# On Search Engine Evaluation Metrics

Inaugural-Dissertation

zur Erlangung des Doktorgrades der Philosophie (Dr. Phil.)

durch die Philosophische Fakultät der

Heinrich-Heine-Universität Düsseldorf

Vorgelegt von Pavel Sirotkin

aus Düsseldorf

Betreuer:

Prof. Wolfgang G. Stock

Düsseldorf, April 2012

> *- Oh my God, a mistake!*
> *- It's not our mistake!*
> *- Isn't it? Whose is it?*
> *- Information Retrieval.*
>
> ***"Brazil"***

# Acknowledgements

*One man deserves the credit,*
*one man deserves the blame…*
　　　　　　　　TOM LEHRER, "LOBACHEVSKY"

I would like to thank my supervisor, Wolfgang Stock, who provided me with patience, support and the occasional much-needed prod to my *derrière*. He gave me the possibility to write a part of this thesis as part of my research at the Department of Information Science at Düsseldorf University; and it was also him who arranged for undergraduate students to act as raters for the study described in this thesis.

I would like to thank my co-supervisor, Wiebke Petersen, who bravely delved into a topic not directly connected to her research, and took the thesis on sightseeing tours in India and to winter beaches in Spain. Wiebke did not spare me any mathematical rod, and many a faulty formula has been spotted thanks to her.

I would like to thank Dirk Lewandowski, in whose undergraduate seminar I first encountered the topic of web search evaluation, and who provided me with encouragement and education on the topic. I am also indebted to him for valuable comments on a draft of this thesis.

I would like to thank the aforementioned undergraduates from the Department of Information Science for their time and effort on providing the data on which this thesis' practical part stands.

Last, but definitely not least, I thank my wife Alexandra, to whom I am indebted for far more than I can express. She even tried to read my thesis, which just serves to show.

As is the custom, I happily refer to the acknowledged all good that I have derived from their help, while offering to blame myself for any errors they might have induced.

# Contents

# 1 Introduction

> *You shall seek all day ere you find them, and when you have them, they are not worth the search.*
>
> WILLIAM SHAKESPEARE,
> "THE MERCHANT OF VENICE"

## 1.1 What It Is All About

The present work deals with certain aspects of the evaluation of web search engines. This does not sound too exciting; but for some people, the author included, it really is a question that can induce one to spend months trying to figure out some seemingly obscure property of a cryptic acronym like MAP or NDCG. As I assume the reader longs to become part of this selected circle, I will, in this section, try to introduce him[1,2] to some of the basic concepts we will be concerned with; the prominent ones are web search and the web search engines, followed by the main ideas and troubles of the queen of sciences which is search engine evaluations.[3] These, especially the last section, will also provide the rationale and justification for this work. The well-disposed specialist, on the other hand, might conceivably skip the next sections of the introduction since it is unlikely that he needs persuading that the field is important, or reminding what the field actually is.

After the introduction, there will be two major parts. Part I is, broadly speaking, a critical review of the literature on web search evaluation. It will explain the real-life properties of search engine usage (Chapter 2), followed by a general and widely used evaluation framework (Chapter 3). Chapters 4 and 5 introduce the two main types of evaluation metrics, explicit and implicit ones, together with detailed discussions of previous studies attempting to evaluate those metrics, and lots of nit-picking comments. Part I concludes with a general discussion of the relationship between explicit and implicit metrics as well as their common problems (Chapter 6).

Part II is where this thesis stops criticizing others and gets creative. In Chapter 7, the concept of relevance, so very central for evaluation, is discussed. After that, I present a framework for web search meta-evaluation, that is, an evaluation of the evaluation metrics themselves, in Chapter 8; this is also where a meta-evaluation measure, the Preference Identification Ratio (PIR), is introduced. Though relatively short, I regard this to be the pivotal section of this work, as it attempts to capture the idea of user-based evaluation described in the first part, and to apply it to a wide range of metrics and metric parameters. Chapter 9 then describes the layout and general properties of a study conducted on this principles. The study is rather small and its results will not always be significant. It is meant only in part as an evaluation of the issues under consideration; at least equally important, it is a proof of concept, showing what can be done with and within the framework. The findings of the study are presented in Chapter 10, which deals with different metrics, as well as with cut-off values, discount functions, and other parameters of explicit metrics. A shorter evaluation is given for some basic implicit measures (Chapter 11). Finally, Chapter 12 sums up the research, first in regard to the results of the study, and then exploring the prospects of the evaluation framework itself. For the very busy or the very lazy, a one-page Executive Summary can be found at the very end of this thesis.

---

[1] "To avoid clumsy constructions resulting from trying to refer to two sexes simultaneously, I have used male pronouns exclusively. However, my information seekers are as likely to be female as male, and I mean no disrespect to women by my usage." (Harter 1992, p. 603)
[2] Throughout our work, I have been quite liberal with footnotes. However, they are never crucial for the understanding of a matter; rather, they provide additional information or caveats which, I felt, would have unnecessarily disrupted the flow of the argument, however little of that there is. Therefore, the footnotes may be ignored if the gentle reader is not especially interested in the question under consideration.
[3] I am slightly exaggerating.

The most important part of this work, its *raison d'être*, is the relatively short Chapter 8. It summarizes the problem that tends to plague most search engine evaluations, namely, the lack of a clear question to be answered, or a clear definition of what is to be measured. A metric might be a reflection of a user's probable satisfaction; or of the likelihood he will use the search engine again; or of his ability to find all the documents he desired. It might measure all of those, or none; or even nothing interesting at all. The point is that, though it is routinely done, we cannot assume a real-life meaning for an evaluation metric until we have looked at whether it reflects a particular aspect of that real life. The meta-evaluation metric I propose (the Preference Identification Ratio, or PIR) is to answer the question of whether a metric can pick out a user's preference between two result lists. This means that user preferences will be elicited for pair of result lists, and compared to metric scores constructed from individual result ratings in the usual way (see Chapter 4 for the usual way, and Section 8.2.1 for details on PIR). This is not *the* question; it is *a* question, though I consider it to be a useful one. But it is a question of a kind that should be asked, and answered.

Furthermore, I think that many studies could yield more results than are usually described, considering the amount of data that is gathered. The PIR framework allows the researcher to vary a number of parameters,[4] and to provide multiple evaluations of metrics for certain (or any) combination thereof. Chapters 9 to 12 describe a study that was performed using this method. However, one very important aspect has to be stated here (and occasionally restated later). This study is, unfortunately, quite small-scale. The main reason for that sad state of things are the limited resources available; what it means is that the study should be regarded as more of a demonstration of what evaluations can be done within one study. Most of the study's conclusions should be regarded, at best, as preliminary evidence, or just as hints of areas for future research.

## 1.2  Web Search and Search Engines

Web search is as old as the web itself, and in some cases as important. Of course, its importance is not self-reliant; rather, it stems from the sheer size of the web. The estimates vary wildly, but most place the number of pages indexed by major search engines at over 15 billion (de Kunder 2010). The size of the index*able* web is significantly larger; as far back (for web timescales) as 2008, Google announced that its crawler had encountered over a trillion unique web pages, that is, pages with distinct URLs which were not exact copies of one another (Alpert and Hajaj 2008). Of course, not all of those are fit for a search engine's index, but one still has to find and examine them, even if a vast majority will be omitted from the actual index. To find any specific piece of information is, then, a task which is hardly manageable unless one knows the URL of the desired web page, or at least that of the web site. Directories, which are built and maintained by hand, have long ceased to be the influence they once were; Yahoo, long the most prominent one, has by now completely taken its directory from the home page (cp. Figure 1.1). Though there were a few blog posts in mid-2010 remarking on the taking down of Yahoo's French, German, Italian and Spanish directories, this is slightly tarnished by the fact that the closure occurred over half a year before, "and it seems no one noticed" (McGee 2010). The other large directory, ODP/DMOZ, has stagnated at around 4.7 million pages since 2006 (archive.org 2006; Open Directory Project 2010b). It states that "link rot is setting in and [other services] can't keep pace with the growth of the Internet" (Open Directory Project 2010a); but then, the "About" section that asserts this has not been updated since 2002. There are, of course, specialized directories that only cover web sites on a certain topic, and some of those are highly successful in their field. But still, the user needs to find those directories, and the method of choice for this task seems to be web search. Even when the site a user seeks is very popular, search engines play an important role. According to the information service Alexa,[5] the global web site of Coca-Cola, judged to be the best-known brand in the world (Interbrand 2010), receives a quarter of its visitors via search engines. Amazon, surely one of the world's best-known web sites, gets about 18% of its users from search engines; and, amazingly, even Google seems to be reached through search engines by around 3% of its visitors.[6] For many sites, whether popular or not, the numbers are much higher. The University of Düsseldorf is at 27%, price comparison service idealo.de at 34%, and Wikipedia at 50%.

---

[4] Such as the cut-off values, significance thresholds, discount functions, and more.

Web search is done with the help of web search engines. It is hard to tell how many search engines there are on the web. First, one has to properly define a search engine, and there the troubles start already. The major general-purpose web search engines have a certain structure; they tend to build their own collection of web data (the index) and develop a method for returning some of them to the user in a certain order, depending on the query the user entered. But most search engine overviews and ratings include providers like AOL (with a reported volume of 1%-2% of all web searches (Hitwise 2010; Nielsen 2010)), which rely on results provided by others (in this case, Google). Other reports list, with a volume of around 1% of worldwide searches each, sites like eBay and Facebook (comScore 2010). And it is hard to make a case against including those, as they undoubtedly do search for information on the web, and they do use databases and ranking methods of their own. Their database is limited to a particular type of information and a certain range of web sites – but so is, though to a lesser extent, that of other search engines, which exclude results they consider not sought for by their users (e.g. spam pages) or too hard to get at (e.g. flash animations). And even if we, now armed with the knowledge of the difference between search engines, agree upon a definition, there is no means of collecting information about all the services out there, though I would venture that a number in the tens of thousands is a conservative estimate if one includes regional and specialized search engines.

---

[5] www.alexa.com. All values are for the four weeks up to October 5th, 2010.
[6] Though some of these users may just enter the Google URL into the search bar instead of the address bar, I have personally witnessed a very nice lady typing "google" into the Bing search bar to get to her search engine of choice. She then queried Google with the name of the organization she worked at to get to its web site.

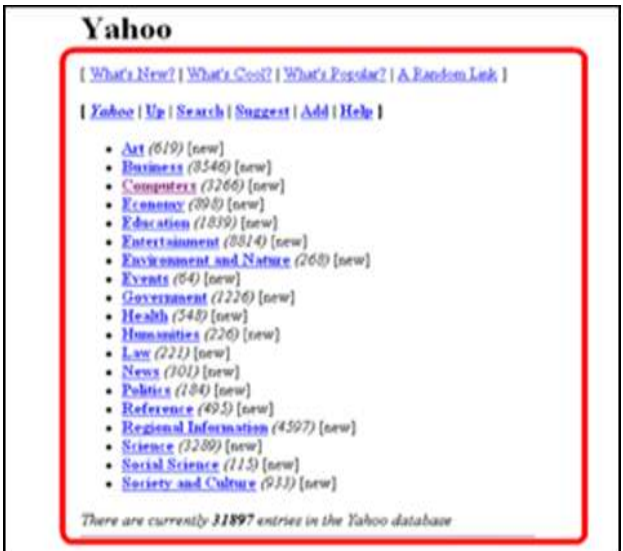 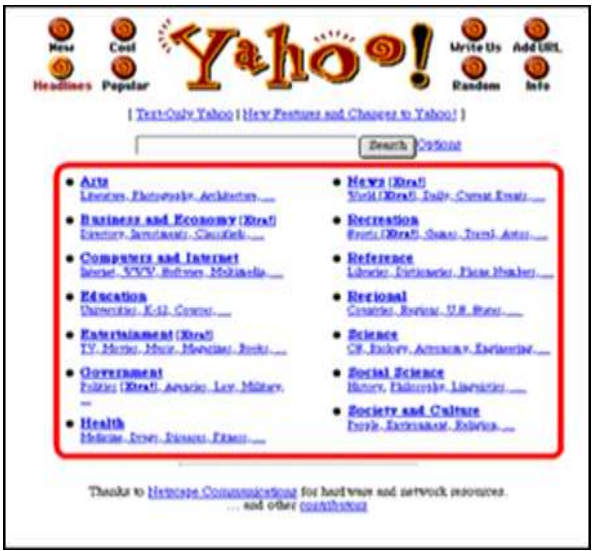
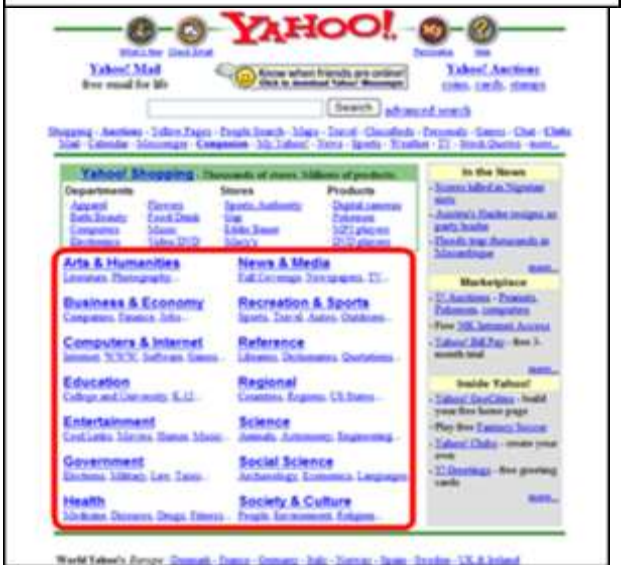 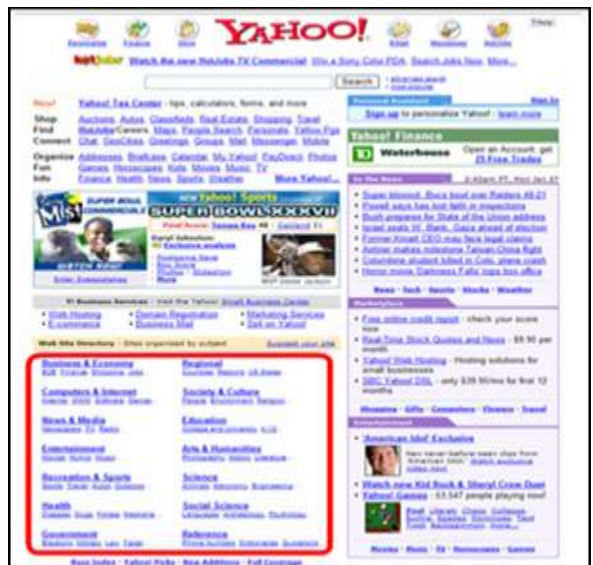
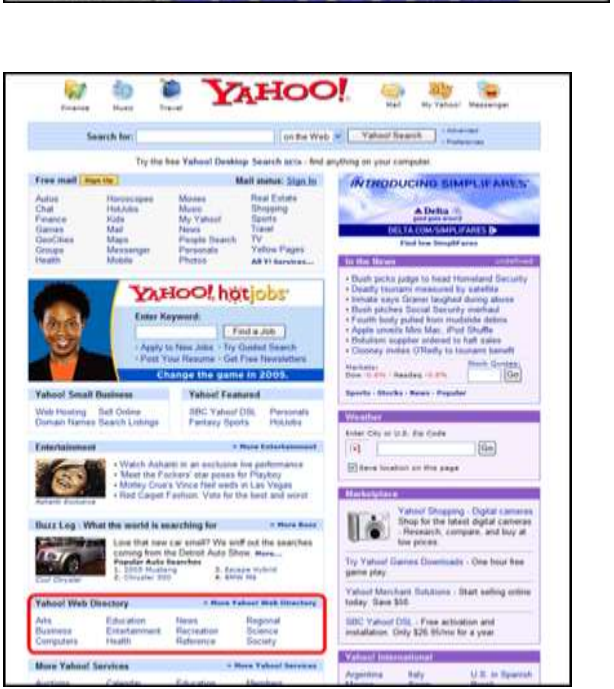 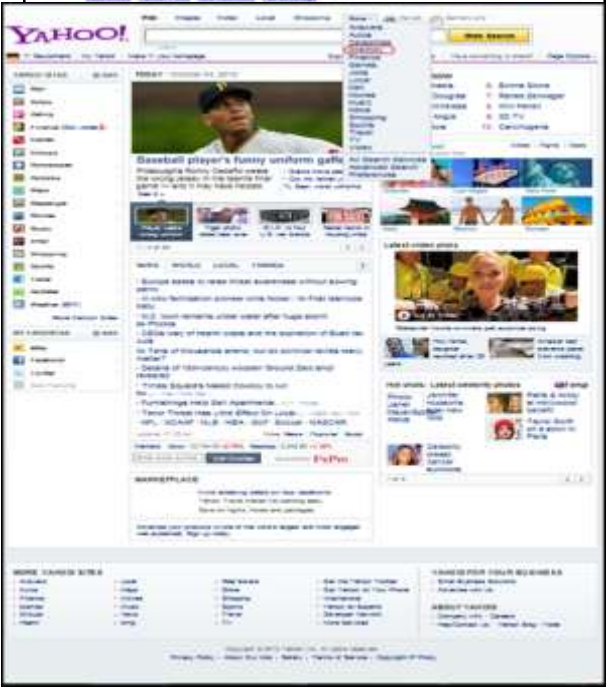

**Figure 1.1. The decline and fall of the Yahoo directory from 1994 to 2010. The directory is marked with red on each screenshot. If you can't find it in the last one, look harder at the fourth item in the drop-down menu in the top-center part of the page, just after „Celebrities". Screenshots of Yahoo homepage (partly from CNET News 2006).**

## 1.3 Web Search Evaluation

Now that we know how important search engines are, we might have an easier time explaining why their evaluation is important. In fact, there are at least three paths to the heart of the matter, and we will take all of them.

Firstly, and perhaps most obviously, since users depend on search engines, they have a lot to gain from knowing which search engine would serve them best. Even if there is a difference of, say, 5% – however measured – between the search engine you are using and a better alternative, your day-to-day life could become just that little bit easier. This is the reason for the popular-press search engine reviews and tests, though they seem to be not as popular these days. Some years back, however, tests like those by on-line publication Search Engine Watch (2002) or German consumer watchdog Stiftung Warentest (2003) were plentiful. However, these tests were rarely rigorous enough to satisfy scientific standards, and often lacked a proper methodology. The few current evaluations (such as Breithut 2011) tend to be still worse, methodologically speaking.

Connected to the first one is another reason for evaluating search engines, which might be viewed as a public version of the former. There is sizeable concern among some politicians and non-government organizations that the search engine market is too concentrated (in the hands of Google, that is); this leads to issues ranging from privacy to freedom of speech and freedom of information. What that has got to do with evaluation? Lewandowski (2007) provides two possible extreme outcomes of a comparative evaluation. Google might be found to deliver the best quality; then it would seem counterproductive to try to diversify the search engine market by imposing fines or restrictions on Google since this would impair the users' search experience. Instead, should one still wish to encourage alternatives, the solution would have to be technological – creating, in one way or another, a search engine of enough quality to win a significant market share.[7] Alternatively, the search engines might be found to perform comparably well, in which case the question arises why Google has the lion's share of the market. It might be because it misuses its present leading position – in whatever way it has been achieved in first place – to keep competition low, which would make it a case for regulators and anti-monopoly commissions, or because of better marketing, or some other reason not directly related to search quality. In any case, the policy has to be quite different depending on the results of an evaluation.[8]

---

[7] Though the French-German Quaero project (www.quaero.org) and the German Theseus research program (www.theseus-programm.de), both touted as "European answers to Google" by the media, develop in directions quite different from a general web search engine.
[8] Though the "winner taxes it all" principle flourishes online because of network effects, and it is unclear whether an attempt to remedy the situation would be sensible (Linde and Stock 2011).

The public might also be interested in another aspect of web search evaluation. Apart from asking "Which search engine is better?", we can also concern ourselves with possible search engine bias. This might be intentional, accidental or an artifact of the search engine's algorithms. For example, preference may be given to old results over new ones or to popular views over marginal opinions. While in some cases this is acceptable or even desirable, in others it might be considered harmful (Diaz 2008).

The third reason web search evaluation is useful is that search engine operators want (and need) to improve their services. But in order to know whether some change really improves the user experience, it is necessary to evaluate both versions. This operator-based approach will be the one we will concentrate upon.

# Part I: Search Engine Evaluation Measures

> *Every bad precedent originated as a justifiable measure.*
> **SALLUST, "THE CONSPIRACY OF CATILINE"**

In Part I, I will do my best to provide a theoretical basis for the study described in Part II. This will require a range of topics to be dealt with – some in considerable depth, and other, less pertinent, ones almost in passing. This will enable us to understand the motivation for the practical study, its position in the research landscape, and the questions it asks and attempts to answer.

The first thing to know about web search evaluation is how web search works, and how people use it; this will be the topic of Chapter 2. Chapter 3 then turns to evaluation; it explains a widely employed IR evaluation framework and its relevance to web search evaluation. Chapter 4 is the largest one in the theoretical part; it describes various explicit metrics proposed in the literature and discusses their merits as well as potential shortcomings. General limitations of such metrics are also discussed. Chapter 5 does the same for implicit metrics, albeit on a smaller scale. The relation between the two kinds of metrics, explicit and implicit, is explored in Chapter 6.

## 2 Search Engines and Their Users

> *People flock in, nevertheless, in search of answers to those questions only librarians are considered to be able to answer, such as 'Is this the laundry?', 'How do you spell surreptitious?' and, on a regular basis: 'Do you have a book I remember reading once? It had a red cover and it turned out they were twins.'*
>
> TERRY PRATCHETT, "GOING POSTAL"

Search engine evaluation is, in a very basic sense, all about how well a search engine serves its users, or, from the other point of view, how satisfied users are with a search engine. Therefore, in this section I will provide some information on today's search engines as well as their users. Please keep in mind that the search engine market and technologies change rapidly, as does user behavior. Data which is two years old may or may not be still representative of what happens today; five-year-old data is probably not reliably accurate anymore; and at ten years, it can be relied upon to produce some serious head-shaking when compared to the current situation. I will try to present fresh information, falling back on older data when no new studies are available or the comparison promises interesting insights.

### 2.1 Search Engines in a Nutshell

Probably the first web search engine was the "World Wide Web Wanderer", started in 1993. It possessed a crawler which traversed the thousands and soon tens of thousands of web sites available on the newly-created World Wide Web, and creating an own index of their content.[9] It was closely followed by ALIWEB, which did have an intention of bringing search to the masses, but no web crawler, relying on webmasters to submit to an index the descriptions of their web sites (Koster 1994; Spink and Jansen 2004). Soon after that, a run began on the search engine market, with names like Lycos, Altavista, Excite, Inktomi and Ask Jeeves appearing from 1994 to 1996. The current market leaders entered later, with Google launching in 1998, Yahoo! Search with its own index and ranking (as opposed to the long-established catalog) in 2004, and MSN Search (to be renamed Live Search, and then Bing) in 2005.

Today, the market situation is quite unambiguous; there is a 360-kilogram gorilla on the stage. It has been estimated that Google handled at least 65% of all US searches in June 2009, followed by Yahoo (20%) and Bing (6%) (comScore 2009). Worldwide, Google's lead is considered to be considerably higher; the estimates range from 83% (NetApplications 2009) up to 90% for single countries such as the UK (Hitwise 2009a) and Germany (WebHits 2009). In the United States, google.com has also been the most visited website (Hitwise 2009b). Its dominance is only broken in a few countries with strong and established local search engines, notably in China, where Baidu has over 60% market share (Barboza 2010), and Russia, where Yandex has over 50% market share (LiveInternet 2010). The dictionary tells us that there is a verb *to google*, meaning to "search for information about (someone or something) on the

---

[9] Though the WWWW was primarily intended to measure the size of the web, not to provide a search feature (Lamacchia 1996).

Internet, typically using the search engine Google" (Oxford Dictionaries 2010; note the "typically"). And indeed, the search engine giant has become something of an eponym of web search in many countries of the world.

The general workings of web search engines are quite similar to each other. In a first step, a web crawler, starting from a number of seed pages, follows outbound links, and so attempts to "crawl" the entire web, copying the content of the visited pages into the search engine's storage as it does so. These contents are then indexed, and made available for retrieval. On the frontend side, the user generally only has to enter search terms, whereupon a ranking mechanism attempts to find results in the index which will most probably satisfy the user, and to rank them in a way that will maximize this satisfaction (Stock 2007).

Result ranking is an area deserving a few lines of its own since it is the component of a search engine which (together with the available index) determines what the result list will be. In traditional information retrieval, where results for most queries were relatively few and the quality of documents was assumed to be universally high, there often was no standard ranking; all documents matching the query were displayed in a more or less fixed order, often ordered by their publication dates. This is clearly not desirable with modern web search engines where the number of matches can easily go into the millions. Therefore, a large number of ranking criteria have been established to present the user with the most relevant results in the first ranks of the list.

One old-established algorithm goes by the acronym tf-idf, which resolves to term frequency – inverted document frequency (Salton and McGill 1986). As the name suggests, this method considers two dimensions; term frequency means that documents containing more occurrences of terms also found in the query are ranked higher, and inverted document frequency implies that documents containing query terms that are rare throughout the index are also considered more relevant. Another popular ranking approach is the Vector Space Model (Salton, Wong and Yang 1975). It represents every document as a vector, with every term occurring in the index providing a dimension of the vector space, and the number of occurrences of the term in a document providing the extension in this dimension. For the query another vector is constructed in the same way; the ranking is then determined by a similarity measure (e.g. the cosine) between the query vector and single document vectors.

These approaches are based only on the content of the single documents. However, the emergence of the web with its highly structured link network has allowed for novel approaches, known as "link analysis algorithms". The first major one was the HITS (Hyperlink-Induced Topic Search) algorithm, also known as "Hubs and Authorities" (Kleinberg 1999). The reason for this name is that the algorithm attempts to derive from the link network the most selective and the most selected documents. In a nutshell, the algorithm first retrieves all documents relevant to the query; then it assigns higher "hub" values to those with many outgoing links to others within the set, and higher "authority" values to those with many incoming links. This process is repeated iteratively, with links from many good "hub" pages providing a larger increase in "authority" scores, and links to many good "authority" pages leading to significant improvements in the "hub" values.[10] HITS is rarely used in general web search engines since it is vulnerable to spam links (Asano, Tezuka and Nishizeki 2008); also, it is query-specific and as such has to be computed at runtime, which would in many cases prove to be too laborious. At one point, a HITS-based ranking was used by the Teoma search engine (see Davison et al. 1999).

The second major link-based algorithm, PageRank (Brin and Page 1998), proved far more successful. It does not distinguish between different dimensions; instead, a single value is calculated for every page. Every link from page A to page B is considered to transfer a fraction of its own PageRank to the linked page; if a page with PageRank 10 has 5 outgoing links, it contributes 2 to the PageRank of every one of those. This process is also repeated iteratively; as it is query-independent, the PageRank values need not be calculated at runtime, which saves time.

It is important to remember that these descriptions are extremely simplified versions of the actual algorithms used; that no major search engine publishes its ranking technology, though brief glimpses might be caught in patents; and that the ranking is always based on a large number of criteria. Google, for example, uses more than 200 indicators, of which the PageRank is but one (Google 2010). The leader of the Russian search engines market, Yandex, claims that the size of their ranking formula increased from 20 byte in 2006 to 120 MB in the middle of 2010 (Segalovich 2010), and to 280 MB in September 2010 (Raskovalov 2010). To put things in perspective, that is approximately 50 times the size of the complete works of William Shakespeare.[11]

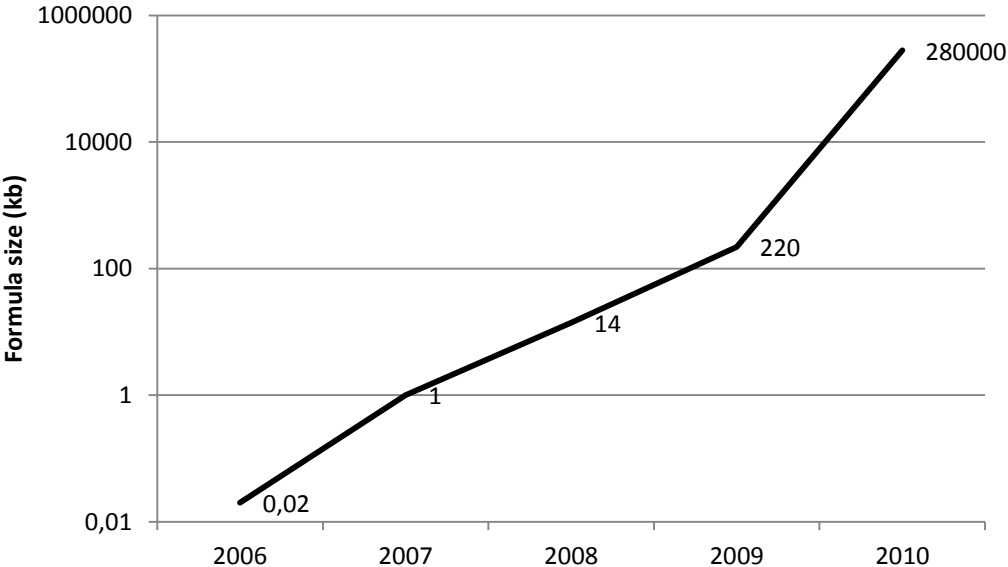

**Figure 2.1. The complication of Yandex ranking formulas (data from Raskovalov 2010; Segalovich 2010). Note that the vertical scale (formula size in kilobyte) is logarithmic.**

---

[10] To allow the results to converge and the algorithm to terminate, the hub and authority values are usually normalized after every iteration.
[11] Uncompressed digital version available at Project Gutenberg (http://www.gutenberg.org/ebooks/100).

Many major search engines offer other types of results than the classical web page list – images, videos, news, maps and so forth. However, the present work only deals with the document lists. As far as I know, there have been no attempts to provide a single framework for evaluating all aspects of a search engine, and such an approach seems unfeasible. There are also other features I will have to ignore to present a concise case; for example, spell-checkers and query suggestions, question-answering features, relevant advertising etc. I do not attempt to discuss the usefulness of the whole output a search engine is able of; when talking of "result lists", I will refer only to the organic document listing that is still the search engines' stock-in-trade.

This is as much depth as we need to consider search evaluation and its validity; for a real introduction to search engines, I recommend the excellent textbook "Search Engines: Information Retrieval in Practice" (Croft, Metzler and Strohman 2010).

## 2.2   Search Engine Usage

For any evaluation of search engine performance it is crucial to understand user behavior. This is true at the experimental design stage, when it is important to formulate a study object which corresponds to something the users do out there in the real world, as well as in the evaluation process itself, when one needs to know how to interpret the gathered data. Therefore, I will present some findings that attempt to shed light on the question of actual search engine usage. This data comes mainly from two sources: log data evaluation (log data is discussed in Chapter 5) and eye-tracking studies. Most other laboratory experiment types, as well as questionnaires, interviews, and other explicit elicitations, have rarely been used in recent years.[12] Unfortunately, in the last years search engine logs have not been made available to the scientific community as often as ten years ago. The matter of privacy alone, which is at any case hotly debated by the interested public, surely is enough to make a search engine provider think twice before releasing potentially sensitive data.[13]

What, then, does this data show? The users' behavior is unlike that experienced in classical information retrieval systems, where the search was mostly performed by intermediaries or other information professionals skilled at retrieval tasks and familiar with the database, query language and so forth. Web search engines aim at the average internet user, or, more precisely, at any internet user, whether new to the web or a seasoned Usenet veteran.

For our purposes, a search session starts with the user submitting a query. It usually contains two to three terms (see also Figure 2.2), and has been slowly increasing during the time web search has been observed (Jansen and Spink 2006; Yandex 2008). The average query has no operators; the only one used with any frequency being the phrase operator (most major engines use quotation marks for that) which occurs in under 2% of the queries, with 9% of users issuing at least one query containing operators (White and Morris 2007).[14] The queries are traditionally divided into navigational (user looks for a specific web page known or supposed to exist), transactional (not surprisingly, in this case the user looks to perform a transaction like buying or downloading) and informational, which should need no further explanation (Broder 2002). While this division is by no means final, and will be modified after an encounter with the current state of affairs in Section 9.2, it is undoubtedly useful (and widely used). Broder's estimation of query type distribution is that 20-25% are navigational, 39-48% are informational, and 30-36% are transactional. A large-scale study with automatic classifying had a much larger amount of informational queries (around 80%), while navigational and transactional queries were at about 10% each (Jansen, Booth and Spink 2007).[15] Another study with similar categories (Rose and Levinson 2004) also found informational queries to be more frequent than originally assumed by Broder (62-63%), and the other two types less frequent (navigational queries 12-15%, transactional queries 22-27%).[16]

---

[12] This is probably due to multiple reasons: The availability (at least for a while) of large amounts of log data; the difficulty in inferring the behaviour of millions of users from a few test persons (who are rarely representative of the average surfer), and the realization that explicit questions may induce the test person to behave unnaturally.

[13] The best-known case is probably the release by AOL for research purposes of anonymized log data from 20 million queries made by about 650.000 subscribers. Despite the anonymization, some users were identified soon (Barbaro and Zeller Jr. 2006). Its attempt at public-mindedness brought AOL not only an entry in CNN's "101 Dumbest Moments in Business" (Horowitz et al. 2007), but also a Federal Trade Commission complaint by the Electronic Frontier Foundation (2006) and a class-action lawsuit which was settled out of court for an undisclosed amount (Berman DeValerio 2010).

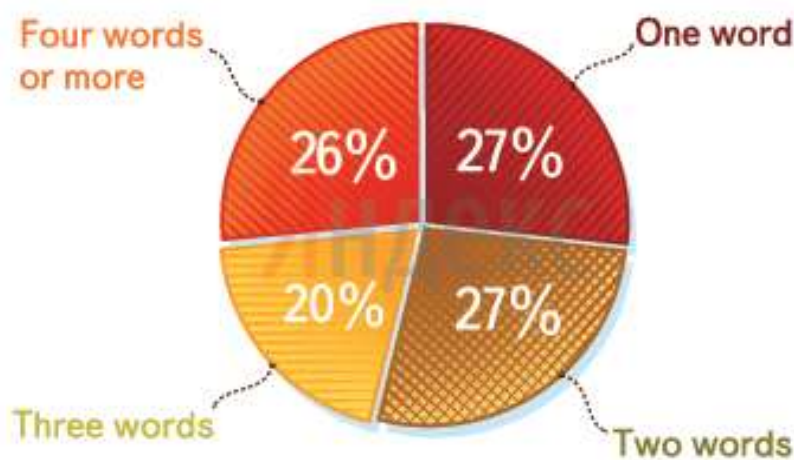

**Figure 2.2. Query length distribution (from Yandex 2008, p. 2). Note that the data is for Russian searches.**

Once the user submits his query, he is presented with the result page. Apart from advertising, "universal search" features and other elements mentioned in Section 2.1, it contains a result list with usually ten results. Each result consists of three parts: a page title taken directly from the page's *<title>* tag, a so-called "snippet" which is a query-dependant extract from the page, and a URL pointing to the page itself. When the result page appears, the user tends to start by scanning the first-ranked result; if this seems promising, the link is clicked and the user (at least temporarily) leaves the search engine and its hospitable result page. If the first snippet seems irrelevant, or if the user returns for more information, he proceeds to the second snippet which he processes in the same way. Generally, if the user selects a result at all, the first click falls within the first three or four results. The time before this first click is, on average, about 6.5 seconds; and in this time, the user catches at least a glimpse of about 4 results (Hotchkiss, Alston and Edwards 2005).[17] As Thomas and Hawking note, "a quick scan of part of a result set is often enough to judge its utility for the task at hand. Unlike relevance assessors, searchers very seldom read all the documents retrieved for them by a search engine" (Thomas and Hawking 2006, p. 95). There is a sharp drop in user attention after rank 6 or 7; this is caused by the fact that only these results are directly visible on the result page without scrolling.[18] The remaining results up to the end of the page receive approximately equal amounts of attention (Granka, Joachims and Gay 2004; Joachims et al. 2005). If a user scrolls to the end of the result page, he has an opportunity to move to the next one which contains further results; however, this option is used very rarely – fewer than 25% of users visit more than one result page,[19] the others confining themselves to the (usually ten) results on the first page (Spink and Jansen 2004).

---

[14] For the 13 weeks during which the data was collected. A study of log data collected 7-10 years earlier had significantly higher numbers of queries using operators, though they were still under 10% (Spink and Jansen 2004).
[15] While the study was published in 2007, the log data it is based is at least five years older, and partly originates in the end-90ies.
[16] Note that all cited statistics are for English-speaking users. The picture may or may not be different for other languages; a study using data from 2004 showed no fundamental differences in queries originating from Greece (Efthimiadis 2008). A study of Russian log data additionally showed 4% of queries containing a full web address, which would be sufficient to reach the page via the browser's address bar instead of the search engine. It also stated that about 15% of all queries contained some type of mistake, with about 10% being typos (Yandex 2008).

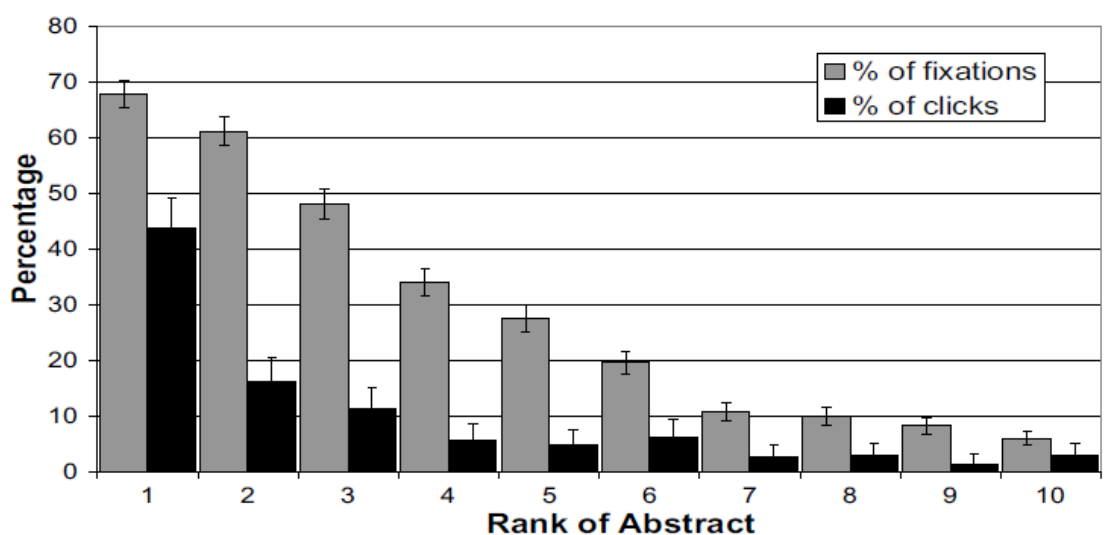

**Figure 2.3. Percentage of times an abstract[20] was viewed/clicked depending on the rank of the result (taken from Joachims et al. 2005, p. 156). Data is based on original and manipulated Google results, in an attempt to eliminate quality bias. Note that the study was a laboratory-based, not log-based one.**

---

[17] Turpin, Scholer et al. (2009) mention 19 seconds per snippet; however, this is to consciously judge the result and produce an explicit rating. When encountering a result list in a "natural" search process, the users tend to spend much less time deciding whether to click on a result.
[18] Obviously, this might not be the case for extra high display resolutions; however, this does not invalidate the observations.
[19] Given the previous numbers in this section, a quarter of all users going beyond the first result page seems like a lot. The reason probably lies in the different sources of data in the various studies; in particular, the data used by Spink and Jansen (2004) mostly stems from the 1990s.
[20] "Abstract" and "snippet" have the same meaning in this case.

The number of pages a user actually visits is also not high. Spink and Jansen (2004) estimate them at 2 or 3 per query; and only about one in 35,000 users clicks on all ten results on the first result page (Thomas and Hawking 2006). This fact will be very important when we discuss the problems with traditional explicit measures in section 4. Most of those clicks are on results in the very highest ranks; Figure 2.3 gives an overview of fixation (that is, viewing one area of the screen for ca. 200-300 milliseconds) and click rates for top ten results. Note that the click rate falls much faster than the fixation rate. While rank 2 gets almost as much attention as rank 1, it is clicked more than three times less often.

# 3    Evaluation and What It Is About

> *We must **decide** what ought to be the case. We cannot discover what **ought** to be the case by investigating what **is** the case.*
> PAUL W. TAYLOR, *"NORMATIVE DISCOURSE"*

The evaluation of a web search engine has the same general goal as that of any other retrieval system: to provide a measure of how good the system is at providing the user with the information he needs. This knowledge can then be used in numerous ways: to select the most suitable of available search engines, to learn about the users or the data searched upon, or – most often – to find weaknesses in the system and ways to eliminate them.

However, web search engine evaluation also has the same potential pitfalls as its more general counterpart. The classical overview of an evaluation methodology was developed by Tague-Sutcliffe (Tague 1981; Tague-Sutcliffe 1992). The ten points made by her have been extremely influential, and we will now briefly recount them, commenting on their relevance to the present work.

The first issue is "To test or not to test", which bears more weight than it seems at a first glance. Obviously, this step includes reviewing the literature to check if the questions asked already have answers. However, before this can be done, one is forced to actually formulate the question. The task is anything but straightforward. "Is that search engine good?" is not a valid formulation; one has to be as precise as possible. If the question is "How many users with navigational queries are satisfied by their search experience?", the whole evaluation process will be very different than if we ask "Which retrieval function is better at providing results covering multiple aspects of a legal professional's information need?". These questions are still not as detailed as they will need to be when the concrete evaluation methods are devised in the next steps; but they show the minimal level of precision required to even start thinking about evaluation. It is my immodest opinion that quite a few of the studies presented in this chapter would have profited from providing exact information on what it is they attempt to quantify. As it stands, neither the introduction of new nor the evaluation of existing metrics is routinely accompanied by a statement on what precisely is to be captured. There are some steps in the direction of a goal definition, but these tend to refer to behavioral observations ("in reality, the probability that a user browses to some position in the ranked list depends on many other factors other than the position alone", Chapelle et al. 2009, p. 621) or general theoretical statements ("When examining the ranked result list of a query, it is obvious that highly relevant documents are more valuable than marginally relevant", Järvelin and Kekäläinen 2002, p. 424). To be reasonably certain to find an explicit phenomenon to be measured, one needs to turn to the type of study discussed in Chapter 6, which deals precisely with the relationship of evaluation metrics and the real world.

The second point mentioned by Tague-Sutcliffe concerns the decision on what kind of test is to be performed, the basic distinction being that between laboratory experiments and operational tests. In laboratory experiments, the conditions are controlled for, and the aim of the experiment can be addressed more precisely. Operational tests run with real users on real systems, and thus can be said to be generally closer to "real life". Of course, in practice the distinction is not binary but smooth, but explicit measures (Chapter 4) tend to be obtained by laboratory experiments, while implicit measures (Chapter 5) often come from operational tests.

The third issue "is deciding how actually to observe or measure [...] concepts – how to operationalize them" (Tague-Sutcliffe 1992, p. 469). This means deciding which variables one controls for, and which are going to be measured. The features of the database, the type of user to be modeled, the intended behavioral constraints for the assessor (e.g. "Visit at least five results"), and – directly relevant to the present work – what and how to evaluate are all questions for this step.

The fourth question, what database to select, is not very relevant for us; mostly, either one of the popular web search engines is evaluated, or the researcher has a retrieval method of his own whose index is filled with web pages, of which there is no shortage. Sometimes, databases assembled by large workshops such as TREC[21] or CLEF[22] can be employed.

More interesting is issue five, which deals with queries. One may use queries submitted by the very person who is to assess the results; in fact, this is mostly regarded as the optimal solution. However, this is not always easy to do in a laboratory experiment; also, one might want to examine a certain type of query, and so needs to be selective. As an alternative, the queries might be constructed by the researcher; or the researcher might define an information need and let the assessors pose the query themselves. As a third option, real queries might be available if the researcher has access to search engine logs (or statistics of popular queries); in this case, however, the information need can only be guessed. The question of information need versus query will be discussed in more detail in Chapter 7 which deals with the types of relevance.

Step six deals with query processing; my answer to this problem is explained in Section 9.1. Step seven is about assigning treatments to experimental units; that is, balancing assessors, queries, and other experimental conditions so that the results are as reliable as possible. Step eight deals with the means and methods of collecting the data (observation, questionnaires, and so forth); steps nine and ten are about the analysis and presentation of the results. These are interesting problems, but I will not examine them in much detail in the theoretical part, while trying to pay attention to them in the actual evaluations we undertake.

Tague-Sutcliffe's paradigm was conceived for classical information retrieval systems. I have already hinted at some of the differences search engine evaluation brings with it, and more are to come when evaluation measures will be discussed. To get a better idea of the entities search engine evaluation actually deals with, I feel this is a good time to introduce our protagonists.

---

[21] Text REtrieval Conference, http://trec.nist.gov
[22] Cross-Language Evaluation Forum, http://www.clef-campaign.org

# 4 Explicit Metrics

*Don't pay any attention to what they write about you. Just measure it in inches.*
ANDY WARHOL

The evaluation methods that are most popular and easiest to conceptualize tend to be explicit, or judgment-based. This is to be expected; when faced with the task of determining which option is more relevant for the user, the obvious solution is to ask him. The approach brings with it some clear advantages. By selecting appropriate test persons, one can ensure that the information obtained reflects the preferences of the target group. One can also focus on certain aspects of the retrieval system or search process, and – by carefully adjusting the experimental setting – limit the significance of the answers to the questions at hand. The main alternative will be discussed in Chapter 5 which deals with implicit metrics.

| *Section* | *Group* | *Metrics* |
|---|---|---|
| 4.1 | Recall/Precision and their direct descendants | *Precision* <br> *Recall* <br> *Recall-precision Graph* <br> *F-Measure* <br> *$F_1$-Measure* <br> *Mean Average Precision (MAP)* |
| 4.2 | Other system-based metrics | *Reciprocal Rank (RR)* <br> *Mean Reciprocal Rank (MRR)* <br> *Quality of result ranking (QRR)* <br> *Bpref* <br> *Sliding Ratio (SR)* <br> *Cumulative Gain (CG)* <br> *Discounted Cumulative Gain /DCG)* <br> *Normalized Discounted Cumulative Gain (NDCG)* <br> *Average Distance Measure (ADM)* |
| 4.3 | User-based metrics | *Expected Search Length (ESL)* <br> *Expected Search Time (EST)* <br> *Maximal Marginal Relevance (MMR)* <br> *α–Normalized Discounted Cumulative Gain (α-NDCG)* <br> *Expected Reciprocal Rank (ERR)* |
| 4.4 | General problems of explicit metrics | |

**Table 4.1. Overview of evaluation metrics described in this section.**

We will first recapitulate the well-known concept of recall and precision and consider the measures derived from it (Section 4.1). Then we will describe some further measures which focus on the system being evaluated (Section 4.2), before turning to more user-based measures (Section 4.3). Finally, we will discuss problems common to all explicit measures and ways suggested to overcome them (Section 4.4). An overview of all described metrics can be found in Table 4.1.

## 4.1 Recall, Precision and Their Direct Descendants

Precision, probably the earliest of formally stated IR evaluation measures, has been proposed in 1966 as part of the Cranfield Project (Cleverdon and Keen 1968). The basic idea is very simple: the precision of a retrieval system for a certain query is the proportion of results that are relevant. A second proposed and widely used metric was recall, defined as the proportion of relevant results that have been retrieved.

$$Precision = P = \frac{relevant\ retrieved\ results}{overall\ retrieved\ results}$$

**Formula 4.1. Precision**

$$Recall = R = \frac{relevant\ retrieved\ results}{relevant\ results\ in\ database}$$

**Formula 4.2. Recall**

The recall and precision measures are generally inversely proportional: if a retrieval system returns more results, the recall can only increase (as the number of relevant results in the database does not change), but precision is likely to fall. This is often captured in the so-called recall-precision-graphs (see Figure 4.1).

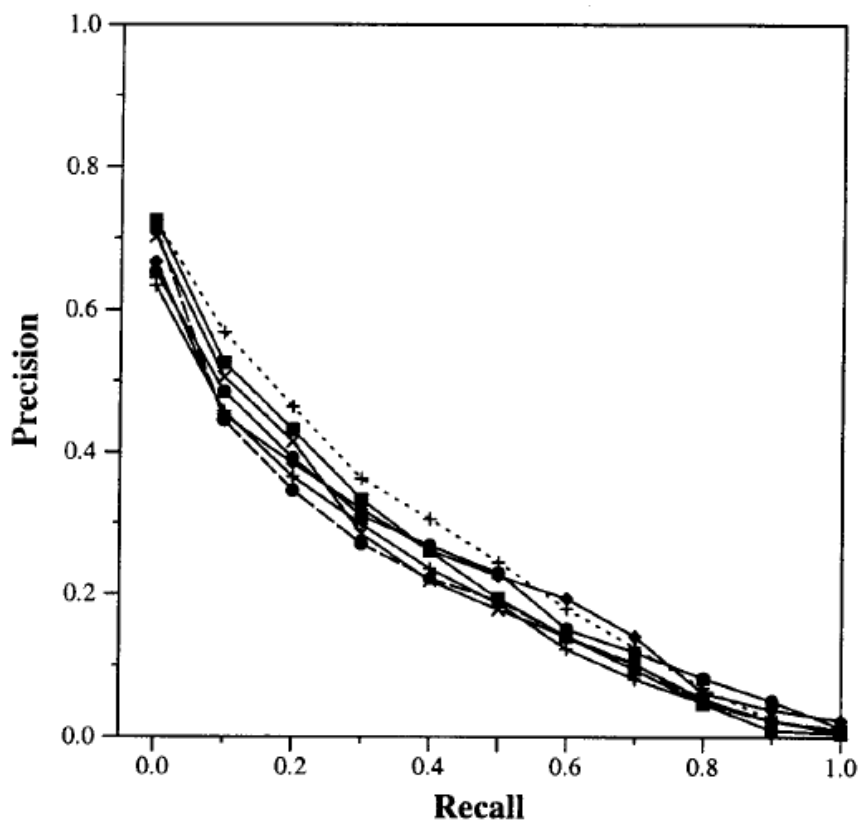

**Figure 4.1. Example of a recall-precision graph showing the performances of different retrieval systems (from Voorhees and Harman 2000, p. 14). The graph visualizes the typical trade-off between recall and precision faced by any IR system. When the recall reaches 1, that is, when every document is retrieved, the precision is at its lowest; when precision is highest, it is recall that plummets. Every system has to find a balance reflecting the expected needs of its users.**

To combine these two aspects of retrieval, the F-Measure was introduced (Van Rijsbergen 1979). It is an indicator for both recall and precision, with the relative importance set by the weighting constant $\beta$. In the special but frequent case of recall and precision being equally important ($\beta=1$) the formula reduces to the so-called $F_1$ score. In this case (or with any $\beta$ close to 1), either low precision or low recall leads to a low overall rating. With a $\beta$ closer to 0, precision has a larger influence on the F-Measure, while a $\beta$ higher than 1 means recall is more important.

$$F = (1 + \beta^2)\frac{PR}{\beta^2 P + R}$$

**Formula 4.3. F-Measure for precision P and recall R. $\beta$ is a constant which can be used to influence the relative importance of recall and precision.**

$$F_1 = 2\frac{PR}{P + R}$$

**Formula 4.4. $F_1$-Measure for precision P and recall R; it corresponds to the F-Measure with $\beta$ set to 1.**

As has often been noted, "since the 1950's the adequacy of traditional measures, precision and recall, has been a topic of heated debates […], and at the same time of very little experimentation" (Su 2003, p. 1175). It was argued that these measures might work well for a relatively small-scale IR system used by information professionals or experienced researchers. But for other uses, their practical and theoretical limitations become apparent. It stands to reason that for large databases with millions of documents recall cannot be easily determined, as it depends on knowing the precise number of relevant documents in the database.[23] Furthermore, as the quantity of returned results increases, it becomes difficult to rate the relevance of each one manually. Especially in web search engines, where there may be millions of relevant results,[24] it is impossible to evaluate each one. The solution comes from studies showing that users look only at a small number of results, with most users restricting themselves to just the first result page (Jansen and Spink 2006). Thus, evaluating all returned results becomes at least unnecessary, and, if one focuses on the user experience, may even be counterproductive.

One further problem remains, however. Precision treats all returned results equally; but as the user's attention decreases, pages in the lower parts of the result list become less important, and may not be considered at all (Jansen et al. 1998). Therefore, most modern metrics, precision-based metrics among them, include some kind of damping constant that discounts the value of later results. One of the most popular measures is the Mean Average Precision (MAP). As its name suggests, it averages mean precisions[25] of multiple queries. In words,

---

[23] This problem was at least partially alleviated by the employment of methods such as Relative Recall (Clarke and Willett 1999). As recall is not a primary issue of the present work, it will not be discussed further.

[24] At least, those are the numbers estimated and displayed by major search engines. In practice, only a subset of those is actually returned in the result list, mostly under a thousand pages. This aggravates the problem, as the results are usually still too many to rate, but now precision cannot be calculated even in theory, lacking the complete list of returned results.

[25] "It means average precisions" would have the words in a more appropriate order, but seems a bit ambiguous.

MAP considers the precision at every relevant result in the list; then, the precision is averaged by dividing the sum of discounted precisions by the total number of relevant results.[26]

$$\text{MAP} = \frac{1}{|Q|} \left( \sum_{Q_i} \frac{1}{|D|} \left( \sum_{r=1}^{|D|} rel(d_r) \frac{\sum_{k=1}^{r} rel(d_k)}{r} \right) \right)$$

**Formula 4.5. MAP formula with queries *Q* and documents *D*. $d_r$ is a document at rank *r*. *rel* is a relevance function assigning *1* to relevant and *0* to non-relevant results.**

An example may be appropriate do demonstrate the calculation of MAP. Given the imaginary relevance judgments in Table 4.2, and supposing no further relevant documents are known to exist, the explicit calculation is given in Figure 4.2. For comparison: The precision for each of the queries is 0.6.

|         | Result 1 | Result 2 | Result 3 | Result 4 | Result 5 |
|---------|----------|----------|----------|----------|----------|
| Query 1 | 1        | 1        | 0        | 1        | 0        |
| Query 2 | 0        | 0        | 1        | 1        | 1        |

**Table 4.2. Imaginary binary relevance for results of two queries.**

$$MAP = \frac{1}{2}\left(\frac{1}{3}\left(\frac{1}{1}+\frac{2}{2}+\frac{3}{4}\right)\right) + \frac{1}{2}\left(\frac{1}{3}\left(\frac{1}{3}+\frac{2}{4}+\frac{3}{5}\right)\right) \approx \frac{1}{2} * 0.92 + \frac{1}{2} * 0.48 = 0.7$$

**Figure 4.2. MAP calculation for values in Table 4.2.**

MAP is one of the most-used metrics and is employed in single studies as well as in large efforts such as TREC (Clarke, Craswell and Soboroff 2009). However, it is also not without its problems. In particular, some studies have detected a lack of correlation between MAP and actual user performance (Hersh et al. 2000; Turpin and Hersh 2001). These studies are based upon data collected in the TREC Interactive Track (see Hersh and Over 2000), which is assessed by asking users to find one or more answers to predefined questions in a given time. The number of answers or *instances* they collect can be used as a standard against which to rate other, more system-centered metrics. The results show that an increase in MAP need not correspond to a significant, or indeed any, increase in user performance. Turpin and Hersh have also looked in more detail at the possible reasons for this discrepancy; their finding is that, while users encounter more non-relevant results, the number of relevant results stays constant. They conclude that the users are not hindered much by a low Mean Average Precision.

The two studies had a relatively small number of queries; furthermore, it could be argued that the setting of a fixed search time significantly distorts the users' real-life behavior and cannot be transferred to other fields such as classical web search. For example, the assessors were not supposed to abandon the search ahead of time if they grew frustrated with the high number of non-relevant results. However, the user model implied by this limit is conceptually not less realistic than MAP which does not assume a user model at all; at the very least, the studies have to be taken seriously and their claims kept in mind.

---

[26] Note that this is not the number of *returned* relevant results but the number of all relevant results in the database, or at least the number of relevant results known to exist.

The findings were also strengthened in a study concerning itself with more simple – and presumably more common – tasks (Turpin and Scholer 2006). A first task was precision-based; it required the users to find a single relevant result for a pre-defined TREC query. The ranking algorithms were selected as to reflect a range of MAP values from 55% to 95%. For every algorithm, the average time elapsed before answer is found and the average percentage of sessions where no relevant document was found were calculated. Neither had any significant correlation to MAP, or to precision at ranks 2, 3, 4 and 10. The only (weak) correlation was observed for precision at rank 1; however, the authors note that this might be an effect of the study design. A second task was recall-based; the users were asked to identify as many relevant documents as possible in a fixed time period. Again, the different ranking algorithms were employed. The only statistically significant differences were observed between the pairs 55% vs. 75% and 65% vs. 75%. However, since the absolute improvement was very low (extra 0.3 documents per session), and the user performance fell again in the conditions featuring MAP values of 85% and 95%, the correlation does not seem impressive. Though some objections against the methodology of the study could be made – the authors mention the different user groups for the TREC-rated documents and the student-performed retrieval task – it clearly adds much doubt to MAPs appropriateness as a measure of probable user performance.

Slightly better correlation between precision and user satisfaction was shown by Kelly, Fu and Shah (2007). They performed a test on four topics from TREC newspaper data. Each subject was given a topic statement and was instructed to construct a query; however, the user was presented with a result list that was fixed in advance and did not depend on the actual query but solely on the topic. There were four result lists; in one condition, they significantly differed in precision, while being similar in MAP (more scattered relevant results versus few relevant results at top ranks), while in another, the precision was equal, but the ranking different (e.g. ranks 1-5 relevant and 6-10 irrelevant, or vice versa). The users were asked to rate single documents as well as the overall performance of every "search engine". The results showed that, with higher MAP as constructed from users' document ratings, the rating of the result list also tend to grow. However, the correlation was only significant for less than half of all users; in fact, users with average ratings for documents as well as result lists were the only group showing this significance. Furthermore, the study has some serious methodological issues. First, there were only four topics, a very small sample. Second, the result lists were not query-specific but topic-specific; this must have lead to different search experiences and different relevance ratings among users. Finally, the result lists were constructed using expected precision values derived from TREC ratings; however, no statistics on their correlation with user assessments of result set quality are provided.

## 4.2 Other System-based Metrics

A simple measure intended for a specific task is the Mean Reciprocal Rank (MRR). Reciprocal Rank (RR) is defined as zero if no relevant results are returned, or else one through the rank of the first encountered relevant result (Formula 4.6). For MRR, the mean of RR values for single queries is taken (Voorhees 1999). MRR assumes that the user is only interested in one relevant result, and anything following that result is irrelevant to his search experience (see Formula 4.7). This seems to be an intuitive assumption for some types of queries, particularly navigational ones.

$$RR = \begin{cases} 0 & \text{if no relevant results} \\ \frac{1}{r} & \text{else} \end{cases}$$

**Formula 4.6. Reciprocal Rank (RR), with *r* the first rank where a relevant result is found.**

$$MRR = \frac{1}{|Q|} \sum_{q \in Q} RR_q$$

**Formula 4.7. Mean Reciprocal Rank (MRR), *Q* being the set of queries and $RR_q$ the Reciprocal Rank measured for query *q*.**

A metric proposed explicitly for web search evaluation is the Quality of result ranking (QRR; Vaughan 2004). It is quite unusual among explicit measures in the way it constructs its rating. First, human judgments are obtained for the search results in the usual way. Preferably, the results will be pooled from several search engines' lists for the same query. In the next step, however, those are used to construct an alternative result list, one that is sorted by user-assigned relevance. This is compared to the original result lists provided by the search engines, and a correlation coefficient is calculated. Vaughan provides an evaluation of her metric. Vaughan applies QRR to three search engines and finds that the results are significantly different. From that, she concludes that it is "able to distinguish search engine performance" (Vaughan 2004, p. 689). This, however, seems to be an example of a vicious circle; the metric is valid because it "recognizes" a difference, which is assumed on the basis of the metric. QRR itself has not been widely used for evaluations; however, it deserves attention for its attempt to detect correlations. We will encounter more correlation-based measures when dealing with implicit measurements; in particular, the notion of Normalized Discounted Cumulated Gain later in this section follows a similar logic.

A measure which has enjoyed some exposure, for example at a few of the TREC tracks, is bpref,[27] so called "because it uses binary relevance judgments to define the preference relation (any relevant document is preferred over any nonrelevant document for a given topic)" (Buckley and Voorhees 2004, p. 26). This is unusual among explicit metrics, as it does not attempt to assign an absolute rating to results, but merely to find pairs where one document should be preferable to another. The metric, given in Formula 4.8, calculates the number of non-relevant documents *n* ranked higher than a relevant document *r*, and averages the number over the first *R* relevant results.[28] Bpref was shown to correlate well (with Kendall τ>90%) with MAP, while being more stable when fewer judgments are available. Also, increases in bpref were correlated to lower times for task completion and higher instance recall in a passage retrieval task, although the connections are not linear, and may not apply to other retrieval tasks (Allan, Carterette and Lewis 2005).

---

[27] The authors write "bpref" with a lower-case "b" except in the beginning of a sentence, where it is capitalized. I do not see any reason not to follow their approach – after all, they should know best.
[28] To ensure that the values fall into the [1..0] interval, *n* is also restricted to the first *R* occurrences.

$$bpref = \frac{1}{R}\sum_{r=1}^{R}(1 - \frac{|n\ ranked\ higher\ than\ r|}{R})$$

**Formula 4.8. Bpref (Buckley and Voorhees 2004). Its distinguishing characteristic is the comparison of results within a result list. For each relevant result *r* (up to a predetermined number of relevant results *R*), the number of non-relevant results *n* above it in the list is calculated, and the results average over the different *r*'s. Thus, every nonrelevant result ranked higher than a relevant one causes bpref to decrease from its maximum value of 1.**

There has also been the rpref metric, introduced mainly to allow for graded instead of binary relevance (De Beer and Moens 2006). However, neither bpref nor its modification will play a significant role in the current evaluation. The reason is that these metrics are explicitly designed for situations where relevance judgments are incomplete. In the study presented in Part II, this problem does not occur, for reasons that will be discussed in the methodological Section 9.1. Additionally, it has been suggested that even in the very circumstances it has been conceived for, bpref performs worse than traditional metrics. When non-judged results are simply disregarded, metrics like Average Precision or NDCG show more robustness and discriminatory power than bpref or rpref (Sakai 2007a).

An early measure comparing a system's actual result list to an idealized output is the Sliding Ratio (Pollock 1968). It is calculated as the ratio of the actual retrieval system's score to an ideal ranking of the same documents for every rank up to a certain threshold; hence the name. In Formula 4.9, *c* is the cut-off value (the number of results considered), *rel($d_r$)* the relevance (or weight) of the document at rank *r*, and *rel($d_{r_{ideal}}$)* the relevance of the *r*-th result in an ideally ranked result list. The simple sample shown in Table 4.3 illustrates that since only the retrieved documents are considered for the construction of the ideal ranking, the SR at rank n is always 1.

$$SR(c) = \frac{\sum_{r=1}^{c} rel(d_r)}{\sum_{r=1}^{c} rel(d_{r_{ideal}})}$$

**Formula 4.9. Sliding Ratio (Korfhage 1997)**

| Rank | Relevance | Ideal relevance at rank | Sliding Ratio |
|---|---|---|---|
| 1 | 1 | 5 | 0.2 |
| 2 | 2 | 4 | 0.33 |
| 3 | 3 | 3 | 0.5 |
| 4 | 4 | 2 | 0.71 |
| 5 | 5 | 1 | 1 |

**Table 4.3. Example of Sliding Ratio calculation**

A measure which has enjoyed wide popularity since its introduction is Discounted Cumulated Gain or DCG for short (Järvelin and Kekäläinen 2000). The more basic measure upon which it is constructed is the Cumulated Gain, which is a simple sum of the relevance judgments of all results up to a certain rank. DCG enhances this rather simple method by introducing "[a] discounting function [...] that progressively reduces the document score as its rank increases but not too steeply (e.g., as division by rank) to allow for user persistence in examining further documents" (Järvelin and Kekäläinen 2002, p. 425). In practice, the authors suggest a logarithmic function, which can be adjusted (by selecting its base) to provide a more or less strong discount, depending on the expectations of users' persistence.

$$DCG_r = \begin{cases} CG_r & if\ r < b \\ DCG_{r-1} + \dfrac{rel(r)}{log_b(r)} & if\ r \geq b \end{cases}$$

**Formula 4.10. DCG with logarithm base *b* (based on Järvelin and Kekäläinen 2002). $CG_r$ is the Cumulated Gain at rank *r*, and rel(r) a a relevance function assigning *1* to relevant and *0* to non-relevant results.**

|  | Result 1 | Result 2 | Result 3 | Result 4 | Result 5 |
|---|---|---|---|---|---|
| **Query 1** | **1** | **1** | **0** | **1** | **0** |
| CQ | 1 | 2 | 2 | 3 | 3 |
| DCQ $log_2$ | 1 | 2 | 2 | 2,5 | 2,5 |
| **Query 2** | **0** | **1** | **1** | **1** | **0** |
| CQ | 0 | 1 | 2 | 3 | 3 |
| DCQ $log_2$ | 0 | 1 | 1,63 | 2,13 | 2,13 |

**Table 4.4. CG and DCG calculation for values from Table 4.2.**

One weak point of DCG is the missing comparability to other metrics as well as between different DCG-evaluated queries. Should one query be "easy" and have more possible relevant hits than another, its DCG would be expected to be higher; the difference, however, would not signify any difference in the general retrieval performance. A measure which indicates retrieval quality independent from the quality of available results (that is, from the "difficulty" of the search task) would be more helpful. To remedy the situation, Normalized DCG (NDCG) can be employed. NDCG works by pooling the results from multiple search engines' lists and sorting them by relevance. This provides an "ideal result list" under the assumption that all the most relevant results have been retrieved by at least one of the search engines. The DCG values of the single search engines can then be divided by the ideal DCG to put them into the [0..1] interval, with 0 meaning no relevant results and 1 the ideal result list.[29] Note, however, that NDCG is quite similar to a version of Sliding Ratio with added discount for results at later ranks.

The authors evaluated the different CG measures (Järvelin and Kekäläinen 2000; Järvelin and Kekäläinen 2002). However, this was not done by comparing the new measure with a standard, or with established measures; instead, it was used to evaluate different IR systems, where one was hypothesized to outperform the others. The CG measures indeed showed a significant difference between the systems, and were considered to have been validated. I think this methodology is not quite satisfactory. It seems that evaluating the hypothesis with the new measure while at the same time evaluating the new measure against the hypothesis may produce a positive correlation without necessarily signifying a meaningful connection to any outside entity. However, a more stringent evaluation of DCG was performed by Al-Maskari, Sanderson and Clough (2007). In the study, Precision, CG, DCG and NDCG were compared to three explicit measures of user satisfaction with the search session called "accuracy", "coverage" and "ranking". The results were mixed. From the overall 12 relations between metric and user satisfaction, only two showed a significant correlation, namely, Precision and CG with the ranking of results. Unfortunately, the authors provide no details as to what the assessors were instructed to rate by their satisfaction measures; it seems possible that, say, a user's perception of accuracy may well be influenced by the result ranking. Still, the study is the only one I know of that directly compares the CG family to user satisfaction, and its results are only partly satisfactory. These results notwithstanding, (N)DCG is conceptually sound, and provides more flexibility than MAP. Since its introduction in 2000, it has become one of the most popular search engine evaluation measures; and we definitely do not mean to suggest throwing it overboard as soon as some initial doubt is cast on its correlation to real-world results.

---

[29] Obviously, if the ideal DCG is zero, the calculation is not possible. However, this is not a large problem, since this value would mean no relevant pages are known, and such a query would probably best be excluded from the evaluation altogether. Alternatively, if relevant documents are supposed to exist, they can be explicitly added into the ideal ranking, either on a per-query basis or as a default baseline across all queries.

$$ADM = 1 - \frac{\sum_{d \in D} |SRS(d) - URS(d)|}{|D|}$$

**Formula 4.11. ADM for retrieved documents D, System Relevance Score SRS and User Relevance Score URS**

A measure which is rarely used for actual evaluation but provides some interesting aspects for the current study is the Average Distance Measure (ADM). It has been introduced explicitly to replace existing measures which were considered to rely too heavily on binary distinctions. The distinction between relevant and non-relevant documents is opposed to documents lying along a scaled or continuous relevance axis, and that between retrieved and non-retrieved documents is to be smoothed by considering the rank of the retrieved result (Mizzaro 2001; Della Mea et al. 2006). The ADM measure is, in short, the average difference between a system's and a user's relevance scores for the documents returned for a query. The interesting characteristic of ADM is that, while it attempts to distinguish itself from precision in some aspects, it is similar in that it is an extremely system-based measure, focusing on single system ratings rather than on user experience. To provide a simple example, a retrieval system providing a bad result list but recognizing it as such can get an ideal ADM (since its evaluation of the results' quality is perfect); one that provides ideal results but regards them as mediocre them performs worse (since it underestimates the result quality).[30] Evaluation results calculated using ADM have been compared to those produced by Average Precision, and found not to correlate with it (Della Mea, Di Gaspero and Mizzaro 2004); however, this merely means that the two measures do not evaluate the same aspects of search engines. Both still may be good at specified but distinctly different tasks.

---

[30] Also, ADM obviously depends on a system's relevance scores for documents, which are not readily available.

## 4.3 User-based Metrics

Another class of metrics goes beyond relevance by employing models of user behavior. The distinction is somewhat vague since any discounted metric such as MAP or DCG already makes some assumption about the behavior, viz. the user being more satisfied with relevant results in earlier than in later ranks. DCG even provides a method for adjusting for user persistence by modifying the logarithm base of its formula. Still, it can be argued that the measures to be presented in this section differ significantly by explicitly stating a user model and constructing a metric to reflect it. The rationale for turning away from purely system-based measures was nicely formulated by Carbonell and Goldstein:

> Conventional IR systems rank and assimilate documents based on maximizing relevance to the user query. In cases where relevant documents are few, or cases where very-high recall is necessary, pure relevance ranking is very appropriate. But in cases where there is a vast sea of potentially relevant documents, highly redundant with each other or (in the extreme) containing partially or fully duplicative information we must utilize means beyond pure relevance for document ranking.[31]
>
> (Carbonell and Goldstein 1998, p. 335)

An early metric focusing on the user's information need is Expected Search Length (ESL). It is defined as the number of non-relevant documents in a weakly ordered document list a user has to examine before finding *n* relevant ones (Cooper 1968). It may be viewed as a more general version of Mean Reciprocal Rank (MRR), which is analogous (or, rather, inversely proportional) to ESL with *n*=1. It should also be noted that the number of retrieved relevant documents is not the only variable; the definition of relevance can also be set so as to model a user's information need. One problem with ESL is the underlying assumption that user satisfaction is primarily influenced by non-relevant results. This is not self-evident; indeed, some studies suggest that at least in some circumstances, the number of non-relevant results plays hardly any role (Turpin and Hersh 2001).

For this and other reasons, ESL was expanded by Dunlop (1997) to an "Expected Search Time" (EST) measure. By estimating the time needed to interact with the search engine (to read an abstract, to read a document and so forth), he comprised a metric for the probable duration of a user's search session before he finds the *n* relevant results he needs. However, further limitations of EST as well as ESL remain; the number of desired relevant documents may vary heavily depending on the information need as well as on the individual user. Also, a user himself probably does not know how many documents he will require; he may prefer all the information to be in a single document, but if it is scattered, he might really want all five or ten relevant documents.

A measure concerned with the diversity of the result list is the Maximal Marginal Relevance, or MMR (Carbonell and Goldstein 1998). It assumes that the user wishes to encounter documents which are as similar as possible to the query, but as dissimilar as possible between each other so as to avoid duplicate information. In short, MMR finds, for any rank, the document for which the difference between query similarity (topicality) and similarity to already retrieved documents (repetition of information) is maximal. A parameter can be used to assign relative importance to the two concepts; if it is set to *1*, the novelty factor disappears, and MMR is reduced to a standard topicality measure. If it is *0*, the topicality is disregarded, and novelty remains the only factor. The concept of marginal relevance is especially interesting since it neatly encompasses the economic notion that only new information is useful (Varian 1999). MMR was conceived as a ranking function, but, it can easily be adapted for evaluation. However, since it requires a sophisticated similarity measurement, it will not be employed in the study presented in Part II (and I will spare you a complicated formula).

---

[31] Here, "relevance" is taken to mean the objective relevance of a single document to a query (topicality). Relevance is discussed in more detail in Chapter 7.

MMR's approach is to start from first principles; an alternative is to take existing measures and expand them by adding a novel user model. One such measure is α-NDCG, which adds to the (normalized) DCG score the concepts of users' information needs as well as novelty and relevance rating details (Clarke et al. 2008). It employs the concept of an "information nugget" which has gained acceptance in several IR subfields (Dang, Lin and Kelly 2006). In the context of α-NDCG, "information nugget" denotes any property of the document which might be relevant to the user. Thus, a document can have no, one or multiple useful information nuggets. The metric then proceeds with some quite complex calculations, determining the probability that a user will find nuggets in the current document as well as the probability for each nugget that it has already been encountered during the search session. These two probabilities then determine the gain at the current rank, which is handled similarly to the gains in usual NDCG. Once again, it is too complex for the present evaluation, and is mentioned mainly because it shows the sophistication of user models that can be employed by current metrics.

An example of a novel evaluation metric with a different agenda is the Expected Reciprocal Rank (ERR). Whereas α-NDCG pays particular attention to the diversity and novelty of the information nuggets, ERR focuses on the documents' cumulated probability of being useful enough to satisfy the user's information need. In practice, this means examining whether the documents earlier in the result list have been useful; if so, the value of a new document is discounted accordingly (Chapelle et al. 2009).

$$ERR = \sum_{r=1}^{n} \frac{1}{r} \prod_{i=1}^{r-1} (1 - \frac{2^{rel(d_i)} - 1}{2^{rel_{max}}}) \frac{2^{rel(d_r)} - 1}{2^{rel_{max}}}$$

**Formula 4.12. ERR formula. For each rank *r*, the probabilities of a relevant result at each earlier rank *i* are multiplied; the inverse probability is used as a damping factor for the gain at the current rank.**

The calculation of ERR shown in Formula 4.12 shows a structure similar to those of traditional measures like MAP, with a summation over document ratings *g* (here as $\frac{2^{g_r}-1}{2^{g_{max}}}$ to indicate a slight probability that even the best-rated result might not satisfy the user completely) and a rank-based discounting function 1/r.[32] The novel part is that every addend is also discounted by a factor indicating how likely it is that the user has already found enough information not to be interested in the current result; the factor is the product of previous results' weights, each subtracted from 1 to provide higher discount for better previous results.

## 4.4  General Problems of Explicit Metrics

There are some problems common to all the explicit ratings described above. This section provides an overview and indicates where solutions may be found.

Firstly, user judgments tend to be hard to obtain, as they require considerable resources. For the quantities of judgments needed for study results to be stable and significant, these resources may surpass the capabilities of single research groups. This is less of a problem for businesses and large collective coordinated such as TREC. However, commercial research does not always find its way into public availability; and the amount of work put into every large-scale evaluation such as TREC means there is likely to be a limited number of methods for which user judgments will be provided, and those methods are likely to be well-established and already tested.

The second issue is the one partially addressed by user-based metrics. The explicit relevance judgments are almost always collected for individual results, and then combined in a certain way to deduce a measure of user satisfaction. The raters do answer the questions asked of them; but it is important to know whether those questions really relate to the point at stake. It is the downside of the precision advantage offered by explicit user judgments: The results are focused, but the focus may be too narrow. It has been argued, for example, that the concentration on individual results is counterproductive and should be replaced or at least supplanted by methods taking into account the user's interaction with the result list as a whole (e.g. Spink 2002; for an overview, see Mandl 2008). Indeed, there are studies that suggest that discounting for rank and even for previously encountered relevant documents does not always alleviate the problems.

One such study tested four models of user behavior as related to position bias of search engines (Craswell et al. 2008). Among them are a baseline model (no positional bias at all; clicks are based only on the snippet quality), an examination model (the probability of a click is based on position-based chance of snippet examination and snippet quality), and a cascade model (clicks are based on the attractiveness of previous snippets and the snippet of the examined document). The models were trained on data acquired from "a major search engine" (the study has been performed at Microsoft), where the results were manipulated as follows: for a subset of users, adjacent results from the first result page were swapped (e.g., the first result with the second or the 9[th] with the 10[th]). The difference in click occurrence was measured, and used to train the individual models. The predictions of these models were then compared to the actual click data.

---

[32] As the authors note, "there is nothing particular about that choice" (Chapelle et al. 2009, p. 624); another function can be easily employed if it promises closer correspondence with the assumed user model.

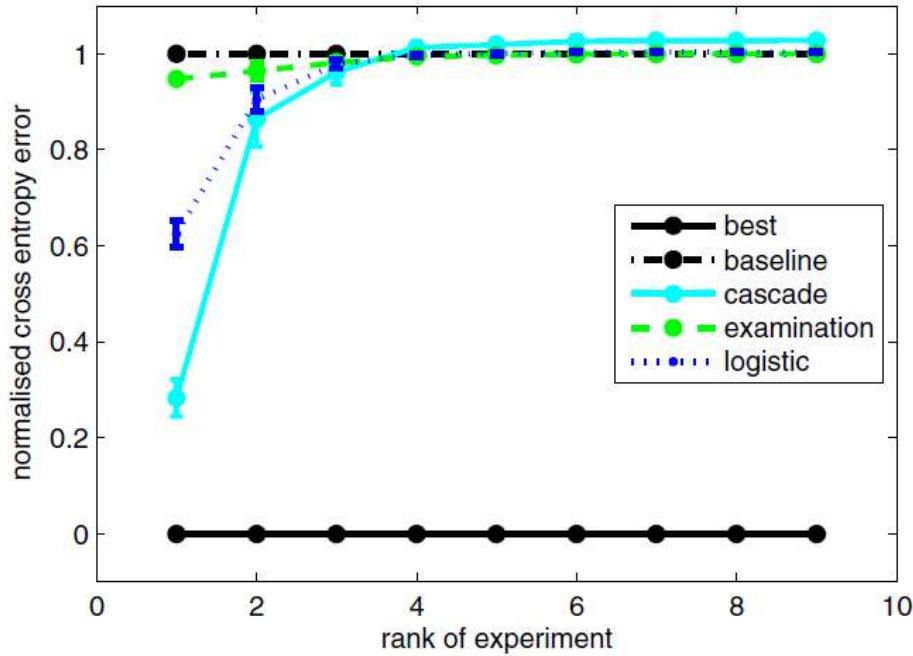

Figure 4.3. Entropy evaluation for different position bias models (Craswell et al. 2008). A value of 0 means a perfect method; 1 is set to the baseline level.

The results, based on over 100,000 data sets, are very interesting. In short, the cascade model performs better than all the others, while the examination model produces results barely above the baseline (see Figure 4.3). This is remarkable, especially given that the cascade model used here was deliberately kept very simplistic and used some crude assumptions (such as there always being one and only one click per query). The obvious conclusion would be that a measure comprised of ratings for single results does not lead to a coherent representation of a user's probable behavior, even if position bias is taken into account. However, the results also show that the strength of the cascade model lies in predicting the user behavior for the very first ranks. At the first rank,[33] its performance is superb; at the second, good; at the third, unremarkable. From the fourth result on, the cascade model actually performs worse than the baseline model, which assumes no positional bias at all. The authors "take this to mean that there is a small amount of presentation bias at lower ranks, and a large amount of variability because clicks are rare" (Craswell et al. 2008, p. 93). This is not the only possible interpretation. Another is that the difference is not primarily one of rank, but of query type. In this reading, the model performs very well on navigational queries, which indeed contain only one result that the user is interested in, as assumed by the model. Furthermore, these results are usually near the top of the result list, in the positions where the cascade model is strongest. Broad informational queries, in contrast, tend to require multiple documents to satisfy the information need. Also, the results for non-navigational queries are considered to be more difficult to rank (Dupret and Liao 2010), so that the ranking list might be worse to begin with, allowing a relatively good performance of the baseline model. If this interpretation is correct, different metrics would be required in order to adequately assess different query types.

---

[33] That is, when the first result is switched with the second; the rank indicates the first of two consecutive swapped results.

Whatever the precise reasons for the cascade model's performance, the overall results of the study seem to indicate that the methods for using explicit relevance ratings to construct result list evaluations, as they are mostly applied today, do not correspond well with real-life user behavior.

There have been further studies indicating the need for more holistic ratings. Ali, Chang et al. (2005) have shown that the correlation between result-based DCG scores and result list scores (on a tertiary scale) is 0.54 for image and 0.29 for news search. While the fields were more specific than general web search, the numbers clearly do not indicate a reliable link between the scores. Instead, the low correlation indicates that a result list might be larger or smaller than the sum of its parts; that is, good individual results might combine into a bad result list, and a result list judged to be of high quality might consist of results which are, on average, considered individually to be low-grade.

A further study examined the correlation between average precision and user success (Al-Maskari et al. 2008). For a number of TREC topics, test runs were performed on three search engines to determine their average precision. Then, users were presented with the best or the worst of the search engines for each topic, and were instructed to find as many relevant documents as possible in 7 minutes. The results showed a strong correlation between average precision and user success metrics (such as the number of retrieved documents) as well as user satisfaction.

However, a number of conceptual issues might be raised. The correlation values are significant; but the increase in average precision that they correlate with is fourfold, which is quite an extraordinary difference. Compared with this distinction, the increase in user success and especially user satisfaction is quite low. When the (absolute or relative) difference between the systems' average precision was reduced, the significance of correlations promptly dropped and all but disappeared when the increase in average precision was at 30%. The topics were taken from TREC, and the "title" and "description" fields were used as queries for precision calculation; however, the users searched with queries of their own, so that the correlation was actually between the precision of one query and the output and user satisfaction of another query. The average precision was very low (0.09 to 0.16 for the three search engines tested); this might be a consequence of measuring it for all results, without a cut-off value.[34] This makes the "good" systems not very good, and the average precision not very realistic with regard to actual user behavior. Furthermore, the interface used by the study does not have much resemblance to the interface of most search engines, providing less transferrable results, and also aggravating possible effects an artificial setting has on user behavior. Perhaps most significantly, the task itself, with its fixed time frame and requirement of relevant document recall independent of the actual novelty or usefulness of documents, seems to be more of a task one would like to measure with average precision than a typical information need of a web searcher. This would also explain why precision@200 was found to correlate better with user's effectiveness than average precision; since the users were instructed to gather as many relevant documents as possible, and given a fixed (and not very small) amount of time, rank effects probably did not play a major role in this evaluation. This long list of possible error sources stresses the necessity of careful consideration of study set-ups.

---

[34] The authors, at least, do not mention a cut-off value, so we assume the average precision was that of all returned results.

One other possible explanation of the different correlations is provided by Smith and Kantor (2008). They studied the influence of degraded search quality on user performance. The degrading was achieved by returning standard Google results for the control group, and Google results starting at rank 300 for the treatment group.[35] The lower quality was confirmed by significantly lower precision and (relative) recall numbers. The users were asked to find as many good sources for the topics as they could. There was no significant difference between the numbers of good results marked by the users, neither in absolute nor in relative terms. The authors conjecture that the users adapt their behavior to the change in result quality; an assumption supported by collected implicit metrics (for example, users working on "bad" result lists tended to pose significantly more queries per topic). Regrettably, the study provides no user satisfaction measures. Those would be quite interesting, as the results seem to suggest that if the correlation between precision and user performance is high, and correlation between precision and user satisfaction is low, as suggested by other studies (e.g. Turpin and Scholer 2006), there might not be any direct relationship between user performance and user satisfaction. Instead, the difference between performance and satisfaction may be caused by user effort (in this case, issuing more queries, and perhaps spending more time considering non-relevant abstracts and documents in search for relevant ones).

One more study which examined the relationship between explicit relevance ratings and user satisfaction has been conducted at Google (Huffman and Hochster 2007). It used 200 random queries from those posed to the search engine, asking raters to infer probable information needs and to provide further information on the query (e.g. navigational versus informational versus transactional). Then, one set of raters was presented with the information need and the Google result list for the query, and asked to perform a normal search. After the user indicated he was done, an explicit satisfaction rating was solicited. A second group of raters was asked to give relevance ratings for the first three documents of each result list. A simple average precision at rank 3 had a very strong Pearson correlation of 0.73 with user satisfaction, which is quite a high number for so simple a metric. A higher correlation was observed for highly relevant results; this means that with really good results in the top 3 ranks, the user will not be unsatisfied. This is not really surprising; however, we have seen that studies do not always support intuitions, so the results are valuable. By manipulating the metric (in ways not precisely specified), the authors managed to increase correlation to over 0.82. The important adjustments were for navigational queries (where the relevance of the first result was given a more important role, and further results were discounted stronger), and for the number of events (such as clicks) during the session.

---

[35] There was also an intermediate condition providing, for different topics, results starting at different ranks, from 1 to 300; the findings for this treatment were mixed, and will not be detailed here.

There might be multiple reasons for the fact that the Huffman and Hochster study (2007) provides a strong correlation between precision and user satisfaction while most other studies do not. First, the queries used and the tasks posed were those of real search engine users, something that other studies rarely featured. Second, the relevance considered was only that of the first three results. Studies have shown that on the web, and especially at Google, the first three or four results are the most often viewed and consequently the most important ones (Hotchkiss, Alston and Edwards 2005). This might mean that the relevance ratings of lower-ranked results do not play an important role for many queries, and as such only complicate the overall picture. It would have been interesting to see how the correlation changed for average precision at, say, 10 results; that is something that will be addressed in the evaluation performed in Part II.

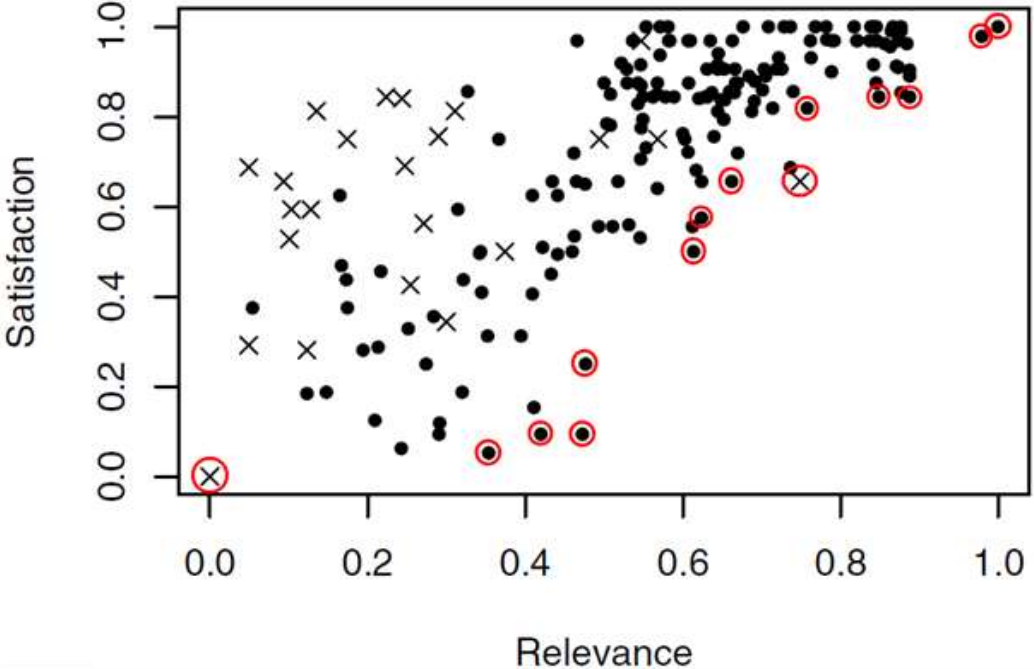

**Figure 4.4. Session satisfaction vs. first-query relevance (modified from Huffman and Hochster 2007, p. 568). Crosses mark misspellings. The red circles indicate data points which do not have any further points with higher relevance *and* lower satisfaction; for all other points, a positive change in relevance might, even within the data set of that study, come with a decline in satisfaction.**

Perhaps more importantly, one has to keep in mind that correlation is not everything. The models described by Huffman and Hochster may help to provide an answer to the question "Will the user be satisfied with his search session, given certain relevance rankings?". But they are probably significantly worse at another task: judging whether a certain change in indexing or ranking procedures improves performance. The authors write: "Relevance metrics are useful in part because they enable repeated measurement. Once results for an appropriate sample of queries are graded, relevance metrics can be easily computed for multiple ranking algorithms" (Huffman and Hochster 2007, p. 571). However, if we look at the relevance-satisfaction-plot (Figure 4.4), we see that there are at most 14 data points (circled in red) which do not have further results right and down from them; that is, where increased relevance might not correspond to decreased satisfaction. If we discard misspellings, this means that only less than 10% of all data points in this sample cannot be actually worsened by a result set with a higher relevance score. Without access to the raw data of the study, it is difficult to tell how well these models predict exact satisfaction values given a relevance rating and – related to that and important in a practical sense – the direction of change in satisfaction given a certain (usually relatively small) change in the relevance ranking. These questions will also be followed up in Part II, where they will be answered based on new data.

Similar results were also produced in a Yahoo study aimed mainly at goal success determination from user behavior (Hassan, Jones and Klinkner 2010). The average precision of the top 3 results had an 80% accuracy in predicting goal success, while DCG offered 79% (here, the considered rank was not disclosed). A Markov Model trained on query reformulation and click data provided 85% accuracy. However, one has to note that this study differed from most others in that it considered success as a function of sessions instead of the more usual queries. Also, success was estimated not by the originators of the information need but by company-internal editors.

# 5  Implicit Metrics

> *The character of the logger's admiration is betrayed by his very mode of expressing it... He admires the log.*
>
> HENRY DAVID THOREAU,
> "THE ALLEGASH AND EAST BRANCH"

The measures described in this section are not acquired by asking the user, as was the case with explicit ones. Instead, they are calculated using search engine logs. Logs are documents automatically created by the search engine's servers which record user actions. Typical contents of a log are a user ID, a session ID, the query, the result list and a timestamp.

The user ID is an identifier which allows tracking the user's progress through different sessions. In the log's original state, this is normally the user's IP address. The session ID is a limited form of a user ID; its precise form depends on the definition of "session" used. This may range from a single query to a timeframe of many hours, during which the user submits multiple queries. The query is just the query in the form submitted by the user; it is paired with the result list. The result lists contains not just the URLs of the results offered to the user; it also shows which of the results have been clicked on. Finally, the timestamp provides information on the exact moment of the query, and – if available – the time of each individual click (Croft, Metzler and Strohman 2010). It is also important to note what the logs of a search engine do not usually contain, not least because much of the research on implicit metrics has used features not readily available. Some of these features are viewing time for a document ("How long did the user stay on a web page he found through his search?"), the amount of scrolling ("How far down the document did the user go?"), or the bookmarking or printing of a document (Kelly and Teevan 2003). Also, the logs do not provide any information as to the users' information needs, a topic that will be discussed in Chapter 7.

It is especially the click data from the result list that provides the basis for analysis. Different explanations have been proposed as to how exactly the click data corresponds to user preferences, and some of those explanations will be presented in this section. As clicks and other log data reflect user behavior, a user model is indispensable in analyzing the data; therefore, those measures lean towards the user-based camp.

Some relatively straightforward metrics are being used for internal evaluations at Google. While not much is known about the details, at least two metrics have been mentioned in interviews: average click rank and average search time (Google 2011d). The smaller (that is, higher up in the result list) the average click rank, and the less time a user spends on the search session, the better the result list is assumed to be. However, the company stops short of providing any clues on how exactly the evaluation is conducted or any evidence they have that these metrics, intuitive as they are, actually represent user preferences.

A general difficulty with interpreting click data as relevance judgments is the absence of any non-click data. While a click on a result can intuitively be interpreted as a positive judgment on the quality of the document, a non-click has little informative value. The user may not have liked the result, or just not seen it. One solution is to relinquish absolute relevance judgments in favor of relative ones (Joachims 2002). The most common way is assuming that a user tends to scan the result list from top to bottom.[36] In this case, all results up to the lowest-ranked click he makes can be considered to have been examined. The logical next step is to assume that every examined but not selected result has been judged by the user to be less relevant than any later result that has been clicked. This model, called "click > skip above" by Joachims, does not make any statements about the relative relevance of clicked results among themselves, or between the non-clicked results among themselves. The preference of selected documents over non-selected but presumably considered ones is the only assertion it makes.

This is not the only possible interpretation of click data. Joachims, Granka et al. (2005) also test further evaluation (or re-ranking) strategies. Assuming that the user learns from results he sees, later clicks would be made on the basis of more available information and thus more meaningful. A possible expression of this is the "last click > skip above" measure, which considers the result last clicked on[37] to be more relevant than all higher-ranked unclicked results. A stronger version of the same assumption is "click > earlier click", interpreting the second of any two clicks as the more relevant (the rationale is that if a highly relevant result has been encountered, it does not pay to click on a less relevant result afterwards). Lastly, noticing that eye-tracking results show the users are likely to examine the abstracts directly above or below a click, they introduce the "click > skip previous" and the "click > no-click next" measures.

The authors also compared their measures to explicit ratings collected for the same data. The results are mixed; the correlations for the "click > skipped above", "last click > skip above" and "click > skip previous" range from 80% to 90% compared to explicit ratings of the descriptions in the result list, and from 78% to 81% for the ratings of the pages themselves. The values for "click > earlier click" and "click > no-click next" are about 15% lower. On the one hand, the values are quite high, especially considering that the inter-judge agreement (and with it the reasonable upper border for possible correlation) was at just 83% to 90%. On the other hand, the variances were very large, from 5% up to 25%. Also, the correlation of the implicit measures with explicit abstract ratings varied strongly between two phases of the study, suggesting a worrying lack of result stability.

All of the described metrics are based on the comparison between clicked and unclicked results. They assume that a click reflects a high quality of the document, while a non-click on a supposedly considered document reflects low or at least lower quality. This approach has, apart from being once again focused on individual results, two main conceptual problems. First, all described methods – apart from "click > no-click next" which has a relatively low correlation with explicit ratings – can only indicate a preference of a lower-ranked result over a higher-ranked one, and "such relevance judgments are all satisfied if the ranking is reversed, making the preferences difficult to use as training data" (Radlinski and Joachims 2006, p. 1407). Thus, the ranking of a result list is never vindicated, and a reversal of a result list can only result in an improved (or at least equally good) result list, as measured by those metrics.

---

[36] Though some studies reported a significant minority (up to 15% of users in a test group) employing "an extreme breadth-first strategy, looking through the entire list [of 25 results] before opening any document" (Klöckner, Wirschum and Jameson 2004, p. 1539), and another group preferring a mixed strategy, looking at a few results before selecting one.

[37] As it is determined by the timestamp, this is not necessarily the lowest-ranked clicked result.

Secondly, clicks are not based on document quality, but at best on the quality of document presentations shown in the result list. The document itself may be irrelevant or even spam, but if the title, snippet or URL is attractive enough to induce a click, it is still considered to be an endorsement.

Dupret and Liao (2010) suggest a metric which does not compare clicked to non-clicked results. Instead, the question the method addresses is whether, after clicking on a result, the user returns to the result list to proceed with his search. If he does not, the document is assumed to have satisfied (together with previously clicked documents) the user's information need. This assumption, as the authors point out, is unrealistic (the user could have abandoned his search effort) – but it provides some practical benefits. The method is not used for creating explicit relevance predictions; but if a function maximizing a posteriori likelihood is added as a feature to the ranking system of "a leading commercial search engine",[38] the NDCG – as compared to explicit relevance judgments – grows by 0,06% to 0,76%, depending on the cut-off value and type of query. This seems to be an improvement over a presumably already sophisticated effort; however, the order of magnitude is not large, and there is a notion that "a gain of a few percent in mean average precision is not likely to be detectable by the typical user" (Allan, Carterette and Lewis 2005, p. 433) which presumably also applies to NDCG.

---

[38] The authors work for Yahoo.

# 6 Implicit and Explicit Metrics

*Just because your ratings are bigger doesn't mean you're better.*
TED TURNER

A major problem with the cited experimental confirmations of log data evaluation is that the confirmation comes from their correlation with measures based on explicit document relevance rankings, whose significance, in turn, is often based on not very persuasive evidence; indeed, sometimes their vindication comes from log data evaluations. Strictly speaking, this would imply only that explicit and log-based ratings correlate with each other, but not necessary with any satisfaction levels on the user side. The claim that implicit and explicit measures do not reflect any user preferences is disproportionate; but the point that there is hardly any validation of their significance outside this vicious circle remains. Another strategy is to compare explicit metrics among each other (e.g. Kekäläinen 2005; Della Mea et al. 2006; Sakai 2007b); I consider this to be valuable, once the validity of the primary metric has been established.

There have been some attempts to break this circle. One interesting method is to create two result lists so that one of them is deemed to be better than the other "by construction" (Radlinski, Kurup and Joachims 2008). In one of the conditions of this study, the authors implemented the "good" result list by calculating the cosine similarity between the query and multiple fields (e.g. title, abstract and full text) of the arxiv.org database of scientific articles. A "worse" result list was created by removing from the score the content of record fields considered to be important, such as title and abstract. Those were still included in the full text and thus matched; but in the "better" list, the inclusion of further fields increased the match count for every word in title and abstract. A "worst" list was created by randomizing the top 11 results of the "worse" list. In a second condition, the "worse" list was constructed from the same "good" list by randomly swapping two of the results in positions 2-6 with two in the positions 7-11. In the "worst" list, four instead of two results were swapped. One of those lists were randomly shown to users of a search engine implemented for the very purpose, and the data from their interaction with it recorded. The data was evaluated using "absolute metrics", that is, metrics attempting to put a concrete number on a user's satisfaction with the result list. Among this metrics were e.g. the Abandonment Rate (queries without clicks on any result), Clicks per Query, and the Mean Reciprocal Rank (one over the mean rank of clicked results). The authors formulated expectations for the measures' behavior when confronted with a worse result list; for example, the Clicks per Query and the Mean Reciprocal Rank were expected to decline. In each of the two conditions, there are three result lists which can be compared to each other; thus, the overall number of comparisons is six. The results are given in Table 6.1; it is obvious that, while the metrics tend to agree with the hypothesized quality difference more than contradict them, the results are by no means conclusive. This is especially true if only statistically significant differences (right column) are considered; of 48 comparisons, only 10 do significantly correlate with the assumed difference, while one contradicts it. The authors draw the conclusion that, while the problems may be "due to measurement noise", the most that can be said for absolute metrics is that the amount of data (from the graphs in the paper, ca. 1400 unique users can be induced) is insufficient for them to reliably measure retrieval quality.

|  | weak ✓ | weak ⚡ | signif ✓ | signif ⚡ |
|---|---|---|---|---|
| Abandonment Rate (Mean) | 4 | 2 | 2 | 0 |
| Reformulation Rate (Mean) | 4 | 2 | 0 | 0 |
| Queries per Session (Mean) | 3 | 3 | 0 | 0 |
| Clicks per Query (Mean) | 4 | 2 | 2 | 0 |
| Max Reciprocal Rank (Mean) | 5 | 1 | 3 | 0 |
| Mean Reciprocal Rank (Mean) | 5 | 1 | 2 | 0 |
| Time (s) to First Click (Median) | 4 | 1 | 0 | 0 |
| Time (s) to Last Click (Median) | 4 | 2 | 1 | 1 |

**Table 6.1. Absolute metric results and their significance with check mark meaning agreement and danger symbol disagreement with hypothesized quality (Radlinski, Kurup and Joachims 2008).**

In the next step, the three result lists of a condition were shuffled into one using a novel method (which, ingenious as it is, does not directly concern us here) so that the resulting result list contains at any point almost an equal number of results from each list in the original order. If we assume that the three lists were completely different (which they were not) and call them A, B and C, than the combined result list might look like A1, B1, C1, C2, B2, A2, A3, B3, C3… As we further assume that users scan the list progressively from top to bottom, a click on C1 without one on A1 or B1 would indicate that the user has seen the first two results but has not found them relevant, in contrast to the third. From this we can in turn follow that the first result of list C is more relevant than those of the other lists. This combined list was also shown to users of the same search engine, with results which were much more encouraging. Without delving too deep into the methodology, the results from better-conceived results lists were preferred in every statistic, and this preference was statistically significant in over 80% of evaluations. The authors conclude that, in contrast to absolute metrics, this so-called "paired comparison tests" deliver stable results and are well suitable for establishing the preferences even of relatively small user groups.

I have described this study in some detail because it addresses a central question of the present work, namely, the validity of evaluation metrics, and, furthermore, it does so in a novel and promising way. Despite, or, perhaps, precisely because of that, it is important to realize that the study is not as conclusive as it seems, and at almost every step objections can be made to the interpretation of its findings. The problems start at the very onset, when two result list triples are constructed. In each triple, there is supposed to be a significant difference in retrieval quality. But does this really follow from the construction of the result lists? The "good" list is the result of a relatively simple cosine similarity measure. We do not know much about the usual retrieval behavior of users who made up the test group; but for many queries, such a ranking algorithm that does not include, for example, any citation data (which a major academic search engine like Google Scholar does quite heavily), may provide a result list where the average decline of result quality from rank to rank may be very shallow. If the relevance of the $10^{th}$-ranked result is, on average, indeed not much lower than that of the first one, then a slight decrease of the importance of some record fields like title and abstract might not lead to a significant decline in perceived retrieval quality. And even a completely random re-scrambling of the first 11 results might not produce large distortions if they are of close enough quality to start with.[39] Of course, that would also hold for the "swap 2" and "swap 4" lists of the second triple. If we consider what might be termed the bibliographic equivalent of navigational queries, that is, search for a concrete document, the situation becomes still more complicated. If the query is not precise enough, the document may land anywhere in the result list, and the "worsening" of the list (which, as we recall, consists of shuffling some or all of the results) may move it up or down. If, on the other hand, the query does precisely identify the document, it is likely to land on top of the "good" result list. It is also likely to land on top of the "worse" result list in the first condition, and of both "worse" and "worst" lists in the second condition, as the top result is never shuffled. Of course, there can be queries which are somewhere between those two states. The point is that the quality difference between the lists might not be as self-evidently significant as suggested by the authors. The statement that "[i]n fact, our subjective impression is that these three ranking functions deliver substantially different retrieval quality […] and any suitable evaluation method should be able to detect this difference" (Radlinski, Kurup and Joachims 2008, p. 44) stands, and, indeed, falls with its first part.

More doubt awaits us at the evaluation of absolute measures. One issue follows directly from the above argument; if the differences in ranking quality are insignificant, than insignificant metric differences may still reflect the situation correctly. The evaluation might also be taken to support the thesis that the "good" result lists are not that good to begin with. The Abandonment Rate for these result lists lies at 68 to 70%; that is, over two thirds of searches with the retrieval quality assumed to be superior do not lead the user to any useful results.[40] A second problem is that once again, the assumptions made by the authors may, but need not be correct. The hypothesized metric changes from one result list to another are founded in sensible logical arguments; but sensible logical arguments could also be made for changes in the other direction. Indeed, the authors inadvertently illustrate this ambiguity in the brief motivations they provide for their hypotheses. For Clicks per Query, the prediction is "Decrease (fewer relevant results)"; for Mean Reciprocal Rank, it is "Decrease (more need for many clicks)" (Radlinski, Kurup and Joachims 2008, p. 46), which, of course, implies an *increase* in Clicks per Query.

These problems should make us view the final evaluation, that involving "paired comparison tests", with caution. Again, one issue stems from the study's difficulties in creating result lists which significantly differ in retrieval quality a priori. The evaluation undoubtedly shows significant correlation of the measure with result list types; but we cannot be certain this correlation is with retrieval quality. Instead, it might correlate with one of a number of other properties, such as the appearance of search keywords in the title.[41] A further problem is the method of assembling a joint result list. As the results from two separate lists are rearranged, de-duplicated, and re-joined, the original ordering and half of the results are lost for every result page. Thus, even if the comparison tests can be used to judge which list has more relevant results,[42] they may well be unable to compare the usefulness of the two original result lists as cohesive, interrelated units. By neglecting the structure of the result list, the evaluation method turns away from user models which can be seen as one of main advantages of log data evaluation.

---

[39] Note that the result quality at any rank is always meant to be averaged over all queries; if a ranking algorithm is not very effective, the relevance of results at any rank will be virtually unpredictable for any single query.

[40] There seems to be some error in the data, which may or may not weaken this line of reasoning. The authors state that the search engine received ca. 700 queries per day, with about 600 clicks on results; this would imply an overall click rate of ca. 0.86. The tables, however, give us an Abandonment Rate of 0.68 to 0.70; and the remaining queries where at least one result was selected (ca. 250 of them) had a click rate of about 0.72. Now, 0.72 is by itself significantly lower than 0.86; but if the queries with no clicks were indeed excluded as stated by the authors, the Clicks per Query metric cannot be lower than 1. Furthermore, 250 queries with a click rate of 0.72 would mean around 180 clicks overall, not 600.

It should be stated once more that the study by Radlinski, Kurup et al. is a promising contribution, and I do not intend to claim its results are wrong. They are, however, based on premises that are less self-evident than the authors assume, and therefore may be not very conclusive.

Another study that addresses, albeit indirectly, the connection between explicit and implicit measures, has been conducted at Microsoft (Dou et al. 2008). The authors once again try to determine relative judgments from click-through data, but their method is slightly different. Instead of calculating the preferences for every session, the clicks on every document are aggregated for every query, and only then the preference is established. That is, if a query $q$ is encountered 10 times in the log data, and document $d_1$ is clicked 4 times, while document $d_2$ is clicked once, a preference for $d_1$ over $d_2$ is assumed.

In a first step, the authors compare such preferences derived from click data with explicit judgments. The latter are graded ratings on a 0..4 scale. However, the prediction accuracy of the implicit judgments is determined by comparing simple preferences, without regard to the actual difference in rating. Still, the results are not very encouraging; the Kendall tau-b measure (Kendall and Smith 1938) shows a correlation of 0.36 in the test condition. This includes pairs where one document received some clicks, while the other was not clicked at all; if those are excluded, the correlation drops to 0.19. Neither condition includes statistics for pairs where none of the documents receive clicks; those obviously do not provide means for determining preferences, and would lower the correlation still further. There are multiple possible reasons for the results. As the authors note, their method is very simple and disregards many known issues such as position bias (Joachims et al. 2005). A more sophisticated approach might bring better results. Another explanation might be a generally low predictive power of click data; a third is that implicit judgments are relevant, but the explicit judgments they are compared to are not.

---

[41] Remember that clicks on a result mean not positive feedback on the document but positive feedback on its representation in the result list. The "good" retrieval function with its focus on matching query keywords with title and abstract can be seen as increasing not necessary the quality of retrieved documents, but the attractiveness of the result list.
[42] That is, either more results which are relevant or results which are more relevant. Clicks provide a binary distinction at best, so we don't know which of these are being measured.

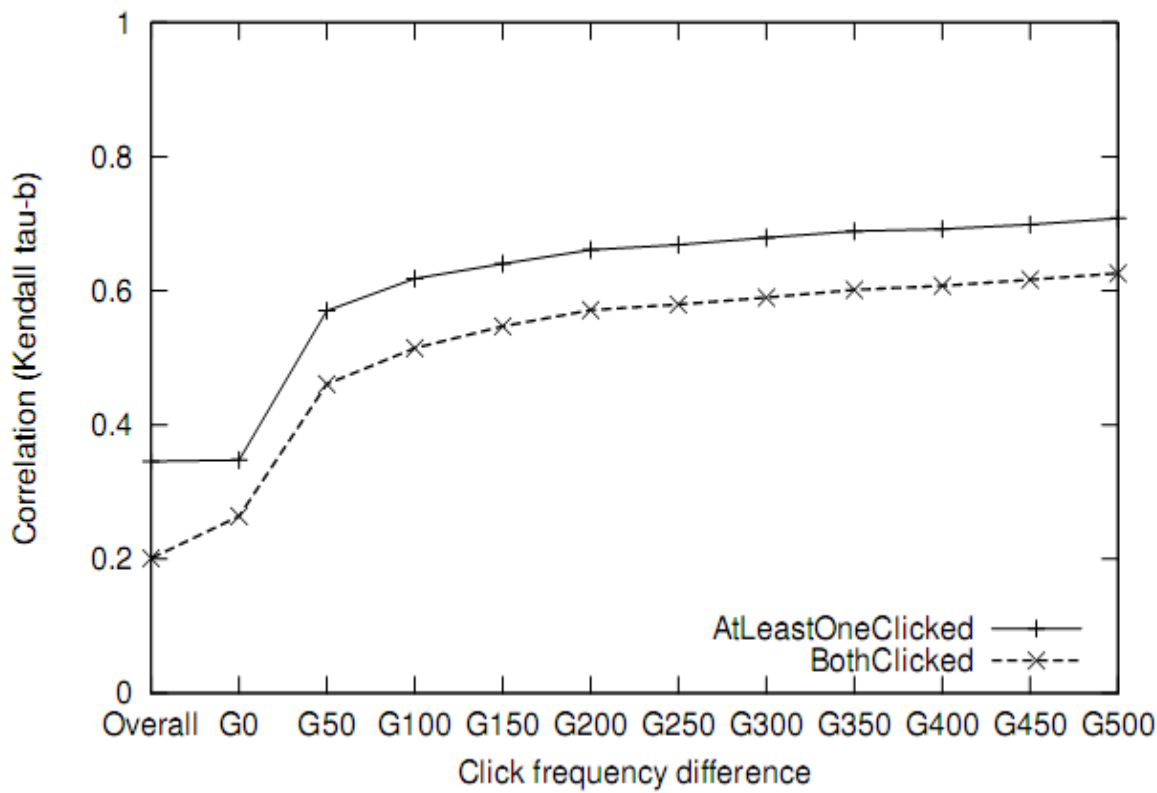

**Figure 6.1. The correlation between preferences as determined by CT-Gn and explicitly stated preferences for different values of *n* (from Dou et al. 2008).**

To improve the results, the authors introduce a more discriminating measure they call CT-Gn (for Click-Through Greater than n). The first evaluation considers only cases where the number of clicks on a pair of results differs by more than a certain number *n*. The correlation statistics for different values of *n* are given in Figure 6.1. The numbers look significantly better; but, as always, there is a trade-off. The higher *n* (and the prediction confidence), the less document pairs with a sufficient click difference there are. Unfortunately, the authors do not provide any numbers; but they may be roughly estimated using the size of the data samples used for training and testing. The training sample consisted of 10,000 queries with ca. 580,000 documents, of which ca. 70,000 have been clicked.[43] The easiest way to get a document pair with *n*≥100 is for one document to have no clicks while the other has 100.[44] This, in turn, means that a query must comprise at least 1% of all stated queries, and in every of this cases a document $d_1$ must be clicked on while a document $d_2$ must not. Obviously, this can produce results for at most 100 queries, and realistically viewed for considerably less. The problem can be mitigated by using significantly larger amounts of data, which are surely available for company researchers. But those logs will probably not be open to the scientific community for scrutiny; and besides, the problem of proportions will emerge. If the amount of log data is significantly higher so that a larger number of queries reach the *n*≥100 level, some queries will have very high frequencies, with tens or perhaps hundreds of thousands of occurrences. For those queries, the $n \geq 100$ border might be too low if considered in proportion to much larger click numbers. The solution to that might be a border line which combines the total number of clicks with the frequency of a query; but that remains to be implemented and tested in detail.

---

[43] This proportion of over 7 clicks per query is extremely high. The usual number given in the literature varies, but generally lies between 0,8 (Joachims et al. 2005) and 2 (Baeza-Yates 2004; Baeza-Yates, Hurtado and Mendoza 2005). However, the 58 documents considered for each query are also a very high number.
[44] To be precise, this would be *n*=101.

There also have been attempts to derive more general connections between log data and explicit measures. One such study looked at the correspondence between a result list's MAP rating and the number and rank of clicked results (Scholer et al. 2008). The result lists were constructed in advance so as to deliver certain levels of MAP, ranging from 55% to 95% (the method previously used by Turpin and Scholer 2006). Then, users were given a recall-based search task and five minutes to complete it by finding as many relevant documents as possible. The results were unambiguous; there was no significant correlation between MAP (or precision at rank 10), whether calculated based on TREC assessments or ratings by users doing the searching, and the average number of clicks or the mean rank of clicked results. Again, there are some methodical issues which can be raised, such as the unrealistic fixed session time; also, the results lists were not directly related to the queries users entered, but rather to the TREC topics from which the task statements were taken. However, the tendency still points to a discrepancy between explicit measures and log data. This may or may not be a problem for web evaluation using click-through measures. With more realistic tasks, which do not have fixed time limits or the goal of finding as many documents as possible,[45] click frequency would not necessarily be expected to be an indicator of retrieval quality. Indeed, for search engines which deliver highly relevant results at the top ranks, the number of required clicks might be expected to drop.

One more Microsoft study has busied itself with the implicit-explicit-measure problem (Fox et al. 2005). The authors collected a variety of data from the company's employees using web search engines; the session measures included data on clicks, time spent on the result page, scrolling data etc., while document measures included time spent viewing a document, the document's position in the result list, the number of images in the document and so forth. Also, during and after each search session explicit feedback was collected on individual results as well as on the satisfaction with the complete session; it was given on a tertiary scale, where the result or session were classified as satisfying, partially satisfying or not satisfying. The collected implicit measures were submitted to a Bayesian model which was used to predict the user satisfaction with individual results or sessions.

The results this model produced were twofold. For explicit measures, the overall prediction accuracy was 57%, which is significantly higher than the baseline method of assigning the most frequent category ("satisfied") to all results (which would provide 40% accuracy). If only results with high prediction confidence are taken into account (those constitute about 60% of all results), the prediction accuracy rises to 66%. The most powerful predictors were the time spent viewing a document (that is, from clicking on a result in the result list to leaving the page) and exit type (whether the user left the document returning to the result page, following a link, entering a new address, closing the browser window etc.). Just these two features provided an overall prediction accuracy of 56%, as compared to the 57% accuracy when using all features; thus, it may be concluded that they carry the main burden of providing evidence for the quality of documents. These findings are highly significant; but it should be noted that they do not provide as much help for search engine evaluation as the numbers suggest. The two main features can be obtained through a search engine log only if the user returns from a document to the search list; for other exit types, the type as well as the time spent viewing the document can only judged if further software is installed on the test persons' computers.[46] Furthermore, if the user does return to the result list, this was found to be an indication of dissatisfaction with the viewed document. Thus, if this feature is available in the logs, it is only in cases where the prediction accuracy is lower than average (ca. 50%, as compared to 57% overall prediction accuracy and 40% baseline). As there are multiple other ways of leaving a document than returning to the result page, the absence of this feature in the logs seems not to be an indicator of user satisfaction. To summarize, the most important features are either not available through the search engine logs, or are available but have lower than usual predictive power. This means that the logs' main advantage, that is, their ready availability and large size, all but disappear for most researchers.

---

[45] Even an informational query has a certain, limited amount of required information; this holds all the more for transactional and especially navigational queries.

A second part of the study concerned itself with the evaluation of complete sessions. Here, the overall prediction accuracy was 70%, as compared to a 56% baseline (again assuming that all sessions produced satisfying experiences). However, the features considered for this model included the explicit relevance judgments collected during the study; if those were excluded to construct a model based entirely on implicit metrics, the accuracy sank to 60% – better, but not much better, than the baseline. On the one hand, this obviously means that the study could not show a sizable improvement of predictive power over a baseline method when using implicit metrics on sessions, even if those metrics included much more than what is usually available from search engine logs.[47] On the other hand, and perhaps more importantly, it gives us two very important results on the relation between explicit measures, which have not been addressed by the authors.

Firstly, the impact of result ratings on the precision of session evaluation suggests that there is indeed a connection between explicit single-result ratings and user satisfaction as measured by the explicit session ratings collected in this study. This is a rare case where result- and session-based measures have been indeed collected side by side, and where an attempt has been made to incorporate the first into a model predicting the second. The "secondly", however, goes in the opposite direction. If we consider that they were aided by other log data as well as usage data not usually available to researchers, and a complex model, the prediction power of explicit result ratings is remarkably low – 70% as compared to a baseline of 56%. Given that the ratings were obtained from the same user who also rated the query, and the result rating was in relation to this query, we can hardly hope for more or better data. This means that either the Bayesian model employed was very far from ideal, or that the correlation of result ratings with user satisfaction with a search session is quite low. Together, the results suggest that, while the perceived quality of individual results has a noticeable influence on the perceived quality of the search experience, this influence is not as high as to be, at least on its own, an accurate predictor of user satisfaction.

---

[46] Fox, Karnawat et al. used a plug-in for the Internet Explorer browser.
[47] As the authors rightly notice, this problem could be mitigated by using the implicit result ratings constructed in the first part of the study. However, it remains to be seen whether this less reliable input can provide the same level of improvement in session-based evaluation.

The concentration of some of the studies presented in this chapter on user satisfaction (or sometimes user success in executing a task) merits a separate discussion. For studies performed within the Cranfield paradigm, the "gold standard" against which a system's performance is measured is the rating for a single document provided by a judge. More concretely, in most studies using TREC data, "gold standard" is associated with a rating made by an expert who is, at the same time, the originator of the query (e.g. Bailey et al. 2008; Saracevic 2008). However, this is only a meaningful standard if one is interested in the relevance to a query of one isolated document. If, instead, the question one wants to ask is about the quality of a result list as a whole (and that is what popular measures like MAP and DCG attempt to answer), it seems prudent to ask the rater for his opinion on the same subject. User ratings of single documents might or might not be combinable into a meaningful rating of the whole; but this is a question to be asked and answered, not an axiom to start from. An evaluation measure is not logically deduced from some uncontroversial first principles; instead, it is constructed by an author who has some empirical knowledge and some gut feelings about what should be a good indicator of search engine quality. If we probe further and ask what constitutes high search engine quality, or what an evaluation standard for evaluation methods should be, we unavoidably arrive at a quite simple conclusion: "By definition, whatever a user says is relevant, is relevant, and that is about all that can be said" (Harter 1992, p. 603).

# Part II: Meta-Evaluation

> *Are you happy? Are you satisfied?*
> **QUEEN, "ANOTHER ONE BITES THE DUST"**

"While many different evaluation measures have been defined and used, differences among measures have almost always been discussed based on their principles. That is, there has been very little empirical examination of the measures themselves" (Buckley and Voorhees 2000, p. 34). Since the writing of that statement, things have changed; the previous chapter contains a number of studies concerning themselves with empirical examination of search engine evaluation measures. However, it also shows that the results of this empirical work are not ones that would allow us to carry on with the methods we have. For most measures examined, no correlation with user satisfaction has been found; for a few, brief glimpses of correlation have been caught, but the results are far too fragile to build upon. What, then, do the search engine evaluation measures measure? This is where the present work enters the stage.

# 7 The Issue of Relevance

> *Some of the jury wrote it down "important", and some "unimportant".*
> **LEWIS CARROLL, "ALICE IN WONDERLAND"**

As I already remarked, it is important to make sure the metric one uses measures the intended quality of the search engine. So, after considering the various evaluation types, it is time for us to ponder one of the central issues of information retrieval evaluation: the concept of relevance. I do not intend to give a complete overview over the history or width of the field; instead, I will concentrate on concepts which are promising for clarifying the issues at hand. This is all the more appropriate since for most of the history of the "relevance" concept, it was applied to classical IR systems, and thus featured such elements as intermediaries or the query language, which do not play a major role in modern-day web search engines.

The approach I am going to embrace is a modification of a framework by Mizzaro (1998). He proposed a four-dimensional relevance space; I shall discuss his dimensions, consider their appropriateness for web search evaluation, and modify his model to fit my requirements.

The first category, adopted from earlier work (Lancaster 1979), is that of the information resource. Three types are distinguished. The most basic is the *surrogate*; this is a representation of a document as provided by the IR system, such as bibliographical data for a book. The next level is the *document* itself, "the physical entity that the user of an IR system will obtain after his seeking of information" (Mizzaro 1998, p. 308).[48] The third level, the most user-centered one, is the *information* the user obtains by examining the provided document.

The second category is the representation of the user's problem; again, it is based on earlier research (Taylor 1967). The original form is called the *Real Information Need (RIN)*. This means, broadly speaking, the real world problem which the user attempts to solve by his search. The user transforms this into a *Personal Information Need (PIN)*; this is his perception of his information need. Then, he formulates his PIN in a *request* (a verbal formulation of the PIN), before formalizing it into a *query* (a form that can be processed by the information retrieval system).

The third category is time. The user's requirements may change over time, whether as a result of changes in the world and therefore in the RIN, or because of the progress of the user's search itself. For example, the user may hit upon a promising branch of information and then wish to encounter more documents from this area.

The fourth category is constituted by what Mizzaro calls *components*. He identifies three of those; *topic* – "the subject area interesting for the user", *task* – "the activity that the user will execute with the retrieved documents", and *context* – "everything not pertaining to topic and task, but however affecting the way the search takes place and the evaluation of results" (Mizzaro 1998, p. 312). The model (without the time category) is shown in Figure 7.1.

---

[48] Obviously, "physical entities" include electronic entities.

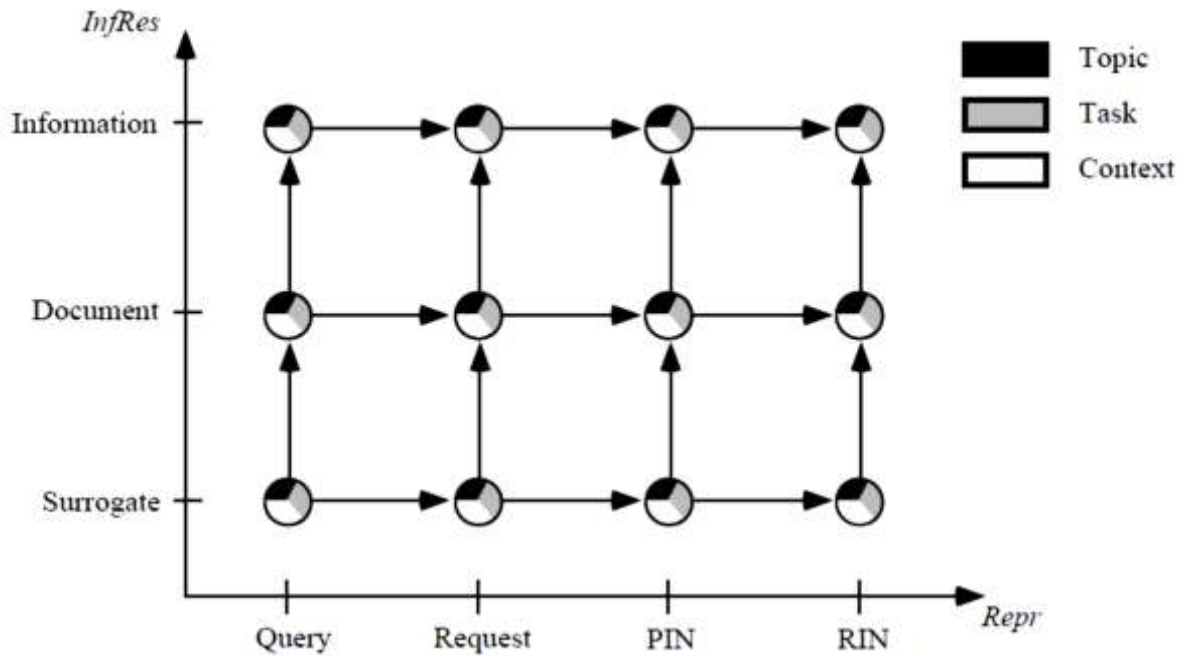

**Figure 7.1. Mizzaro's relevance model without the temporal dimension (Mizzaro 1998, p. 314).**

For the purposes of this study, I propose some changes to this model. First, as mentioned above, some of its aspects are not appropriate for modern web search. Second, the present work concerns itself only with a certain part of web search evaluation, making it possible to omit parts of the model, while acknowledging their importance for other aspects of evaluation. To distinguish my model from others, I will call it the Web Evaluation Relevance model, or WER model for short.

First, let us address the question of representation. The possibility of differences between RIN and PIN is undeniable; as an extreme example, the user can just misunderstand the task at hand. But, while "find what I mean not what I say" has been mentioned quite often[49] (it refers to inducing the PIN from the query), "find what I need not what I want", its equivalent for the RIN-PIN-distinction, does not seem to be very relevant to the field of Information Retrieval.[50] Granted, there are cases when a user learns during his search session, and recognizes that his information needs are different from what he assumed. But most of them lie outside the scope of this work. The easy cases might be caught by a spell checker or another suggestion system. In difficult cases, the user might need to see the "wrong" results to realize his error, and thus those results will still be "right" for his RIN as well as PIN. All in all, it seems justifiable in these circumstances to drop the distinction and to speak, quite generally, of a user's information need (IN). The request, on the contrary, has its place in search evaluation. If the raters are not initiators of the query, they might be provided with an explicit statement of the information need. This is obviously more close to a user's situation than a rater provided with just a query. Thus, the query is on the system-based end of the scale, while IN is user-based, with the request in between.

---

[49] It has even been the title of at least two articles (Feldman 2000; Lewandowski 2001).
[50] Although Google, for one, has had a vacancy for an autocompleter position, with a "certificate in psychic reading strongly preferred" (Google 2011a; Google 2011b).

Next, let us consider the information resources. Surrogates and documents are obviously the most frequently used evaluation objects. Lately, however, there has been another type: the result set (studies directly measuring absolute or relative result set quality include Fox et al. 2005; Radlinski, Kurup and Joachims 2008). Different definitions of "result set" are possible; we shall use it to mean those surrogates and documents provided by a search engine in response to a query that have been examined, clicked or otherwise interacted with by the user. Usually, that means all surrogates up to the one viewed last,[51] and all visited documents. The evaluation of a result set as a whole might be modeled as the aggregated evaluation of surrogates as well as documents within their appropriate contexts; but, given the number of often quite different methods to obtain set judgments from individual results, the process is nothing like straightforward. Thus, for our purposes, we shall view set evaluation as separate from that of other resources, while keeping in mind the prospect of equating it with a surrogate/document metric, should one be shown to measure the same qualities. It is clear that the user is interested in his overall search experience, meaning that set evaluation is the most user-centered type. Surrogates and documents, however, cannot be easily ordered as to the closeness to the actual user interest. While the user usually derives the information he needs from documents, for simple, fact-based informational queries (such as "length of Nile") all the required data may be available in the snippet. Furthermore, the user only views documents where he already examined the surrogate and found it promising. Therefore, I will regard surrogates and documents as equidistant from the user's interests.

I do not distinguish between components. Topic and task are parts of the user's information need, and as such already covered by the model.[52] I also do not use context in the way Mizzaro did. Instead, I combine it with the *time* dimension to produce quite another *context*, which I will use meaning any change in the user's searching behavior as a result of his interaction with the search engine. In the simplest form, this might just mean the lesser usefulness of results on later ranks, or of information already encountered during the session. On a more complex level, it can signify anything from the user finding an interesting subtopic and desiring to learn more about it, to a re-formulation of the query, to better surrogate evaluation by the user caused by an increased acquaintance with the topic or with the search engine's summarization technique.

---

[51] The result list being normally examined from the top down in a linear fashion (Joachims et al. 2005).
[52] One might reverse the reasoning and *define* the information need as the complex of all conceivable components. There may be many of those, such as the desired file type of the document, its language, and so forth. These are all considered to be part of the user's information need.

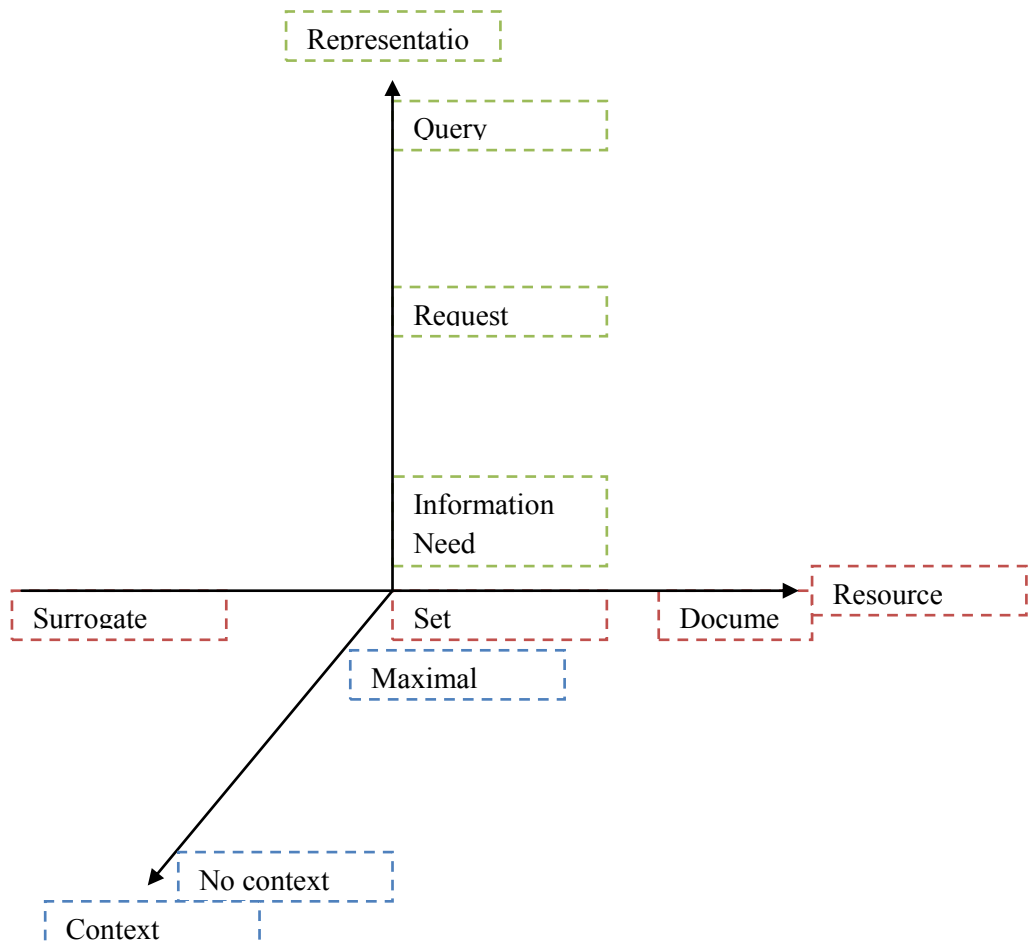

**Figure 7.2. The Web Evaluation Relevance (WER) model. The point at the intersection of the three axis indicates maximal user-centricity.**

Using this model, we now can return to the metrics described above and categorize them. Precision, for example, is document-based (though a surrogate precision is also conceivable) and context-free; the representation type may be anything from a query to the information need, depending on whether the originator of the query is identical with the assessor. MAP is similar, but allows for a small amount of context – it discounts later results, reflecting the findings that the users find those less useful, if they find them at all. Click-based measures, on the other hand, are based on the information need and incorporate the entire context, but regard the surrogate (since the relevance of the document itself is not assessed).

What is the WER model good for? We can postulate that the more user-centric metrics are better at discerning a user's actual search experience. For me, it stands to reason that a user's satisfaction with a result list as a whole provides a more appropriate view of his web search session than his satisfaction with individual results; or that a result list should be, if possible, evaluated with regard to the information need rather than the query itself.[53] Similarly, metrics that are considering more aspects of context might be supposed to be better. However, this last assertion is more problematic. We can hypothesize about the superiority of some context-sensitive user model, but, as the variety of user models employed in evaluations suggests, they cannot all be appropriate in all circumstances. Therefore, this is another issue awaiting testing: do metrics which employ a more user-centric model of relevance generally perform better than those that do not?

---

[53] Of course, there are cases when a query is obviously and blatantly inappropriate for an information need (e.g. a "me" search to find information about yourself). But these queries do occur, and all search algorithms are in the same situation with regard to them.

The discussions in the earlier sections indicate that user-based measures can generally be expected to be closer to the users' real satisfaction than system-based ones. It is not very hard to obtain very user-centered measures; all one has to do is to ask the user about his satisfaction after a search session. However, this metric has one very important disadvantage: it is hard to use to *improve* the search engine, which, after all, is what evaluation should be about. If a user states he is unsatisfied or hardly satisfied with a search session, nothing is learned about the reasons of this dissatisfaction. One possible way to overcome this problem is to submit the ratings, along with a list of features for every rated result list, to an algorithm (e.g. a Bayesian model) which will try to discern the key features connected to the user's satisfaction. However, this seems too complex a task, given the amount of data that can possibly influence the results; and we do not know of any attempts to derive concrete proposals for improvement from explicit set-based results alone.

Document-based measures have, in some ways, the opposite problems. They are "recomposable under ranking function changes" (Ali, Chang and Juan 2005, p. 365), that is, it can be directly calculated how the metric results will change if the retrieval mechanism is adjusted in a certain way. However, this comes at a price. As factors like already encountered information or, for that matter, any distinction on the *context* dimension, are generally not accounted for in a document-based metric, its correlation with user satisfaction is by no means self-evident. Of course, the distinction is not binary but smooth; as metrics become less user-centered, they become more usable for practical purposes, but presumably less reliably correlated to actual user satisfaction, and vice versa. This is one of the reasons why understanding the connections between different measures is so important; if we had a user-centered and a system-centered measure with a high correlation, we could use them to predict, with high confidence, the effect of changes made to a retrieval algorithm.

An additional complication is introduced by the abstracts provided by the search engines on the result page. The user obviously only clicks on a result if he considers the abstract to indicate a useful document. Therefore, even the best document-based metric is useless if we do not know whether the user will see the document at all. This might suggest that, for a complete evaluation of a search engine to be both strongly related to real user satisfaction and helpful for practical purposes, all types should be used. The evaluation of user behavior confirms this notion. "User data […] shows that 14% of highly relevant and 31% of relevant documents are never examined because their summary is judged irrelevant. Given that most modern search engines display some sort of summary to the user, it seems unrealistic to judge system performance based on experiments using document relevance judgments alone. Our re-evaluation of TREC data confirms that systems rankings alter when summary relevance judgments are added" (Turpin et al. 2009, p. 513).

# 8 A Framework for Web Search Meta-Evaluation

*Create your own method. Don't depend slavishly on mine. Make up something that will work for you!*

CONSTANTIN STANISLAVSKI

## 8.1 Evaluation Criteria

As discussed in the beginning of this work, how to judge evaluation metrics is an intricate question which is nevertheless (or rather: precisely for that reason) vital to the whole process of evaluation. To get a meaningful and quantifiable answer, one has to ask the question with more precision.

Mostly, studies are not interested in an absolute quality value of a result list. Rather, the relevant question tends to be whether one result list has a higher quality than another. It can be asked to compare different search engines; or to see whether a new algorithm actually improves the user experience; or to compare a new development with a baseline. The common feature is the comparison of the quality of two result lists.

But how to get such a judgment? I opted for the direct approach of asking the users. This has some disadvantages; for example, the user might not know what he is missing, or might misjudge the quality of the results. These problems reflect the difference between real and perceived information need (as discussed in detail by Mizzaro (1998)). However, as discussed in Chapter 7, a judgment based on a real information need (RIN) would be more problematic. There is no obvious way to assess RIN. Expert judgments, which are often used as part of the Cranfield paradigm, are not just laborious. They may work if the information needs are carefully selected to be judged by certain specialists; if the study goes for real-life information needs, this becomes unfeasible, even if descriptions of the personal information needs are available. For some queries, like the one looking for the definition of the term "panic attack", expert judgment can be expected to work. But even for science-based queries like the one looking for information about language acquisition, they would be problematic, as linguists passionately disagree about which answers are correct. More obviously, a potential visitor looking for information about a city might not notice a web site's omission of an attractive sight, but he is surely a better judge of whether the information he sees is interesting and useful to him than any expert. The argument is similar for transactional queries, where an expert might know whether the price is too high, but probably not the searcher's exact requirements and preferences. In short, to provide "good" assessments, an expert would have to interview the searcher extensively, and then – at least in the case of web search, where no-one is likely to be aware of all potentially useful sites – to conduct research of his own before rating results provided by the search engine under review. And even then, the judgments will not reflect real usefulness if, say, the searcher would not have come to grips with a site's interface, or failed to find the relevant information on a page.[54]

---

[54] And, of course, expert evaluations tend to be on the expensive side. Still, even when the resources are available, one has to carefully consider whether experts are the best solution.

While the question of user versus expert evaluation might be debatable, the goals which one attempts to reach by doing an evaluation have to be clearly defined. Thus, the question the following evaluation is meant to answer is:

**Mission Statement**

If real web users with real information needs could be presented with one of two possible result lists, how well is a particular metric suited to determine which result list would be preferred by this particular user, given explicit relevance judgments for the individual results?

## 8.2 Evaluation Methods

Some of the main decisions when evaluating a metric's relationship to user judgments concern the input, the metric parameters and the threshold and cut-off values.

Some metrics (like precision) were constructed for use with binary relevance, and this is the judgment input they are mostly provided with. However, it has been suggested that scaled ratings might be more useful (Järvelin and Kekäläinen 2002; Zhou and Yao 2008). In the present study, scaled ratings have been used for the simple reason that they can be easily reduced to binary, while the opposite process is not yet within the reach of mathematicians. The complications arise when one considers the different possibilities of conflating the ratings; the two probably most obvious are mapping the lowest rating to "irrelevant" and all others to relevant, or the mapping of the lower half of the ratings to "irrelevant" and the higher half to "relevant". I approach this issue in the way which we consider to be least assuming (or rather to make as little assumptions as possible); I use multiple methods separately, and look at which provide better results, that is, which are closer to the users' real assessments.

$$DCG_r = \begin{cases} CG_r & \text{if } r < b \\ DCG_{r-1} + \dfrac{rel(r)}{\log_b(r)} & \text{if } r \geq b \end{cases}$$

**Formula 8.1, alias Formula 4.10. DCG with logarithm base *b* (based on Järvelin and Kekäläinen 2002).**

The second issue are metric parameters.[55] Some metrics, DCG perhaps most prominently among them, have elements that are supposed to be changed to reflect the assumptions about user behavior or user aims. To illustrate that, we will reproduce the DCG formula as Formula 8.1, and quote once more, now more extensively, from the authors of DCG:

> A discounting function is needed which progressively reduces the document value as its rank increases but not too steeply (e.g., as division by rank) to allow for user persistence in examining further documents. A simple way of discounting with this requirement is to divide the document value by the log of its rank. For example log2 2 = 1 and log2 1024 = 10, thus a document at the position 1024 would still get one tenth of it face value. By selecting the base of the logarithm, sharper or smoother discounts can be computed to model varying user behavior.

---
[55] That is, the parameters of metrics; this has nothing to do with the conversion of pints into liters.

(Järvelin and Kekäläinen 2000, p. 42)

I see two problems with this excerpt. The first is the authors' statement that a division by rank would be too steep a discount. The question not explicitly considered is: Too steep for what? It seems logical to assume that the benchmark is user behavior, or user preference. But, as the authors indicate in the last sentence quoted above, user behavior varies. To make a point: as described in Section 2.2, hardly any users of web search engines go beyond the first result page, that is, the first 10 results. Surely, for them a good result at rank 1 is more useful than ten good results soon after rank 1000 – though the authors' example suggests they would be equal for the purposes of DCG.[56] The authors themselves do not provide any information on which users they mean, and whether there has been any research into what would be the most appropriate discounting function for those users. It is at least conceivable that division by rank, dismissed by Järvelin and Kekäläinen, might yet work well; or that some more complex function, reflecting the viewing and clicking patterns measured for real people, might work better. The second problem is that, though the authors bring up base 2 as an example, and call for adjustments wherever necessary, $b$ is routinely set to 2 in studies. Thus, even the adjustments actually suggested are not made, and I am not aware of any work looking either at the appropriateness of different discounting functions, or even different logarithm bases, for any scenario. I do not claim that the widely used parameters are inappropriate in all or even any particular circumstances; I just point out that, as far as I know, no-one has tried to find that out. Again, my own approach is trying to preselect as little as possible. Obviously, one cannot test every function and base; however, I will attempt to cover a variety of both.

Which discount functions should one examine? Obviously, there is an indefinite number of them, so I will concentrate on a small sample which attempts to cover the possible assumptions one can make about user behavior. It should not be forgotten that user behavior is the raison d'être of discount functions; they are supposed to reflect the waning perseverance of the user, whether from declining attention, increasing retrieval effort, or any other influence that causes the marginal utility[57] of results to decline. The question relevant to discounting is: How steep does the discount have to be to correctly account for this behavior? Therefore, the considered discount functions have to cover a wide range of steepness rates. The ones I consider are, in approximate order of increasing steepness:

- **No discount**. The results on all ranks are considered to be of equal value. This is a special case since the absence of a discount in NDCG makes it analogous to classical precision. Both metrics consist of a normalized total sum of individual result ratings.

---

[56] The definition of a "good result" is, in this case, not important, since the argument holds regardless of what is considered "good" or "relevant".

[57] Defined as "the additional satisfaction or benefit (utility) that a consumer derives from buying an additional unit of a commodity or service. The concept implies that the utility or benefit to a consumer of an additional unit of a product is inversely related to the number of units of that product he already owns" (Encyclopædia Britannica 2011). In the case of IR, it is almost universally assumed that the user gains less and less after each additional result, at least after some rank.

A small disparity in absolute scores results from different normalizing, which leads to different absolute numbers and thus to different performances at different thresholds. This similarity will be discussed in more detail in Section 10.2.

- **log$_5$**. A shallow function. The logarithm is not defined mathematically for ranks less than the base (five, in this case); taking the lead from standard DCG usage, results up to and including rank 5 are not discounted at all.
- **log$_2$**. The standard discount function of DCG. Starts discounting at rank 3.
- **Square root**.
- **Division by rank**. A function used – among others – by MAP.
- **Square rank**. A very steep discounting function; the third result carries just one ninth of the first result's weight.
- **Click-based discount**. The data for this function comes from Hotchkiss, Alston and Edwards (2005), who have looked at the click frequencies for different result ranks. They found click rates for the second result are only a small fraction of those for the first result. This function is unusual in that it does not decreasing monotonically; instead, the click rates rise e.g. from rank 2 to 3 and from 6 to 7. The authors explain the result with certain user and layout properties; for example, most display resolutions allow the user to see the first six results on the screen, and after scrolling the seventh result can become first. For a more detailed discussion, I refer to the original study. This function was chosen to represent a more user-centered approach not just to evaluation, but also to the calculation of the metric itself.[58]

Figure 8.1 gives a graphic overview of the different discount functions. For rank one, all functions offer the same discount (or rather no discount at all).

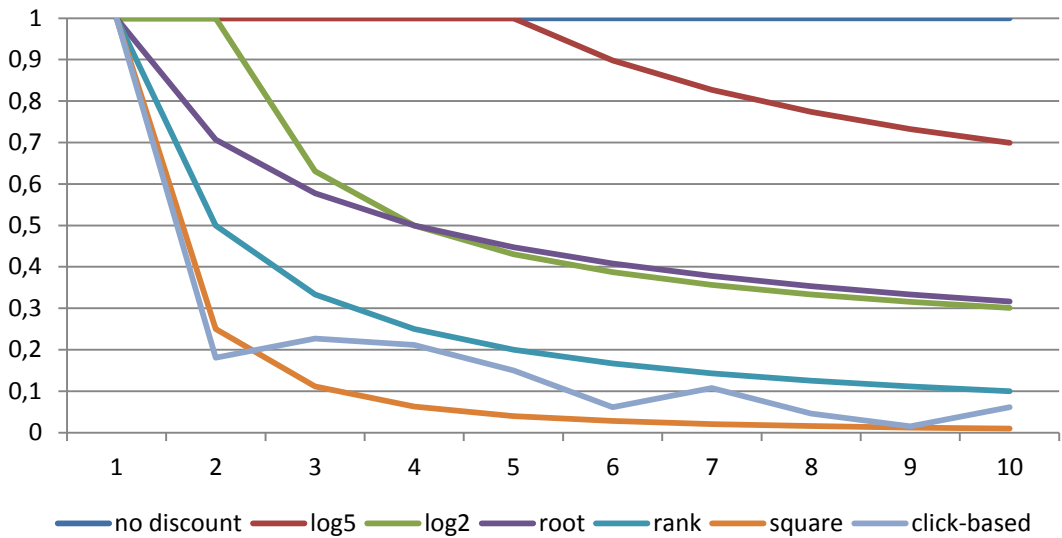

**Figure 8.1. The discount rates of different functions. The Y-axis shows the weights assigned to the ranks shown on the X-axis.**

---

[58] Of course, the data from the (Hotchkiss, Alston and Edwards) is not perfect, and might not be representative for general click behavior (if such a generalization exists at all); I use it as an example of a log-based discount function.

The third question concerns threshold values. When considering the relationship between individual metrics and explicit preference judgments, we want to look at whether, for a pair of result lists, both methods pick the same "preferable" result list. However, result lists can also be equally good (or bad). With explicit preference judgments, there are no problems; users can just indicate such a situation. For classical metrics, however, one has to decide when the quality of two result lists is to be considered similar. Are, for example, result lists with a MAP of 0.42 and 0.43 equally good? It can be argued that a user is unlikely to recognize such a minute difference. Where, then, to draw a line? Once more, I do not make any a priori judgments; instead, we will try out different thresholds and evaluate their suitability.

The forth problem are cut-off values. It is a commonplace that more data is better, but one cannot possibly hope to evaluate all the estimated 59 thousand results for "why a duck?",[59] and even the 774 results actually returned will be more than any but the most determined (and well-paid) rater can manage. On the other hand, research has shown that people would not be likely to go to result №775 even if they could (see Section 2.2). And even if they would, the marginal gain from evaluating the additional results would not necessarily justify the extra effort. Thus, there might be multiple points of interest between the extremes of evaluating only the first result and evaluating everything out there. There might be a minimal number of evaluated results for a study to have any validity; there probably is a point beyond which nothing at all will be gained by evaluating more results; and there can also be multiple points where the marginal gain from rating the next result does not warrant the pains. I speak of multiple points, as different studies have different objectives, for which different cut-off values might be more or less appropriate; but at the very least, there should be a span in which this "usefulness peak" falls for most studies.

The four problems outlined above are aggravated by the fact that they might be interconnected. Perhaps a dual logarithm-based calculation of DCG works well with binary judgments, while other judgment scales require different discount functions. While it is possible to conduct *all* of the necessary evaluations in one study, I will not attempt to do that. The amount of data would probably need to be larger than that which the current study could muster to provide robust results; and the interpretation and presentation would be a challenge in itself. Such an evaluation would have to consider scale versus discount versus threshold versus cut-off value versus the metric itself versus the actual main criterion of the present study, user preference or satisfaction. This six-dimensional matrix is something I happily leave to the researchers to come, content with smugly pointing out that the method introduced in this study seems well suited for the task. Nevertheless, the issues cannot be completely ignored; for this reason, I attempt to combine evaluations of these features on a smaller scale whenever it seems possible and appropriate.

To evaluate the different metrics and their possible parameters, two different methods are possible. The first is statistical correlation in its various forms. In short, they indicate how strong the link between different data sets is. This method would bring with it some methodological problems. For example, most popular correlations measure the connections between two data sets; however, for the preference evaluation, we have three (preference judgments and the metric values for two different result lists). For example, the user might have indicated a preference for result list 1 over result list 2, with the former's precision at 0.6 and the latter's at 0.3. Of course, the two metrics might be conflated into a single number; however, it is unclear how best to do that. Do we take the difference, the ratio, or some other measure of discrepancy? Another problem is that statistical correlation are significant only in a mathematical sense.

---

[59] For Google (http://www.google.com/search?q=%22why+a+duck%22) on July 8th 2011.

### 8.2.1 The Preference Identification Ratio

The second method of evaluating metrics against preference judgments is based on the notion that correlation, while well-tested and statistically meaningful, does not necessarily provide an immediate grasp of what the results actually mean for the algorithm or the user. Let us assume we have a Pearson correlation of 0.73 between average precision at rank 3 and user satisfaction (as was the case in Huffman and Hochster 2007). What does it tell us? What do we gain if we use this particular metric? What if we can increase the correlation to 0.82 (as Huffman and Hochster did)? How much good does it do to the end user? The answers to these questions are far from clear. The most we can say is whether the correlation is statistically significant, and that a higher correlation is, all other things being equal, generally preferable.

To provide a better grasp of metric evaluation, I propose a meta-evaluation metric. "Meta-evaluation" means that it is not a metric for testing the performance of a search engine, or any other retrieval system; rather, its aim and proper use is to offer an intuitive but well-defined[60] indicator of the usefulness of a metric which *is* directly concerned with a search engine's performance. This metric, which I called the Preference Identification Ratio (or PIR for short), is concerned with the following situation:

Imagine that we have two different result lists that we could present to the user for his query (as a result of different index sizes, ranking algorithms, or any other reasons). We also have explicit relevance judgments for the individual results which appear in the result lists. If we use a certain metric (say, Precision) to calculate the quality of the two result lists, and present the users with the one scoring higher, how likely is it, on average, that the user would really prefer it to the lower-scoring one? To put it more simply: How good is a metric at picking out the result list the user would have preferred?

The formula is probably more complex than the concept itself, so here is an example. Let us assume five queries with the characteristics shown in Table 8.1. The preliminary step is to identify the set of queries $Q$ for which a user preference judgment between the result lists actually exists. In the sample data, this excludes $q_2$ and leaves us with $q_1$, $q_3$, $q_4$, and $q_5$. The reason for excluding non-discriminating queries is that, whatever metric we use and whichever result list scores higher, the user is no better or worse off since he regards both result lists as equally good. It may be argued that not recognizing the similarity of the result lists' quality is detrimental since it would suggest a change of algorithm which would not provide a better experience to the user. This may indeed lead a search engine operator to waste resources unnecessary; however, PIR focuses on the user, and since he is indifferent to which result list he sees, the non-discriminating queries are disregarded.[61]

---

[60] It is unfortunate that there is a "but" rather than an "and" between "intuitive" and "well-defined"; but after some thought, I concluded that the overlap of the two attributes cannot be taken for granted.

$$PIR = 0.5 + \frac{\sum_{q \in Q}(pref(m_{q1} - m_{q2}) * pref_{user})}{2|Q|}$$

**Formula 8.2. The Preference Identification Ratio (PIR) indicates which percentage of users will enjoy an improved experience if the choice between two result lists is made based on their score according to a certain metric ($m_{q1}$ and $m_{q2}$ for the two result lists). $pref_{user}$ is the actual user preference with $p_q$=1 if the first result list is preferred, and -1 if the second is. Q is the set and |Q| the number of all queries where the user has judged to be one result list better than the other. The division by *twice* the number of queries and the addition of 0.5 bring the score into the 0..1 range.**

$$pref(x) = \begin{cases} 1, & x > t \\ 0, & |x| \leq t \\ -1, & x < -t \end{cases}$$

**Formula 8.3. This is an auxiliary formula which defines a preference value as 0 if a variable does not exceed a certain threshold *t*, and as 1 or -1 if it is larger than *t* or smaller than -*t*, accordingly. *pref(x)* is similar to the sign function.**

| Query | Precision of Result List 1 | Precision of Result List 2 | User Preference |
|---|---|---|---|
| $q_1$ | 0.4 | 0.7 | RL2 |
| $q_2$ | 0.5 | 0.4 | - |
| $q_3$ | 0.5 | 0.4 | RL2 |
| $q_4$ | 0.8 | 0.4 | RL1 |
| $q_5$ | 0.6 | 0.4 | RL1 |

Table 8.1. Sample data for a PIR calculation

| Query | Precision difference ($m_{q1}$-$m_{q2}$) | $pref(m_{q1}$-$m_{q2})$ | $pref_{user}$ | $sgn(m_{q1}$-$m_{q2})*p_q$ | PIR addend (t=0.15) | PIR addend (t=0.35) |
|---|---|---|---|---|---|---|
| $q_1$ | -0.3 | -1 | -1 | 1 | 1 | 0 |
| $q_3$ | 0.1 | 1 | -1 | -1 | 0 | 0 |
| $q_4$ | 0.4 | 1 | 1 | 1 | 1 | 1 |
| $q_5$ | 0.2 | 1 | 1 | 1 | 1 | 0 |

Table 8.2. Sample PIR calculation

Now that we have cleansed our data set, the first step is to calculate the difference between the metric scores (which are precision scores in this case). This exercise of figuring out how to perform this task is left to the reader, though the results are provided in Table 8.2. The next step is determining the preference indicated by these scores; this is done by the *pref* function, which is calculated according to Formula 8.3. Basically, it is defined as 1 if the argument is positive (that is, the first result list has a higher metric score), -1 if it is negative (the second result list having the higher score), and 0 in case it is zero (or lies below a certain threshold *t*; more on that below). Then, the user preferences are converted into a similar format, with $pref_{user}$ set to 1 if the first result list was judged to be better, and to -1 in the opposite case.[62] After that, the two numbers calculated last (the metric preference *pref* and user preference $pref_{user}$) are multiplied. The result of all those difficult manipulations is a 1 if the metric scores come to the same conclusion as the user, that is, if they pick out the same result list as being better; a -1 if they pick different result lists; and a 0 if the metric scores are similar for both result lists.

---
[61] Section 12.2.2.1 has more to say on this topic.
[62] If you are asking yourself whether $pref_{user}$ is set to 0 if there is no preference, you might want to re-read the previous paragraph.

We can now proceed with the final step by adding together the numbers (1+(-1)+1+1=2), dividing the result by double the number of our queries (2*4=8), and adding 0.5, which gives us a PIR of 0.75. Since three of four queries where the user actually has a preference would deliver the preferred result list when using precision to choose, the PIR value of $^3/_4$ might not come as a shock. Indeed, that is what PIR does – indicating what ratio of queries will get the preferred result list. Generally, if a metric picks out the wrong result lists in most cases, its PIR would be below 0.5. However, as guessing randomly (or assigning all result lists the same score) would provide a PIR of 0.5, this can be considered the baseline. To sum up, the PIR scale goes from 0 to +1, but any metric with a PIR lower than 0.5 is doing something really wrong; so, for all practical purposes, a PIR value of 0.5 is as low as it gets. The division by 2|Q| instead of Q and the subsequent addition of 0.5 to the result serve only to put the PIR results from the conceivable but unintuitive -1..+1 to an immediately understandable and common 0..1 scale.

$$PIR_{t=0} = 0.5 + \frac{\sum_{q \in Q} sgn(m_{q1} - m_{q2}) * p_q}{2|Q|}$$

**Formula 8.4. The simplified version of the PIR calculation given in Formula 8.2, with the threshold *t* set to 0. *sgn* is the sign function which returns 1 if the input is positive, -1 if it is negative, and zero if it is zero.**

There is one additional consideration that has been left out. It can be argued that not all result lists with different precision scores should be regarded as different in quality. One might think, for example, that the user does not really distinguish between results with precision scores of 0.48 and 0.51. To account for that, the preference calculation contains the threshold *t*. If *t* is set to zero, any difference between result list scores is considered significant; this is the route that has been taken when discussing the example up to now (the official form for this simplified metric I call $PIR_{t=0}$ is given in Formula 8.4). But what happens if any difference of less than, say, 0.15 between the precision scores of the two result lists is considered not significant? In this case, the calculation has one additional step. After arriving in the same way as before at the data presented in Table 8.2, we compare the absolute value of the precision difference with the chosen threshold. If the difference does not exceed the threshold, the query's addend value for the final calculation is set to zero; this is the case with $q_3$. The PIR for *t*=0.15 is, then, (1+0+1+1)/8+0.5, or 0.875. In words, there was one result where the difference in precision scores was rather small and pointed to the wrong result list; by requiring any score difference to meet a minimal threshold, the risk of making the wrong judgment is reduced for the close calls. The higher the threshold is set, the less probable both wrong and correct predictions become, as the number of disregarded queries increases; for *t*=0.35, our example's PIR drops to 0.625, and if the threshold is higher than the full span of possible metric scores (in most cases this would be *t*≥1), PIR would always be 0.5, regardless of the input.

So, once more: What does PIR tell us? One interpretation is that it shows how much a metric can help improve user experience. It is a commonplace that one cannot improve the experience for all users, at least if the system is as large as most web search engines; Google "could make 500 [queries] great, and only one worse, but you are never gonna make *all* of them better" (Google 2011d). If the PIR of a metric is 0.75, it means that for 75% of users who actually have a preference that preference can be correctly determined. The other 25% would get the result list they consider to be worse. Alternatively, a PIR of 0.75 could mean that the preference can be determined correctly for 50% of users, while for the other 50%, the metric (wrongly) indicates that the result lists are equally good. In this case, the users might get any of the result lists. The numbers could also be different, as long as the sum of "good" preference prediction and half the "equal" predictions totals add up to the PIR. Put another way, the PIR, as its full name (Preference Identification Ratio, in case you already forgot) suggests, shows how well a metric can identify user preferences.

PIR, as most other measures, is more useful in comparisons. If viewed for just one metric, it can be considered a comparison with the baseline of 0.5 (which might be achieved, for example, by guessing at the better result list randomly). Then, it would show to what extent the metric closes the knowledge gap between guessing and knowing the user's preferences. If PIR is used for inter-metric comparison, it shows which metric will give users more of their preferred result lists, and how often.

### 8.2.2 PIR Graphs

In order to provide as much information as possible, I have used graphs that might be considered to be a bit on the crammed side. However, I think that with a proper explanation, they are quite readable. Here, then, is the explanation. I will not provide any interpretations of the results at this point since those will be given in more detail in the next section. The aim of this graph and the next one is only to provide a guide on how to read them.

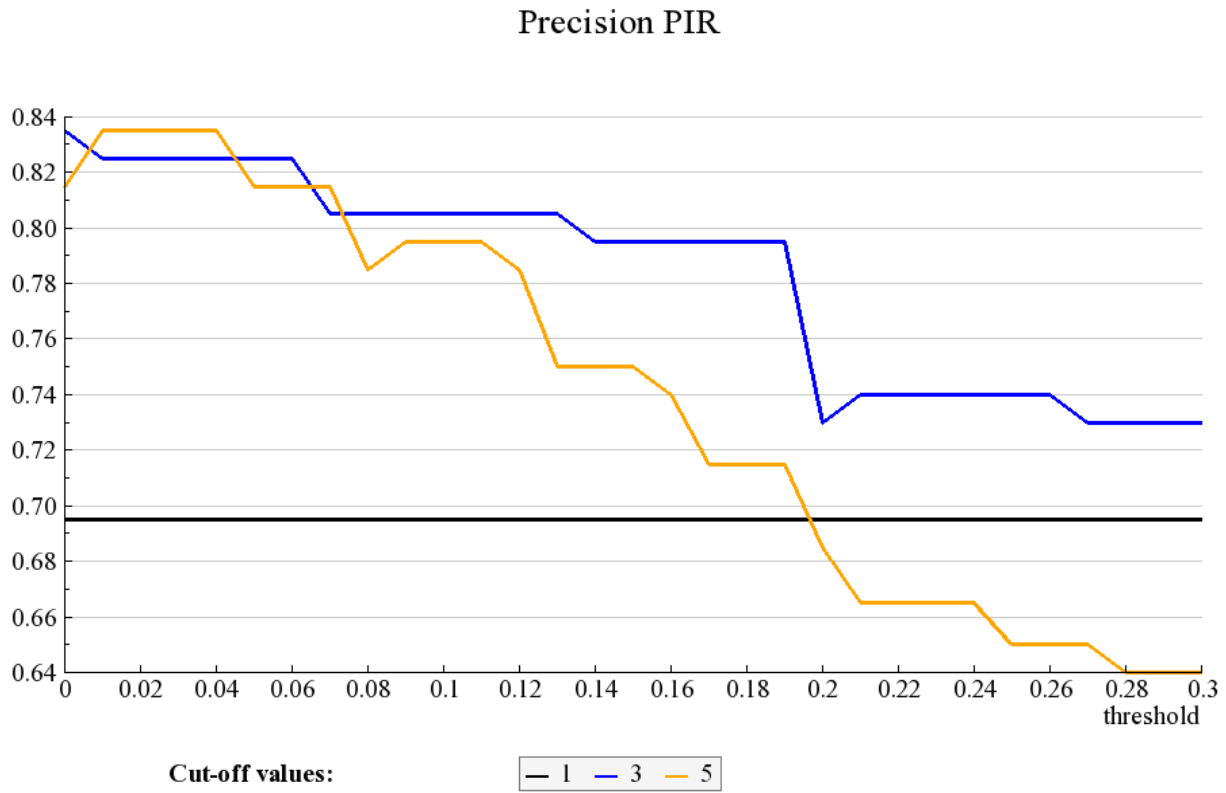

**Figure 8.2. Precision PIR (on Y-axis) depending on threshold value (X-axis) and cut-off value (coloured lines). Note that the Y scale does not start with zero and does not end with 1; rather, it is dynamic and graph-specific. This means different graphs can be compared directly as to the form of single plot lines, but a comparison in absolute PIR terms requires a careful look at the numbers on the Y-axis rather than just a glance on two graphs side-by-side.**

Figure 8.2 shows a Precision PIR graph. Let us first consider the blue line. From the legend at the bottom you can see that it is the curve (or perhaps jagged straight) for precision at the cut-off value of 3. That is, for the purposes of this evaluation, the judgments for any result beyond the third rank are ignored; the retrieval systems might just as well have returned only three results each. On the horizontal axis, the threshold values are given. If we look up the PIR value at zero threshold, we see that it is 0.835. This means that if we consider only the first three results of every result list, and if we regard any difference in Precision, however small, to be significant, for about five of six queries Precision will correctly pick out the result list preferred by the user, while once every six queries, it will pick out the less preferred. If we move right to the threshold of 0.05, we will consider only result lists where the precision scores differ by at least 0.05 to be of significantly unequal quality. This reduces the risk of accidentally regarding a close call as marking a preference and getting it wrong; but it also reduces the chance of getting it right. The larger the threshold value, the more cases are considered to be "close", and the less material the metric uses. Thus, when the threshold value is high enough (1 is the extreme), the PIR will always be 0.5, as no judgments will be made, and whether the better or the worse result list will be returned will be a matter of chance. The black line, indicating a cut-off value of one (only the first result is considered), is straight; that is, the threshold values up to 0.30 do not change the PIR. The reason is that we have only one judgment on what is in this case a six-point scale between 0 and 1 (0.0, 0.2, 0.4, 0.6, 0.8, 1.0), and the likelihood of just the two first-rank results being that one rating step apart is relatively small. With even larger thresholds, there would be differences; also, other ratings or other metrics might well provide for a different graph.

I have two final remarks on this graph. One concerns the threshold scale itself; the highest value you will find displayed on graphs is always 0.30, as there was no single evaluation where the PIR improved in any way when the threshold exceeded this value. The second is to inform that in the evaluation proper, all cut-off values from 1 to 10 will be displayed;[63] in the sample graph, simplicity rather than completeness was called for.

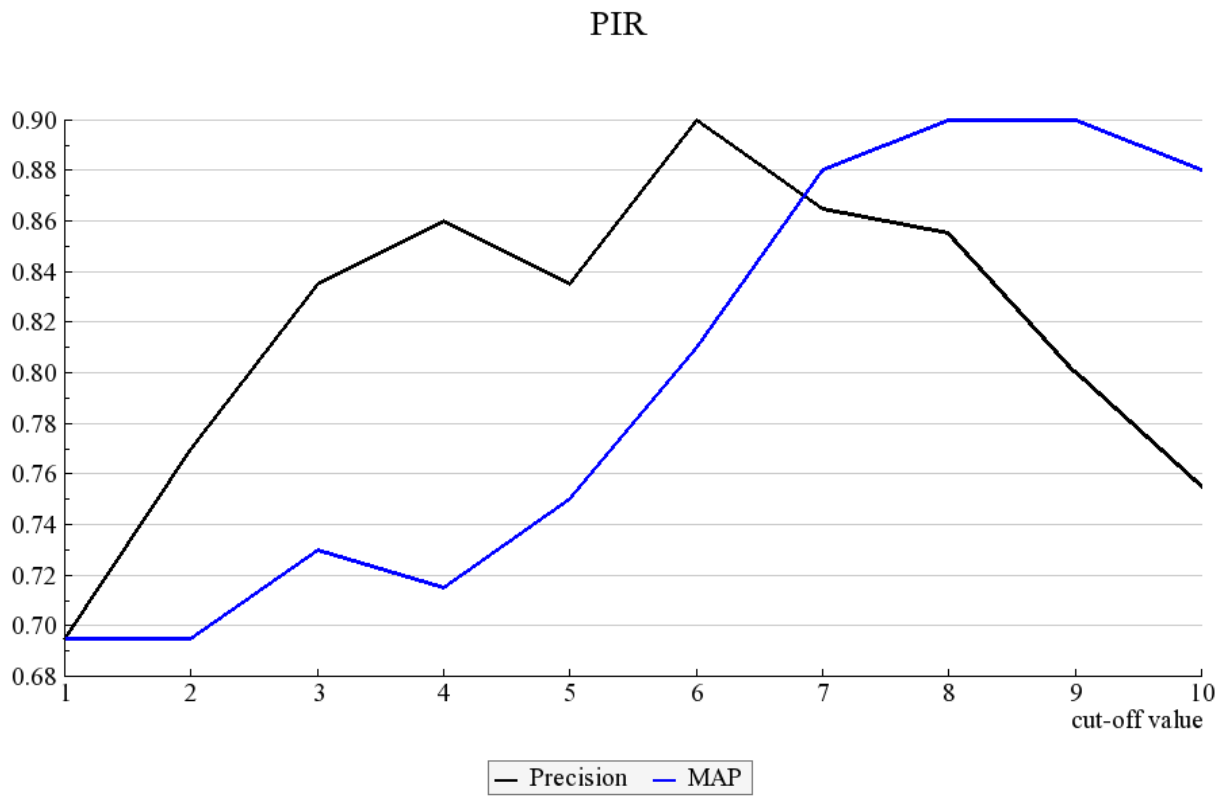

**Figure 8.3. PIR comparison graph. The PIR values are again displayed on the Y-axis. Here, the thresholds are used on a "best-for-cut-off" basis; for each cut-off value, the best-performing threshold is chosen.**

Figure 8.3 shows a graph comparing two PIR plots. Here, the cut-off values are displayed on the X-axis. The graph does not show any threshold information; the threshold values used will be declared with each graph. If we again consider the Precision plot, we can see that it starts with a PIR of 0.695; that is, if only one judgment is used for each result list, Precision will correctly identify the better list seven times out of ten. If you compare this to Figure 8.2, you will see that the value comes directly from the horizontal line showing the stable PIR value for Precision@1 there. If you follow the Precision plot to the right, you will see that the PIR slowly rises, peaking at 0.90 at a cut-off value of 6, and then declines again.[64]

---

[63] The reasons for choosing 10 as the cut-off value will be explained in Section 9.1; basically, the reason lies in the limited availability of resources.

The source of values for cut-offs greater than one is probably not quite self-explanatory. While the Precision PIR at cut-off one was stable (at least, up to 0.30) with regard to threshold values, this is not the case for other cut-offs. The question that arises is: Which threshold do we use to represent Precision in this inter-metric comparison? There are two answers that will be used in the evaluation. One is to use the best possible threshold values for each cut-off; this is the method employed in Figure 8.3. For a cut-off value of 3, for example, we look at Figure 8.2 and see that the PIR is highest (0.835) when the threshold is zero; so 0.835 is the value we use for the metric comparison. For a cut-off value of 5, precision peaks (or rather plateaus) at thresholds 0.01 to 0.04, also with a value of 0.835; so in the current graph, Precision PIR returns to 0.835 at cut-off 5. This method has two disadvantages; firstly, the threshold values used cannot be deduced from the graph;[65] and secondly, it almost certainly overestimates the metrics' success. Consider that in order to be realistic, we have to have enough data to reliably discern the most successful threshold value for each metric and cut-off value. The finer the threshold steps, the less significant any difference found is likely to be. In Figure 8.2, you can see that for the cut-off value of 5, the 0.01 steps might be just about fine enough to capture the ups and downs of the PIR; but at the same time, most differences between neighbouring data points are not statistically significant. While PIR with the threshold set at 0.2 is significantly lower than that for $t=0.1$, the PIR rise from 0.785 to 0.795 between $t=0.08$ and $t=0.09$ is not, at least in this study. If we make the steps larger, we will, in most cases, see the same picture – the PIR tends to be largest with the lowest thresholds, and to decline as the threshold increases. Therefore, the second method of determining which threshold values to use for inter-metric PIR comparison is using $t=0$. As you will see for yourself on the evaluation graphs, this is the most obvious candidate if you are looking for a constant threshold value; it has the maximal PIR in most of the cases (see the discussion in Section 12.2.2). Those two methods constitute convenient upper and lower boundaries for the PIR estimate since the former relies too much on threshold "quality" for individual metrics and cut-off values, while the latter does not make any use of it at all.

---

[64] Why this happens will be the topic of much discussion in the evaluation section itself.
[65] Unless an additional threshold value is given for each data point, which would make the graphs harder to read than Joyce's *Ulysses*.

# 9 Proof of Concept: A Study

*A good definition of fundamental research will certainly be welcomed: let us see whether we can invent one. We have to begin, of course, by defining research. Unfortunately the concept of research contains a negative element. Research is searching without knowing what you are going to find: if you know what you are going to find you have already found it, and your activity is not research. Now, since the outcome of your research is unknown, how can you know whether it will be fundamental or not?*

K. K. DARROW,
"PHYSICS AS A SCIENCE AND AN ART"

In this section, the layout of the study will be explained. After that, some results will be provided which do not directly concern the main questions of the present study, but are interesting in themselves or may provide useful information for other issues in evaluation or IR studies. Furthermore, many of these findings will help to put the main result into the proper context.

## 9.1 Gathering the Data

The study was conducted in July 2010. 31 Information Science undergraduates from Düsseldorf University participated in the study. The users were between 19 and 29 years old; the average age was 23 years. 18 participants (58%) were female, while 13 (42%) were male. The raters were also asked to estimate their experience with the web in general and with search engines in particular on a scale from 1 to 6, 1 signifying the largest possible amount of experience, and 6 the lowest.[66] The overwhelming majority considered themselves to be quite to very experienced (Figure 9.1); this should not be surprising since their age and choice of study indicate them to be "digital natives" (Prensky 2001).

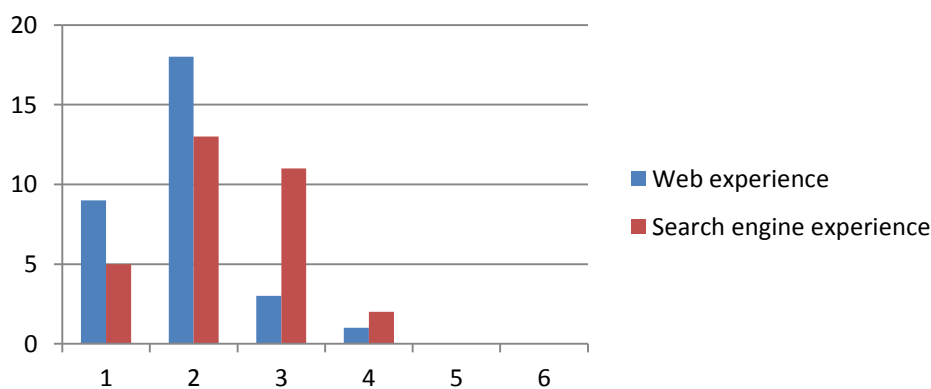

**Figure 9.1. Self-reported user experience, 1 being highest and 6 lowest**

---

[66] The scale is familiar to German students as it is the standard grading system in German education. The scale is explained in more detail later in this section.

The study was conducted entirely online using a web-based application designed and implemented by the author for this very purpose. Each user logged on to the system using a unique ID. Both the front-end and back-end were written in PHP, with data stored in a MySQL database. The present section contains screenshots of the front-end (web interface used by the participants) to help the gentle reader visualize the procedures (apropos visualization: some of the visualizations in this work have been created directly from the back-end using the JpGraph PHP library; others were prepared and created using Microsoft Excel). As a result of this set-up, the raters could participate in the study from any computer with internet access. After a supervised introductory and practice session, they were instructed to proceed with the study on their own schedule; the only constraint was that all tasks were to be finished within one week. If any questions or other issues arose, the participants were encouraged to use the contact form, the link to which was provided on every page.

The raters were asked to submit queries for information needs they had at the time or in the recent past, together with a description of the information need itself[67]. They also were to indicate the language they would prefer the results to be in; the choice was only between English and German. Queries aimed at other languages were not accepted; as all students, through their background, could be expected to be proficient in German and English, queries issued by a rater could be understood (at least insofar the language is concerned) by another.

**Figure 9.2. Query/information need entry form**

The query was then forwarded, via the BOSS ("Build your Own Search Service") API,[68] to the now-defunct Yahoo search index.[69] The top 50 results were fetched; in particular, the result URL, the page title, and the query-specific snippet constructed by Yahoo were obtained. The snippet and URL contained formatting which was also query-specific and mostly highlighted terms contained in the query. Then, the source code of every document

---

[67] Of course, if the user had already attempted to satisfy the information his information need before the study; this would influence the judgments. This would not make a large difference, as most students stated they provided a current information need; furthermore, as will become clear later in this section, every result set was rated by the originator as well as by others. Still, this constitutes one of possible methodological problems with the present evaluation.
[68] http://developer.yahoo.com/search/boss/
[69] http://www.yahoo.com; the Yahoo portal and the search engine obviously still exist, but now they use Bing technology and index.

was downloaded; this was later used to ensure the documents had not changed unduly between evaluations.[70]

As one of criteria for evaluation was to be user preferences, multiple result lists were needed for each query to provide the user with a choice. From the 50 results obtained for each query (no query had fewer than 50 hits), two result lists were constructed. One contained the results in the original order; for the other, they were completely randomized. Thus, the second result list could theoretically contain the results in the original order (though this never happened).

After this, the evaluation proper began. After logging into the system, the raters were presented with a User Center, where they could see whether and how many queries they needed to submit, how many evaluations they had already performed, what the total number of evaluations to be performed was, and what the next evaluation task type would be (Figure 9.3). With a click on the relevant link, the rater could then commence the next evaluation task.

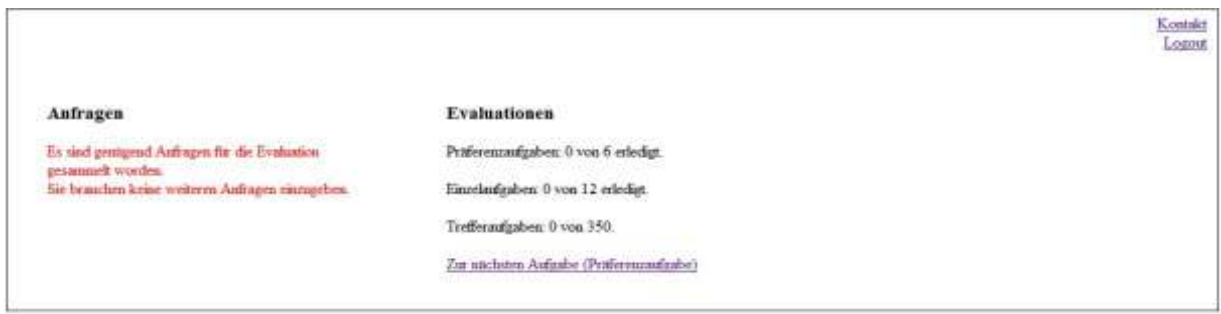

**Figure 9.3. User Center.**

The first type of evaluation was preference evaluation. The user was presented with a query as well as an information need statement and desired language on top of the page. The main part of the page consisted of two columns, each containing one version of the result list for the current query. The left-right-ordering of the result lists was randomized to exclude a bias for one side. Also, the result lists were anonymized, so that it was not discernible that they were provided by Yahoo, to exclude potential search engine bias (Jansen, Zhang and Zhang 2007). Instead, the design of each column was based on that used by Google for the presentation of its result list.[71] This did not include any branding or advanced search features, but rather the then-current styling and colors of the result presentation. Each result column contained 10 results (consisting of title, snippet and URL with term highlighting as provided by Yahoo); below the last results, the raters were provided with a link to navigate to the next result page containing results ranked 11-20 on the corresponding result lists.[72] At the bottom of the page the users were provided with a box. It contained the question "Session completed?" and three links, the left to indicate that the left-side result list was considered better, the right to indicate the preference for the right-side result list, and the central link to signal that the result lists were considered to be of similar quality. This box did not scroll with the page, but was always visible on the bottom of it, so that the rater could at any moment complete the session by clicking on one of the links (the page layout can be seen in Figure 9.4).

---

[70] A small content variation (up to 5% of characters) was tolerated, as many documents contained advertisements or other changeable information not crucial to the main content.
[71] http://www.google.com/search?name=f&hl=en&q=the+actual+query+we+pose+is+not+important%2C+rather%2C+the+aim+of+this+reference+is+to+demonstrate+the+design+of+a+Google+result+page
[72] Obviously, if the user was on page 2 to 5, he could also return to a previous page. The number of pages was limited to five.

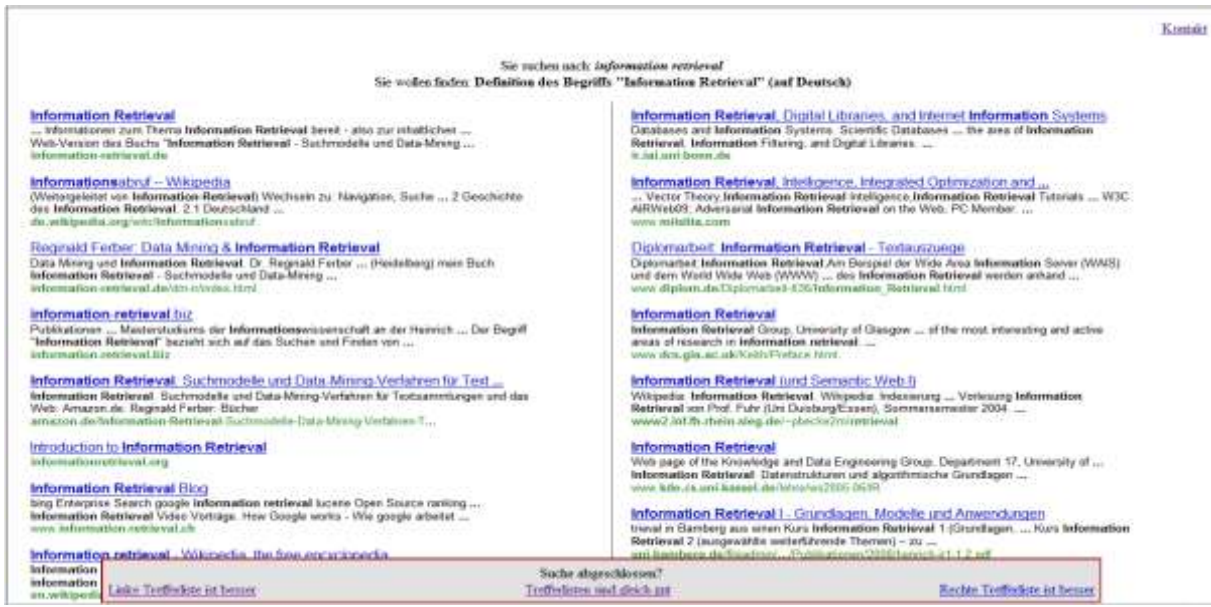

**Figure 9.4. Preference evaluation.**

The raters were instructed to use the result page as they liked, attempting to satisfy the given information need. They were also told to end the session when they felt continuing would not be worthwhile, either because they already found what they sought for or because they thought it unlikely that the rest of the results contained good documents. Simplifying somewhat, the session was to end when the results were thought to be good enough or else too bad. The end of the session was marked by a click on one of the three options described above, marking the user's preference or the lack thereof. This click took the rater back to the User Center. There have been reports of search engines using such side-by-side result lists for their internal evaluations; later, this was confirmed, at least for Google (see Figure 9.5).

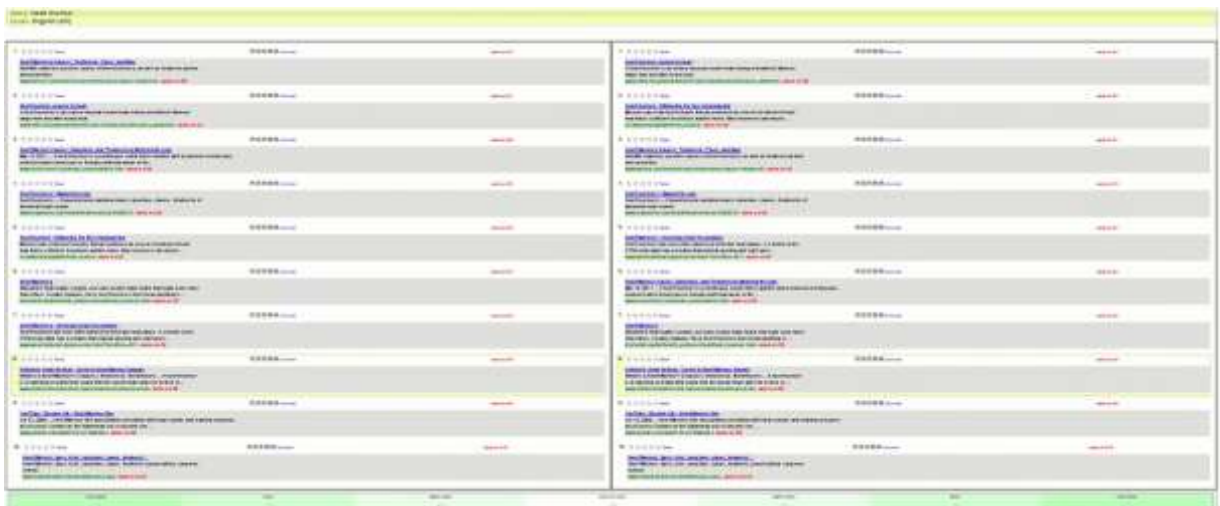

**Figure 9.5. A side-by-side evaluation screen used by Google (Google 2011c). Unfortunately, the screenshot is only available from an online video, hence the poor image quality.**

In the second type of evaluation, the aim was to get an assessment of user satisfaction. The procedure was similar to that employed for preference evaluation, but the raters were provided with just one result list, as is usual for search engines. The list could be either the "original" or the "random" one; the distribution of different result list types was randomized here as well. A user only evaluated one result list type for a query (if he had already worked with a side-by-side evaluation for a certain query, his satisfaction evaluation would surely not be realistic any more, as he knew the result list and individual results already). The only other difference to the preference evaluation was the feedback box which featured only two links, one indicating satisfaction with the search session and one stating dissatisfaction. The users were instructed to judge satisfaction on a purely subjective level. The layout can be seen in Figure 9.6. Again, a click on one of the "Session end" links redirected the rater to the User Center.

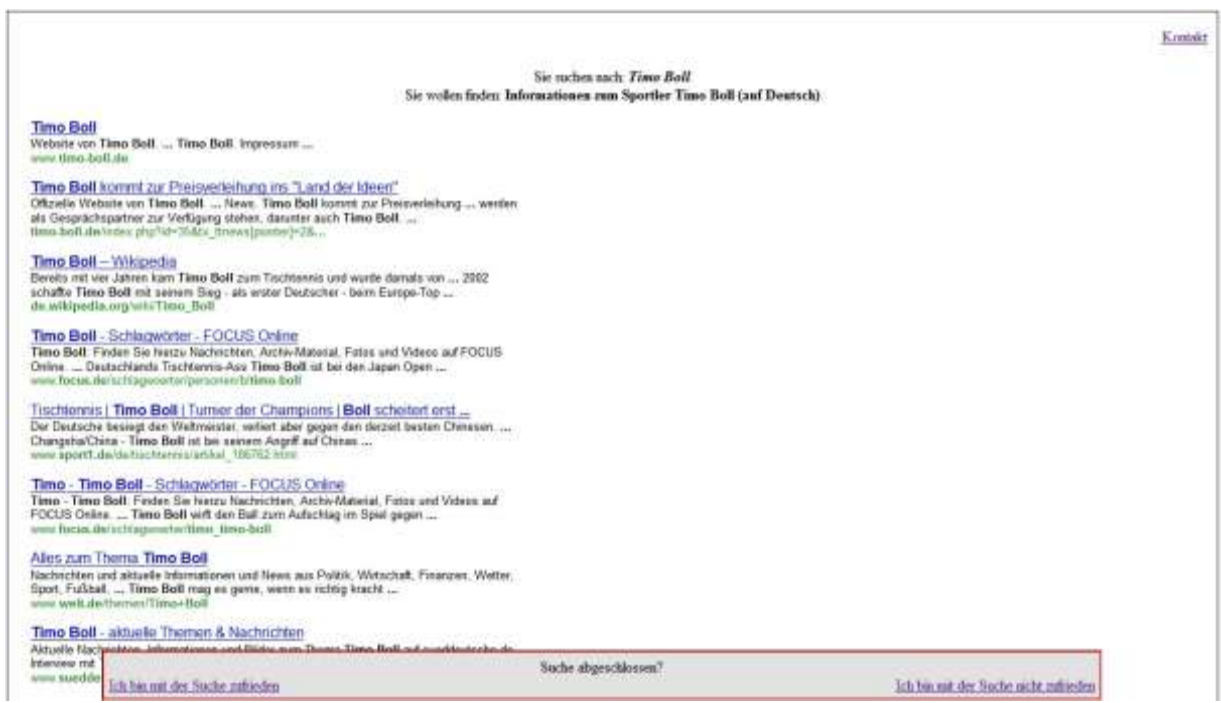

**Figure 9.6. Satisfaction evaluation.**

For both preference and satisfaction evaluation, log data was recorded. It included the session start timestamp (the time when the result page was first shown to the user), the clicks the user performed (in particular, result clicked on and the time of the click), and the session end timestamp (the time when the user clicked on a link in the "session finished" box) as well as the user ID. Also, for every session the query and information need and the display mode of the result lists were included.

The third type of evaluation was result evaluation. Here, the rater was presented with the query and information need and just one result description, again consisting of title, snippet and URL with the appropriate highlighting and kept in the Google layout. After the result description, the user was confronted with two questions. The first, which he was to ponder before actually clicking on the result link, was whether the result seemed relevant for the query under consideration, or, put in another way, whether he would click on this result given the query and information need. After that, the user was to click on the result and examine the actual document. The pertinence of the document to the query was then to be rated on a scale from 1 to 6, with 1 being the best imaginable result and 6 a result not providing any value to the rater. This scale, while unusual in web search evaluation, has the advantage of being very well-established in Germany, as it is the standard academic grading scale for education in schools as well as at universities. Users generally have problems in differentiating between neighboring grades; the employment of a scale they already know how to use and interpret was meant to reduce the uncertainty and the decrease of judgment quality associated with it. For evaluation purposes, the scale was internally converted into the usual 0..1 range (with a rating of 1 corresponding to 1.0, 2 to 0.8, and so forth). After selecting a grade, the user submitted the form and was automatically provided with the next result to judge.

**Figure 9.7. Result evaluation.**

After all the required evaluations have been performed, the raters received the message that they had completed the evaluation. The requirements included 6 preference evaluation judgments, 12 satisfaction evaluation judgments and up to 350 result evaluation judgments.

Overall, the 31 users contributed 42 queries, which they evaluated by providing 147 preference judgments, 169 satisfaction judgments, and 2403 individual result ratings.

## 9.2 The Queries

The raters contributed a total of 42 queries. There were no restrictions placed on the kind of query to be entered since the intention was to gather queries that were as close to possible to those employed for real-life information needs. In particular, this means no query type was predetermined. However, the detailed information need statements allowed for an unequivocal identification of the original intent behind the queries. 31 queries were informational, 6 were transactional and 2 were navigational, according to Broder's widely employed classification (Broder 2002). Three further queries do not fit easily into any of the traditional categories (see Table 9.1).

| Query | Information need[73] | Query type |
|---|---|---|
| **korrellation** | Answer to the question how to spell the German word for "correlation" | Factual |
| **Fussballweltmeisterschaft 2014** | Looking for the country hosting the 2014 Football World Cup | Factual |
| **Flughafen Düsseldorf** | I wanted to see at which rank the site www.clickundflieg.com will appear | Meta |

Table 9.1. Queries not covered by the traditional categories

Two queries are similar to what has been described as "closed directed informational queries" (Rose and Levinson 2004), and commonly called "factual", as they are generally geared towards finding a specific fact (Spink and Ozmultu 2002). This indicates a query aimed at finding not as much information as possible on a certain topic, but rather a (presumably definitive and singular) answer to a well-defined question. Another example would be a query like "How long is the Nile". I would like to further narrow down the definition of such queries to distinguish them from informational ones and highlighting special features particular to them. These features are:

- The query can easily be reformulated as a question starting with "who", "when", or "where".[74]
- The originator of the query knows exactly what kind of information he is looking for.
- The answer is expected to be short and concise.
- The answer can be found in a snippet, eliminating the need to examine the actual results.
- One authoritative result is enough to cover the information need.
- There are many pages providing the needed information.

The last three properties can be used to derive some interesting predictions for this type of queries. If all the needed information can (and, ideally, would) be found on the result page of the search engine, the sessions with the highest user satisfaction will tend to have low durations and no clicks – features also typical of "abandoned sessions". In other words, sessions without any clicks are typically assumed to have been worthless for the user (Radlinski, Kurup and Joachims 2008), while for factual queries, the opposite might be true: the absence of a click can indicate a top result list. Furthermore, the last two points indicate that, while factual queries resemble navigational ones in that a single result will satisfy the user's information need, they are unlike navigational queries in that there are many possible results which might provide the fact the user is looking for.

---

[73] Here and further, information needs are translated from German where needed.
[74] "What" and "how" are more complex cases. Obviously, my own example ("How long is the Nile") starts with "how". Mostly, questions aiming for a concise answer ("how long", "how big", "how old" etc.) are quite typical for factual queries, whereas open-ended, general questions ("how do you ...") are not. A similar division holds for "what", with "What is the capital of Madagascar" being a prototypical factual query, but not "What is the meaning of life". In general, the factuality of a query is more meaningfully defined by the next two points of expected preciseness and brevity of the answer, with the question word being, at best, a shortcut.

Another type of information need not covered by traditional classifications is represented by what I will call "meta-queries". The single example is provided by a query intended to check the position of a certain page in the result list. The exceptional property of these queries is that a result list cannot be "good" or "bad". Whether the result sought in that example is in first position, somewhere in the tail or not in the result list at all – the information need is satisfied equally well. Another example would be the searches done for the comic shown in Figure 9.8. Here, the user was looking for the number of Google hits for a particular query. The number of returned results and the results themselves do not have any effect on user satisfaction; more than that, it is unclear how any single result can be relevant or non-relevant to that query. The information the user is looking for is not found on any web page except for the search engine's result page; it is either exclusively a result of the search algorithm (as in the case of the query aiming to find out the rank of a certain page in Google), or else a service provided by the search engine which can be described as "added value" compared to the information found on the web itself (as with submitting a query to determine the popularity of a certain phrase). The search, in short, is not a web search about real life; it is a search engine search about the web.[75]

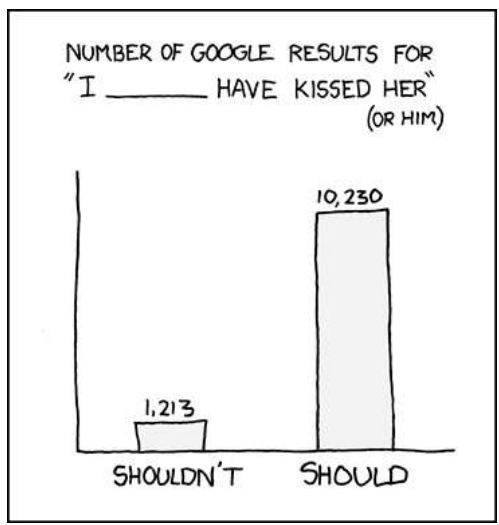

**Figure 9.8. The queries submitted to research this *xkcd* comic (Munroe 2008) fall in the meta-query category.**

There is only one meta-query in the present study, so that it will not play a major role. However, it would be interesting to determine how frequent this type of query is in real life, and particularly whether it occurs often enough to warrant a special consideration in further studies.

---

[75] It is also interesting to note that the web comic shown in Figure 9.8 explicitly states that the relevant information is the number of Google hits, not the number of web pages it is supposed to represent, and not the popularity of the phrases implied by the number of web pages, and also not the general attitudes of the population to be derived from the popularity. This is in no way to criticize the *xkcd* comic, which is generally very much in tune with the general attitudes of the population (at least, to judge from what can be seen online), but rather to point out the extent to which this population is relying on the notion that Google is an accurate reflection of the web, or even of a large part of the modern world.

| Query type | Number of queries | Average terms per query |
|---|---|---|
| **Transactional** | 6 | 2,67 |
| **Informational** | 31 | 2,45 |
| **Factual** | 2 | 1,5 |
| **Navigational** | 2 | 1,5 |
| **Meta-query** | 1 | 2 |
| *Overall* | *42* | *2,38* |

**Table 9.2. Average query length. Only the first two types were present in significant numbers**

Table 9.2 shows the average query length for the different kinds of queries. Leaving aside the small amount of queries that leaves factual, navigational and meta-queries a statistically insignificant minority, we see that the numbers are well in accord with the two to three term average found in other studies (Jansen and Spink 2006; Yandex 2008). Also in accord with other studies (White and Morris 2007), there were no operators used, which was to be expected for a relatively small number of queries.

There are surely not enough factual, navigational and meta-queries to allow for their meaningful evaluation on their own, and probably also not enough transactional queries.[76] For this reason, in considering the study's results, I will evaluate two sets: the informational queries on their own, and all the queries taken together.

## 9.3 User Behavior

Another set of data concerns user behavior during the search sessions. An interesting question is how long it takes for the raters to conduct a session in a single result list or side-by-side condition, and decide upon their satisfaction or preference judgment. Figure 9.9 shows the session length for single result list evaluation. Almost half of all sessions were concluded in less than 30 seconds, and around 80% took up to 90 seconds. Less than 5% of queries took more than 5 minutes. The situation is somewhat different in the side-by-side evaluation (Figure 9.10); here, about 50% of all sessions were completed in up to a minute, and about 75% in up to three and a half minutes. Here, 10% of sessions took more than 6 minutes; it has to be noted, however, that a number of sessions have extremely long durations of up to 24 hours. These are cases where the raters started a session, but seem to have temporarily abandoned the task. These sessions are included in the evaluation, as all of them were later correctly finished with a satisfaction or preference judgment.[77] Also, this might well be in line with actual user behavior since a user can start a search session but be distracted and come back to an open result list page on the next day. However, these outliers greatly distort the average session time, so that median numbers are more meaningful in these cases. Those are 32 seconds for single result lists and 52 seconds for side-by-side result lists. While the raters – understandably – needed more time to conduct a query with two result lists, the session duration does not seem so high as to indicate an excessive cognitive burden.

---

[76] Although there have been studies published with less than six different queries.
[77] These might be considered to consist of multiple sub-sessions, with significantly lower overall durations. However, since in this study there was no clear way of determining the end of the first sub-sessions and the start of the last ones (and since these sessions were rare), I opted to keeping them with their full duration.

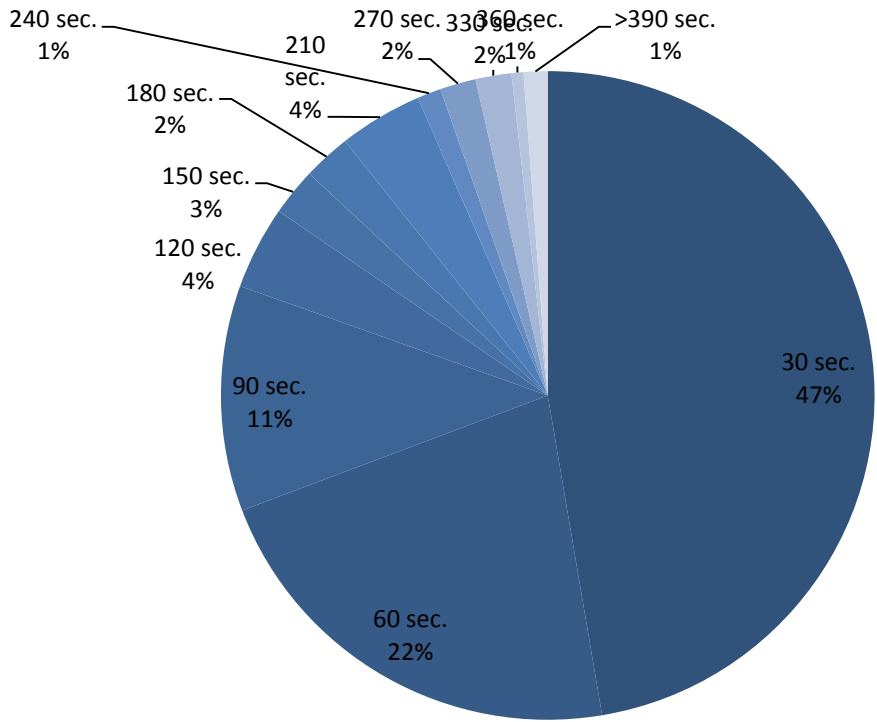

**Figure 9.9. Session duration for single result list evaluation (from 169 sessions).**

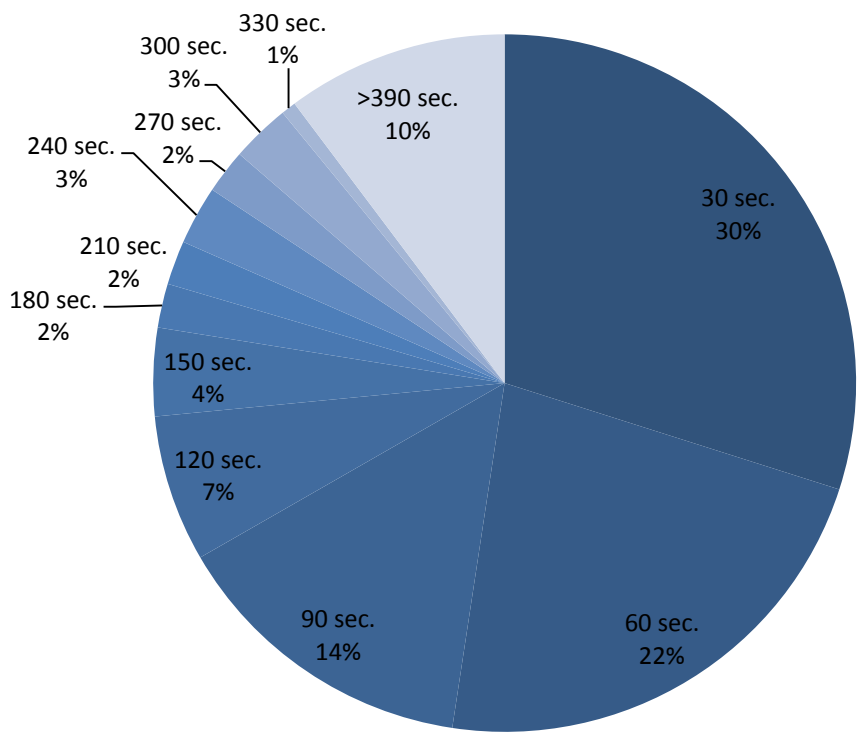

**Figure 9.10. Session duration for side-by-side result list evaluation (from 147 sessions).**

Another question concerns the click behavior. While I will evaluate some possible correlations between user behavior and explicit judgments in Chapter 11, some aspects are once again interesting enough for a general discussion.

The first notable result is the large number of sessions where no results are clicked on. As Figure 9.11 indicates, in both single and side-by-side evaluations, almost half of all sessions end without any result being selected. Of the sessions that did have clicks, most had a very small number, and the sessions with one to three clicks made up 75% and 63% respectively (of those with any clicks at all). The graph for non-zero click sessions can be seen in Figure 9.12. Side-by-side evaluations had generally more clicks than single ones, with the average being 2.4 and 1.9 clicks, respectively. With usual click numbers per session given in the literature ranging from 0.3-1.5 (Dupret and Liao 2010) to 2-3 (Spink and Jansen 2004), these results seem to fall within the expected intervals.

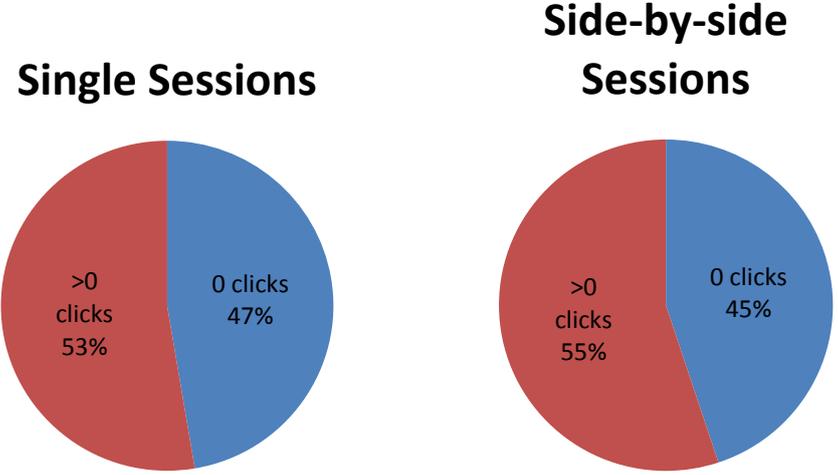

Figure 9.11. Zero click sessions.

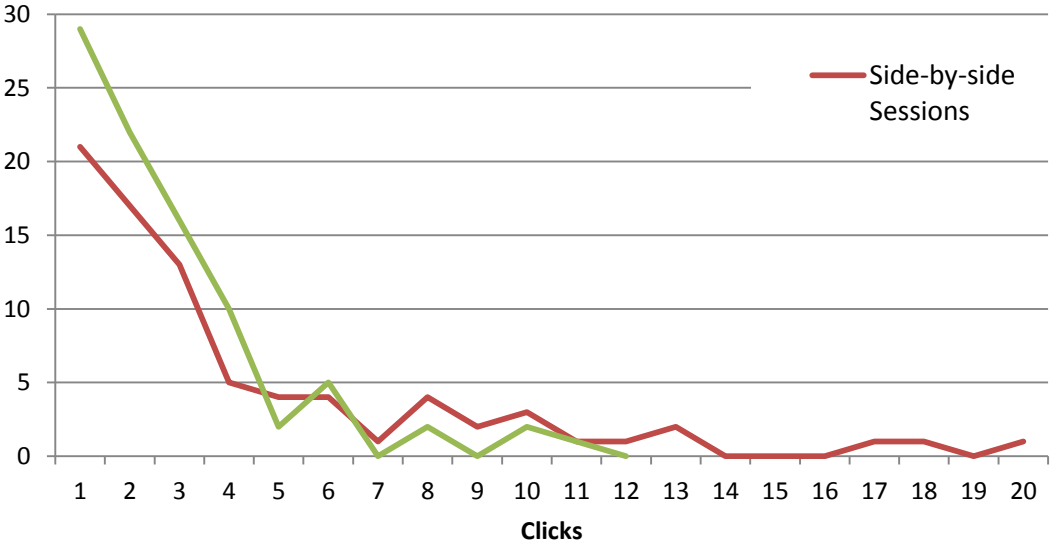

Figure 9.12. Clicks per session. No session had more than 20 clicks.

These numbers indicate that, while raters click more often in side-by-side settings, the click frequency is not as high as to suggest a total change of search habits. With more results to choose from, it is only to be expected that more results are chosen for closer inspection; but since the users have been instructed to abandon the session whenever they felt probable improvements weren't worth the extra effort, they had more chances of seeing good results and satisfying their information needs early on, causing the increase in clicks to be a moderate 0.5 per session.

Another question which can be approached with the help of click data is whether the raters came to grips with the – admittedly unusual – concept of side-by-side evaluation of result lists. As Figure 9.13 clearly shows, the click rates for result lists presented on the left and on the right are very similar; the overall click number is 183 for the left result lists and 166 for the right; the average click rank is 3.4 versus 3.5; and the statistical difference between the result lists is notable, but not significant ($p>0.1$, $W=19$, Wilcoxon signed-rank test). Moreover, the random presentation order of the result lists ensured the lack of any systematic bias towards result lists on either side. Figure 9.14 shows three sample click patterns for the side-by-side condition.

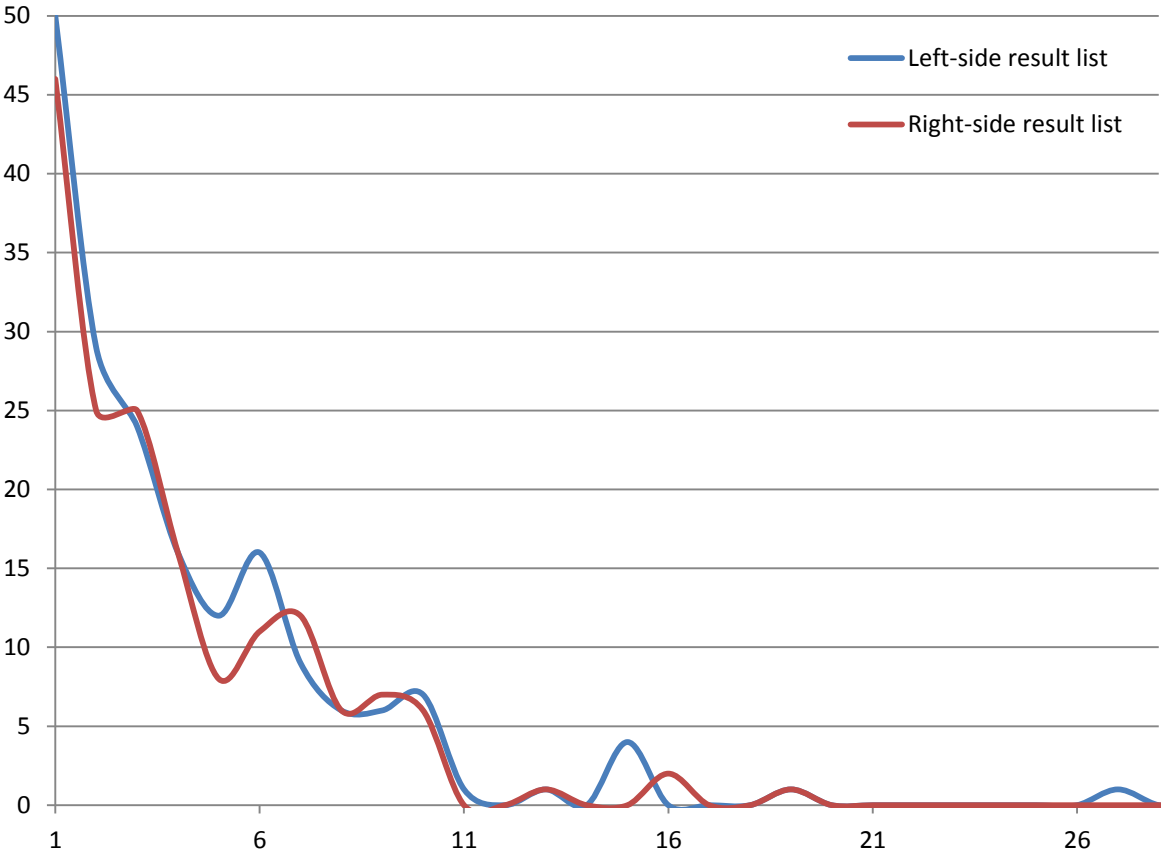

**Figure 9.13. Click distribution for left-presented and right-presented result lists.**

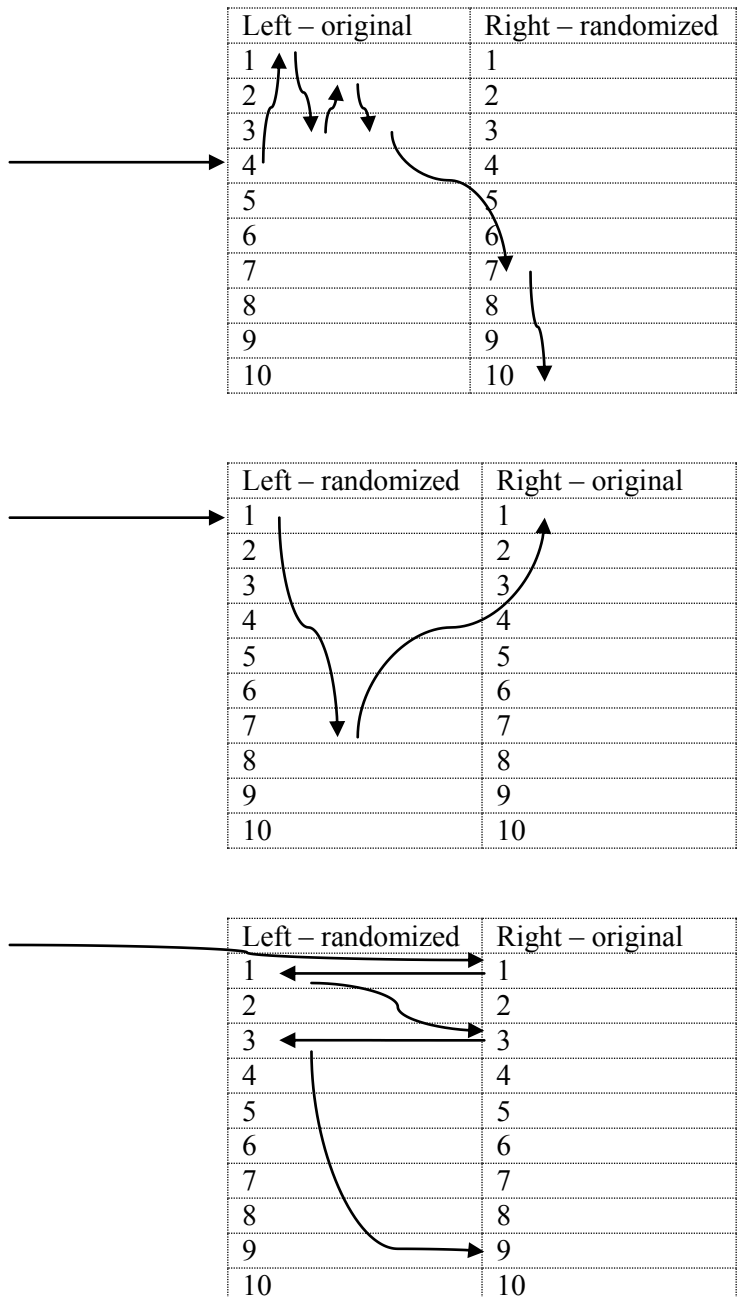

**Figure 9.14. Sample click trajectories. The arrow coming from the left shows the first click, further arrows indicate later clicks.**

## 9.4 Ranking Algorithm Comparison

Apart from evaluating user behavior, it seems interesting to evaluate the performance of the two result list types. Obviously, the original ranking would be expected to perform much better than the randomized list in preference as well as satisfaction. When I submitted a preliminary paper on this topic (Sirotkin 2011) to a conference, an anonymous reviewer raised a logical issue: "It seems obvious that users always prefer the ranked lists. I would be interested in knowing whether this was indeed the case, and if so, whether this 'easy guess' could have an influence on the evaluation conducted." We can approach the matter of rankings coming from the log statistics; namely, with an evaluation of clicks depending on the ranking type. As Figure 9.15 shows, the proportions of clicks (as indicated by the curve shapes) are quite similar. While the randomized ranking receives fewer clicks (247 versus 350) and has a slightly lower mean clicked rank (3.5 versus 3.1), its inferiority as expressed in the click behavior is by no means overwhelming. However, the data does support the notion of positional bias (see e.g. Craswell et al. 2008; Keane, O'Brien and Smyth 2008) since randomizing result order means there is no quality advantage for earlier ranks.[78]

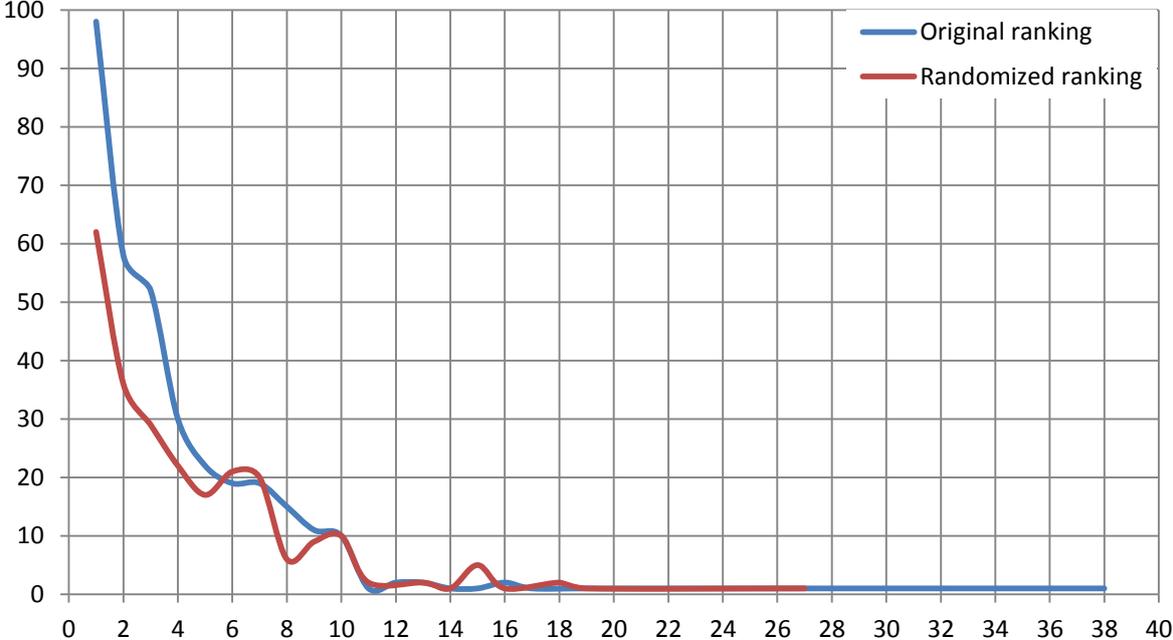

**Figure 9.15. Click ranks for original and randomized rankings.**

Obviously, clicks are not the best indication of the raters' evaluation of the two result list types – especially if we actually asked them to explicitly state their satisfaction and preference levels. Figure 9.16 shows the user preference for the original and randomized result lists; and the results are not what was expected by the reviewer (and myself). Of course, the original result list is preferred most of the time; but in over a quarter of all cases, the lists were deemed to be of comparable quality and in almost 15% of judgments, the randomized result list was actually preferred to the original. The reasons for this are not quite easy to fathom. One simple explanation would be that the quality of the original list was just not good enough, so that a randomization might create a better sequence of results about one in six times. However, a detailed examination of queries where the randomized result list is preferred at least by some users shows another pattern. Almost all of those queries are informational and transactional and stated quite broadly, so that the number of possibly interesting hits can go into the thousands or tens of thousands; some examples are shown in Table 9.3. The case of the query "korrellation", though not representative statistically, may nevertheless be typical. It is a factual query, for which any web site with an authoritative feel mentioning the word will probably be not only highly relevant, but also sufficient. The result list preference might be down to which of the two first-rank results has a more authoritative feel to it, or displays better context in its snippet.

---

[78] Not that this notion needs much support, mind you; it is universally accepted.

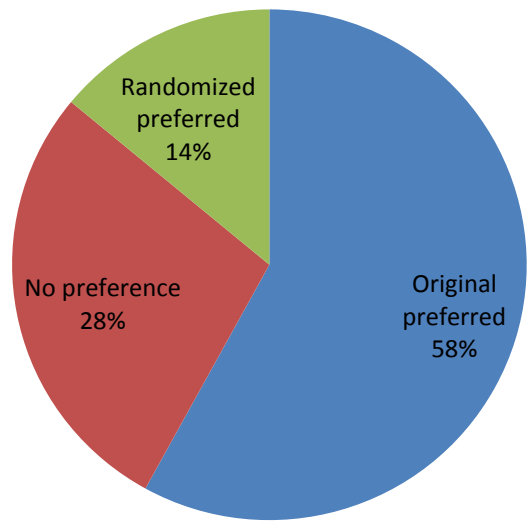

**Figure 9.16. Result list preferences**

| Query | Information need | Query type | Original result list preferred | Result lists similar | Randomized result list preferred |
|---|---|---|---|---|---|
| **sony vx1000** | Finding offers for a DV camcorder not manufactured any more | Transactional | 0% | 50% | 50% |
| **NRW-Kredit Student** | Information on NRW federal student loans | Informational | 50% | 0% | 50% |
| **korrellation** | Answer to the question how to spell the German word for "correlation"[79] | Factual | 25% | 25% | 50% |
| **günstige mobiltelefone** | Cheap mobile phone offers | Transactional | 25% | 50% | 25% |

**Table 9.3. Sample queries with randomized result list preference.**

The difference between result list types becomes smaller still if we switch from preference to satisfaction (shown in Figure 9.17). While the original result lists had a higher proportion of queries which satisfied most raters (average satisfaction of over 0.5 was 80% versus 64%), the randomized option had a higher proportion of average-quality result lists. Result lists which entirely failed to satisfy the users were equally uncommon (17% and 18%, respectively). This can be taken to mean that, while the original result list is indeed more satisfactory, many (or, rather, most) users will find satisfactory results even if they have to look longer for them in a randomized list; the average user satisfaction was 0.79 for the original and 0.65 for the randomized lists. We take this to be a result of the same kind as that found by researchers who observed that "searchers using […] degraded systems are as successful as those using the standard system, but that, in achieving this success, they alter their behavior" (Smith and Kantor 2008, p. 147). The results might also be connected to the fact that binary satisfaction was used for the evaluation; as is the case with other measures like relevance, the threshold for a good rating is probably quite low when the user is presented with only two options. This, in its turn, suggests that retrieval system operators might want to concentrate on improving the weak points in their systems rather than try to further enhance their strong sides, as users seem to be satisfied with a relatively low standard. Of course, this is only one possible interpretation which needs to be thoroughly tested in larger studies before any real practical recommendations can be made on its basis.

---

[79] In case you wonder: it's spelled "Korrelation".

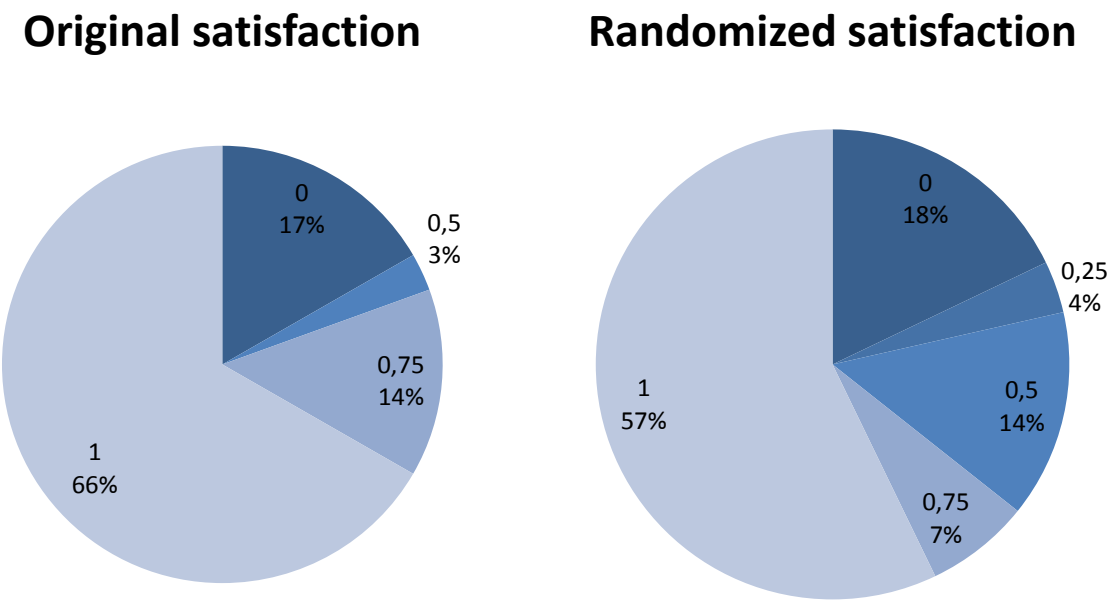

**Figure 9.17. User satisfaction with the original and randomized result list. The numbers indicate the percentage of queries for which mean user satisfaction lies in the bands 0, 0.01-0.25, 0.26-0.50, 0.51-0.75 and 0.75-1.**

Finally, I want to look at the relevance distributions for the two result list types. Figure 9.18 shows the macro-relevance; that is, the relevance ratings for individual results are averaged, and then these average ratings are averaged again for each rank. The first thing to catch the beholder's eye is the narrow range of relevance scores. The best score (original result lists at rank one) is just below 3; all other ratings fall in the strip between just below 3.5 and just below 4.5. That is, all positions but one have relevance scores lower than the theoretical average of 3.5. This fits well with the notion described above, namely, that it does not require top-relevance results (or even *a* top-relevance result) to make a user's search experience satisfactory. The second issue with this graph regards the shapes of the two lines. The randomized result list has values in the range from 4.2 to 4.6, which do not seem to depend in any meaningful way on the rank. This is entirely expected since the results have been randomized. It can be also reasoned that since the randomized ratings stem from results anywhere in the original top 50, the score of 4.2 to 4.6 is also the average relevance of those first fifty results. The relevance ratings for the original result lists do show a clear direction; they start off (relatively) high and decline towards the later ranks. However, from rank 5 on, when the average relevance falls below 4, the decline is not very pronounced, and the values behave similar to those for randomized result lists. The simplest interpretation of this data is that the ranking algorithm is really effective only for the first four or five ranks, and after that does no better than chance.

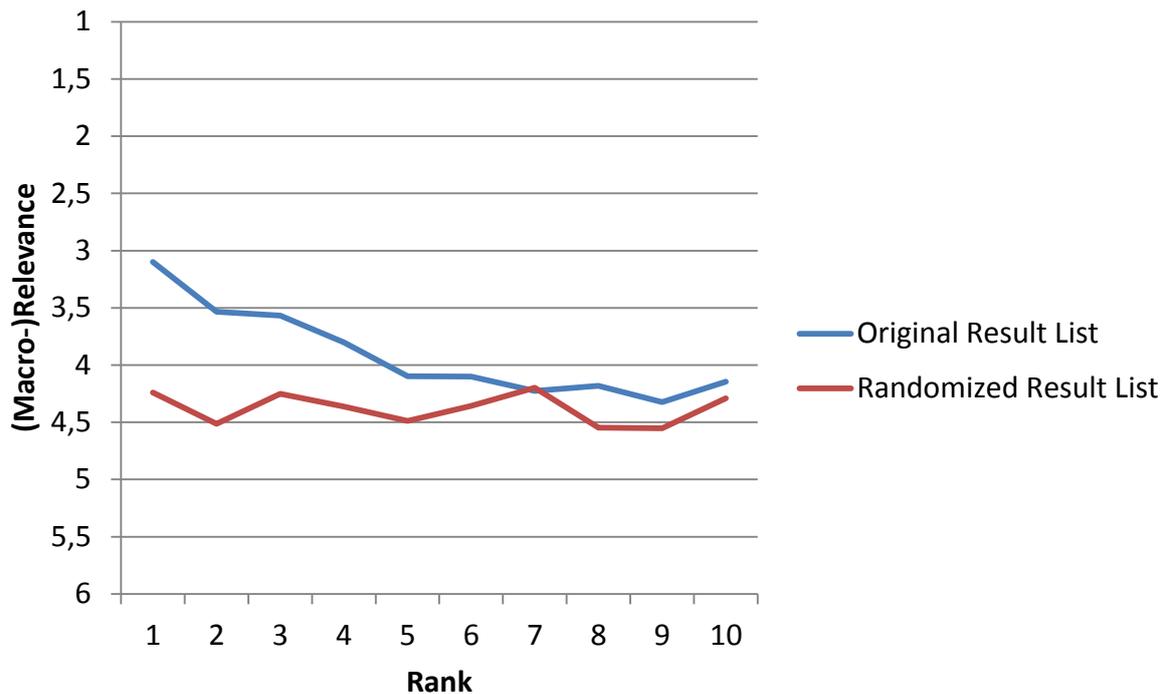

**Figure 9.18. Macro-Relevance at individual ranks for the original and randomized result list. The randomized list relevance stays fairly stable, with fluctuations which can be attributed to chance.**

Figure 9.19 and Figure 9.20 provide a more detailed view of the original and randomized result list, accordingly. They show the precise distribution of relevance rankings by rank. In the original result list, we see that the share of high-relevance results declines towards the end. However, there are some details that need to be kept in mind. Firstly, the decline in highest-relevance results (relevance score 1) more or less ends at rank 7, staying much the same after that. Secondly, it is mostly that relevance band (and to a lesser extent relevance score 2) that declines in frequency. The relative frequencies of average-relevance scores (3 and 4) stay more or less the same, while the lower scores' frequencies increase, as the counterpart to the decline of high-relevance ratings. These tendencies correspond well to the macro-relevance evaluation described in the previous paragraph.

The relevance evaluation for the randomized result lists are, as could be expected, less telling. There are no obvious trends depending on the rank. It is perhaps worth noting that here, even more than when evaluating the original result lists, we see stable relative frequencies for mid-way relevance ratings. Almost all the changes occur in the highest and lowest ratings.

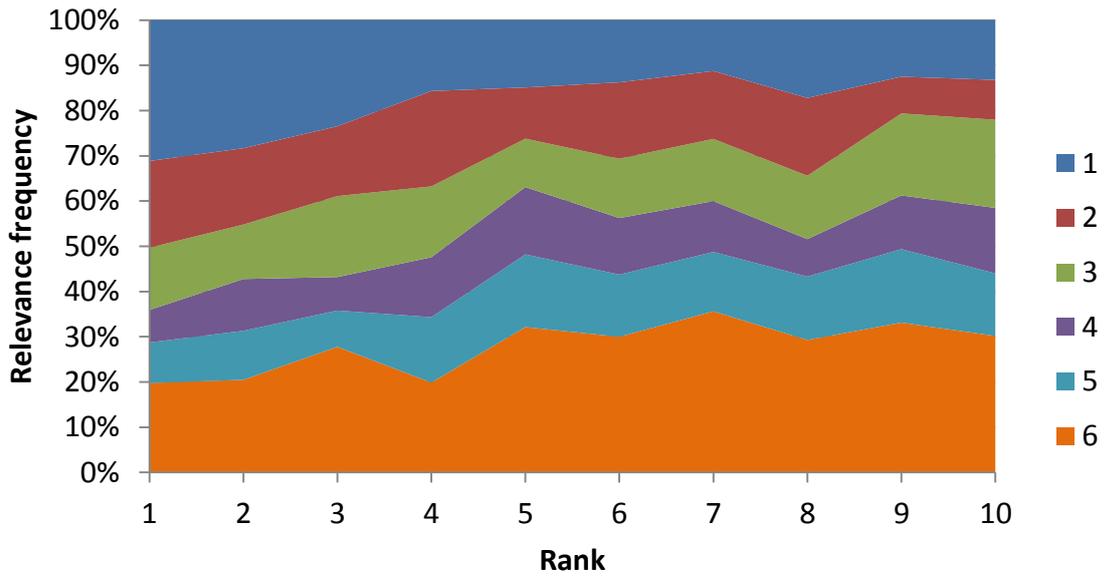

Figure 9.19. Micro-relevance for the original result list.

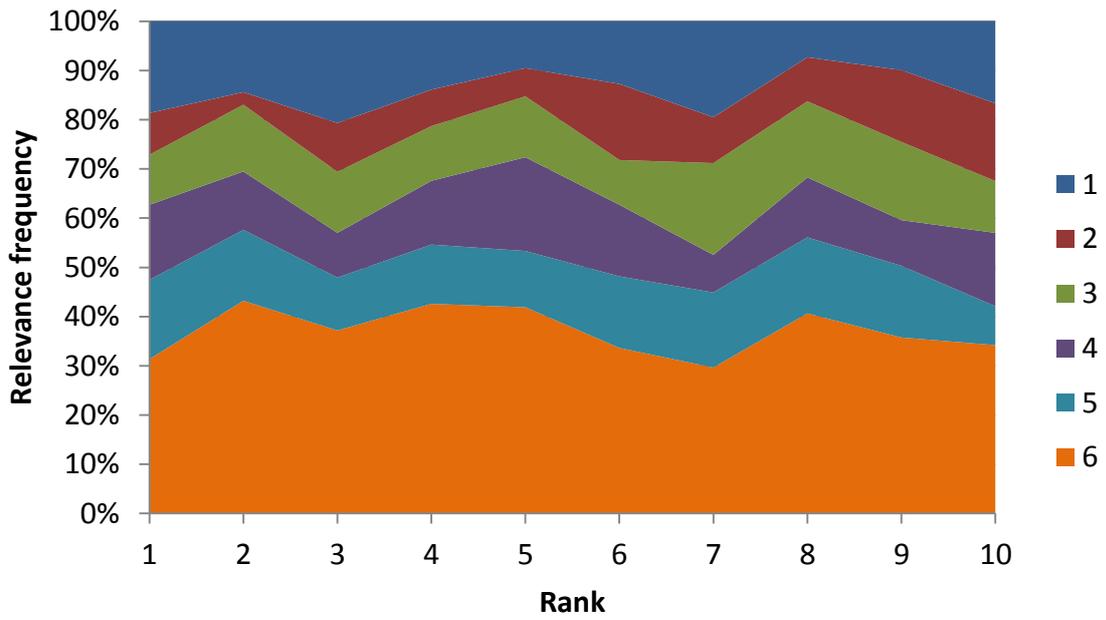

Figure 9.20. Micro-relevance for the randomized result list.

# 10 Explicit Metrics

*The average man's judgment is so poor he runs a risk every time he uses it.*
EDGAR WATSON HOWE, "ROSICRUCIAN DIGEST"

In this section, I will describe how well many of the different explicit metrics described in Section 4 perform. We will start with considering the effect of different thresholds and discounting functions (combined with cut-off value changes) on a metric's performance (Sections 10.1-10.4). Then we will compare the different metrics themselves (Section 10.5). After that, the influence of using different raters for different judgments will be considered (Section 10.6). Finally, we will look at the effects of choosing different judgment scales, bearing in mind the previous findings (Section 10.7).

## 10.1 (N)DCG

I shall start with (N)DCG, which provides a somewhat unusual standard discount rating; while most other metrics normally discount by rank, the authors of DCG use the dual logarithm as their denominator of choice, while indicating that it might not always be the best option. I will use NDCG instead of DCG. As the PIR is always concerned with aggregating individual queries, both metrics offer the same relative scores; however, NDCG falls into the more familiar and uniform 0..1 range, while DCG scores can potentially be arbitrarily high.

Figure 10.1 to Figure 10.7 show our first real PIR graphs. Each Y-axis shows the PIR scores, and is scaled from 0.5 (the baseline, equivalent to guessing) to 1 (every single user preference can be correctly induced from the metric in question). The figures show PIR values for NDCG using the seven discount functions listed above. There are quite a few interesting observations to be made about the differences between as well as the similarities among the graphs; however, most of those are better viewed on a comparison graph, which is presented later. I will start with the discussion of the properties of Figure 10.1 (showing NDCG PIR scores without any discounting), and work down from there; this will hopefully make for a more clear picture.

Probably the first thing an observer sees is that the graph is not very intuitive. However, concentrating on specific questions, one can read quite a lot. My primary objective in this particular evaluation was to see what threshold value performed best for each NDCG discount function. This means we should look at all lines to see whether we can find regularities in their runs. The first impression is that, while the lines go up and down a bit, in general the PIR values for smaller thresholds tend to be higher, and those for the largest threshold (0.30) are almost universally the lowest of all.[80] Table 10.1 makes this situation immediately comprehensible; it shows different average measures for the thresholds which have the highest PIR scores. Five out of seven median values and six out of seven mode values have a threshold of 0 as the best-performing one; and four out of seven average scores are below the lowest non-zero threshold, 0.01.[81] While the 0.01 threshold itself might perform slightly better, the zero threshold provides a useful counterpart to the best-threshold approach for PIR comparison as discussed in Section 8.2.1. It makes the evaluation simpler by leaving thresholds entirely out of the equation; and if it underestimates best possible PIR scores (which might be achieved by using, say, a threshold of 0.01, or perhaps some even smaller number), it provides a lower boundary as opposed to the best-threshold approach's upper boundary for PIR scores. In later sections, we will see that this holds true not just for (N)DCG, but also for most other metrics.

---

[80] Keep in mind that values larger than 0.30 are not pictured, since they failed to produce any increase in PIR.

|  | No discount | log5 | log2 | Root | Rank | Square | Click-based | Overall |
|---|---|---|---|---|---|---|---|---|
| **Median** | 0.01 | 0.01 | 0 | 0 | 0 | 0 | 0 | **0** |
| **Mode** | 0 | 0.01 | 0 | 0 | 0 | 0 | 0 | **0** |
| **Mean** | 0.026 | 0.024 | 0.001 | 0.007 | 0.006 | 0 | 0.013 | **0.011** |

Table 10.1. The thresholds with the highest PIR scores for NDCG with different discount functions.

The next graphs show other discounts, which get progressively steeper. You will note that the curves follow this trend and become steeper themselves; that is, larger threshold values have more rapidly declining PIR scores for steeper discounts.

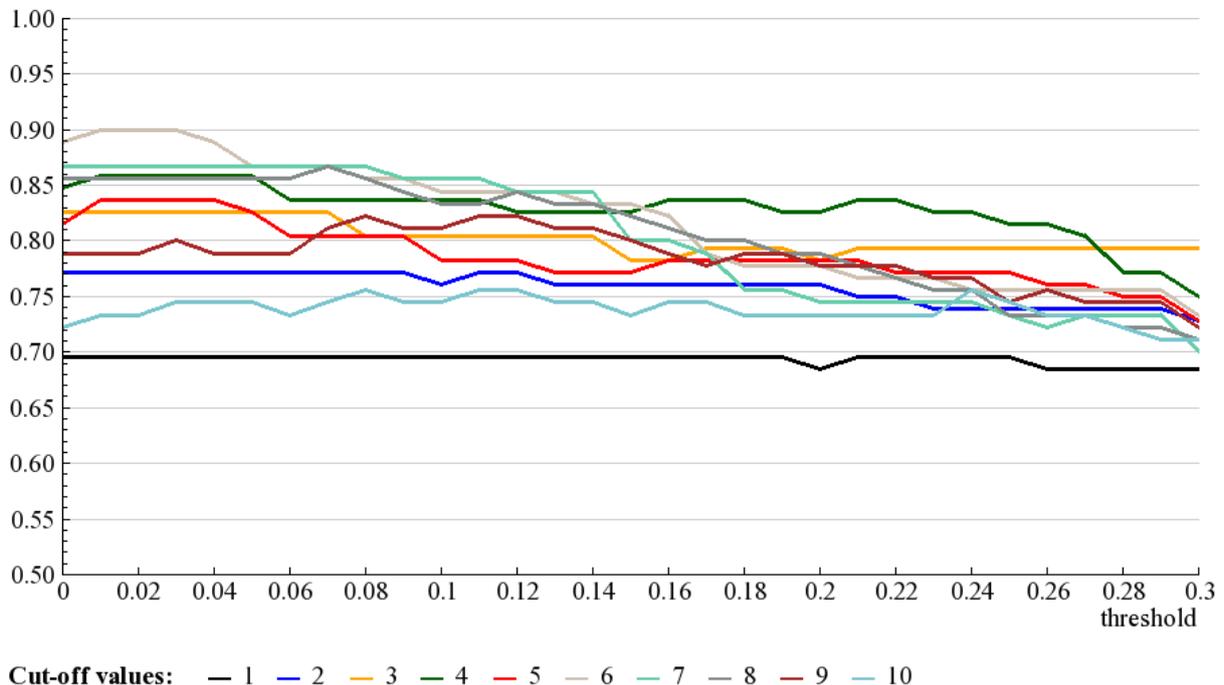

Figure 10.1. NDCG PIR values without any discount for results at later ranks.

---

[81] The numbers are on the low side since in the cases when multiple thresholds have the same PIR, the lowest one has been used for the calculation. I feel this method is acceptable since the important point is that, in this situation, lower thresholds rarely produce *lower* PIR scores, not that they always produce *higher* ones.

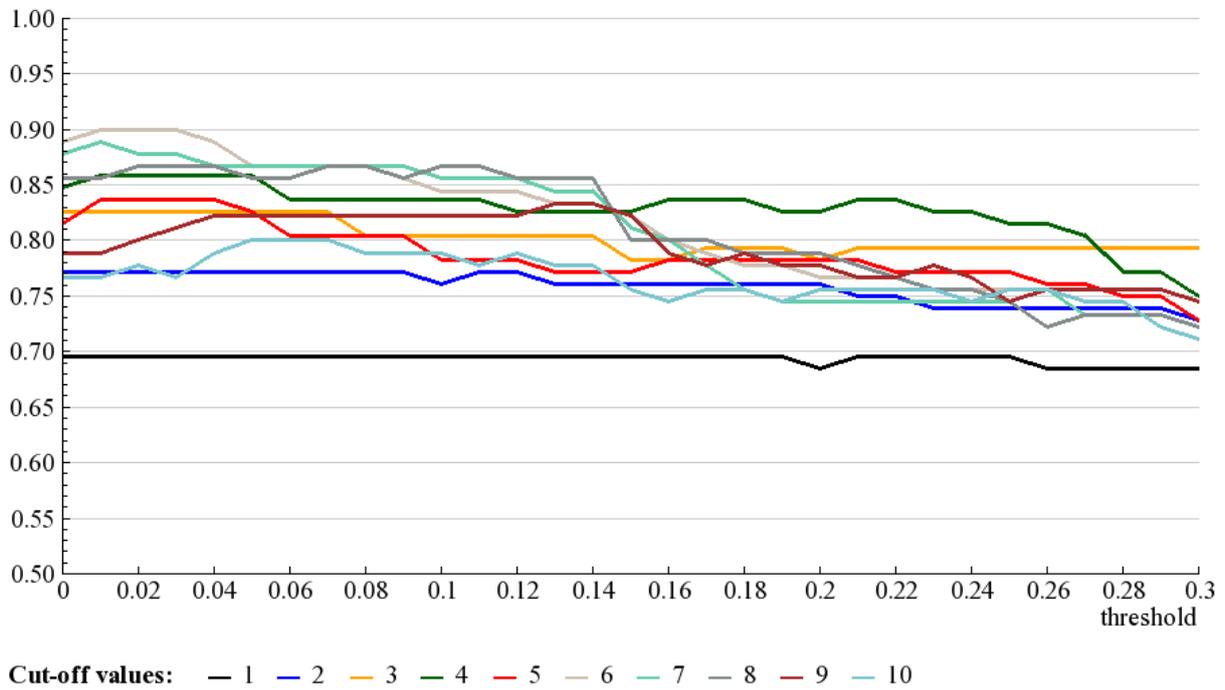

Figure 10.2. NDCG PIR values discounted by $\log_5$ of result rank.

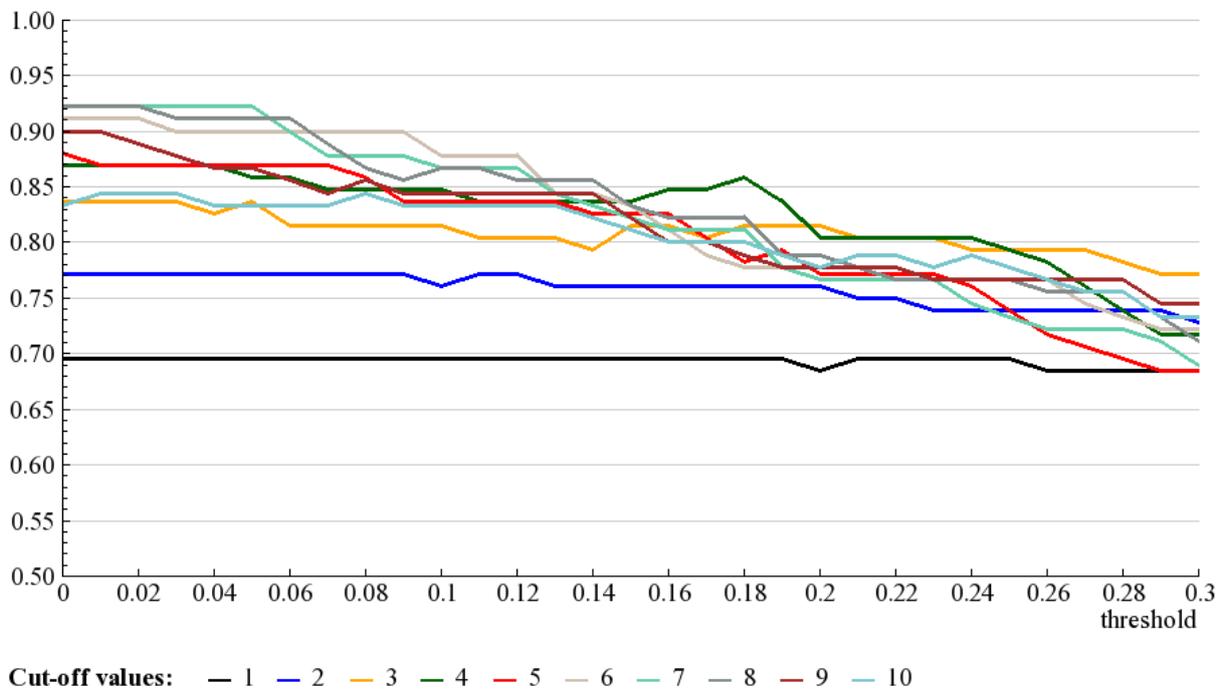

Figure 10.3. NDCG PIR values discounted by $\log_2$ of result rank.

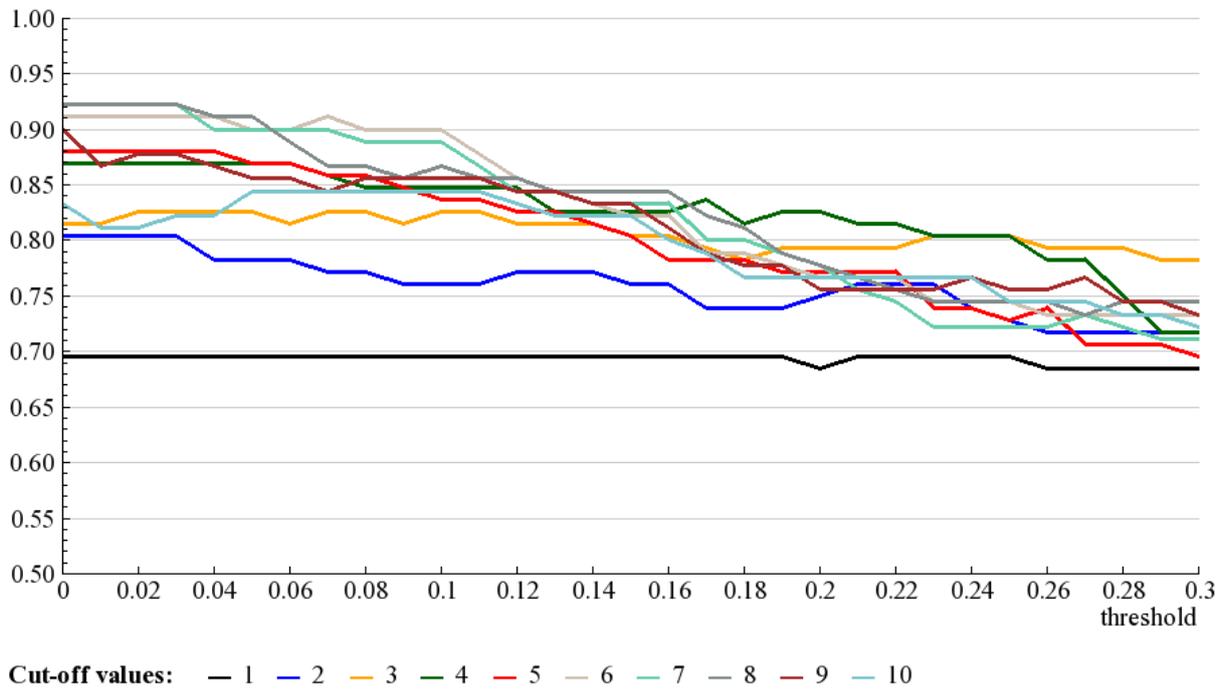

**Figure 10.4. NDCG PIR values discounted by the square root of result rank.**

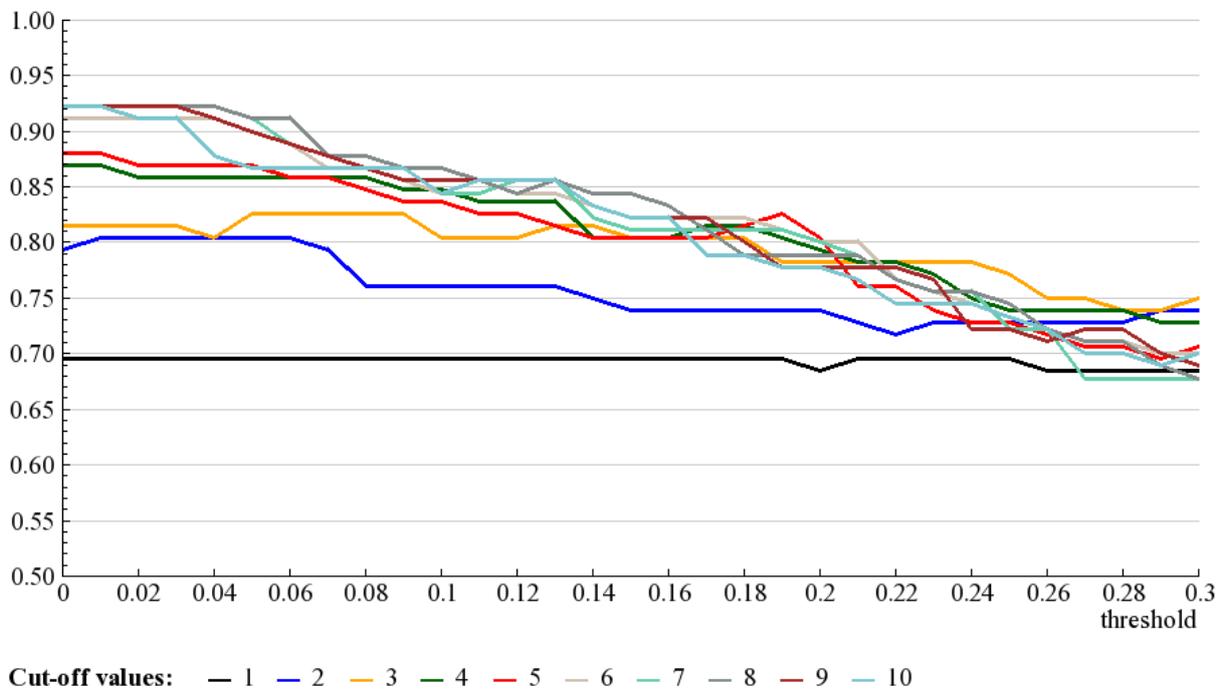

**Figure 10.5. NDCG PIR values discounted by the result rank.**

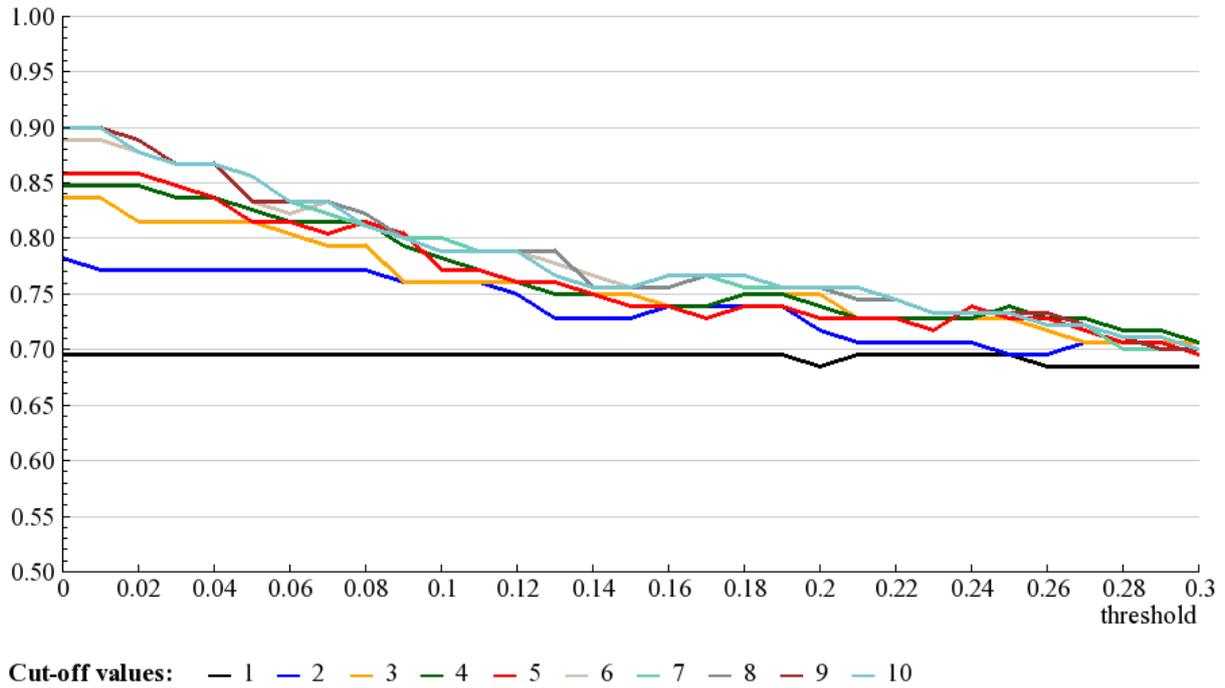

**Figure 10.6. NDCG PIR values discounted by the result rank squared.**

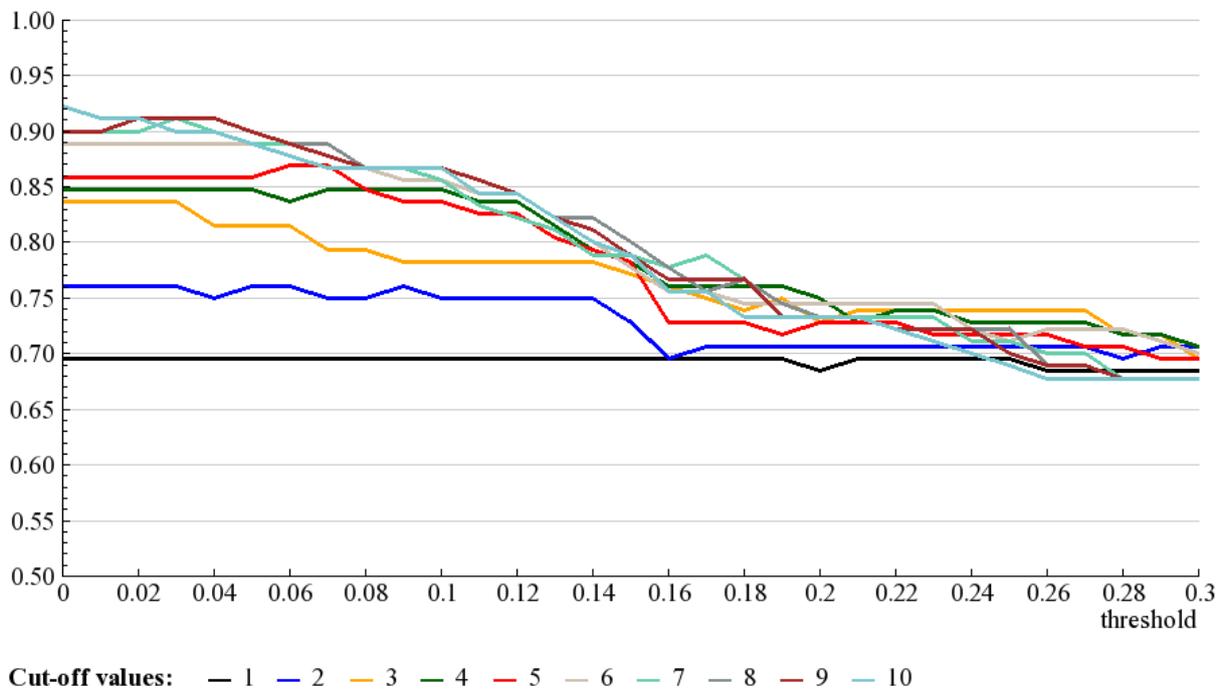

**Figure 10.7. NDCG PIR values discounted by the frequency of clicks on results of a particular rank according to Hotchkiss, Alston and Edwards (2005).**

There are more things to be seen in the graphs; but they concern not the changes brought on by different thresholds; rather, they are about the influence of cut-off values and discount functions. Thus, they are more visible and better discussed in the context of inter-metric PIR evaluation, as shown in Figure 10.8 and Figure 10.9.

The first point I would like to make about these two graphs is their similarity. On first glance, it might not even be obvious that they are indeed different. For this reason, I will discuss the best-threshold results, but everything can be equally well applied to the *t=0* graph.

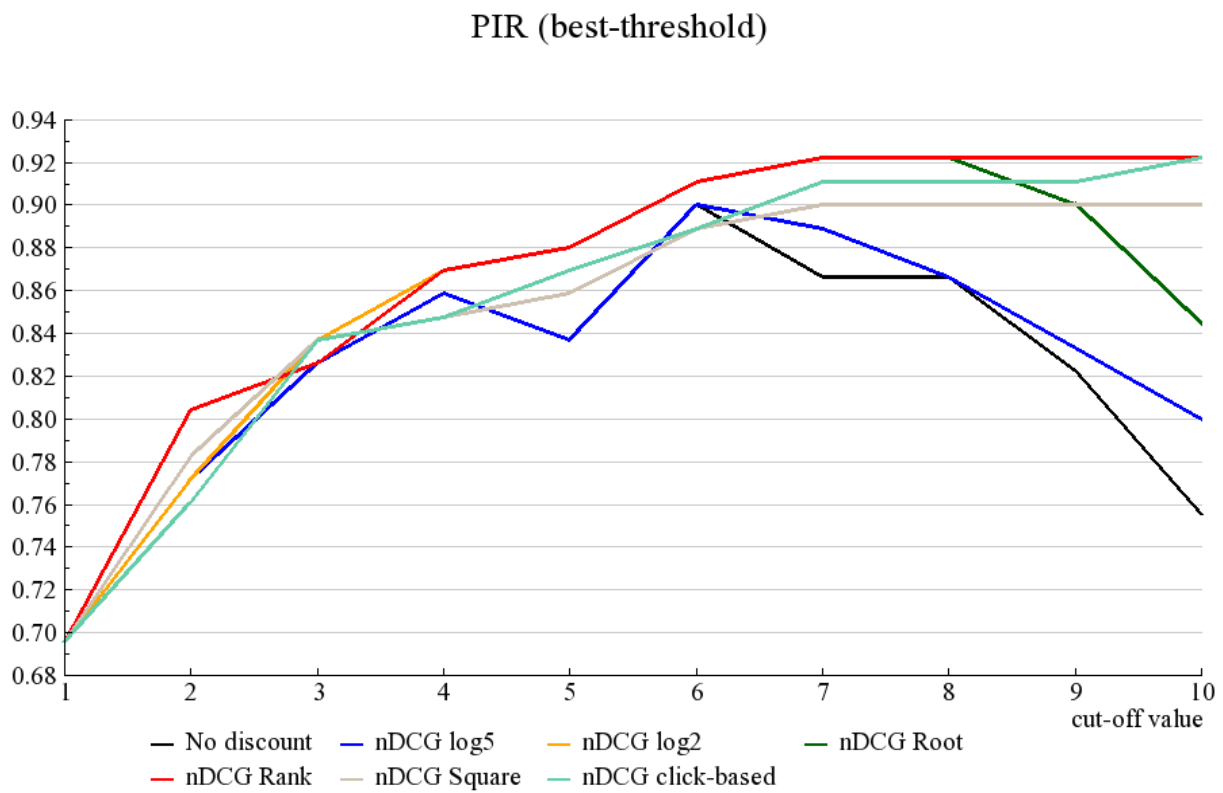

**Figure 10.8. NDCG PIR scores for different discount functions, with the best-threshold approach. As many of the discount functions often produce identical PIR scores, a description is needed. Due to the discount function definitions, the "No discount", "log$_5$" and "log$_2$" conditions produce the same results for cut-off values 1 and 2, and "No discount" and "log$_5$" have the same values for cut-off values 1 to 5. Furthermore, the "Root" and "Rank" conditions have the same scores for cut-off ranks 1 to 8, and "Square" and "Click-based" coincide for cut-off values 3 to 4. "log$_2$" has the same results as "Root" for cut-off values 4 to 10 and the same results as "Rank" for cut-off values 4 to 8.**

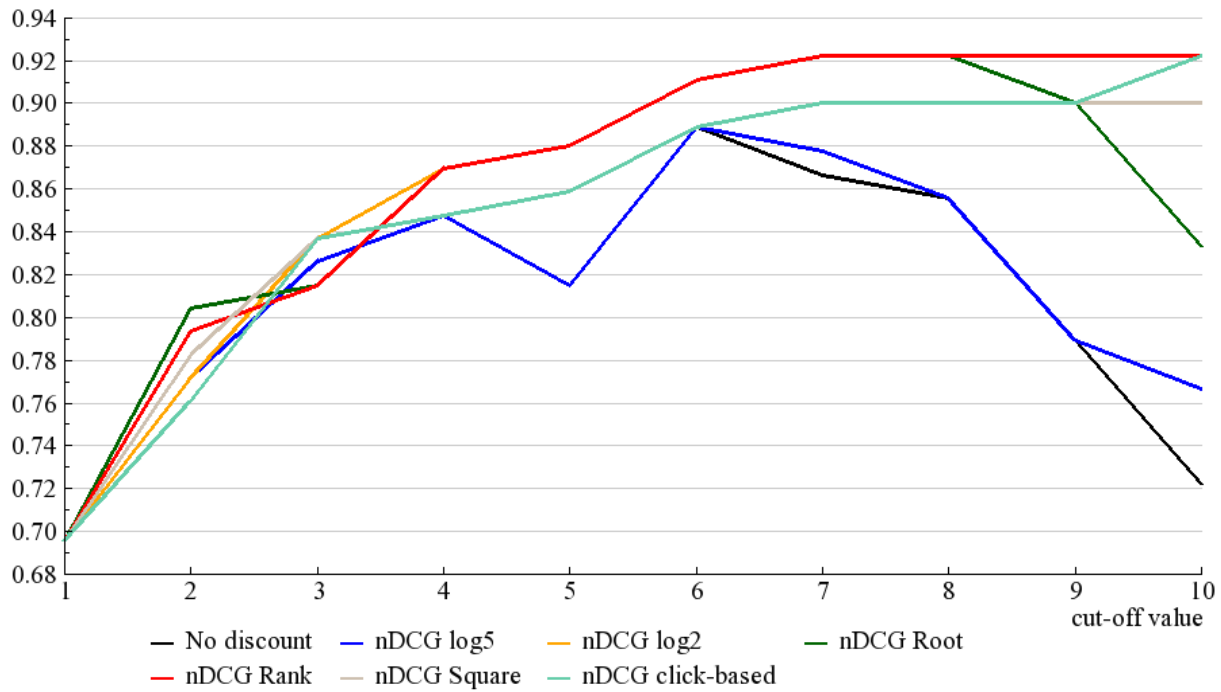

Figure 10.9. NDCG PIR scores for different discount functions, with the threshold values set to zero. As many of the discount functions often produce identical PIR scores, a description is needed. Due to the discount function definitions, the "No discount", "$\log_5$" and "$\log_2$" conditions produce the same results for cut-off values 1 and 2, and "No discount" and "$\log_5$" have the same values for cut-off values 1 to 5. Furthermore, the "Root" and "Rank" conditions have the same scores for cut-off ranks 3 to 8, and "Square" and "Click-based" coincide for cut-off values 3 to 9. "$\log_2$" has the same results as "Root" for cut-off values 4 to 10 and the same results as "Rank" for cut-off values 4 to 8.

The second issue concerns the overall absolute values. The best PIR scores are slightly higher than 0.92; this means that for over 92% of all sessions the user would get the result list he preferred if the result list choice is based on the NDCG scores. This is a surprisingly high value. However, there are a few factors that put the numbers in perspective. While for 92% of sessions the users get their preferred result lists, for 8% they get a result list they consider inferior. In other words, one in twelve sessions still results in a choice of a result list which the user finds worse. Furthermore, the relevance ratings on which the NDCG calculations are based come from the same user as the preference ratings with which they are compared. Having the user rate single results for the very query he is interested in is a luxury which no search system provider is likely to be able to afford. This and other interpretations will be discussed in more detail in Section 10.6.

The third issue is the general shape of the different discount function curves. All curves start at the same PIR score of approximately 0.7; as no discount is ever applied at rank 1, the PIR scores for a cut-off value of 1 will always be the same for all discount functions. They rise when the cut-off value is increased; no surprises there – the amount of information on which to base NDCG ratings increases, and with it the quality of preference predictions. What *is* surprising, however,[82] is that the highest PIR scores are reached by rank six or seven by every discount function except for the click-based one. After that, the scores stay the same, and often even decline, sometimes quite significantly (for example, from a peak PIR score of 0.9 at rank 6 for the "No discount" function to just over 0.75 at rank 10). This implies that an increased amount of information can actually lead to lower-quality metric performance. We will see this trend in most other metrics as well; the possible reasons and implications will be discussed in detail in Sections 10.5 and 12.2.5.

The final issue is the one which the graphs were originally intended to clarify: How do the different discount functions perform in comparison to each other? Again, one thing that catches the eye is the relative similarity of the graph lines. The PIR scores peak between 0.9 and slightly over 0.92, and up to a cut-off value of 8, there are no PIR differences larger than 0.055 (to put flesh on the numbers once again: this is equivalent to ca. 5% of sessions displaying the better result list instead of the worse one). This difference is not negligible, of course, but neither is it gaping. Only after that, when the "No discount" and "$\log_5$" functions' PIR scores decline sharply at cut-off values 9 and 10, are the differences becoming really large.

A further property of the graphs that has at first surprised me was the large number of identical data points for different discount functions (for example, NDCG Square and Click-based coinciding at cut-offs 3, 4, and 6), necessitating the detailed verbal descriptions found in Figure 10.8 and Figure 10.9.[83] However, this becomes less surprising when one considers that the discount functions start out with the same performance at cut-off 1, and then improve when for a number of sessions, the better result list can be identified correctly. The improvement in PIR score is then a multiple of one over the overall number of sessions with stated result list preferences. If, for example, we had 50 sessions where a user has indicated a preference for one of the possible result lists, the PIR would always be a multiple of $^1/_{50}$ or 0.02. If we furthermore consider that the PIR baseline is 0.5, there are only 25 possible PIR scores left. Therefore, it is not really surprising that different discount functions might have the same score. Also, the functions with the same scores tend to be ones with relatively similar discounts ("Square" and "Click-based", for example, are the two functions which discount most strongly).

Overall, the "No discount", "$\log_5$" and "Square" discounts can be said to perform worse, while "Root", "$\log_2$" and "Rank" do better, the "Click-based" function falling in between. For the purposes of later inter-metric comparison, "$\log_2$" will be used for two reasons. It is one of the more successful functions; though it performs slightly worse that "Rank" and "Root" at cut-off 2, it is better at cut-off 3 and equally good up to cut-off 8. We will have to bear in mind, though, that its PIR scores decline at cut-offs 9 and 10, falling almost 0.08 points below "Rank" at 10. The reason to nevertheless pick it rather than "Rank" for future prominence is its role in the literature in the past and present. As discussed in Section 8.2, $\log_2$ is routinely and universally used in (N)DCG evaluations, mostly without any thought given to the matter. As such, NDCG with $\log_2$ discount is a metric successful (in terms of PIR) as well as widely employed, and as such a good candidate to represent the NDCG family.

---

[82] For me, at least. Perhaps *you* suspected all along that the results would turn out to be just like this.
[83] And in many of the upcoming graphs; though I omit them later when we get a feel for the graphs and what the course of each line is.

An issue with the data requires more graphs. As stated in Section 9.2, only 31 of the 42 evaluated queries were informational, with the others classified as transactional, navigational, factual, or meta-query. While the latter categories contain too few queries to be evaluated on their own, they might have disturbed the picture if their PIR score patterns look significantly different from those of informational queries. Therefore, it is important to examine the issue.

Luckily, Figure 10.10 ($\log_2$ PIR threshold graph for informational queries only) and Figure 10.11 (comparison of PIR scores for different discount functions for informational queries only) show such a strong similarity to their respective counterparts (Figure 10.3 and Figure 10.8) that it is unnecessary to reproduce all the graphs in an "informational-only" form.

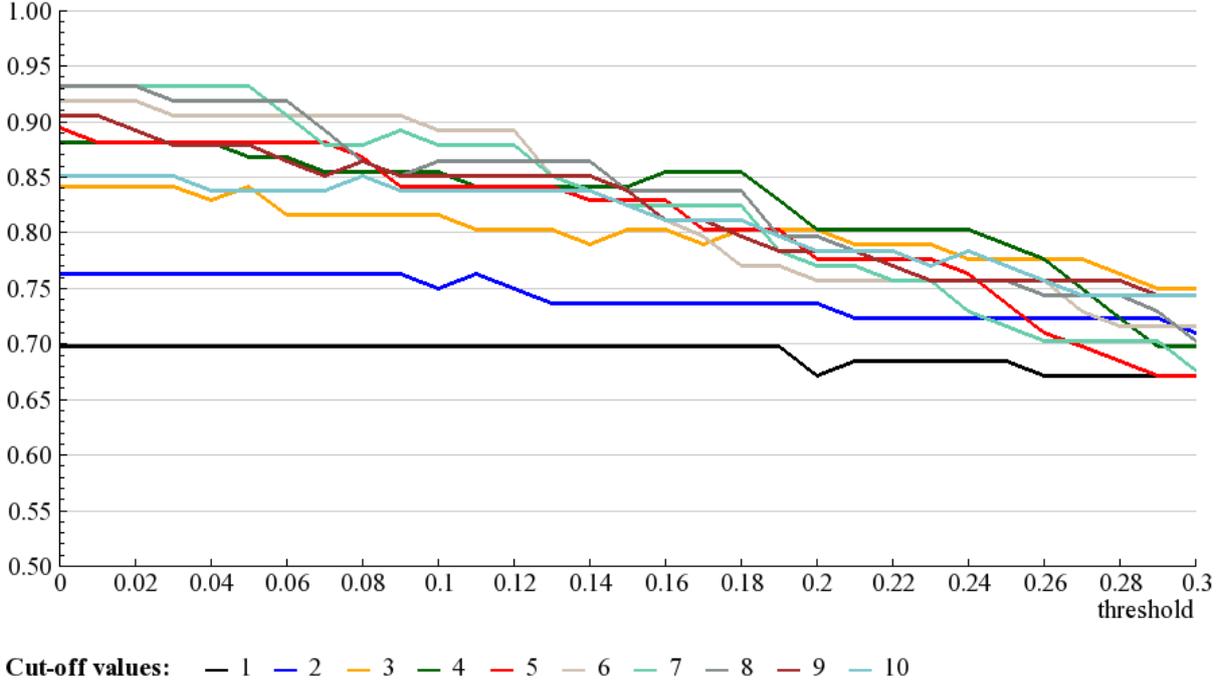

**Figure 10.10. NDCG PIR values for informational queries only, discounted by $\log_2$ of result rank.**

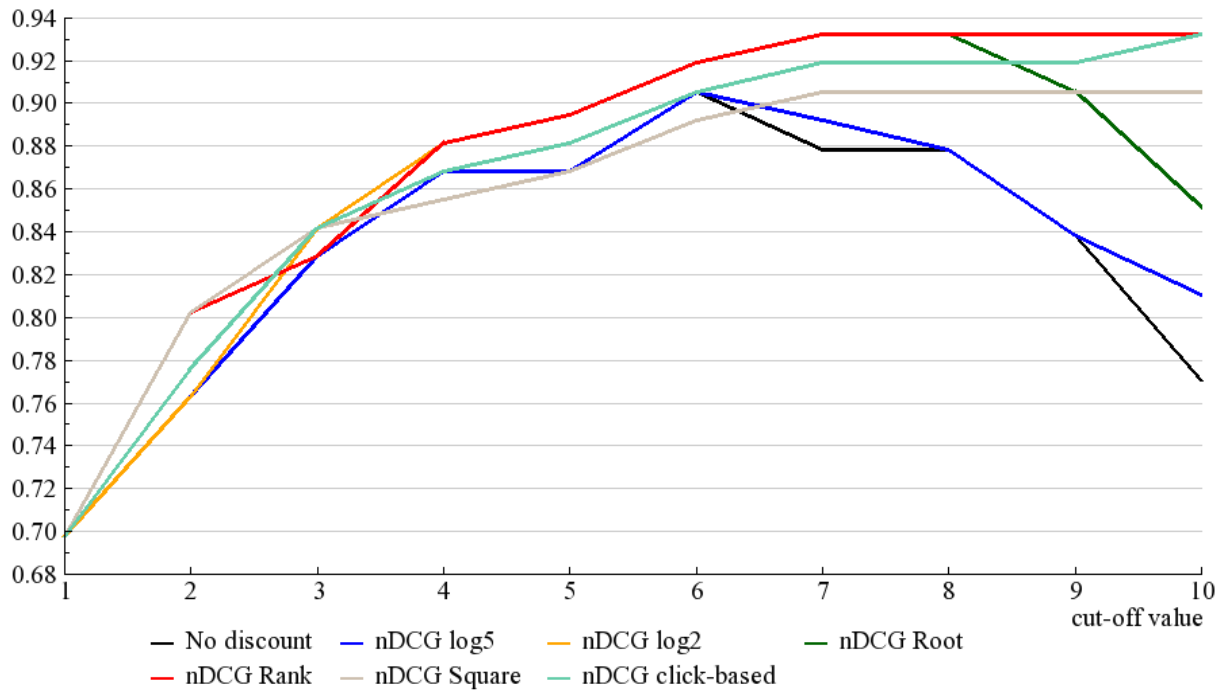

Figure 10.11. NDCG PIR scores for informational queries only, with the best-threshold approach. As many of the discount functions often produce identical PIR scores, a description is needed. Due to the discount function definitions, the "No discount", "$\log_5$" and "$\log_2$" conditions produce the same results for cut-off values 1 and 2, and "No discount" and "$\log_5$" have the same values for cut-off values 1 to 5. Furthermore, the "Root" and "Rank" conditions have the same scores for cut-off ranks 1 to 8, and "Square" and "Click-based" coincide for cut-off values 3 to 4. "$\log_2$" has the same results as "Root" for cut-off values 4 to 10 and the same results as "Rank" for cut-off values 4 to 8.

## 10.2 Precision

As mentioned in Section 8.2, classical Precision has quite a lot of common with a no-discount version of NDCG (alternatively, one might say standard NDCG has a lot in common with Precision with a $\log_2$ discount function). Both metrics rely on simple addition of relevance scores, which are then normalized. The main difference is that Precision, in dividing the sum of scores by their number, has its highest possible score of one in a situation where all results are perfectly relevant; NDCG, instead, considers the situation where all known relevant results are presented in the optimal order to be the ideal. Therefore, Precision tends to have lower absolute scores than NDCG.

At first glance, Figure 10.12, showing the PIR scores of classical Precision depending on cut-off and threshold values, looks significantly different from its NDCG counterpart (Figure 10.1). And the difference becomes extreme when one compares the two measures with steep discounting functions, such as the squared rank (Figure 10.13 versus Figure 10.6). While the no-discount versions appears basically similar, though with a more marked decrease in PIR scores at higher threshold values, the strongly discounted Precision graph shows not just a strong PIR decline – every line showing PIR scores for cut-off values of 5 and above hits the 0.5 baseline. This means that at this point, Precision performs no better than chance in selecting the preferred result list. However, this is easily explained by the differences in absolute values mentioned above. If, for example, the NDCG scores for all queries tend to lie in the interval between 0.2 and 0.9, while the Precision scores are between 0.1 and 0.5, then the maximum difference is 0.7 and 0.4 respectively. With the Precision differences spread over a shorter score span, the same threshold steps will provide less fine distinctions between result lists; in the above example, any threshold above 0.4 will fail to distinguish between any result lists. The effect is much stronger for high-discount versions, where the discounts reduce the score range even further. One could change this situation by choosing a finer threshold step for Precision, and selecting a different denominator for discounted Precision metrics to compensate; however, I think that this would combat a non-existent problem.

To see why the threshold issues do not matter a lot, we will examine two Precision PIR graphs comparing different discount functions, just as we did with NDCG. Figure 10.14, using the best-threshold approach, is very similar to its NDCG counterpart (Figure 10.8). There are a few minor differences (mostly some PIR scores being slightly lower in the Precision graph), but the overall picture is clearly the same. And when we compare the graphs where threshold values of 0 are used (Figure 10.15 and Figure 10.9), we see that they are completely identical. This should not really surprise; if no threshold is used, any difference between two result lists is considered significant, and the magnitude of absolute scores does not matter anymore. For all future purposes, then, Precision and NDCG can be regarded as the same metric with different discount functions. I will still refer to them by their familiar names, but the connection should be kept in mind. For the later inter-metric comparison, I will also show classical Precision as a separate metric for the simple reason that it is an all-time favourite, and is still used in evaluations. It will be interesting and instructive to see how it performs against more modern metrics.

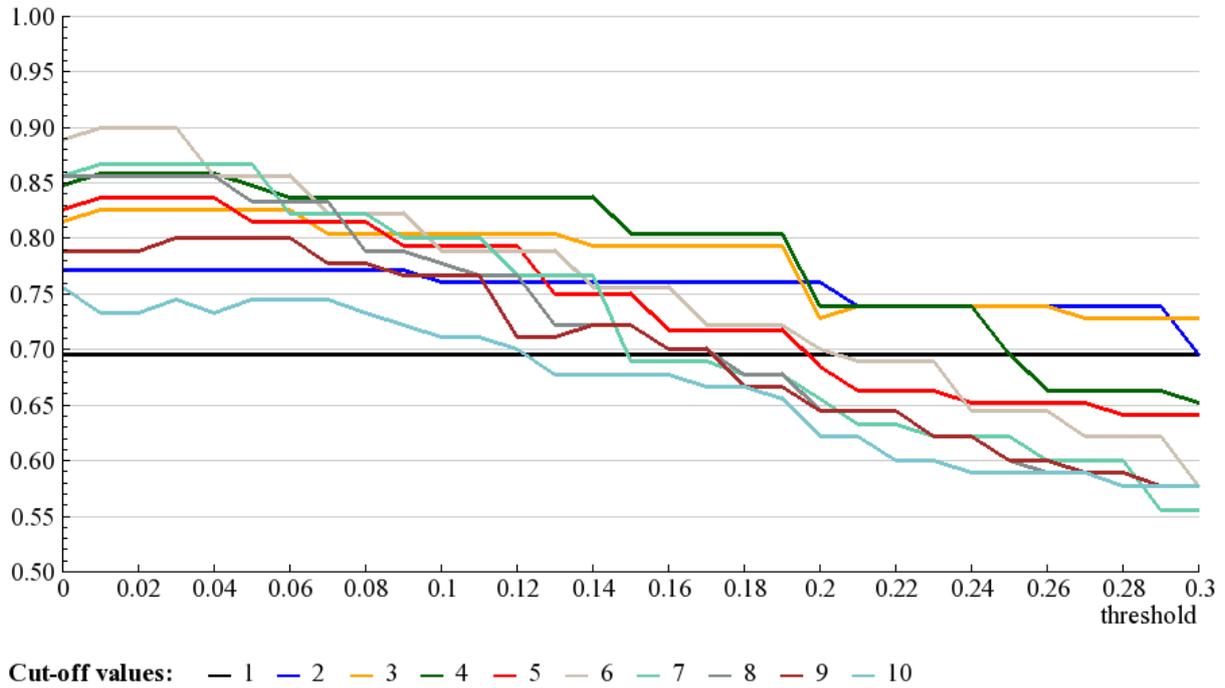

**Figure 10.12. Precision PIR values without any discount for results at later ranks.**

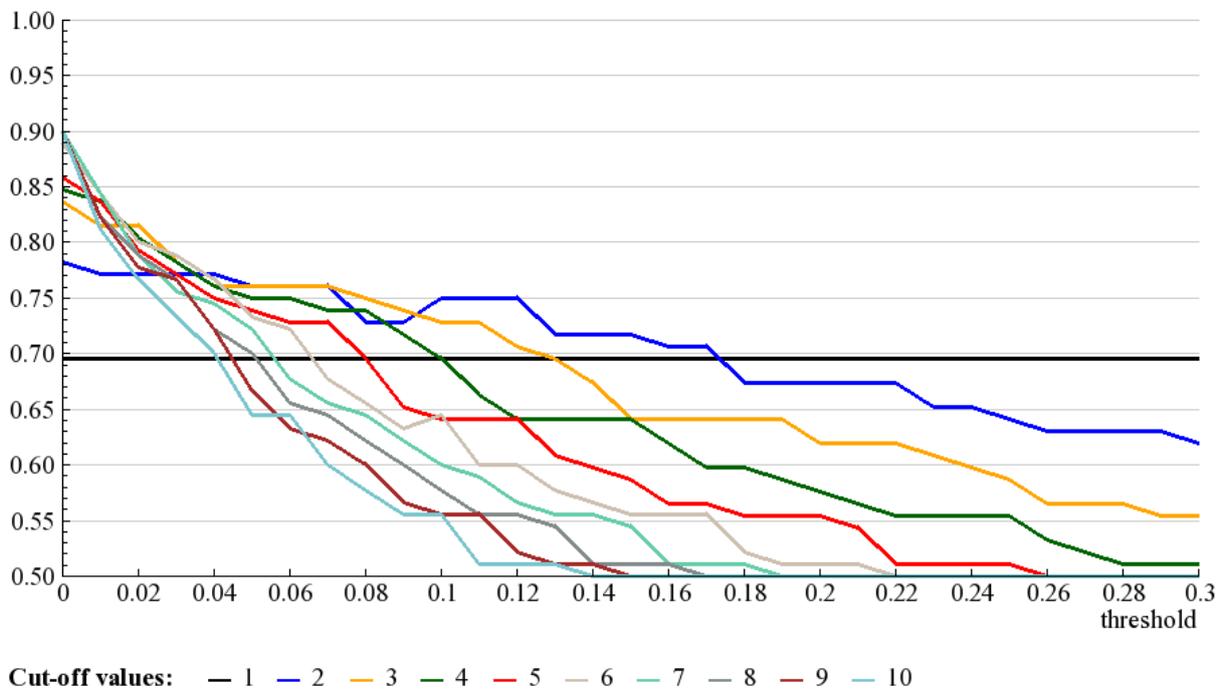

**Figure 10.13. Precision PIR values discounted by the result rank squared.**

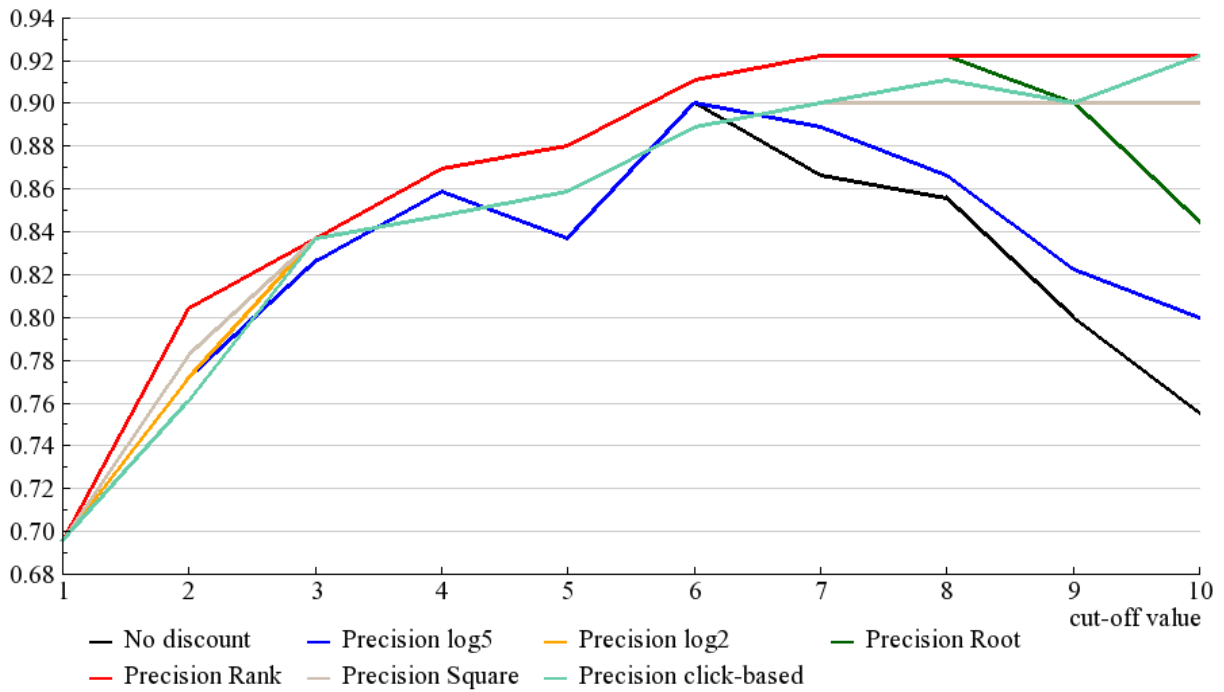

**Figure 10.14. Precision PIR scores for different discount functions, with the best-threshold approach.**

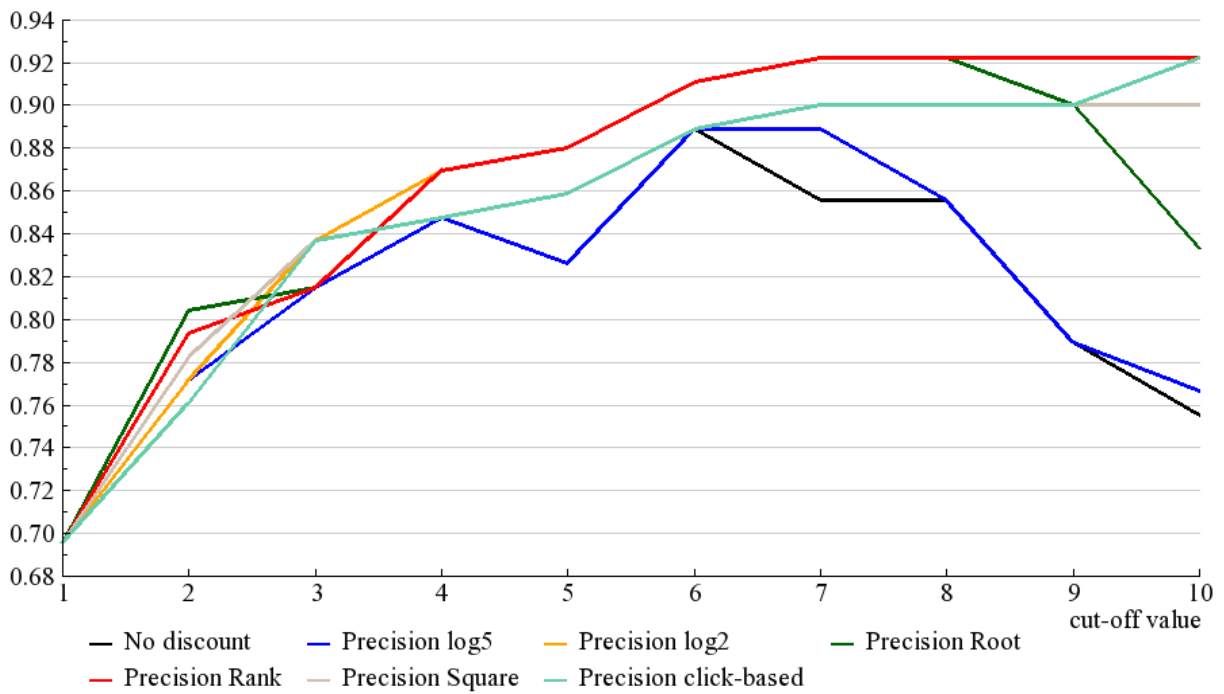

**Figure 10.15. Precision PIR scores for different discount functions, with the threshold values set to zero. The graph is identical to the one in Figure 10.9 (NDCG PIR scores with *t*=0).**

## 10.3 (Mean) Average Precision

With NDCG and precision dealt with for the time being, let us turn to the third of the most popular metrics: Mean Average Precision.[84] I will present it in the now familiar way by introducing threshold graphs and then following up with discount-based comparisons.

$$\text{MAP} = \frac{1}{|Q|} \left( \sum_{Q_i} \frac{1}{|D|} \left( \sum_{r=1}^{|D|} rel(d_r) \frac{\sum_{k=1}^{r} rel(d_k)}{r} \right) \right)$$

**Formula 10.1. MAP formula with queries *Q*, relevant documents *R* and documents *D* (at rank *r*). *rel* is a relevance function assigning *1* to relevant and *0* to non-relevant results, or, in this study, one of six values in the range from 1 to 0.**

Like NDCG, MAP also has a built-in discount factor. Consider the MAP calculation, presented here as Formula 10.1. The discount function is present in the AP formula as the denominator *r*, signifying the rank of the result. It is this denominator that we will change in order to test different discount functions; the modified Formula 10.2 shows the discount function replacing the rank as denominator.

$$\text{MAP} = \frac{1}{|Q|} \left( \sum_{Q_i} \frac{1}{|D|} \left( \sum_{r=1}^{|D|} rel(d_r) \frac{\sum_{k=1}^{r} rel(d_k)}{disc(r)} \right) \right)$$

**Formula 10.2. Modified MAP formula with queries *Q*, relevant documents *R* and documents *D* (at rank *r*). *rel* is a relevance function assigning *1* to relevant and *0* to non-relevant results, or, in this study, one of six values in the range from 1 to 0. *disc(r)* is a discount function depending on the rank *r*.**

Note that it is not the relevance of later results that is being discounted, but rather the importance of later addends. In a way, the formula is not really using Precision any more; the addends (which are later averaged) may range from a simple sum of relevance judgment sums (if no discount is used) to a sum of cumulative gains discounted, for later addends, by the square of their rank. If the function discounts by rank, it is obviously the same as standard MAP.[85] Also note that for *disc(r)<r*, MAP does not necessarily fall into the [0..1] range. This is not a problem for the present study for two reasons. Firstly, this would only have a significant effect for consistently high relevance ratings on all ranks, which hardly ever happens. Second, even where MAP scores do exceed 1, they will do so for both result lists of a comparison, thereby preserving the preference.[86]

The two graphs showing MAP by threshold are somewhere between those for NDCG and those for Precision. Both Figure 10.16 (no discount) and Figure 10.17 (discount by rank, that is, traditional MAP) show a marked PIR decline at higher thresholds, though it is not as steep as that of Precision PIR graphs. Otherwise, they are rather unspectacular, which is why I omit the detailed threshold graphs for the other five discount functions and proceed straight to the comparative graphs of the different discount metrics.

---

[84] As mentioned before, for our purposes (without cross-query comparison), MAP is the same as Average Precision. I will use the more familiar term MAP throughout this work.
[85] Well... Not quite. In this first batch of evaluations, a six-point relevance scale is used. While this usually only influences the metrics by providing a more faceted relevance distribution, AP is different. The relevance function occurs twice in Formula 10.2; the second time it is as a multiplier for the entire precision fraction. Thus, its influence is higher than for other metrics. This issue will be dealt with in Section 10.7, which concerns itself with relevance scales.
[86] The only problem could be threshold steps, which might be too small to reach an optimum level. However, the graphs in this section clearly show that this is not the case.

With MAP, many results are analogous to those of NDCG. The graphs using the $t=0$ and the best-threshold approach are quite similar; and the peak scores lie at about 0.92 once again. The tendency of PIR scores to decline after a peak has been reached is less pronounced; it only happens at the very last rank, and only with two curves. These two discount functions, "No discount" and "log5", are also the best-performing ones. The equivalent of traditional MAP, the "Rank" discount, lies in the middle, and the only discount function it constantly outperforms is the "Square" discount. All in all, the tendency in this case seems to be for more steeply discounting functions to perform worse, while the tendency seemed to go in the opposite direction in NDCG evaluations.

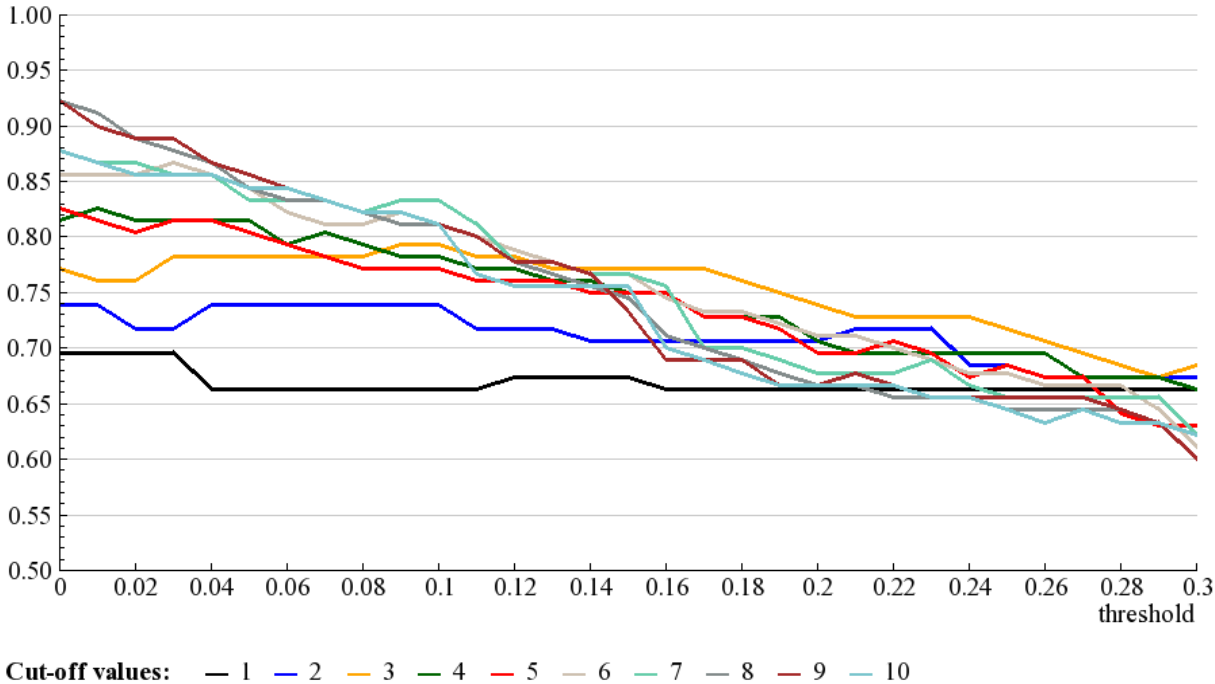

**Figure 10.16. MAP PIR values without any discount for results at later ranks.**

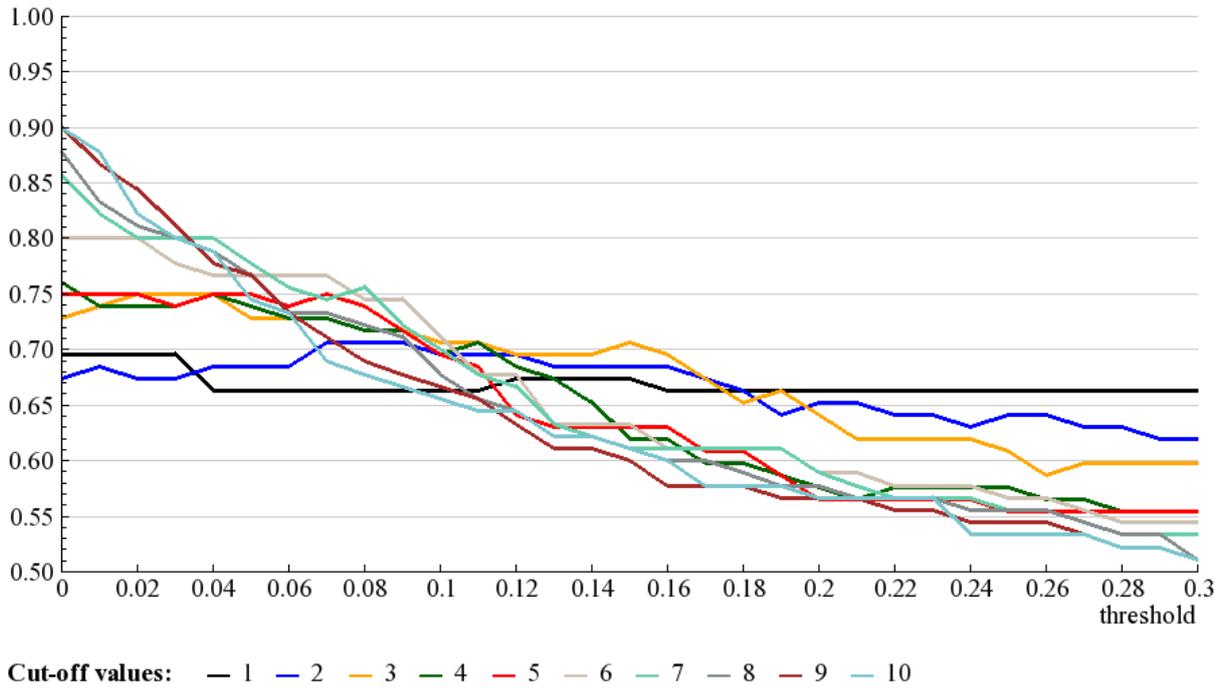

Figure 10.17. MAP PIR values discounted for the rank of the result (as in the traditional MAP calculation).

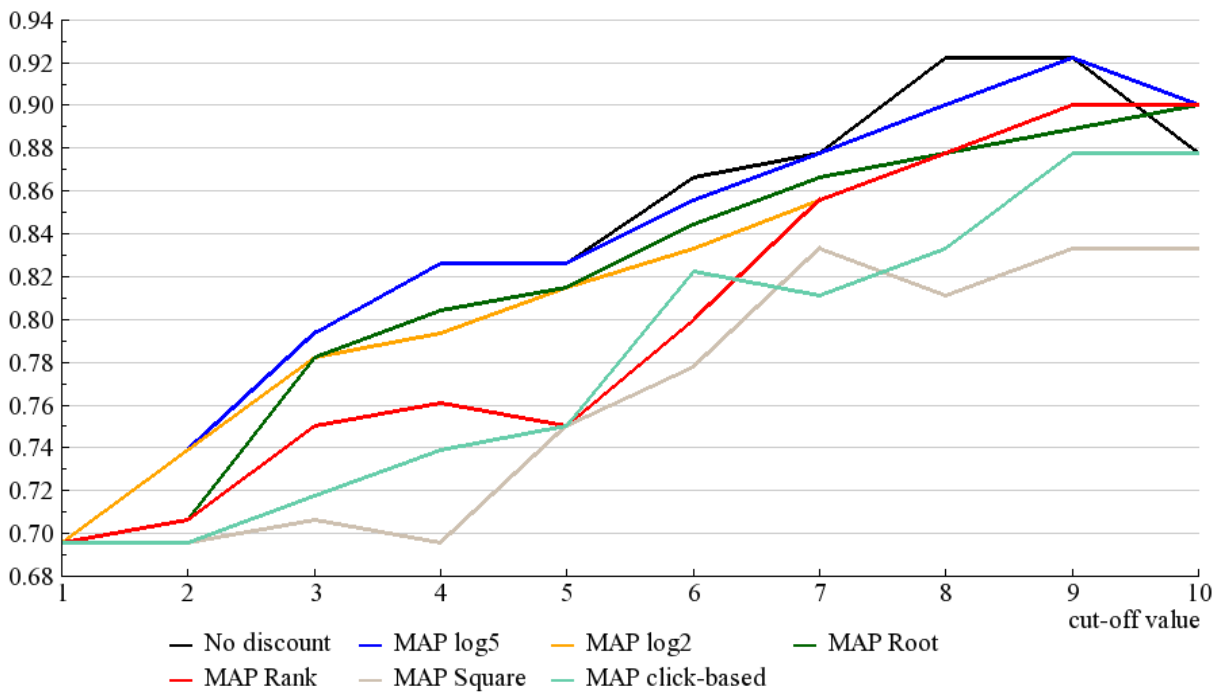

Figure 10.18. MAP PIR scores for different discount functions, with the best-threshold approach. MAP Rank is the traditional MAP metric. As many of the discount functions often produce identical PIR scores, a description is needed. Due to the discount function definitions, the "No discount", "$\log_5$" and "$\log_2$" conditions produce the same

results for cut-off values 1 and 2, and "No discount" and "log$_5$" have the same values for cut-off values 1 to 5. Also, "log$_2$" has the same scores as "Rank" for cut-offs 7 and 8, and the same as "Root" for cut-off values 8 to 10.

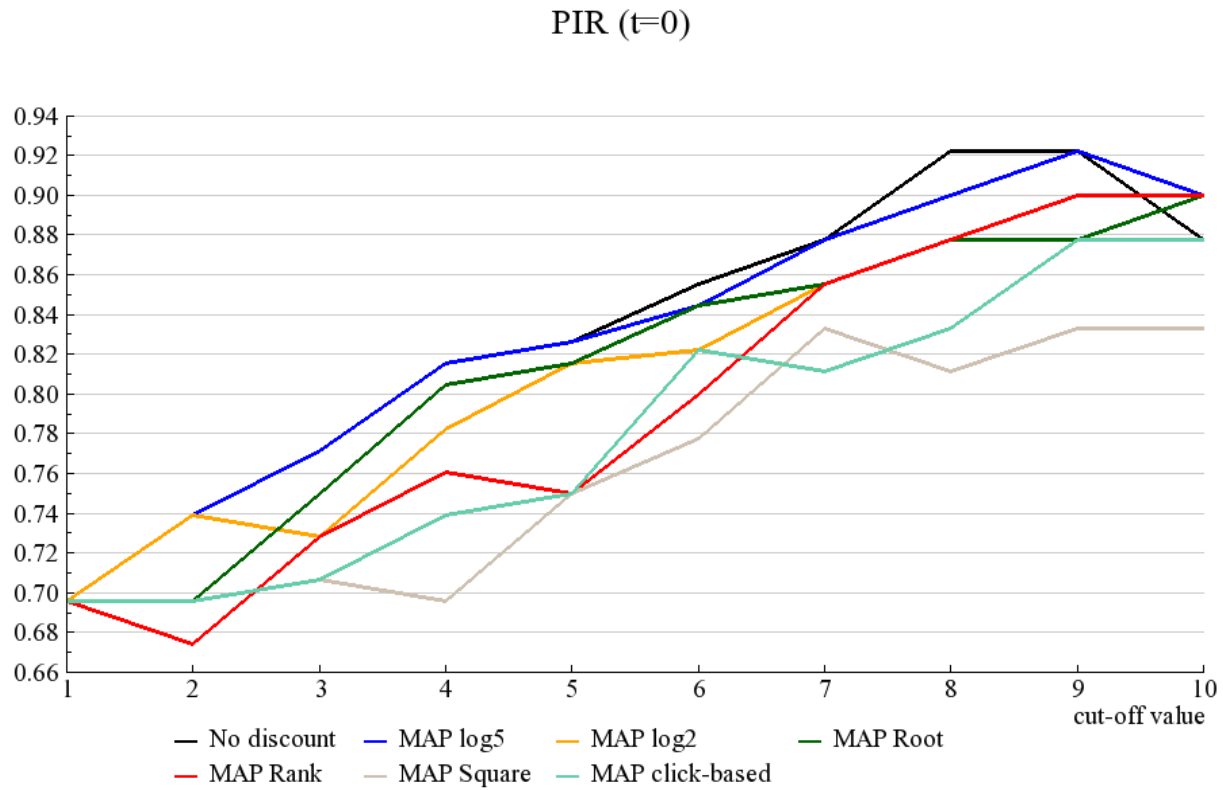

**Figure 10.19.** MAP PIR scores for different discount functions, with the threshold values set to zero. Again, MAP Rank is the traditional MAP metric.

For later inter-metric comparison, two discount functions are of interest. One is "Rank"; it is the default, widely employed version of MAP, which makes it important. The other is "No discount", which significantly outperforms "Rank", at some cut-offs by as much as 0.1 PIR points. Its calculation is given in Formula 10.3; in practice, it is really just a mean of relevance sums, devoid of any explicit discount.

$$MAP_N = \frac{1}{|D|} \sum_{r=1}^{|D|} rel(d_r) \sum_{k=1}^{r} rel(d_k)$$

**Formula 10.3.** MAP calculation for the "No discount" version.

As with NDCG, I would like to compare these results with those for informational queries only, shown in Figure 10.20. The difference to Figure 10.18 (MAP PIR for all types of queries) can be discerned with the naked eye; while the overall shapes of the individual lines are similar, they lie consistently higher for the informational queries. The reason for this is not quite clear. As returning to the graph the few navigational, factual and meta-queries does not change the scores in any way, the only remaining influence are the transactional queries. The main difference between informational and transactional queries is that, while with the former the user is generally supposed to have an information need to be satisfied by multiple sources, the "transactional need" in the latter case is often satisfied by the first good result. By counting relevant results multiple times (as part of precisions for all following ranks), MAP might be more sensitive to changes where one result may make a big difference.[87]

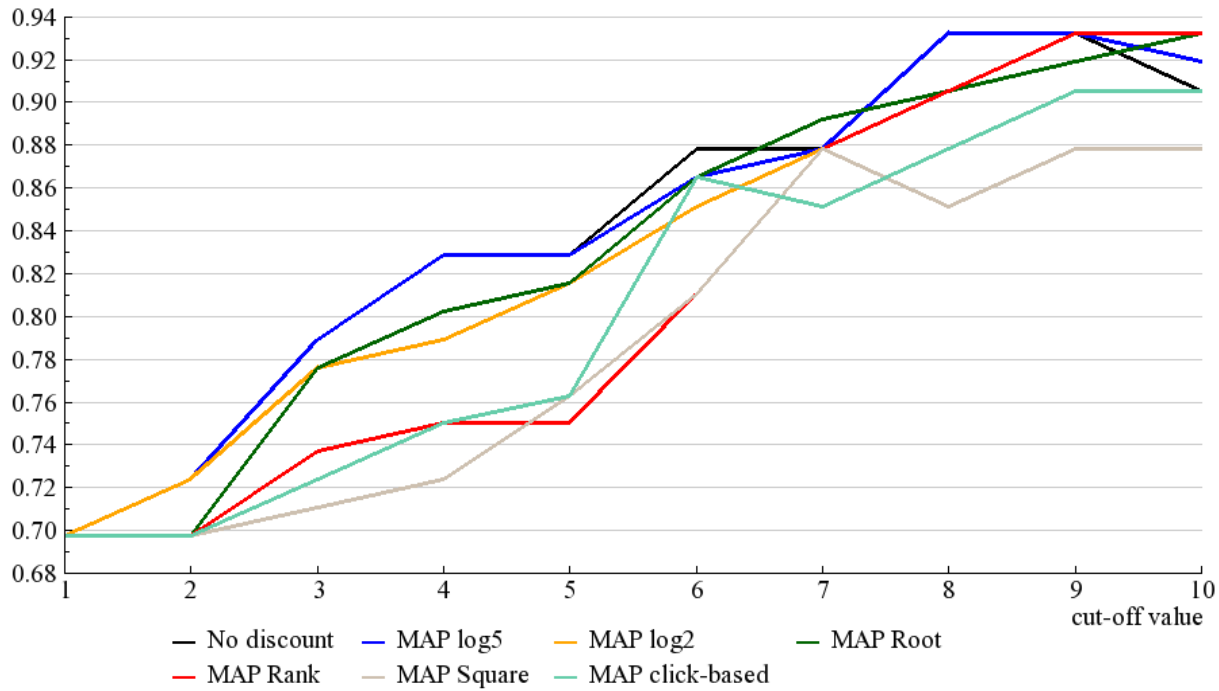

**Figure 10.20. MAP PIR scores for informational queries only, with the best-threshold approach.**

## 10.4 Other Metrics

With the more popular metrics safely tucked, we can now turn towards others, which have not been used as extensively in evaluations. These metrics sound promising on paper, and it is not easy to tell whether they have not been used more often because other researchers found some intrinsic flaws in them, or because they did not see why they should abandon more familiar metrics, of perhaps just because of some unlucky circumstances.

$$ERR = \sum_{r=1}^{n} \frac{1}{disc(r)} \prod_{i=1}^{r-1} (1 - \frac{2^{rel(d_i)} - 1}{2^{rel_{max}}}) \frac{2^{rel(d_r)} - 1}{2^{rel_{max}}}$$

**Formula 10.4. ERR formula. For each rank *r*, the probabilities of a relevant result at each earlier rank *i* are multiplied; the inverse probability is used as a damping factor for the gain at the current rank.**

The first of these metrics is Expected Reciprocal Rank (ERR). To recapitulate briefly, its outstanding feature is a built-in user model which assumes that the more relevant results the user found already, the less important the relevance of further results becomes; Formula 10.4 shows the ERR calculation with a flexible discount function *disc(r)*. As I have often emphasized the importance of a user-based approach in Part I, ERR makes for a promising candidate. How, then, does it performs in terms of PIR?

---

[87] This is only a possible explanation, and a post-hoc one at that. Theoretically, it could be argued with equal logic that the increased influence of any single relevant result is most visible in cases where there are many such relevant results, as the scores add up faster. Clearly, more research is needed on this topic.

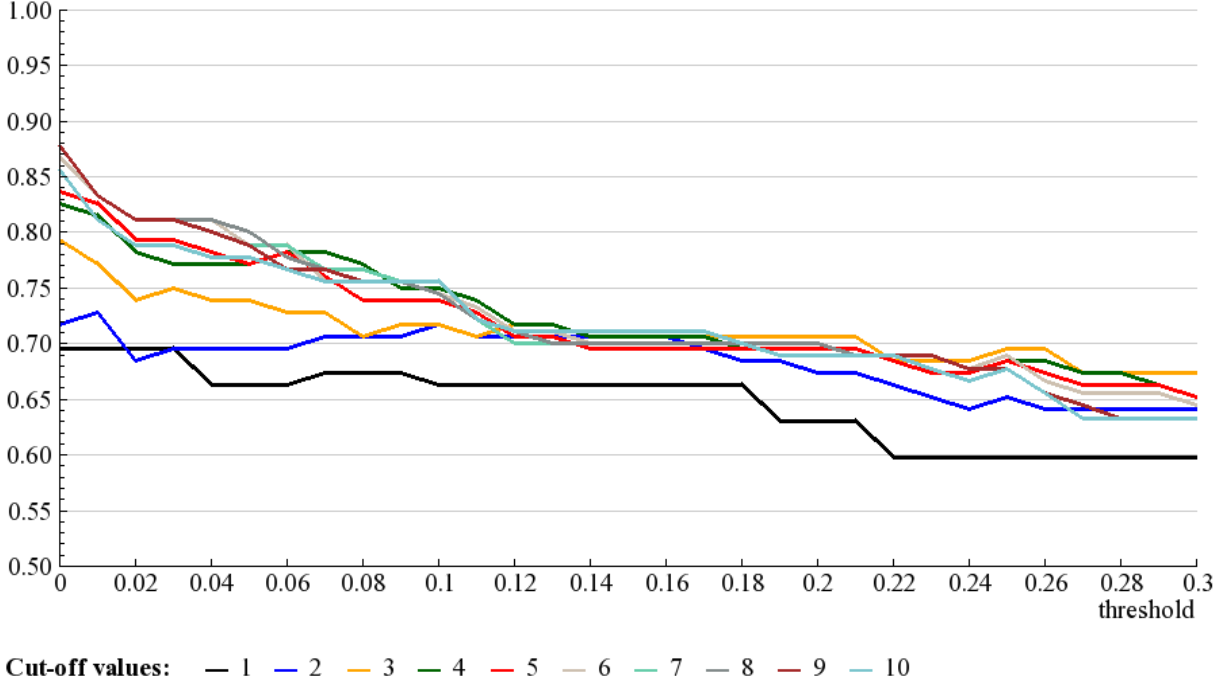

**Figure 10.21. ERR PIR values with discount for result rank.**

The threshold-dependant graph in Figure 10.21 (discounted by rank, which corresponds to the standard ERR metric formula) is typical for ERR, so it will be the only one of its kind shown. It has a by now surely familiar pattern of PIR score decline with increasing threshold values, which is steeper than in NDCG but not as steep as in MAP.

In the comparative graph (Figure 10.22),[88] ERR also offers a picture similar to what we have seen with the previous metrics. All discount functions have their peak at rank 6 or 7, and their scores start to decline at the last ranks. However, it is harder to discern a pattern regarding the different discounts' relative performance. While high-discount functions performed best in NDCG, and low-discount functions in MAP, there is no clear regularity in ERR. While most low-discount functions do perform worse than average, the $\log_2$ discount has some of the highest scores (at least up to rank 6). The rank-based discount (used in standard ERR) lies quite reliably in the middle of the group over all cut-off ranks.

---

[88] The $t=0$ graph is omitted, since it shows the same features as previous $t=0$ graphs, being almost the same as the corresponding best-threshold graph, with minimally lower scores.

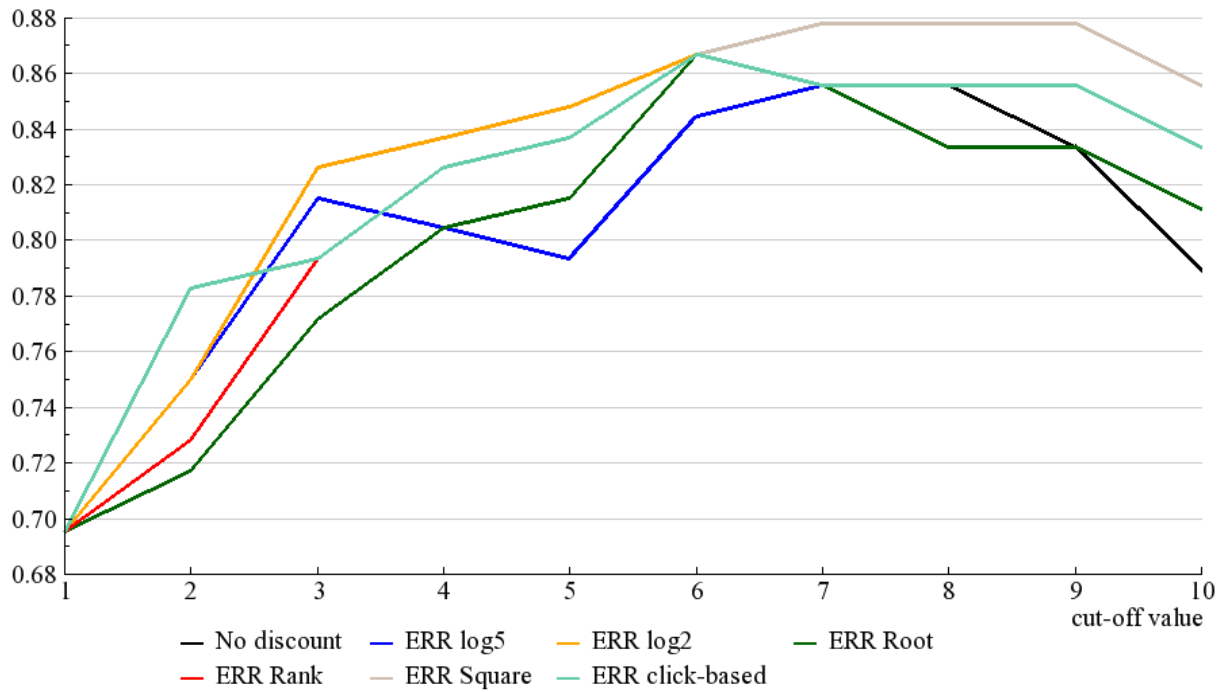

**Figure 10.22. ERR PIR scores for different discount functions, with the best-threshold approach. ERR Rank is the traditional ERR metric. As many of the discount functions often produce identical PIR scores, a description is needed. Due to the discount function definitions, the "No discount", "$\log_5$" and "$\log_2$" conditions produce the same results for cut-off values 1 and 2, and "No discount" and "$\log_5$" have the same values for cut-off values 1 to 5. On the non-self-evident side, "No discount" and "$\log_5$" coincide further (up to rank 9), while "Rank" and "Click-based" values are the same from rank 4 onwards. "Square" and "Click-based" scores are the same for ranks 1-6. Also, "$\log_2$" and "Click-based" coincide at ranks 6 to 10.**

As with MAP, ERR PIR scores increase if only informational queries are evaluated (Figure 10.23). Here, the argument that the metric could have been expected to perform better for this type of query could be made as well as with MAP; after all, ERR discounts for previously found results, and informational queries are the ones where multiple results can be expected to be needed. However, as well as with MAP, the logic might have run the opposite way had the results been different.[89]

---

[89] After all, the discount for highly relevant previous results is warranted if it means later results are not very useful anymore – which is precisely the case with navigational, fact-based and (perhaps to a lesser extent) transactional queries.

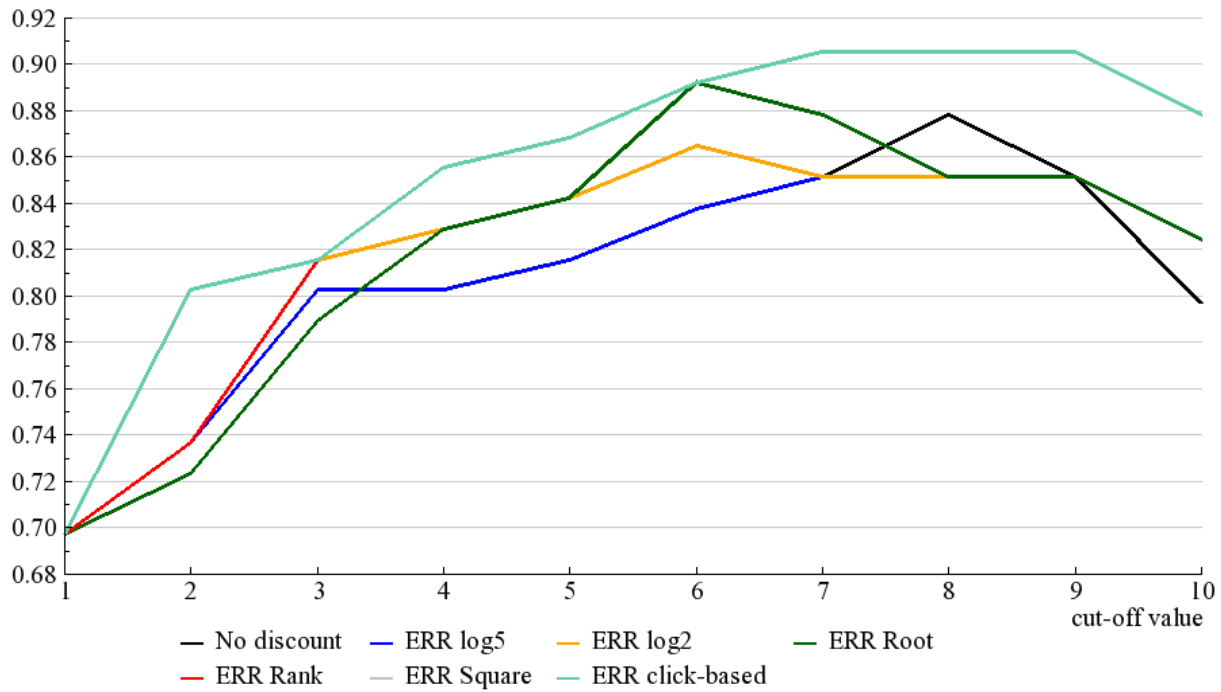

**Figure 10.23. ERR PIR scores for informational queries only, with the best-threshold approach.**

Another little-used metric is Mean Reciprocal Rank (MRR), defined as one over the first rank at which a relevant result is found. The discount comparison graph (Figure 10.24) differs from the ones we have encountered before in a few respects. First, it covers a narrow range of PIR scores; the difference between the highest and the lowest is below 0.15 points. Secondly, the scores are very low in absolute terms; the peak is below 0.65, and the lowest scores lie barely above the baseline of 0.5. Thirdly, while most previous graphs showed PIR scores rising (mostly until ranks 6-8) and then declining, MRR PIR scores hardly rise at all. The high-discount functions' scores rise from 0.63 at rank one by just about 0.03 points; the low-discount functions' scores decline sharply after rank one. And finally, the Root, Rank, Square and Click-based discounts have precisely the same PIR scores.

Fortunately, there is an obvious explanation for all of these phenomena. In preference terms, MRR selects the better result by looking at the result lists and picking the one where the first relevant result lies in an earlier rank. Therefore, it is irrelevant how strong the discount is, as long as there is any discount at all. When there is none (as in the $\log_2$, $\log_5$ and especially no-discount conditions), the discriminatory power of the metric actually diminishes; the no-discount condition at cut-off rank 10, for example, can only differentiate between two result lists if one of them has at least one relevant result in the top ten ranks and the other has none. The problem is aggravated by the broad definition of "relevant" used in the current batch of evaluations; any result which is not completely useless counts. We will examine the influence of different definitions of "relevance" in Section 10.7.

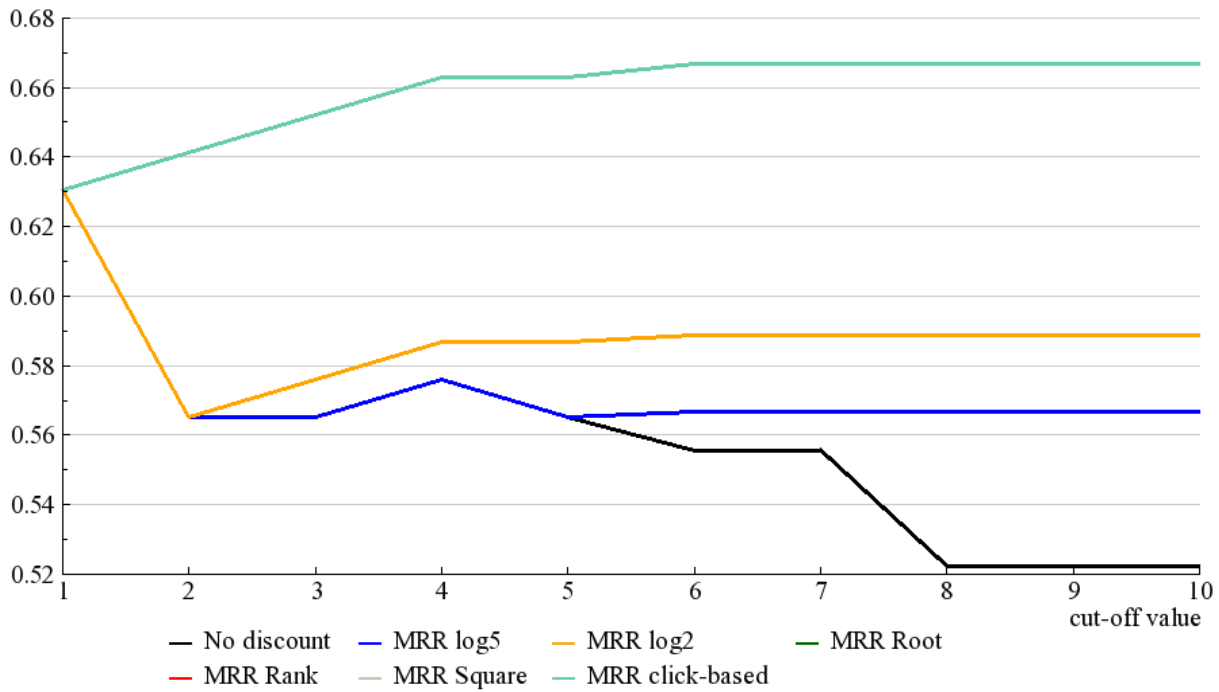

Figure 10.24. MRR PIR scores for different discount functions, with the best-threshold approach. MRR Rank is the traditional MRR metric. As many of the discount functions often produce identical PIR scores, a description is needed. Due to the discount function definitions, the "No discount", "log$_5$" and "log$_2$" conditions produce the same results for cut-off values 1 and 2, and "No discount" and "log$_5$" have the same values for cut-off values 1 to 5. "Root", "Rank", "Square" and "Click-based" have the same scores at all cut-off ranks.

As mentioned in Section 4.3, a more general analogue of MRR is the Expected Search Length (ESL), defined as the number of non-relevant results a user has to examine before finding *n* relevant ones. With non-binary relevance, and a need for the metric's scores to fall into the usual 0..1 range as well as to be usable with a discount function, a modified definition is needed. It is introduced in Formula 10.5, which attempts to capture a relatively simple concept. Unfortunately, it seems to loose quite a lot of simplicity, but hopefully none of the concept.[90]

$$ESL = 1 - \frac{r_n - \sum_{i=1}^{r_n} \frac{rel(i)}{disc(i)}}{c}$$

**Formula 10.5. Expected Search Length adapted for graded relevance and discount.** *r* is the rank at which the sum of single result relevance scores reaches a threshold *n*. The relevance scores are obtained by dividing the user relevance score for rank *i* by a discount *disc(i)* dependant on the same rank. *c* is the cut-off value used.

The numerator reflects the number of non-relevant results a user has to examine before satisfying his information need. For this, we take the number of all results examined (the rank *r*) and subtract the relevant results. This number is then divided by the cut-off value to normalize it since the worst case is having nothing but non-relevant results. In this case, the fraction is 1, while in the optimal case (e.g. a perfectly relevant result at first rank for a required cumulative relevance threshold of 1) it is 0; therefore, the fraction is subtracted from 1 to revert the score.[91]

---

[90] The metric given in Formula 10.5 might more appropriately be called "*Normalized* Expected Search Length". However, for simplicity's sake I will continue to refer to it as ESL.

However, as we have a multi-graded relevance scale, instead of simply counting results which have any relevance, I opted for using the sum of single relevance scores. In this way, a result list which provides perfectly relevant results until the user's information need is satisfied still receives an optimal score of 1, one without any relevant results receives a score of 0, and one providing only results with relevance 0.5 would also get an ESL score of 0.5. That is, at least, if no discount is used; otherwise, the score is lower. It might not be intuitive (or advisable) to use a discount for ESL, as the rank $r_n$ already provides some discount – if good results appear earlier, the information need is satisfied faster. However, a few good results up front and one at the end are presumably better than all good results being stuck at the lower end of the result list. Another objection against using a discount is the very nature of ESL, which assumes rather plausibly that the user looks for information until he is satisfied, and then stops. In this case, he either sees the result or he does not; thus, either no discount should apply (before $r_n$), or the results should not be regarded at all (after $r_n$). However, this model does not take into account that the user may abandon his search before fully satisfying his information need; from this viewpoint, a discount is a factor incorporating the increasing likelihood of query abortion at later ranks. Ultimately, as usual, it will be up to the evaluation to show which method works best; and we always have the no-discount function to fall back on if discounts turn out to be a bad move.

---

[91] For any discount function (other than "No discount", that is), and cut-off values larger than 1, the maximum final score is actually lower than 1. However, as with the opposite case we have seen in the MAP calculation, this would only be a problem if the threshold steps were wrong (in this case, too coarse). However, there is no evidence of that problem in the data.

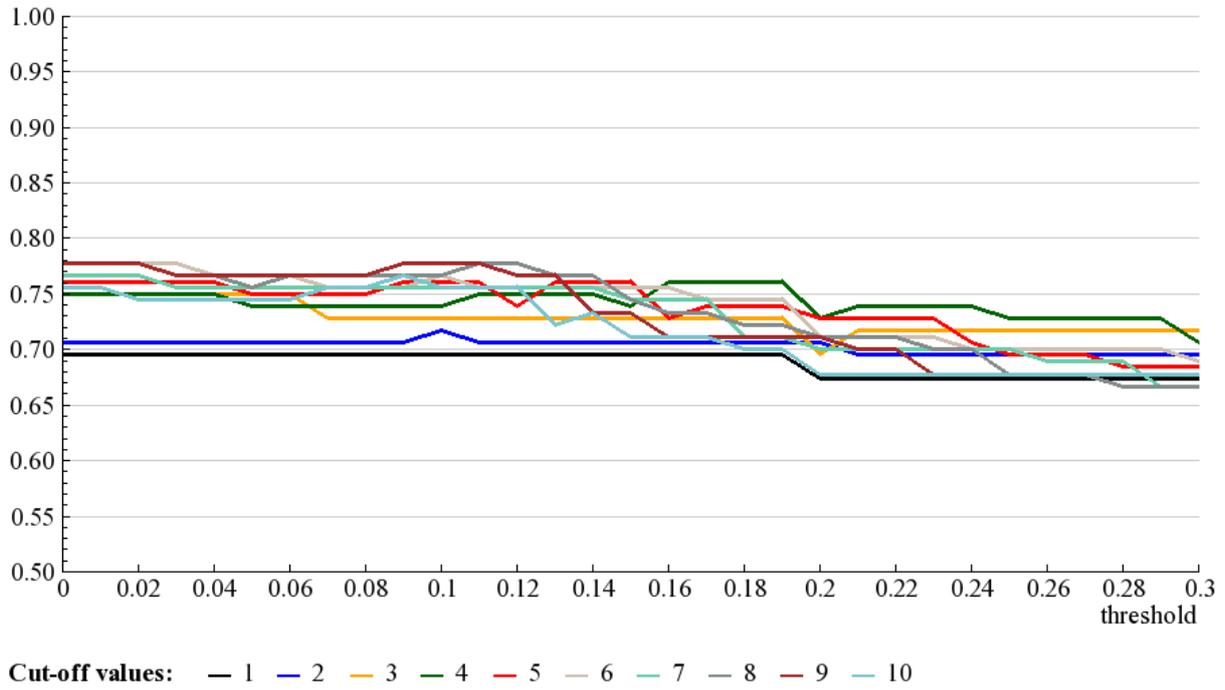

**Figure 10.25.** ESL PIR values without any discount for results at later ranks with required cumulated relevance *n*=1.

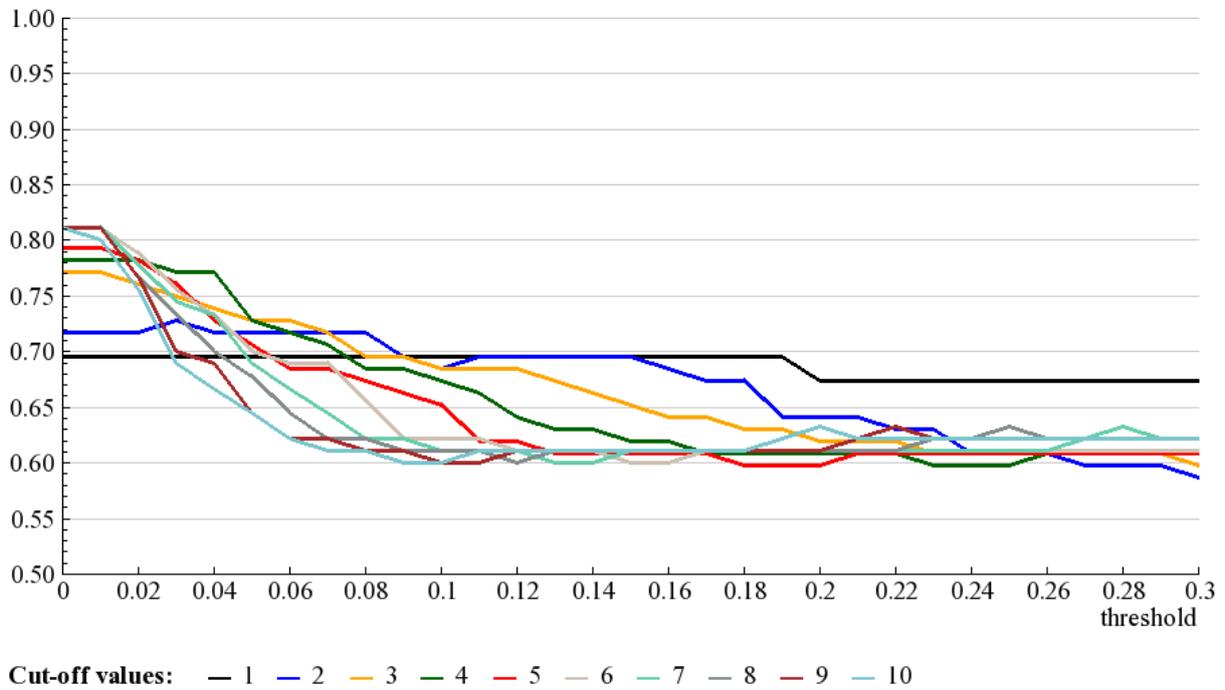

**Figure 10.26.** ESL PIR values discounted by the frequency of clicks on results of a particular rank with required cumulated relevance *n*=1.

Figure 10.25 and Figure 10.26 show a marked difference when compared to graphs we have encountered before, the individual scores at any cut-off value show much less variability at different threshold values. In the no-discount condition, the difference between the highest PIR score of the best-performing metric and the lowest PIR score of the worst-performing metric is just about 0.1. In the most divergent case, with the discounts being click-based, this number rises to just over 0.2. Remarkably, while the decline occurs at low threshold values and proceeds steeply for higher cut-off values, once the line reaches a PIR score of about 0.6, its fall stops. The steepest fall is seen in the no-discount condition for a cut-off rank of 10. The PIR scores fall rapidly from ca. 0.81 for a threshold value of 0 to ca. 0.61 for a threshold value of 0.08 – but then they stay at or even above this boundary up to the maximum threshold of 0.30. The probable reason for this is that the most significant factor in ESL calculation is $r_n$, the rank at which the predetermined cumulated relevance threshold is attained. That, in its turn, depends mainly on the cumulative relevance $n$, which will be considered in more detail below.

The inter-discount comparison in Figure 10.27 shows the "Rank" function performing best, with "No discount" (which is as close as we have to the "classical" ESL, considering the number of changes made to the formula) having the lowest scores. Generally, functions with lower discounts perform worse, though the "Square" and "Click-based" functions lie slightly below the maximum.

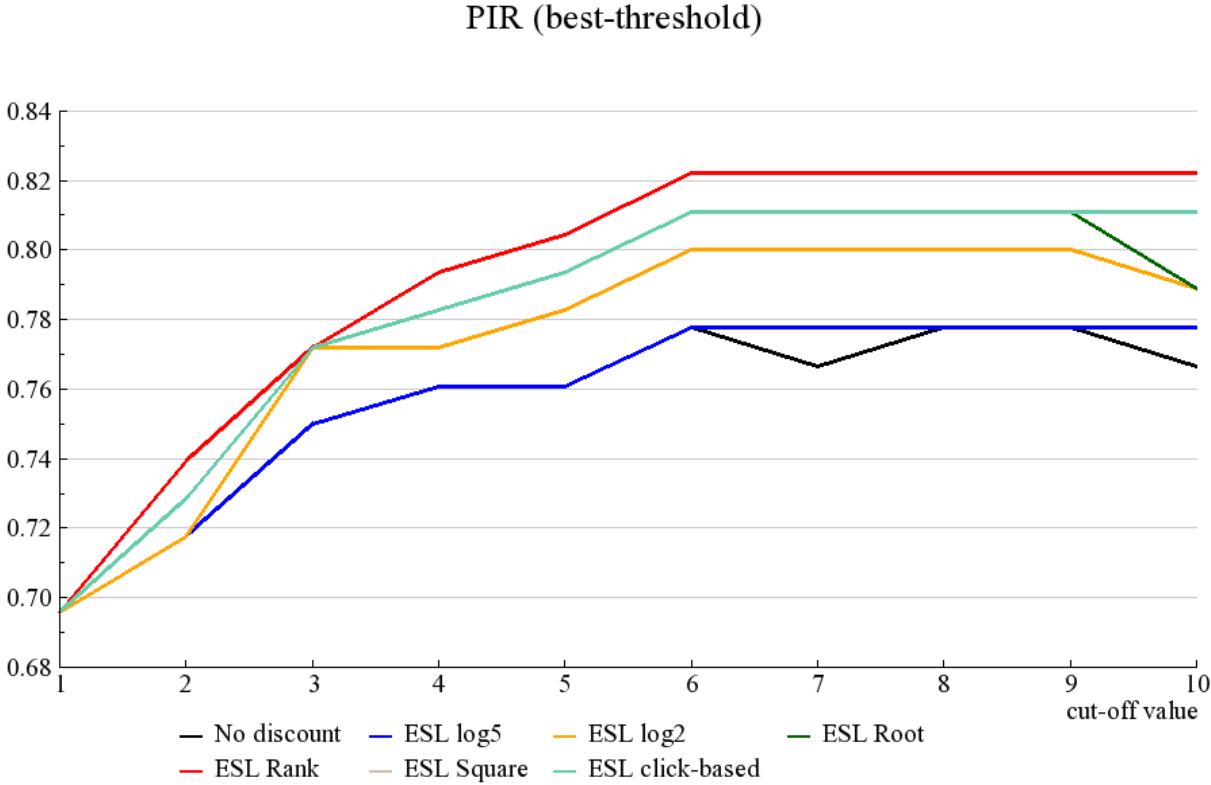

**Figure 10.27. ESL PIR scores for different discount functions with required cumulated relevance *n*=1 and the best-threshold approach. Per definition, the "No discount", "log$_5$" and "log$_2$" conditions produce the same results for cut-off values 1 and 2, and "No discount" and "log$_5$" – for cut-off values 1 to 5. "Root" scores coincide with those of "Rank" for cut-offs 1-3, and with those of "click-based" for 3-9. The "click-based" line is identical with "Square".**

However, the discount function is not the only way in which the ESL results can be modified; it has a parameter for regulating the required cumulated relevance, *n* (used in $r_n$). In the evaluation above, *n*=1 was used, analogous to looking for one top-quality result. It is then a logical next step to ask whether other values of *n* might produce better results. Figure 10.28 through Figure 10.31 show the PIR graphs for *n*=0.5 to *n*=2.5 (with *n*=1 in Figure 10.27). A tendency is easy to discern; higher values of *n*, that is, assuming that the user wants more information, lead to higher PIR scores. Again, low-discount functions tend to perform worse; and again (and even more in these functions), the PIR scores decline at the later cut-off values. For values of *n* higher than 2.5, the peak values stay the same, but scores tend to decline faster at later ranks. From these graphs, the "Rank" function is the obvious best candidate for inter-metric comparison since it generally performs as well as, and often better than other discount functions in terms of PIR.

As for scores depending on query type, Figure 10.32 and Figure 10.33 show PIR graphs for informational queries only. The general shape of the curves is the same; however, the absolute values are higher. This is especially visible for lower cumulative relevance values (for *n*=1, the peak PIR scores rise by about 0.03).

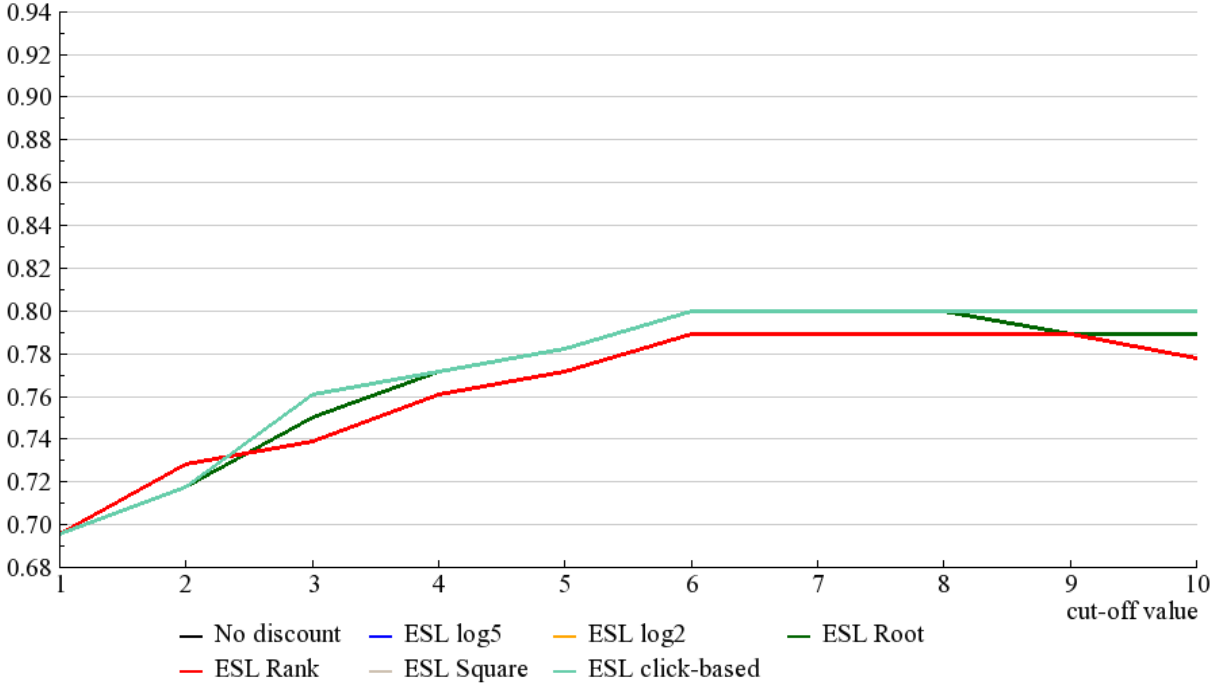

**Figure 10.28. ESL PIR scores for different discount functions with required cumulated relevance *n*=0.5, with the best-threshold approach.**

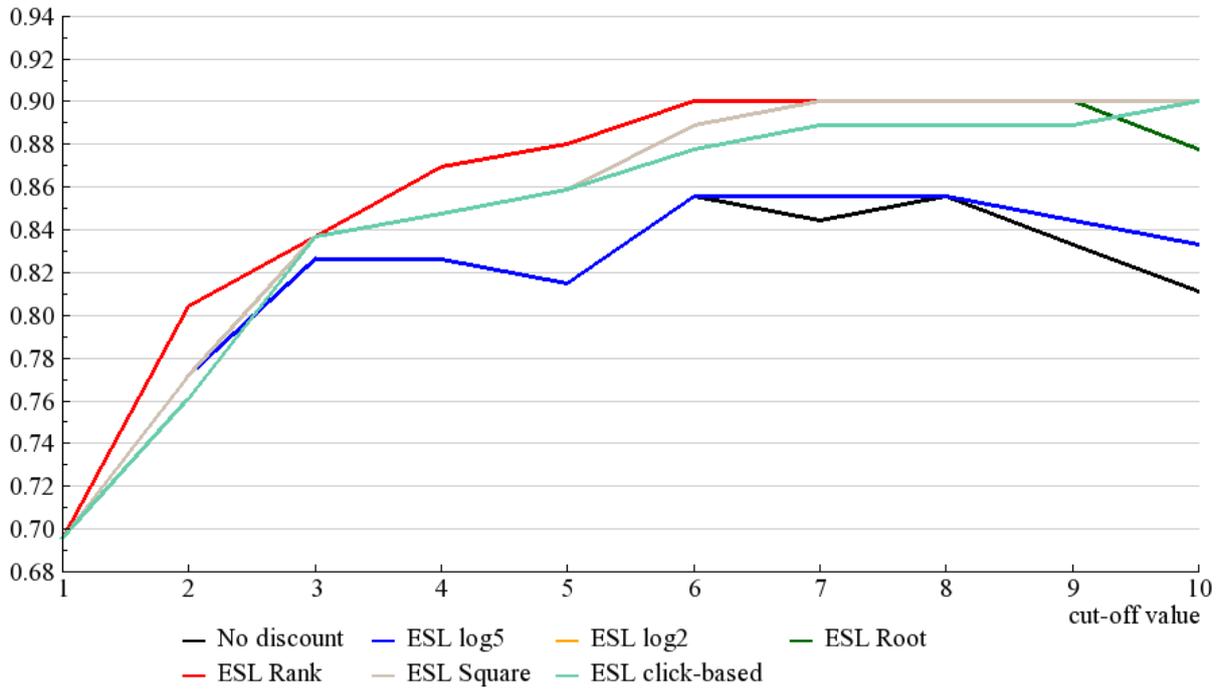

**Figure 10.29.** ESL PIR scores for different discount functions with required cumulated relevance *n*=1.5, with the best-threshold approach.

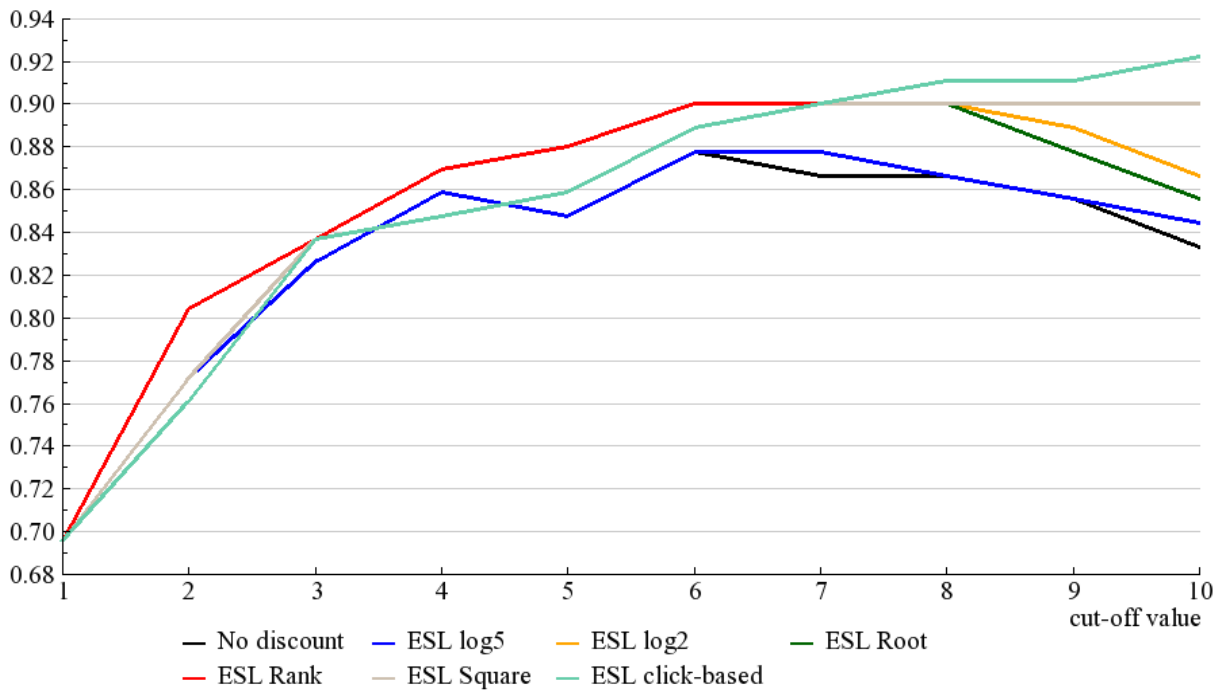

**Figure 10.30.** ESL PIR scores for different discount functions with required cumulated relevance *n*=2, with the best-threshold approach.

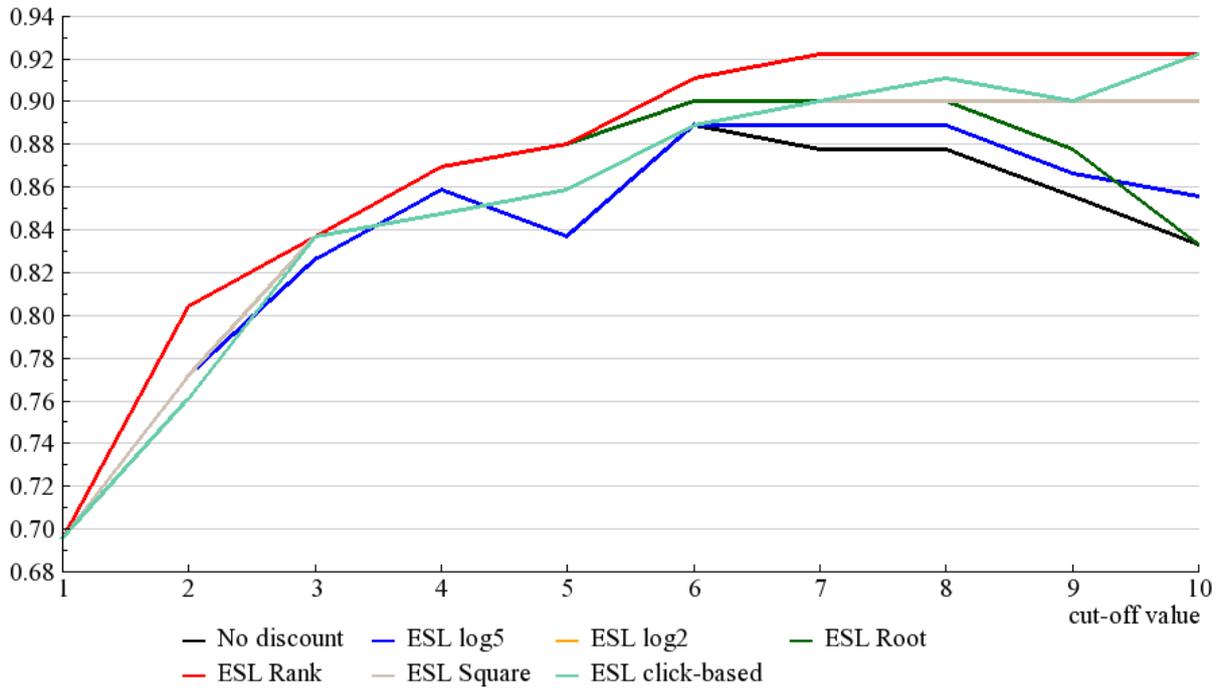

**Figure 10.31.** ESL PIR scores for different discount functions with required cumulated relevance *n*=2.5, with the best-threshold approach.

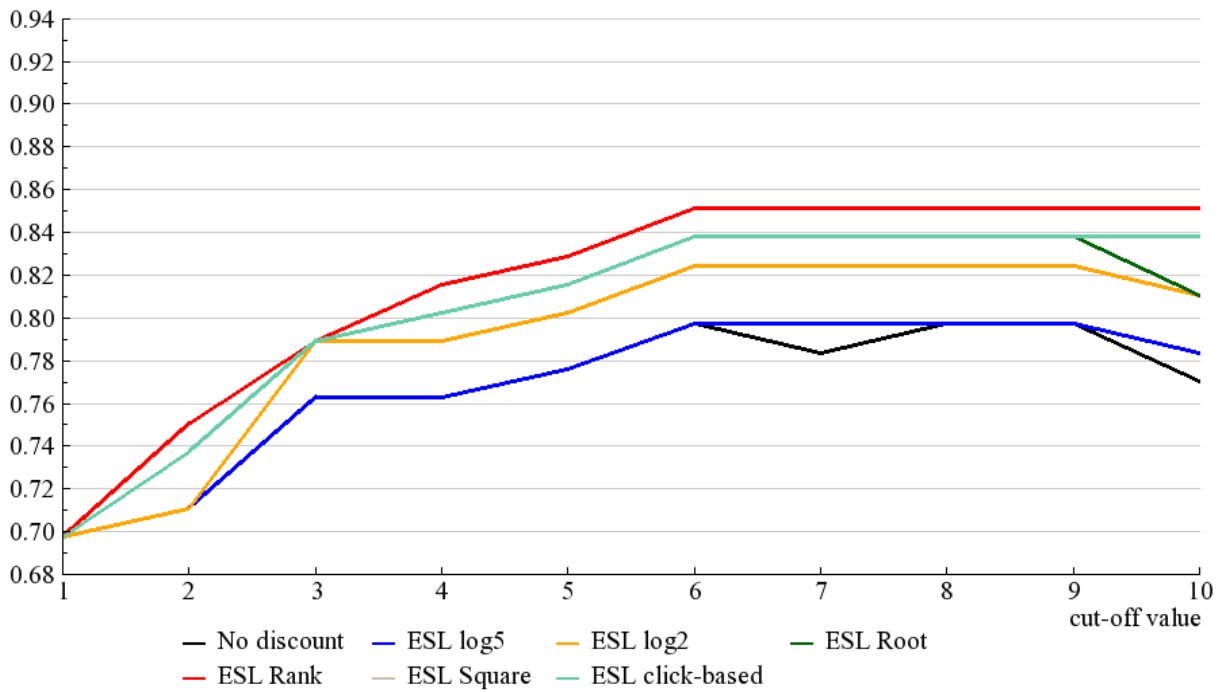

**Figure 10.32.** ESL PIR scores for informational queries only, cumulated relevance *n*=1 with the best-threshold approach.

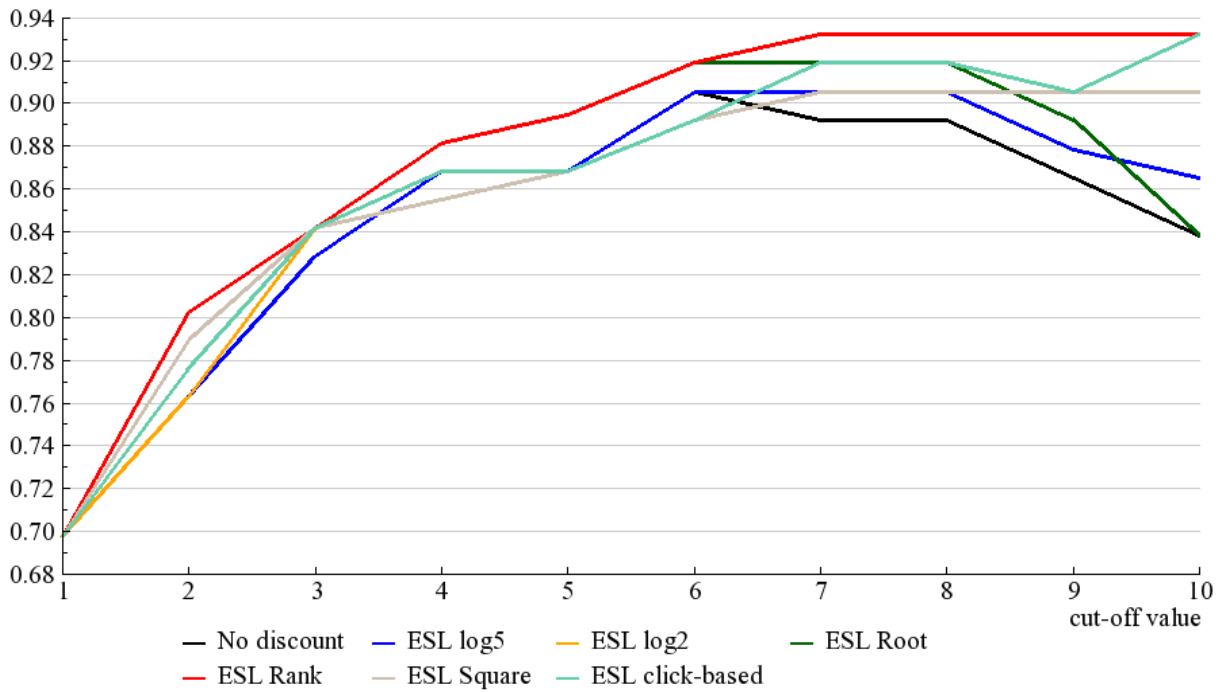

**Figure 10.33.** ESL PIR scores for informational queries only, cumulated relevance *n*=2.5 with the best-threshold approach.

## 10.5 Inter-metric Comparison

Now we come to the first iteration of the answer to a question which had set us off in first place: Which metric is best at predicting user preferences? To provide a visual answer, Figure 10.34 shows the PIR performance all the different metrics discussed in the preceding sections at a glance. In most cases, only one line is presented for each metric; however, MAP has two lines, one for the universally used discount function (by rank) and one for the best-performing (without any discount).

The graph provides many instantly visible results. MRR lies far below all the other metrics; it is better than chance at recognizing preferred result lists, but does so only in about $^2/_3$ of all cases, picking the worse result list one-third of the time. Because of this vast discrepancy, I will ignore MRR for the rest of this section when discussing the performances of individual metrics. All other metrics have peak PIR scores in the range from 0.88 to 0.92; nevertheless, many of them differ significantly.

The two metrics performing worst are ERR and traditional (that is, discounted by rank) MAP. ERR has the lowest peak value, 0.88. While MAP (Rank) reaches 0.90, it does so only at cut-off rank 9. Up to cut-off rank 7, it is the worst-performing metric. Somewhat surprising are the PIR values for classical Precision. While it does not attain the highest scores, it does very well up to rank 6; after that, though, the PIR declines sharply, to hardly 0.75 towards cut-off rank 10. The no-discount version of MAP takes off rather slowly, but overtakes Precision by cut-off rank 7 and reaches a PIR score of 0.92 at ranks 8 and 9.

The best-performing metrics are NDCG (with the standard discount function $\log_2$) and ESL (with rank-based discount and a cumulative relevance of 2.5). NDCG rises just a bit slower at the start, and declines after cut-off rank 8 while ESL stays stable, but over the middle distance, their scores are the same.

Apart from individual metrics' scores, there are two important questions to be raised about this data, both of which have been mentioned in the NDCG discussion in Section 10.1. The first of those is the shape of the individual curves, while the second concerns absolute values.

None of the metrics in Figure 10.34 starts with high PIR values at low cut-off ranks. Instead, there are two basic forms; either the scores rise up to a point, and then stay stable (MRR, MAP (Rank), and ESL (Rank, $n$=2.5) ), or they rise to a point, plateau, and the decline again (Precision, NDCG). MAP (No discount) and ERR (Square) only decline at the cut-off 10, the former more, the latter less strongly. The broad pattern that can be discerned is one of discount steepness; the metrics with steeper discounts tend to keep their high scores, while those with low or no discounts tend to do worse starting with cut-off rank 6 to 9. The no-discount MAP and ERR (Square) metrics do not quite conform with this observation, but they do not depart from it strikingly, either. But is there, I hear you cry out, any possible reason for this strange phenomenon? Indeed, there might well be.

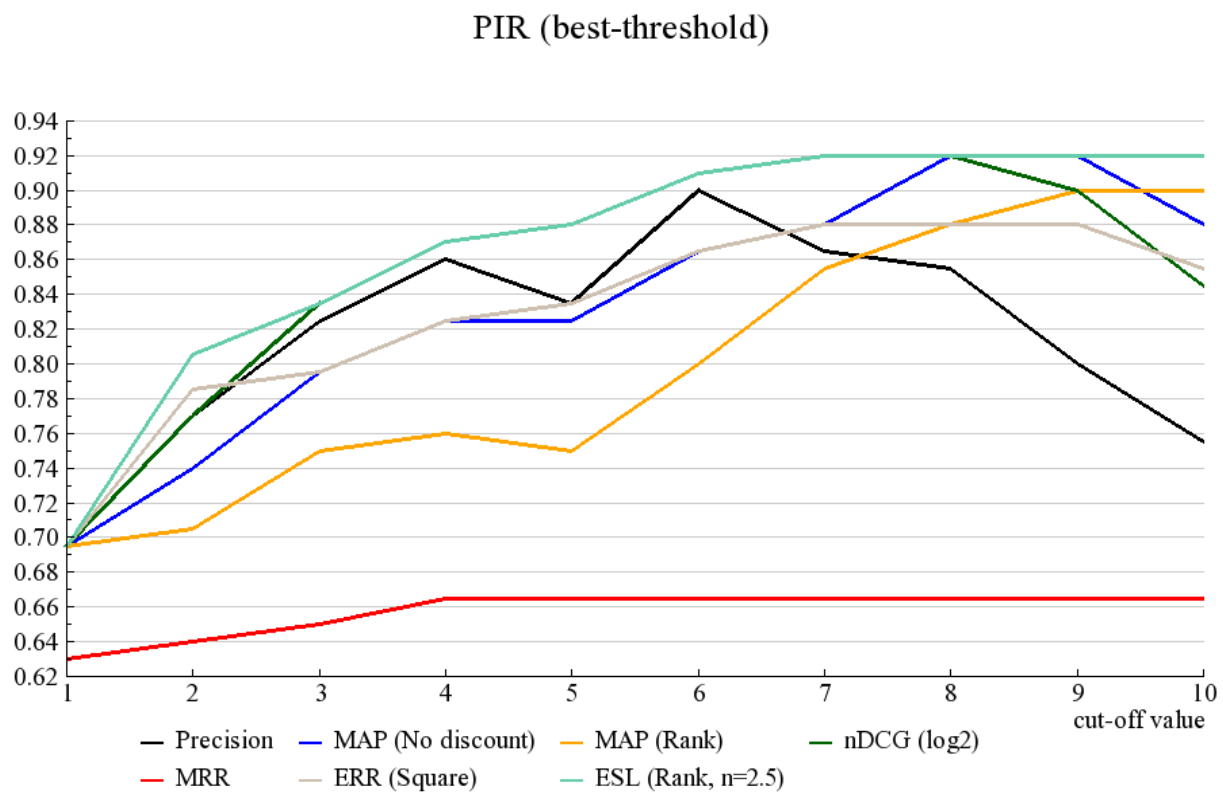

Figure 10.34. Inter-metric PIR comparison using the best-threshold method. NDCG ($\log_2$) and ESL (Rank, $n$=2.5) have the same values for cut-off ranks 3 to 8.

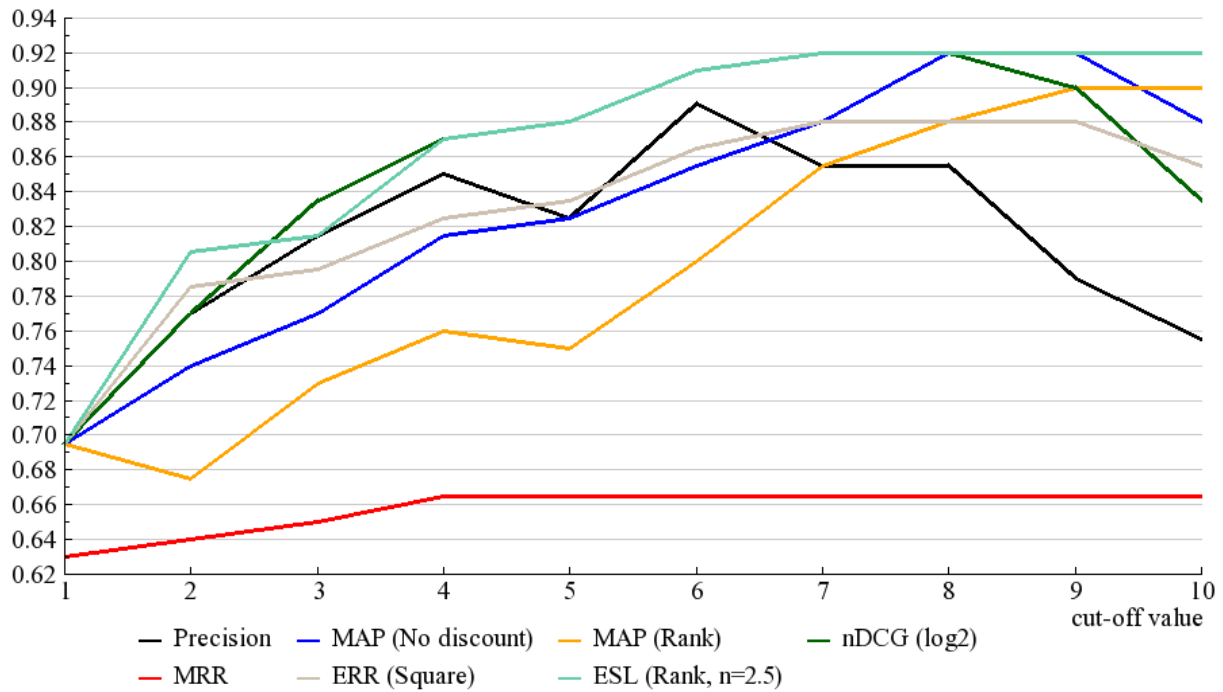

Figure 10.35. Inter-metric PIR comparison using the zero-threshold method. NDCG ($\log_2$) and ESL (Rank, $n$=2.5) have the same values for cut-off ranks 4 to 8.

Usually, it is assumed that it would be best to evaluate all available results if possible. In web search engines, this possibility rarely occurs, and the researchers settle for the second-best option – evaluating as many results as they can afford. But this may turn out to be a relic of good old days of professional databases used by professional researchers. Consider Precision, the most venerable of metrics. In olden days, the perfect solution would have been to evaluate all results returned by the information retrieval system. But, as study after study has shown, modern-day search engine users do *not* pay attention to all documents. Their attention dwindles rapidly, so that in most search sessions, a result at rank ten is unlikely to even be considered, much less clicked on. This means that if one result list has a top-quality page at rank 10, while a second has a completely irrelevant result, the user is unlikely to care (or even notice). Precision, however, regards this difference to be important – in fact, as important as a similar difference at rank 1. What does this mean for the relative evaluation of the two result lists? If the result at a later rank does not influence user preference, but is counted towards the metric score, the score may go up or down independent of the preference rating. And an indiscriminate fluctuation of scores pushes the PIR to its baseline of 0.5, which, as discussed in Section 8.2.1, constitutes precisely the case where the user preference and metric score are independent.

More generally, this line of reasoning suggests that, while later ranks have the potential to improve preference identification, this diminishes the farther away one gets from the first few results; the chance of leading the metric off course with a result ignored by the user will, in contrast, rise. If the metric takes too little notice of lower-ranked results, it might miss an opportunity to fine-tune; if it takes too much notice, it can be quite certain to destroy the fine tuning sooner or later by adjusting for a seemingly relevant result which is not even examined by the user. In short, a metric has to discount for results at later ranks in just the right way.

The result of this view on discounting is that considering only a small part of results returned by a search engine might not be a necessary evil, saving resources by diminishing the quality of the study. It might instead be a way to save resources while *improving* the study quality – a trade-off most researchers would not consider too disadvantageous.

The second important question is that of the metrics' absolute values. As has been already mentioned in Section 10.1, the scores seem quite high. There might be multiple explanations for that; for example, metric quality, query type, threshold advantages, lack of non-preferable result lists, and unrealistic rater availability, which I will discuss in the following paragraphs.

First, let us call to mind what the numbers actually mean. The "good" metrics (that is, every metric but MRR) peak at PIR scores of 0.88 to 0.92, so I will use a value of 0.9 as an average. A PIR score of 0.9 means that, given a query, two result lists one of which is preferable to the user, and explicit result judgments by this user, a metric will correctly pick the preferred result list nine times out of ten; one in ten result lists will be the less desirable one. Alternatively, the metric might pick the preferred result eight times out of ten, and return no preference for the other two results (leading to guessing and a preference recognition rate of 50% for those results).

The most straightforward explanation is that the numbers simply reflect the high quality of the metrics. After all, many of these metrics have been used extensively, often for decades, and there is no compelling intrinsic reason to assume they have to be poor predictors of user preference. However, previous studies examining the relationship between these metrics and user satisfaction and quality have produced results which are hard to reconcile with these high scores (e.g. Turpin and Hersh (2001) or Al-Maskari, Sanderson and Clough (2007); see Section 4 for detailed discussions of these studies).

The next possibility is that the mix of different query types used for the evaluation pushed the scores higher than they would be otherwise, assuming that different metrics should be used for different query types. We can check this by comparing Figure 10.34 with its informational-only counterpart (Figure 10.36), as we did before. The results for informational queries show slightly higher PIR scores than, but no major differences from, the all-query results.

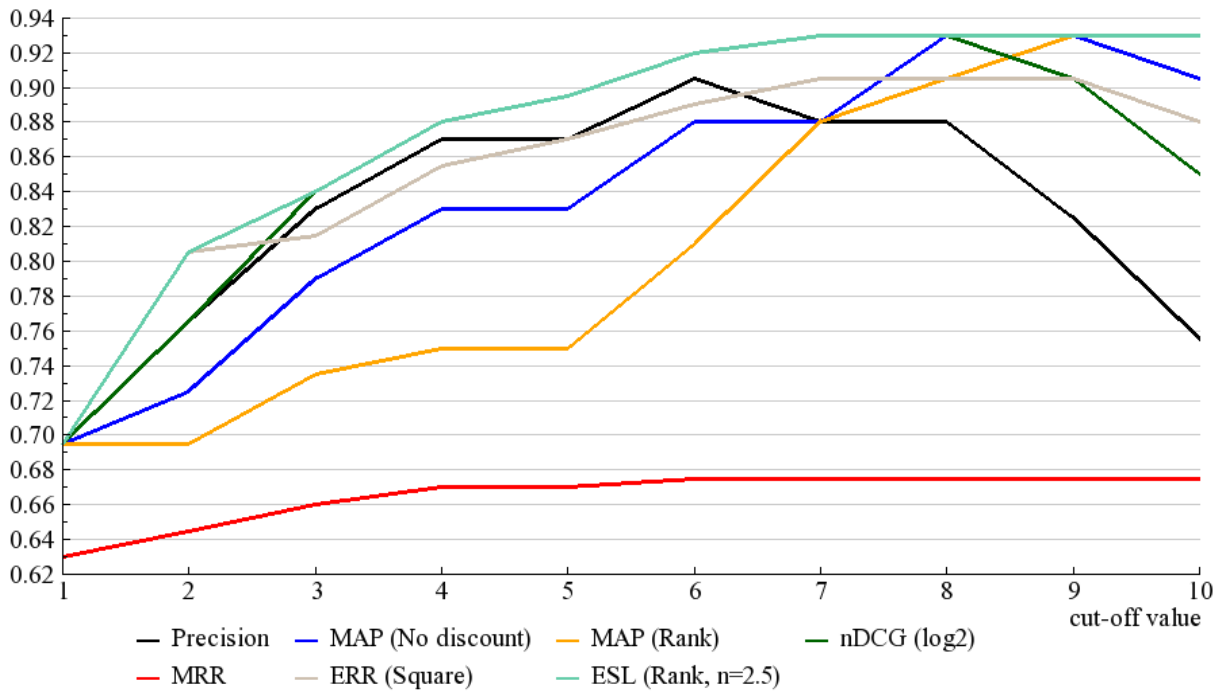

**Figure 10.36. Inter-metric PIR comparison using the best-threshold method and only informational queries. NDCG (log$_2$) and ESL (Rank, *n*=2.5) have the same values for cut-off ranks 3 to 8. Again.**

Another conceivable reason for the high scores is the best-threshold approach; by cherry-picking the best thresholds, the results are likely to be on the upper bounds of what is possible to determine on a relatively modest data set. This possibility can also be easily tested. In Figure 10.35, the zero-threshold approach is used instead of best-threshold, and the results are so similar that it takes a bit of time to find the differences.[92] The peak values remain practically unchanged; and so we can shelve this possibility for good.

A third option for doubting the high values comes from the pre-selected nature of result lists used for this evaluation. As explained in Section 8.2.1, result list pairs where the user does not have a preference have been excluded from the study. The reason for this is simple: if the user does not care which result list he gets, the metric can neither improve nor degrade the user experience. On the downside, this means that PIR measures only a specific sub-part of a metric's ability to correctly predict the user's judgments. The question of judging metric quality for other, more detailed evaluations will be discussed in Section 12.2.2.1. But we can already hypothesize the results in this case; as we have seen, most high PIR scores are reached at very low thresholds. However, low thresholds mean that we judge almost any difference in metric scores to be important; and the likelihood of assuming a user preference when there is none is quite considerable in these cases.

A final reason for possibly inflated scores is grounded in the relevance ratings and preference judgments used. As it is probably the most complex one, it has a section of its very own.

---

[92] Yes, there are some. I leave it to the reader to find them as a home assignment.

## 10.6 Preference Judgments and Extrinsic Single-result Ratings

As mentioned before, the results discussed in the previous section were all based on a comparison of result lists preferred by users and ratings of single results by the very same users. As was also briefly mentioned before, this is an unrealistic situation. If a user has already seen the individual results, he is unlikely to be interested in seeing result lists containing links to those results, even with the added benefit of abstracts. If an evaluation effort uses any single result based metric to determine the better result list, its ratings will not come from precisely the people for whom the result list is intended. Accordingly, a more realistic estimate of a metric's usefulness will come from the comparison of user preferences with metric scores based on *other* users' ratings. The calculation is straightforward; the standard PIR is used, as it has been stated in Formula 8.2, but with the relevance rankings of individual results being the average relevance ratings of all users except for the user whose result list preference is being used.

As in the same-user rating condition, I will start with threshold graphs. Only a few will be used, as the graphs show many of the same properties. One difference, however, can be seen on graphs such as Figure 10.37 (NDCG with $\log_2$ discount) and Figure 10.38 (NDCG with square-rank discount). When compared to their same-user-rating counterparts (Figure 10.3 and Figure 10.6), they show much more shallow curves for individual cut-off ranks. This is especially true for higher-discount functions; while the PIR scores diminish rapidly with increasing thresholds in the same-user condition, here, they stay relatively stable or even increase slightly at threshold values up to 0.1, and decline relatively slowly after that.

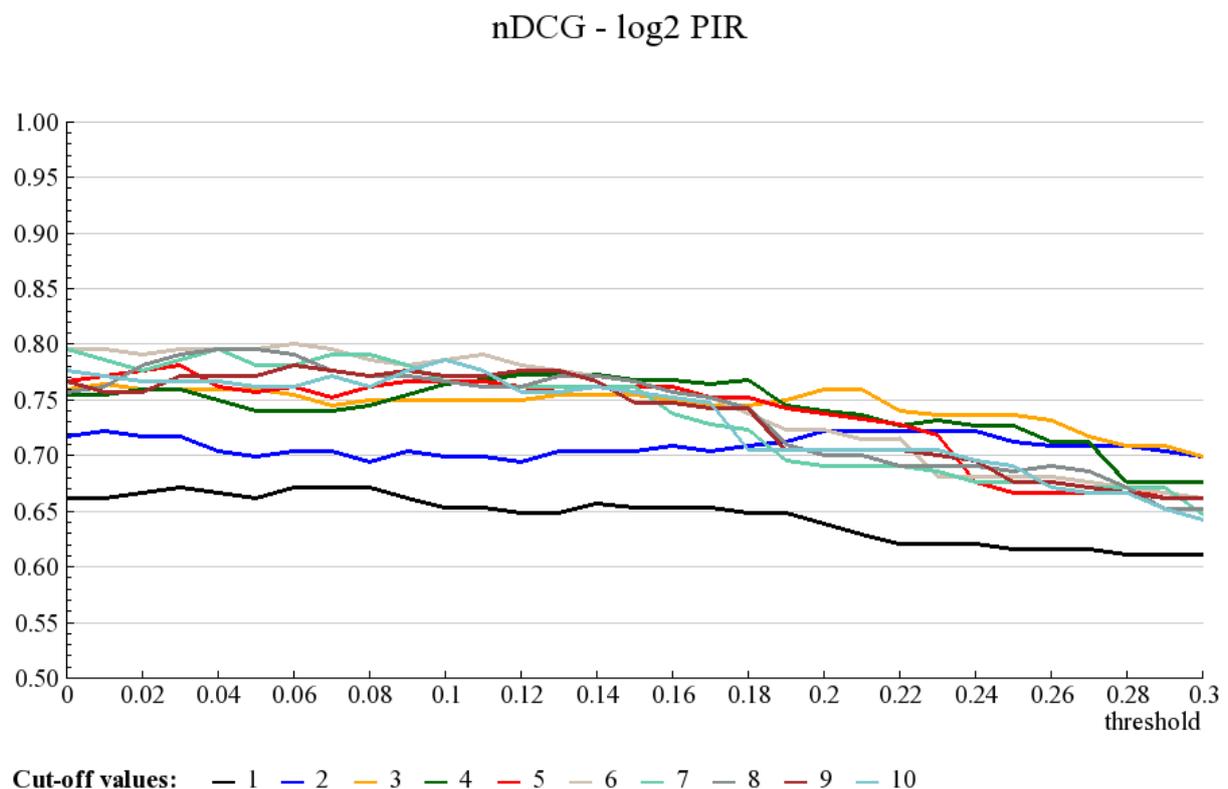

Figure 10.37. NDCG PIR values discounted by $\log_2$ of result rank. Result and preference ratings come from different users.

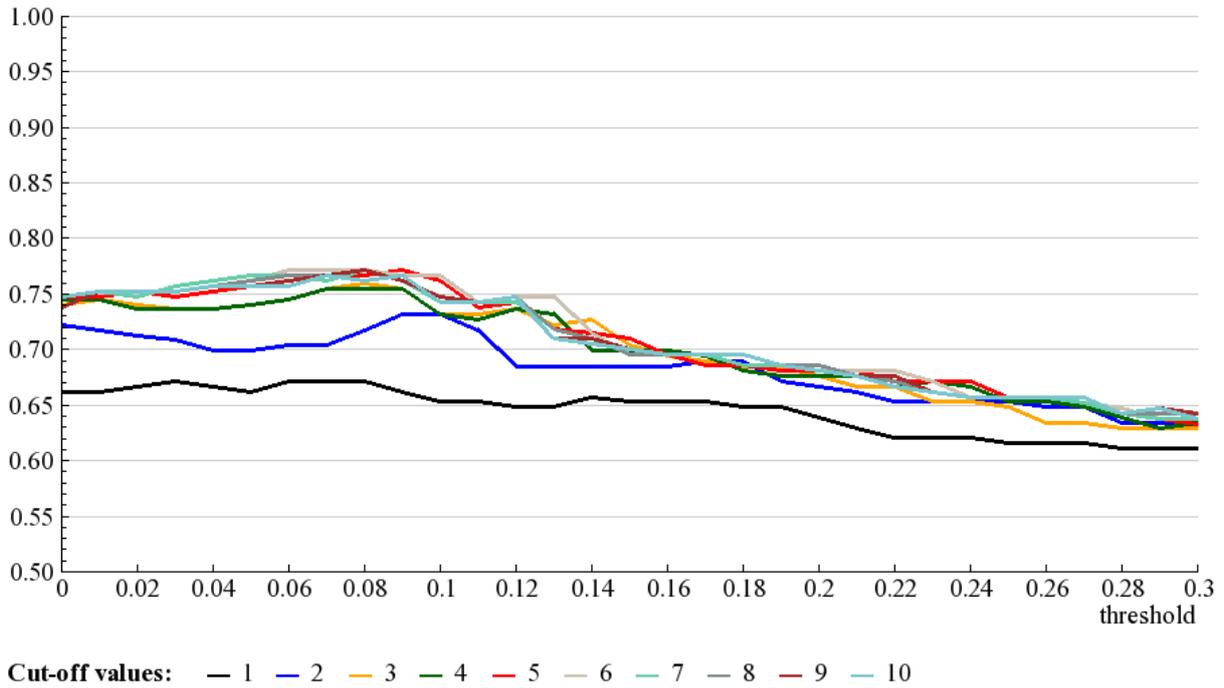

**Figure 10.38. NDCG PIR values discounted by the square of the result rank. Result and preference ratings come from different users.**

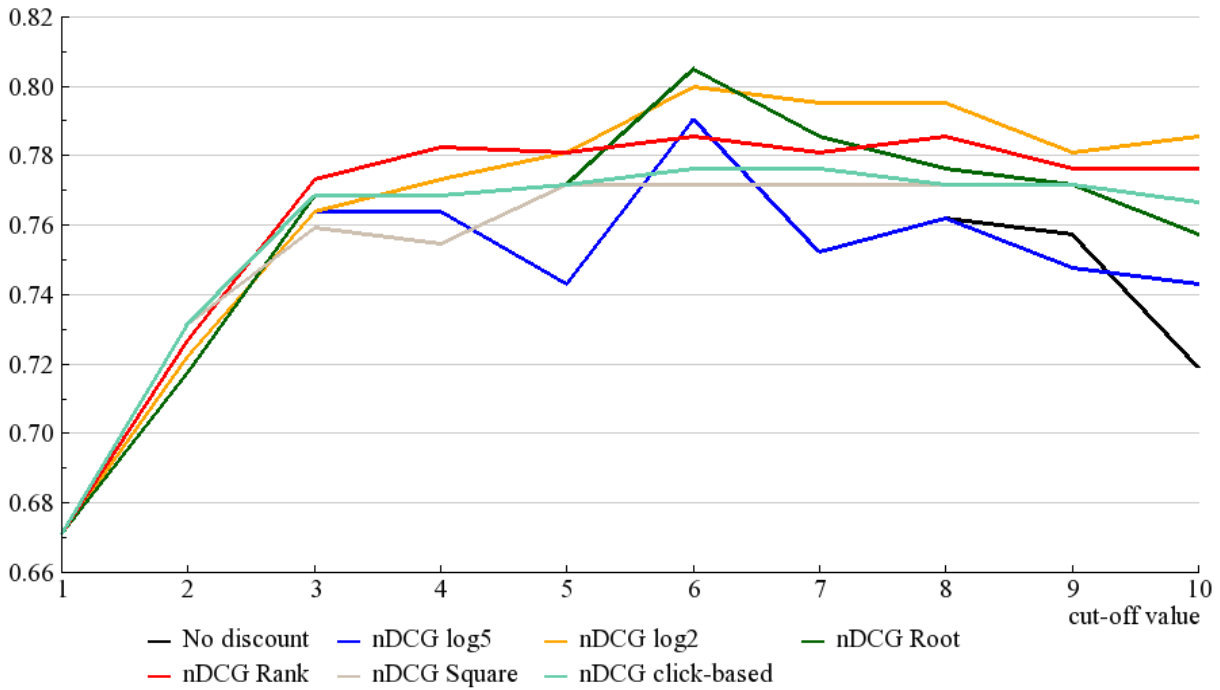

**Figure 10.39. NDCG PIR scores for different discount functions, with the best-threshold approach. Result and preference ratings come from different users.**

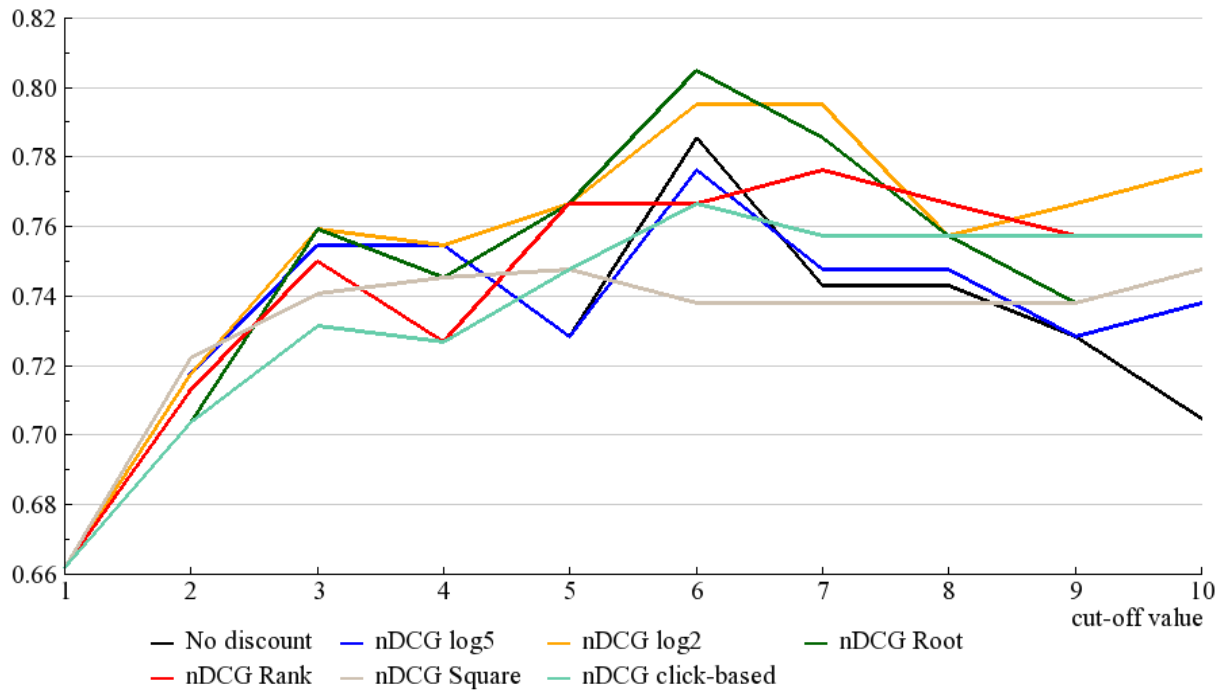

**Figure 10.40. NDCG PIR scores for different discount functions, with the zero-threshold approach. Result and preference ratings come from different users.**

This change has predictable implications for the PIR comparison graphs using different threshold approaches (since *t*=0 does not necessarily provide scores close to the best-threshold method, as it did almost universally in the condition with same-user ratings). While the general shapes of the curves in Figure 10.39 and Figure 10.40 are still quite similar, the differences are immediately visible. The differences between different discount functions are larger with the zero-threshold approach, as are the score amplitude and score fluctuations for individual functions. The best-threshold approach produces more stable results, but it also shows some changes. The PIR scores peak earlier, mostly at cut-off rank 6; the declines after the peak are less steep and less universal; and because of the stronger score fluctuations, it is harder to determine which metrics perform better. The low-discount functions (no-discount and $\log_5$) seem to perform worst if viewed over all cut-offs; but their peak score is higher than that of the high-discount functions (rank square, click-based). Arguably, the average-discount functions ($\log_2$, root, rank) perform the best; for this reason, I will keep the NDCG $\log_2$ as NDCG reference metric.

A similar picture arises for MAP. The zero-threshold graph (Figure 10.42) appears even more chaotic than that for NDCG. Scores still tend to peak around cut-off 6, but there are some large increases at ranks 7 or 8. The best-threshold evaluation (Figure 10.41) is again slightly more stable, but still hard to read. The low-discount functions tend to perform better; the "No discount" function can still be considered to perform best in the best-threshold case, while falling behind for zero-threshold; the "Rank" discount, equivalent to traditional MAP, still performs rather poorly.

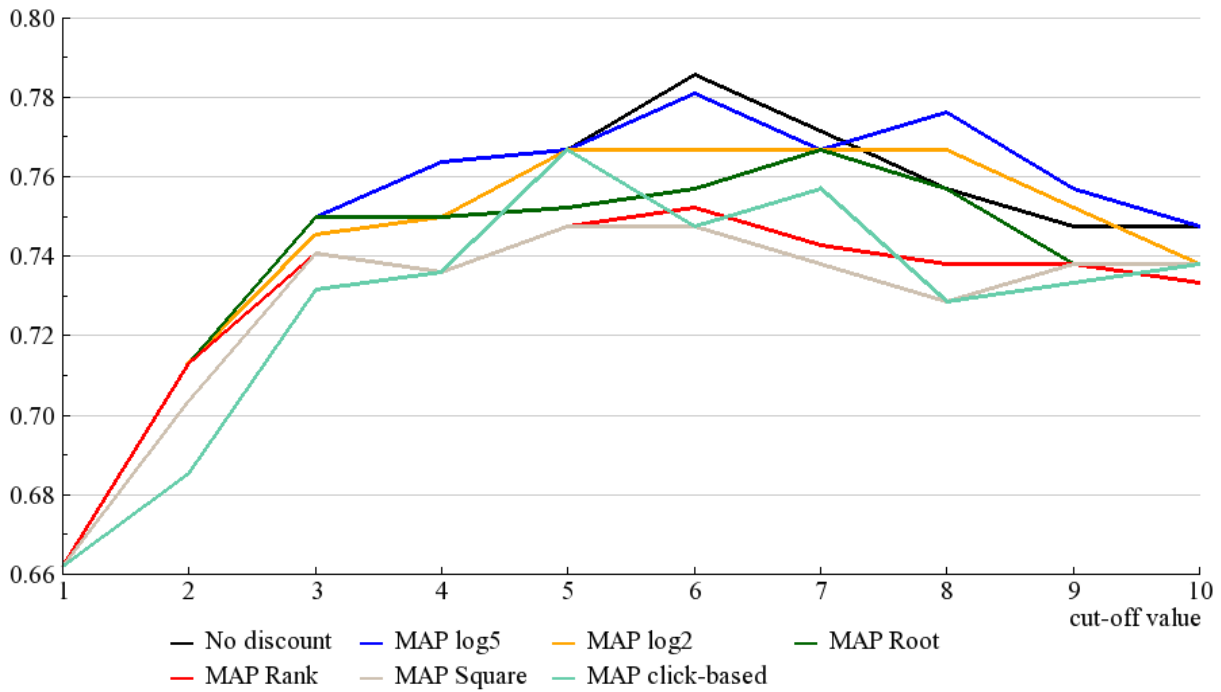

**Figure 10.41. MAP PIR scores for different discount functions, with the best-threshold approach. Result and preference ratings come from different users.**

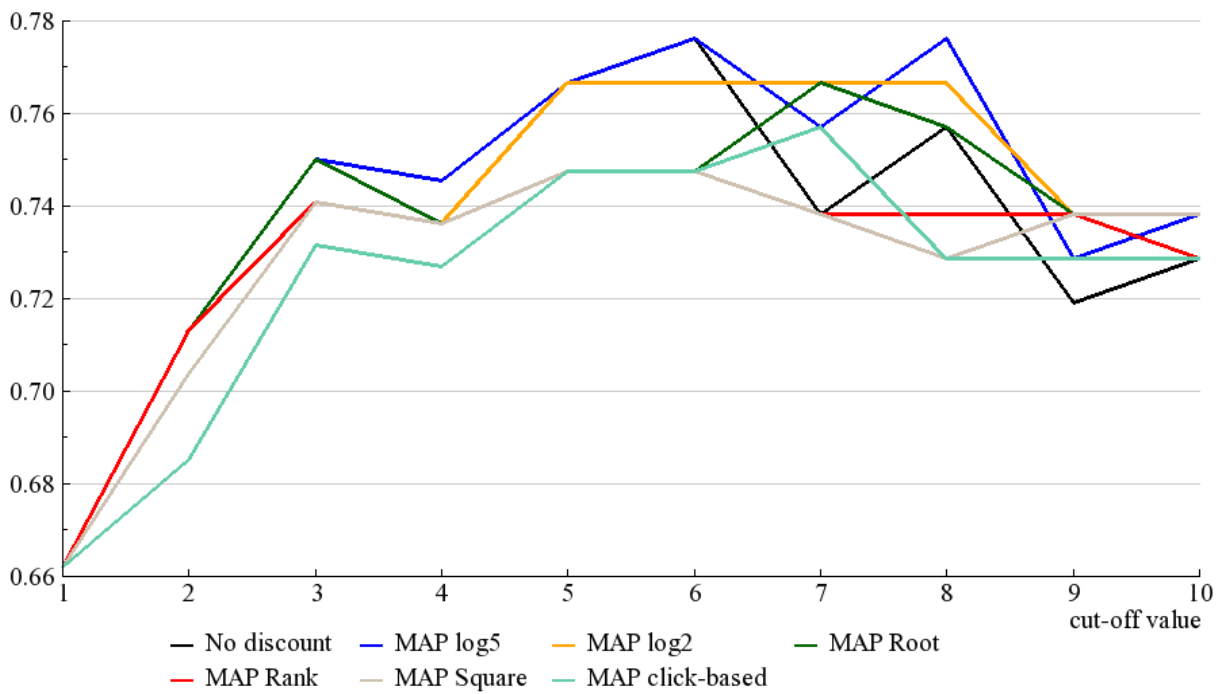

**Figure 10.42. MAP PIR scores for different discount functions, with the zero-threshold approach. Result and preference ratings come from different users.**

Our next test case is ESL. The shapes of the curves in Figure 10.43 and Figure 10.44 show less difference to their counterparts with same-user ratings. The latter graph, showing the PIR scores for the cumulative relevance *n*=3, is somewhat more jagged, and more lines show a slight decline at later ranks (*n*=3 produced better results than *n*=2.5 in this condition, and will be used for the purposes of this section). Nevertheless, both graphs are much more orderly than MAP or even NDCG. The scores obtained with the zero-threshold approach are less stable for *n*=3 (Figure 10.46), but otherwise both these results and those for *n*=1 (Figure 10.45) are quite similar to the results of the best-threshold approach.

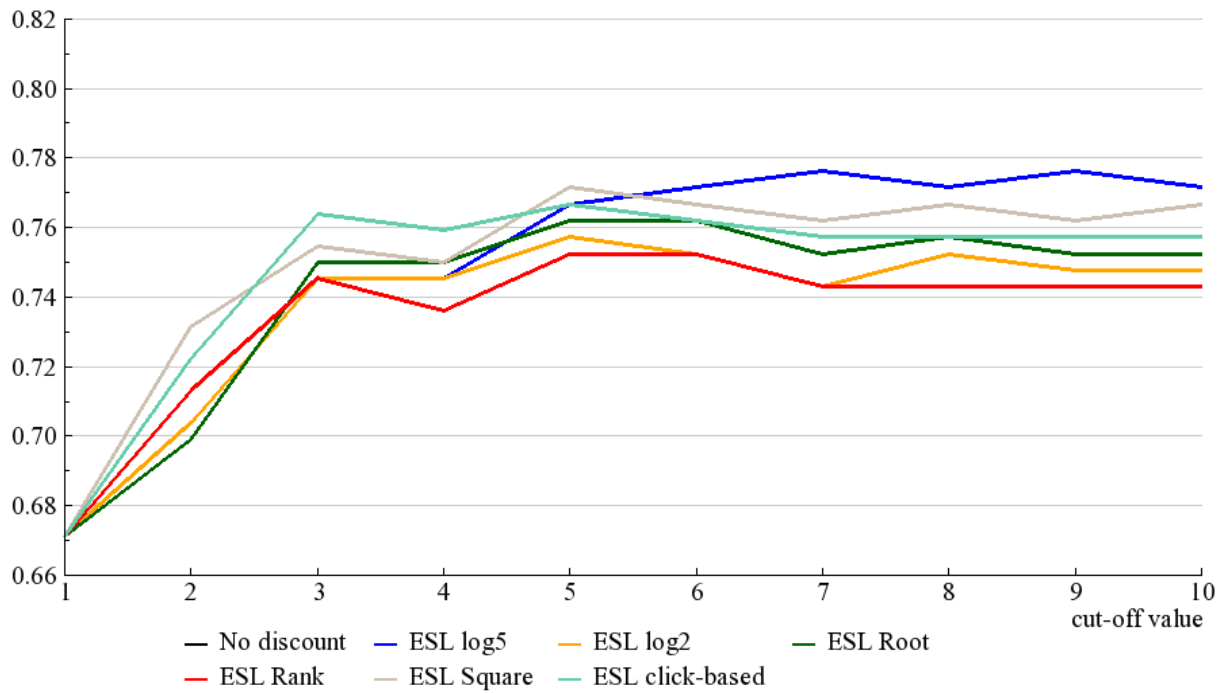

**Figure 10.43. ESL PIR scores for cumulated relevance *n*=1 with the best-threshold approach. Result and preference ratings come from different users.**

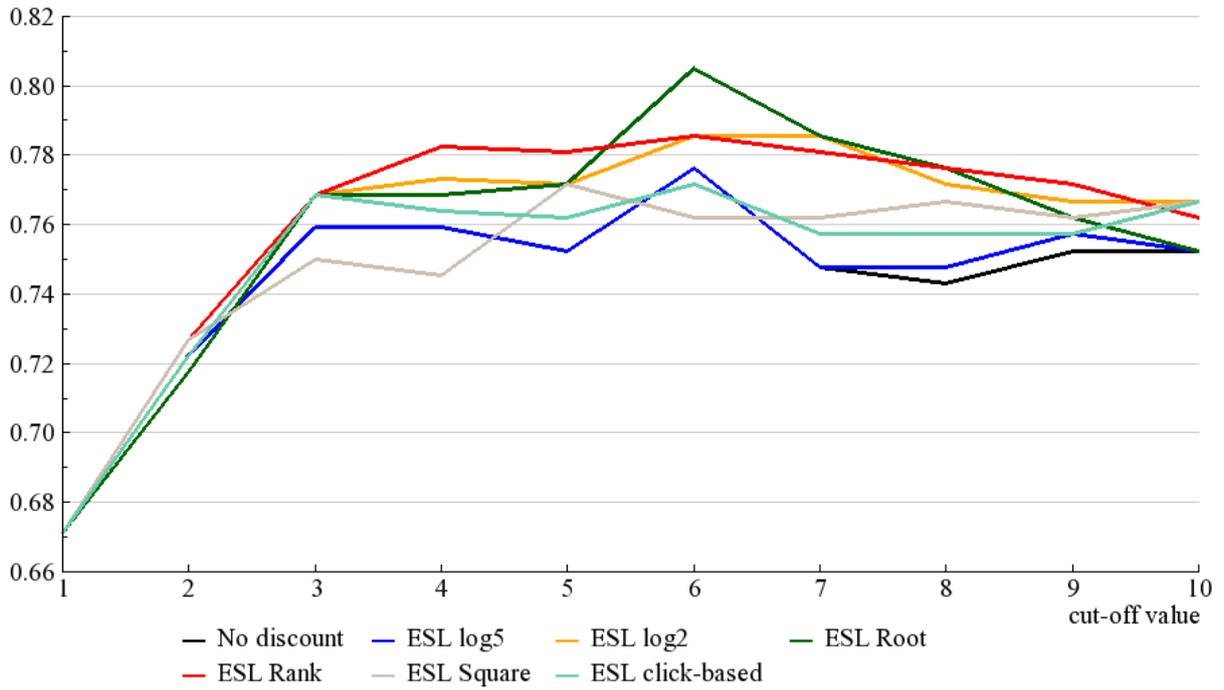

**Figure 10.44.** ESL PIR scores for cumulated relevance *n*=3 with the best-threshold approach. Result and preference ratings come from different users.

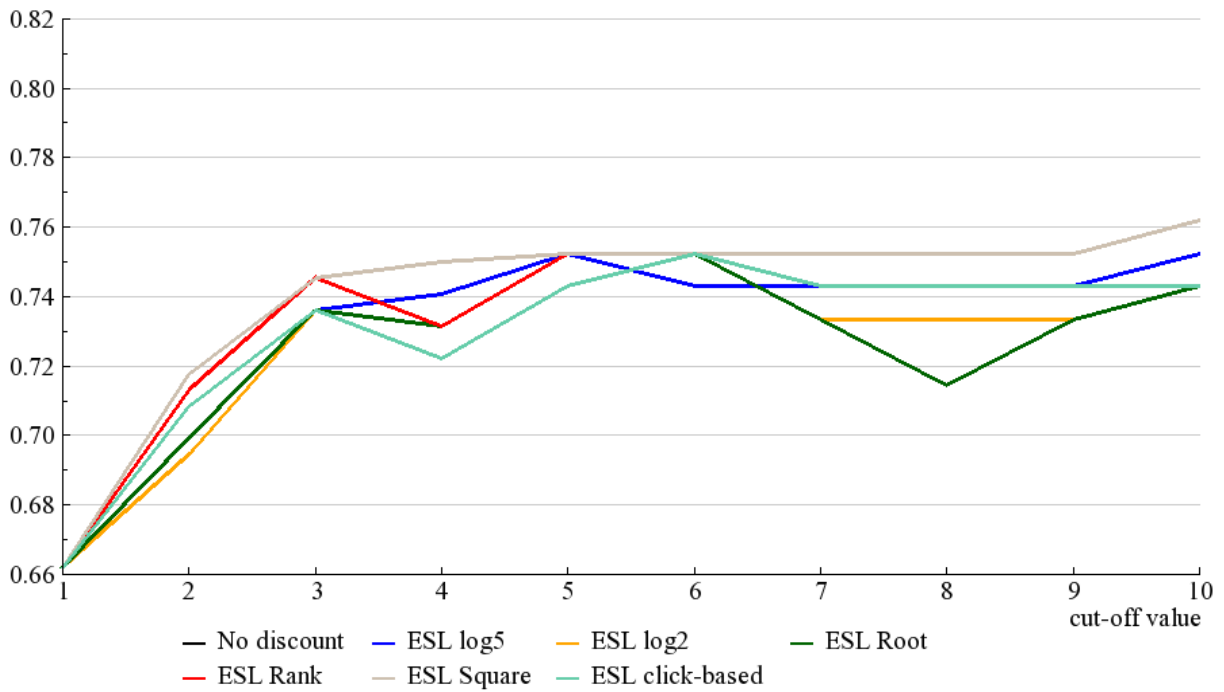

**Figure 10.45.** ESL PIR scores for cumulated relevance *n*=1 with the zero-threshold approach. Result and preference ratings come from different users.

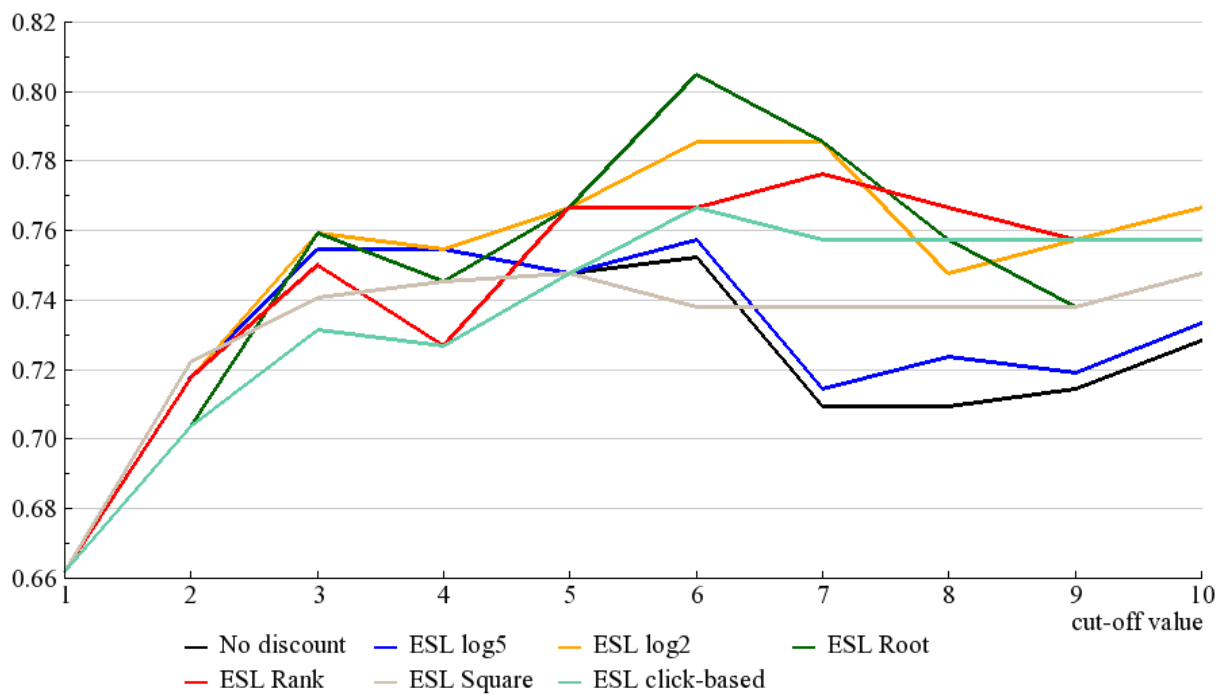

**Figure 10.46. ESL PIR scores for cumulated relevance *n*=3 with the zero-threshold approach. Result and preference ratings come from different users.**

The final question regards the relative and absolute PIR performances of the individual metrics. As Figure 10.47 shows, they are significantly different from those obtained with same-user ratings. The runs of the individual curves are different, as are the ranking of individual metrics and the absolute values in terms of PIR.[93]

When the individual result ratings and preferences came from the same users, the PIR curves were relatively stable, mostly rising continuously towards their peak PIR value around cut-off ranks 7 or 8, and possibly declining after that (Figure 10.34). However, when we remove the luxury of same-user ratings, the picture changes. The metrics mostly peak at cut-off rank 6, with a few scores peaking as early as rank 5 (ERR and traditional, rank-discounted MAP). Both of these metrics, as well as precision, also experience a dip in PIR scores at rank 4. Only ERR, with its square discount used, manages to stay at its (relatively low) peak score until cut-off rank 10; all other metrics' scores fall towards the latter ranks, sometimes quite significantly. Again, the no-discount metrics (Precision and MAP) have the largest declines in PIR towards later cut-off ranks.

The relative performance of the different metrics changed as well, though not as strikingly. The log$_2$-discounted NDCG is still arguably the best-performing metric, although ESL (with *n*=3 and discounted by rank) has fractionally higher scores at early ranks. But traditional

---

[93] MRR is omitted from this and all further evaluations where single result relevance ratings and preference judgments come from different users. It consistently lies far below the other metrics, and does nothing to aid our understanding while distracting from more important matters (and stretching the Y-scale on the graphs).

MAP, which, at least at later ranks, had high PIR scores in the same-rater condition, performs poorly throughout the current evaluation. And Precision is even more variable than before; it has a PIR score of 0.80 at cut-off rank 6, tying with NDCG for the highest overall PIR score and beating traditional MAP (which is also at its peak value) by over 0.05. At cut-off rank 10, however, Precision not only falls over 0.06 PIR points behind NDCG, but also falls behind all other metrics, including MAP.

Finally, the absolute scores also change, not quite unexpectedly. If the same-user approach produced PIR scores up to 0.92, meaning that over nine out of ten users will get their preferred result list based on individual-result rating, the different-user approach brings the top scores down to 0.80. This means that, even if we determine the ideal threshold value and the ideal rank, one out of five queries will result in an inferior result list being presented to the user. And if we do not have the ideal thresholds for each cut-off value, the results are slightly worse and more complex. Figure 10.48 shows the results of the zero-threshold approach; while the absolute peak values decline insignificantly (from 0.80 to 0.775), the individual metrics' scores fluctuate more. If we pair off two metrics at a time, we will find only a few pairs where one metric performs better than the other at all ranks. The scores increase slightly if only informational queries are evaluated (Figure 10.49).

It is not quite easy to explain these results. As a possible explanation for the general PIR decline at later ranks, I previously offered the diminished chance of the results playing a role in the user's assessment, and the influence of the relevance rating on the metric therefore changing PIR for better or worse by chance, bringing it closer to the baseline of 0.5. This could also explain why the PIR scores rise at cut-off rank 6 for many of the metrics. Some studies suggest that the result in the sixth rank is clicked on more often than the previous one, thus raising its importance and usefulness (Joachims et al. 2005). Others suggest a possible explanation; the user studies results immediately after those visible on the first screen in more detail since his scrolling down means he has not satisfied his information need yet (Hotchkiss, Alston and Edwards 2005). This transition might well happen immediately before or after cut-off rank 5. But it is unclear to me why this would only be the case if relevance and preference judgments come from different users. If the reason for the stagnation or decline of PIR scores lay in general user behavior, it should be the same across methods. The differences between the results of same-user and different-user evaluations have to stem from the only point of departure, that is, the raters determining the relevance of single results. This does produce some effects; for example, as the relevance ratings of different users are averaged, the relevance ratings change from a six-point to a more faceted scale. This leads to fewer instances of precisely the same PIR scores for different metrics. Obviously, the decline in absolute PIR scores can also be attributed to the different raters; the relevance of individual results for one person can be assumed to have a more loose connection with another's preference than with his own. However, I see no obvious explanation of the distinctive declines of PIR scores at cut-off ranks 4 or 5 and their subsequent higher rise at rank 6 in terms of different users providing the relevance ratings.

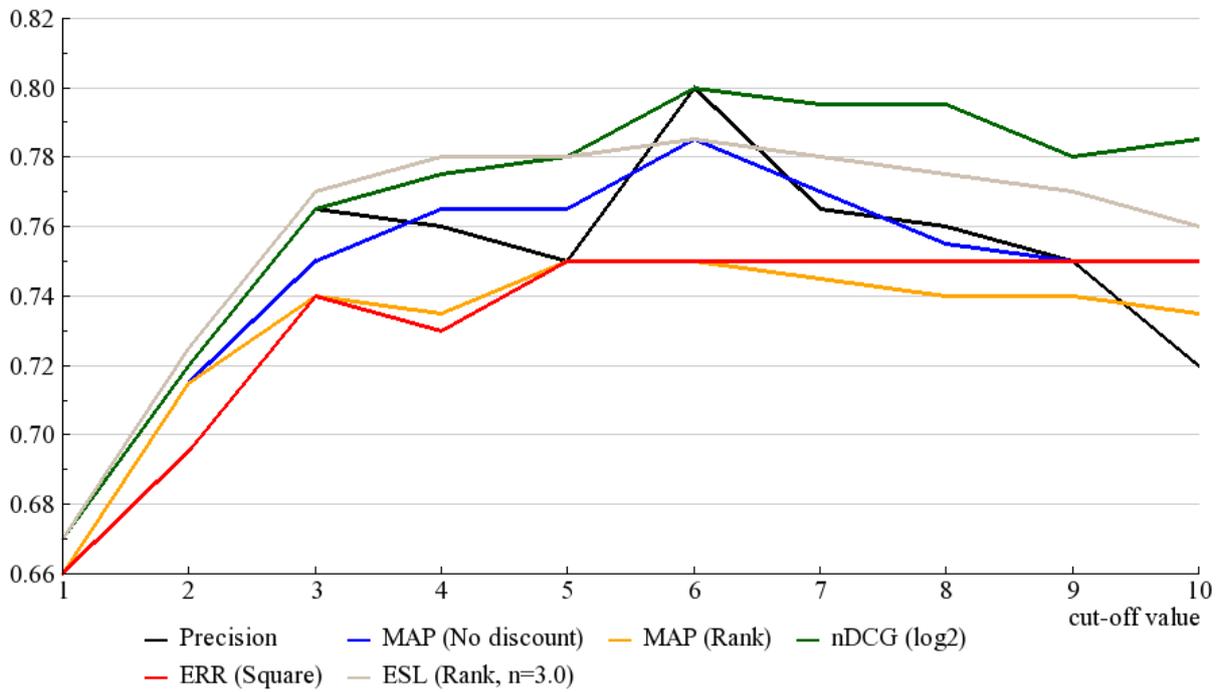

**Figure 10.47. Inter-metric PIR comparison using the best-threshold method. Result and preference ratings come from different users.**

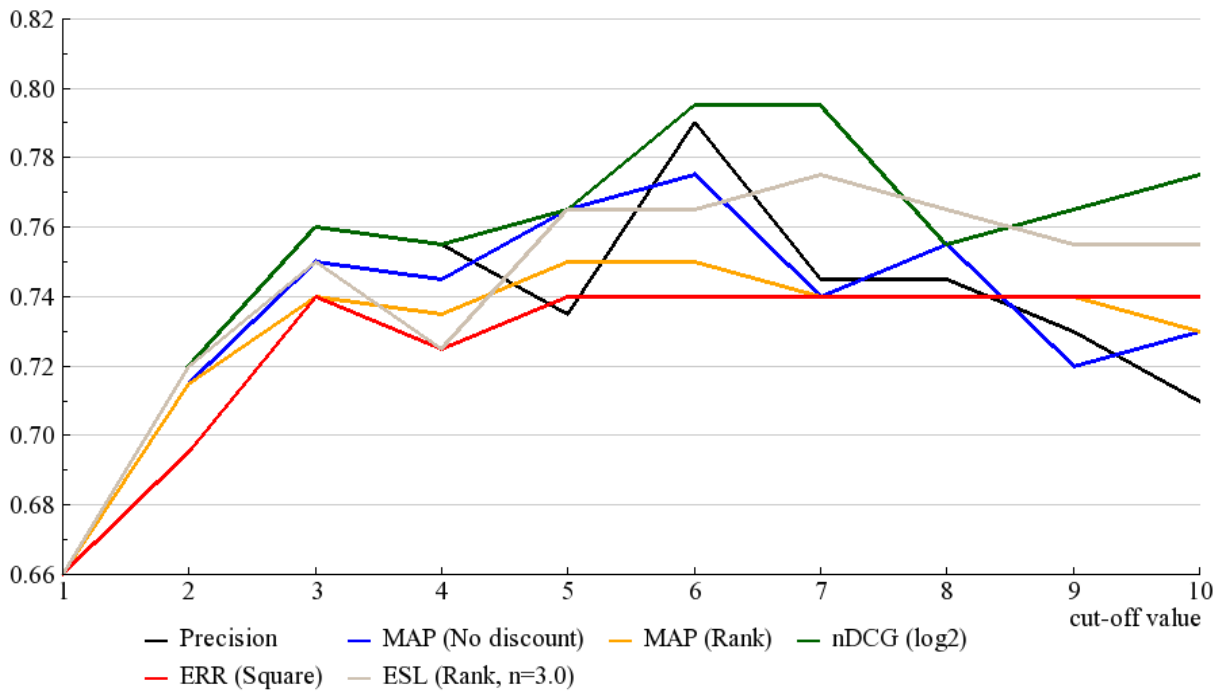

**Figure 10.48. Inter-metric PIR comparison using the zero-threshold method. Result and preference ratings come from different users.**

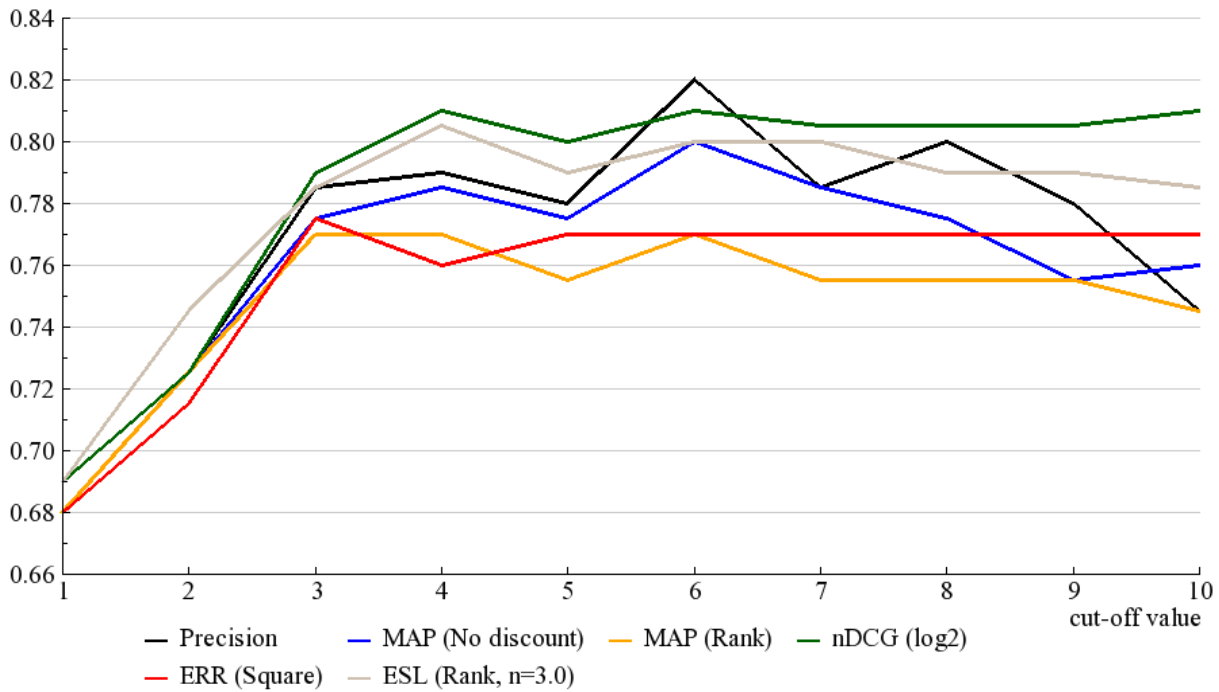

**Figure 10.49.** Inter-metric PIR comparison using the best-threshold method and only informational queries. Result and preference ratings come from different users.

## 10.7 PIR and Relevance Scales

As mentioned in Section 8.2, one more question arising with regards to the evaluation is whether the six-point scale used to rate relevance for individual results uses too fine a distinction. The most popular scale is still often binary (Macdonald et al. 2010), and occasionally a three-point scale is used (Chu 2011). Perhaps the metrics I have evaluated are not well suited for graded relevance, having been conceived for use with binary or three-point relevance scales?[94] The way to deal with this question will be, of course, consistent with the rest of this study: I will test different configurations and see which relevance scales produce what kind of results.

The two relevance scales will be two-level and three-level relevance. While four-level and five-level relevance scales are quite conceivable (and occasionally used, see Gao et al. (2009) or Kishida et al. (2004) for some examples), the differences between a five-point and a six-point scale (and with them the need for an evaluation) are not as large as between two-point and six-point scales. Moreover, it is more difficult to convert judgments from a six-point to a four- or five-point scale in a meaningful way. The same is true for more detailed relevance ratings.

For both possible scales there are also questions of conversion. If we want to employ a binary scale, what would we want the rater to regard as relevant? There are three obvious approaches. One is to rate as relevant anything having even a slightest connection with the information need; another to rate as irrelevant anything not regarded as optimal for the query; and the third to draw the line in the middle and using the categories "rather relevant than irrelevant" and vice versa (or perhaps "rather relevant" versus "not too relevant"). This would correspond to regarding as relevant the ratings 1 to 5, 1, and 1 to 3, respectively.[95] For brevity, the three binary relevancies will be referred to as $R2_5$, $R2_1$ and $R2_3$, the "2" designating binarity and the subscript showing the worst ratings still regarded to be relevant. For the three-point scale, the two possibilities of conflating the ratings are either in equal intervals (1 and 2 as "highly relevant" or 1; 3 and 4 as "partially relevant or 0.5; and 5 and 6 as "non-relevant" and 0), or regarding 1 as highly relevant (1), 6 as non-relevant (0), and everything in between as "partially relevant" (0.5). The two methods will be called $R3_2$ and $R3_1$ respectively, with the "3" showing a three-point scale and the subscript designating the worst rating still regarded to be highly relevant.

---

[94] Although DCG, at least, was constructed with graded relevance input being one of its explicitly stated advantages (Järvelin and Kekäläinen 2000).

How, then, will the PIR performance change if we use different ratings scales? I will again start with NDCG, and proceed through the individual metrics, giving fewer details when the tendencies start to repeat.

**10.7.1 Binary Relevance**

Figure 10.50 and Figure 10.51 show that the difference between cut-off ranks becomes lower for no-discount and rank-discounted NDCG when binary $R2_3$ precision is used; especially with higher discounts, the scores after cut-off rank 3 get very close to each other. The picture is similar for $R2_5$ (Figure 10.52) and $R2_1$ (Figure 10.53).

---

[95] There are also other possibilities, like considering 1-4 to be relevant, since those are the "pass" grades in the German school system, or considering 1-2 to be relevant, since these are the grades described as "very good" or "good", respectively. I do not wish to complicate this section (further) by considering all possibilities at once, instead leaving these particular evaluations to future studies.

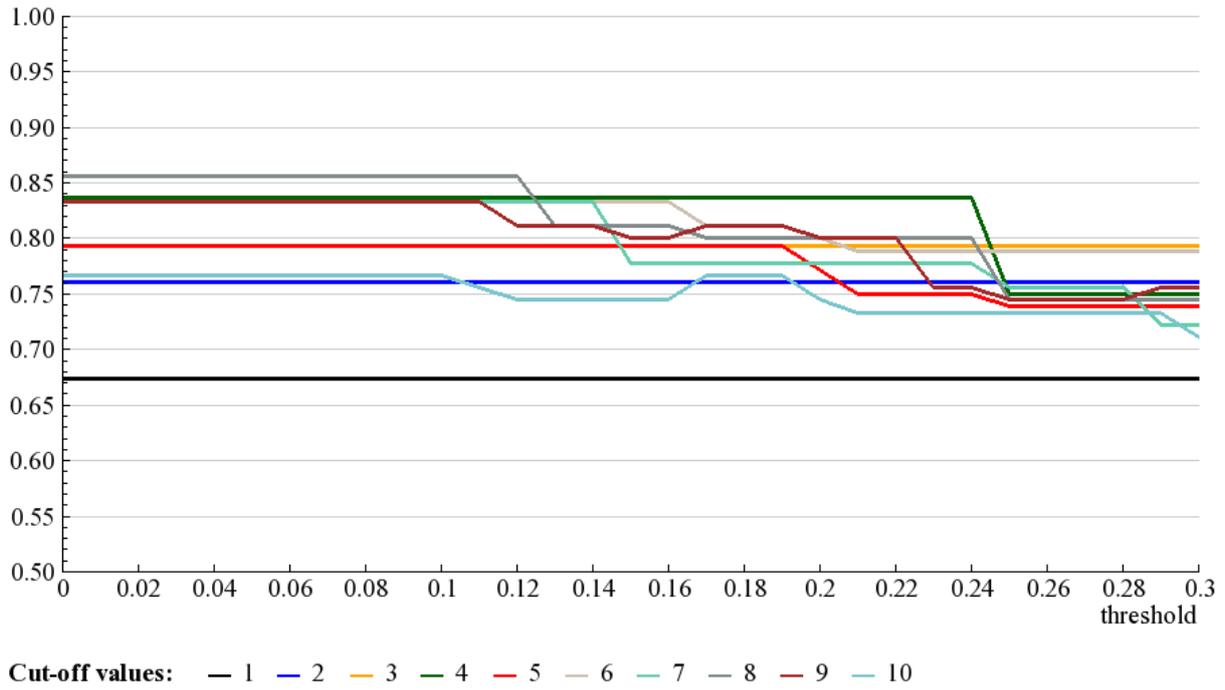

**Figure 10.50.** NDCG PIR values without any discount for results at later ranks. R2$_3$ relevance (1,2,3 as relevant, 4,5,6 as non-relevant) is used.

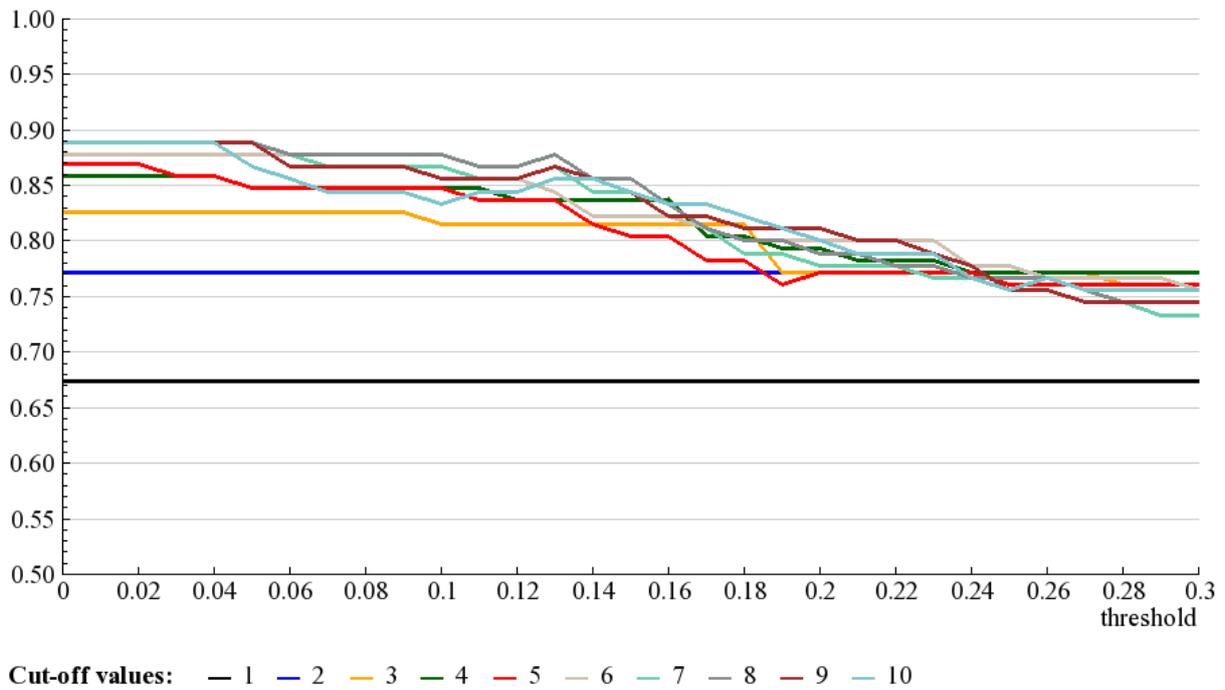

**Figure 10.51.** NDCG PIR values discounted by result rank, with R2$_3$ relevance used.

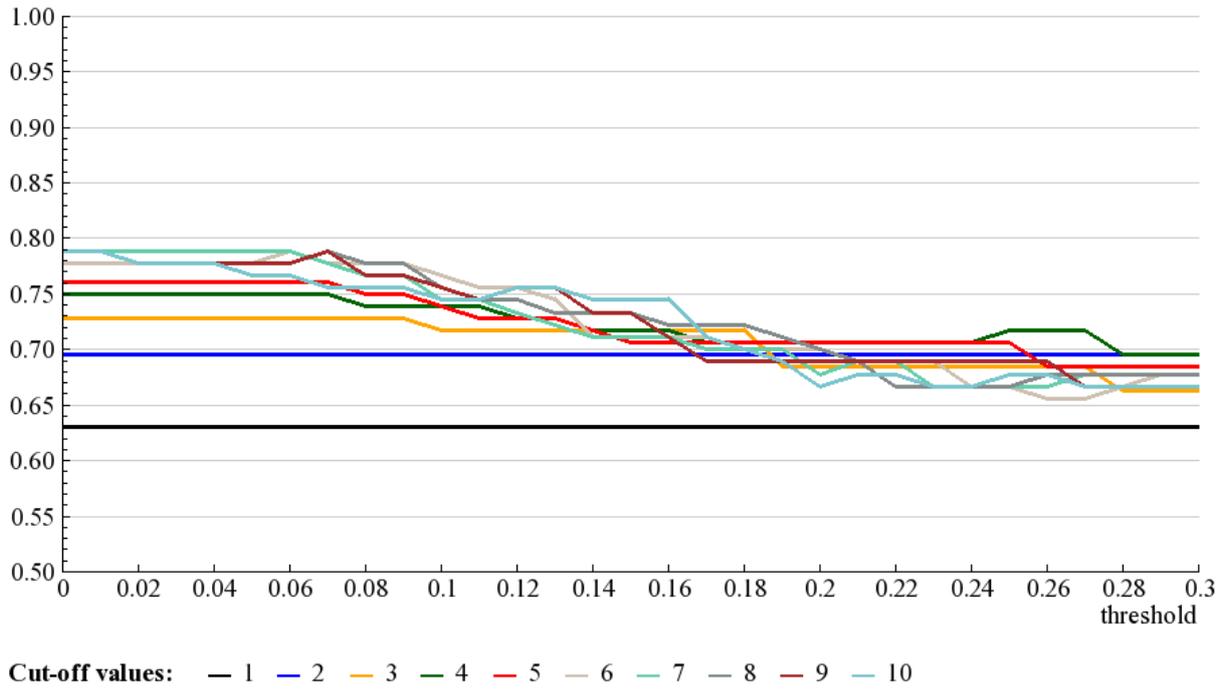

Figure 10.52. NDCG PIR values discounted by result rank, with $R2_5$ relevance used.

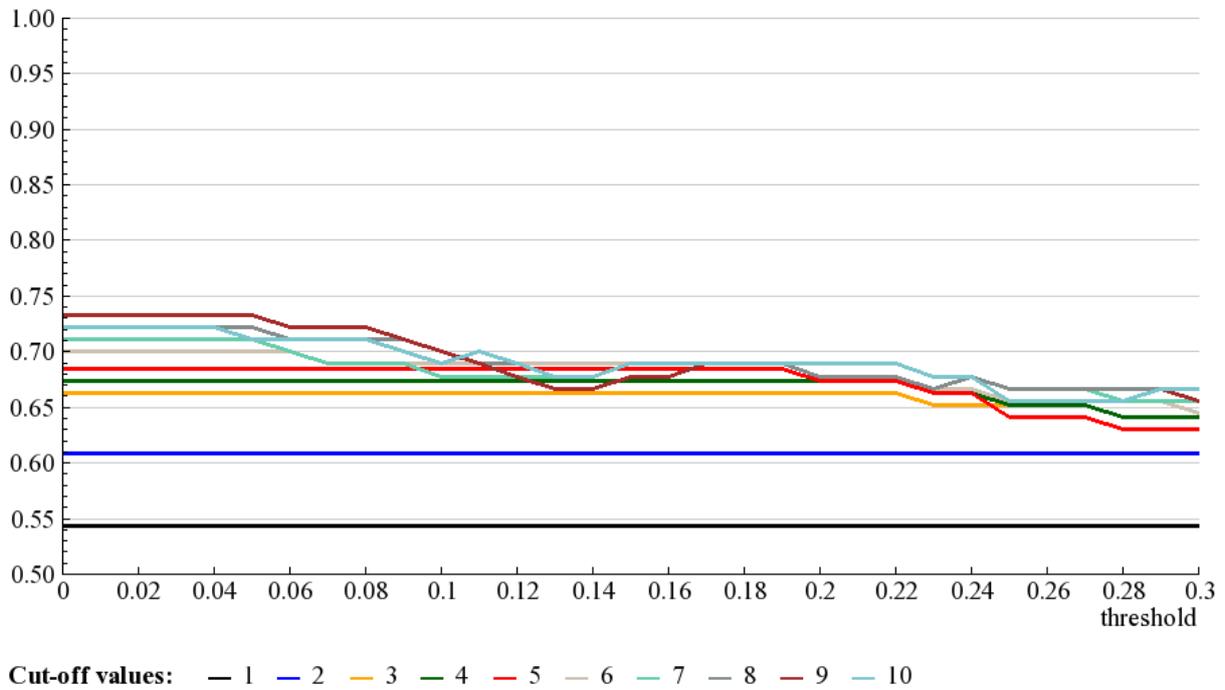

Figure 10.53. NDCG PIR values discounted by result rank, with $R2_1$ relevance used.

Of more immediate interest are the relative performances of differently discounted NDCG metrics. Again, we see that the performances (in terms of PIR) become more similar. As in the six-level relevance evaluation, the performance in the $R2_3$ evaluation (Figure 10.54) is dependent on the discount strength; the low-discount functions (no discount, $\log_5$) perform worse than the higher-discount ones. In the other two binary relevance evaluations, the individual lines are still closer to each other. With $R2_5$, even the differences between low-discount and high-discount functions' PIR scores become less pronounced (Figure 10.55), and in the $R2_1$ evaluation, the difference between the best-performing and the worst-performing function at any particular cut-off value almost never exceeds 0.02 PIR points (Figure 10.56).

The $\log_2$-discounted NDCG metric, the one actually used in evaluations, does not perform as well as with the six-point relevance range. It does have the highest scores with $R2_1$ relevance, but is inferior to other discounts with $R2_3$ and $R2_5$. Nevertheless, as no other discount consistently outperforms $\log_2$, I shall stick with it as the NDCG reference.

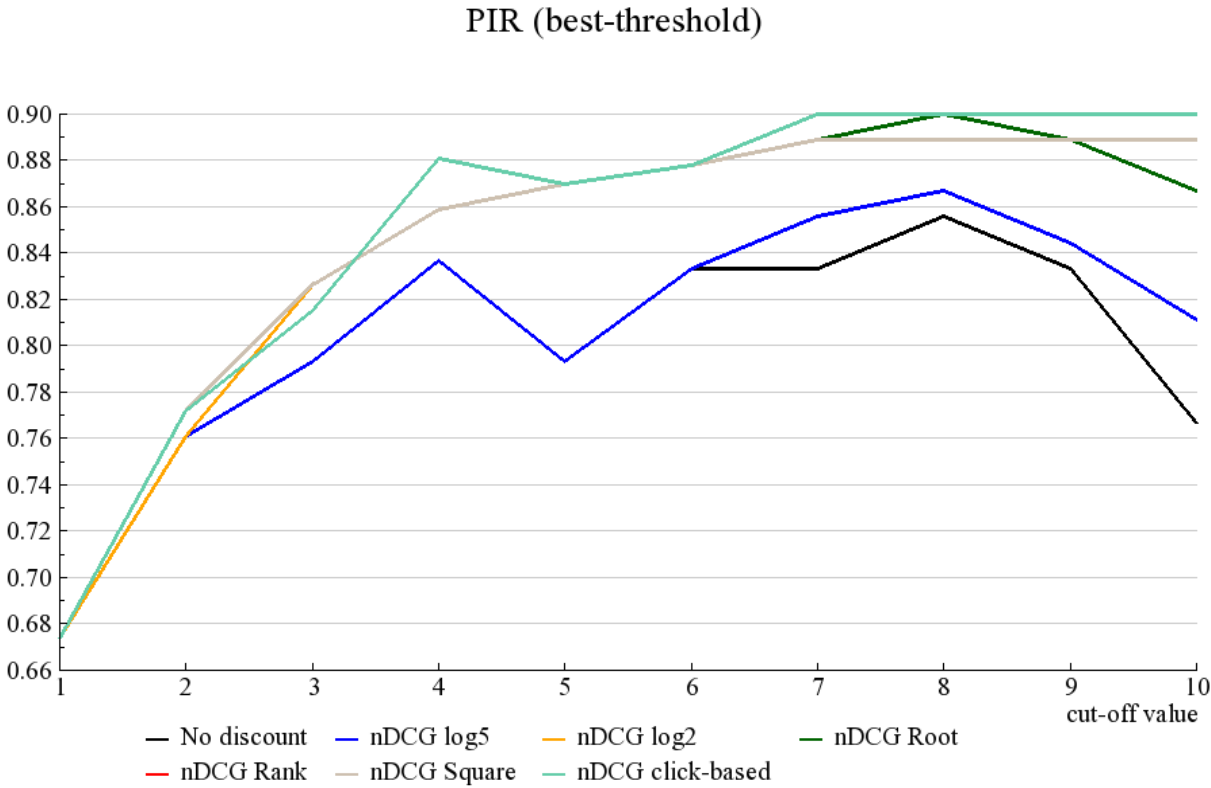

**Figure 10.54. NDCG PIR scores for different discount functions, with the best-threshold approach using $R2_3$ relevance.**

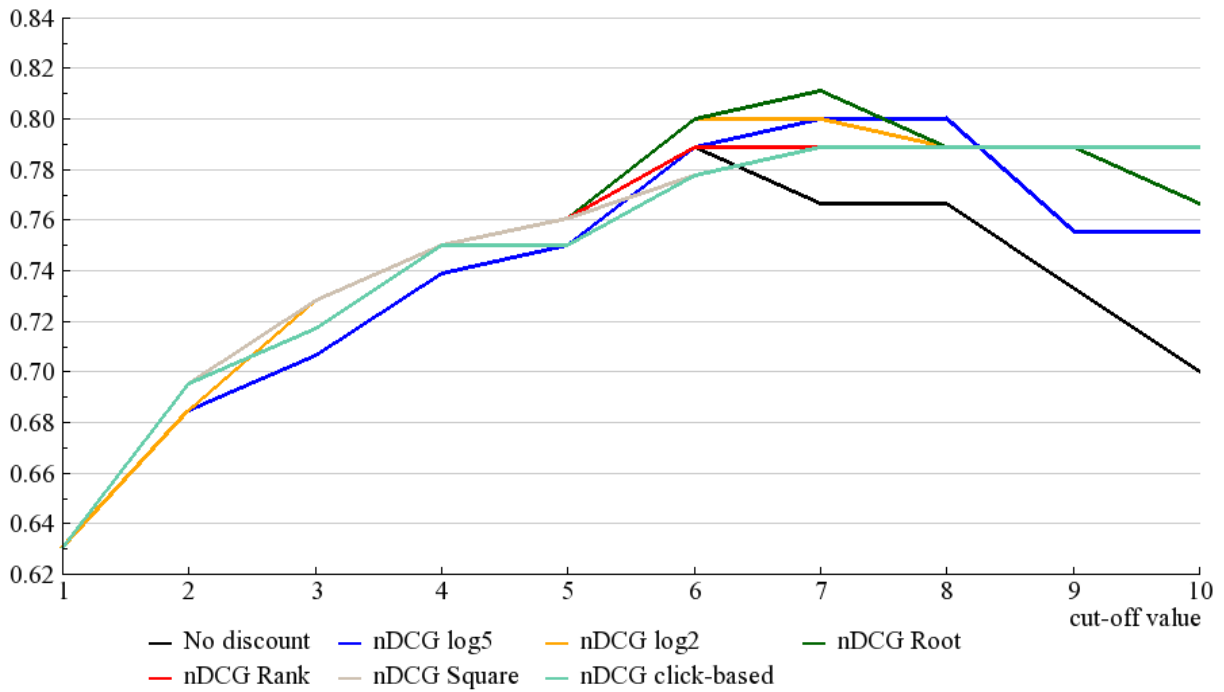

Figure 10.55. NDCG PIR scores for different discount functions, with the best-threshold approach using $R2_5$ relevance.

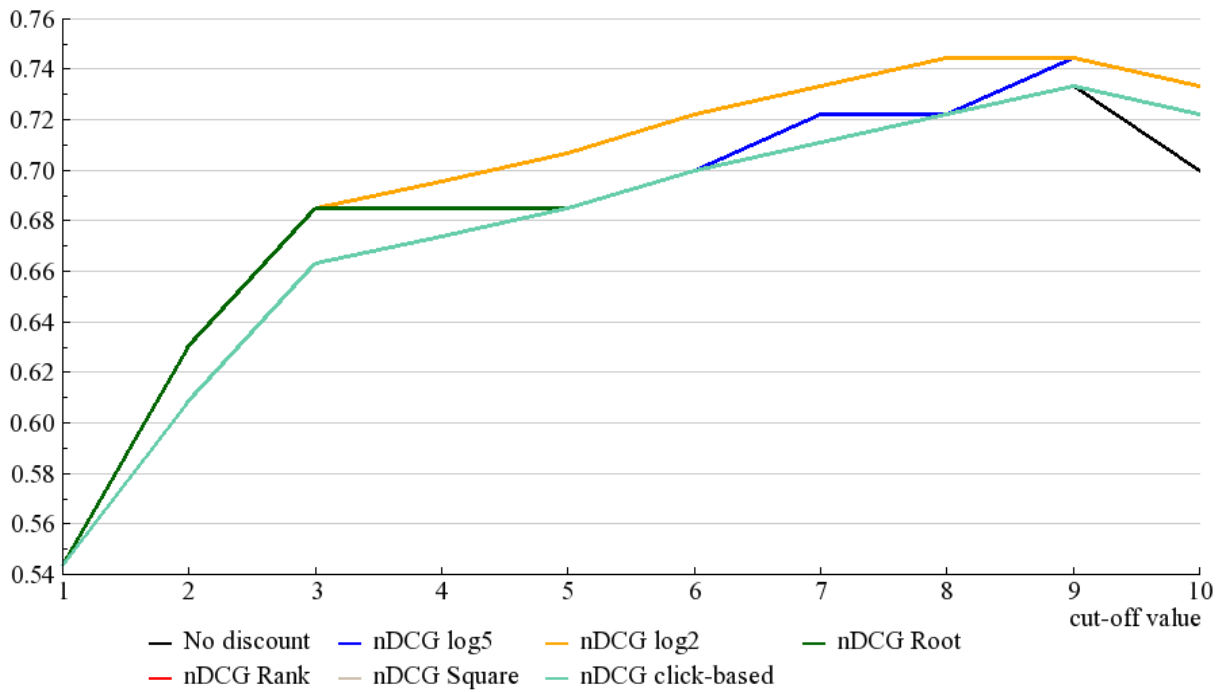

Figure 10.56. NDCG PIR scores for different discount functions, with the best-threshold approach using $R2_1$ relevance.

As the results for individual MAP-based discount functions are similar in kind to those of NDCG, I will not bore you with more threshold graphs. The discount comparisons, however, are a different matter. The graph for $R2_3$ relevance (Figure 10.57) is not dissimilar to that for six-point relevance (Figure 10.18). Though there are some dips, the lines rise quite steadily throughout the cut-off values. The PIR scores for $R2_5$ (Figure 10.58) relevance rise faster, with most discount functions reaching their peak values by cut-off rank six to eight. For $R2_1$, the scores fluctuate still more (Figure 10.59).

As in the six-point relevance evaluation, the no-discount function performs best in all three conditions (closely followed by $\log_5$ discount), while traditional MAP scores (discounted by rank) are at best average, compared to other discount functions.

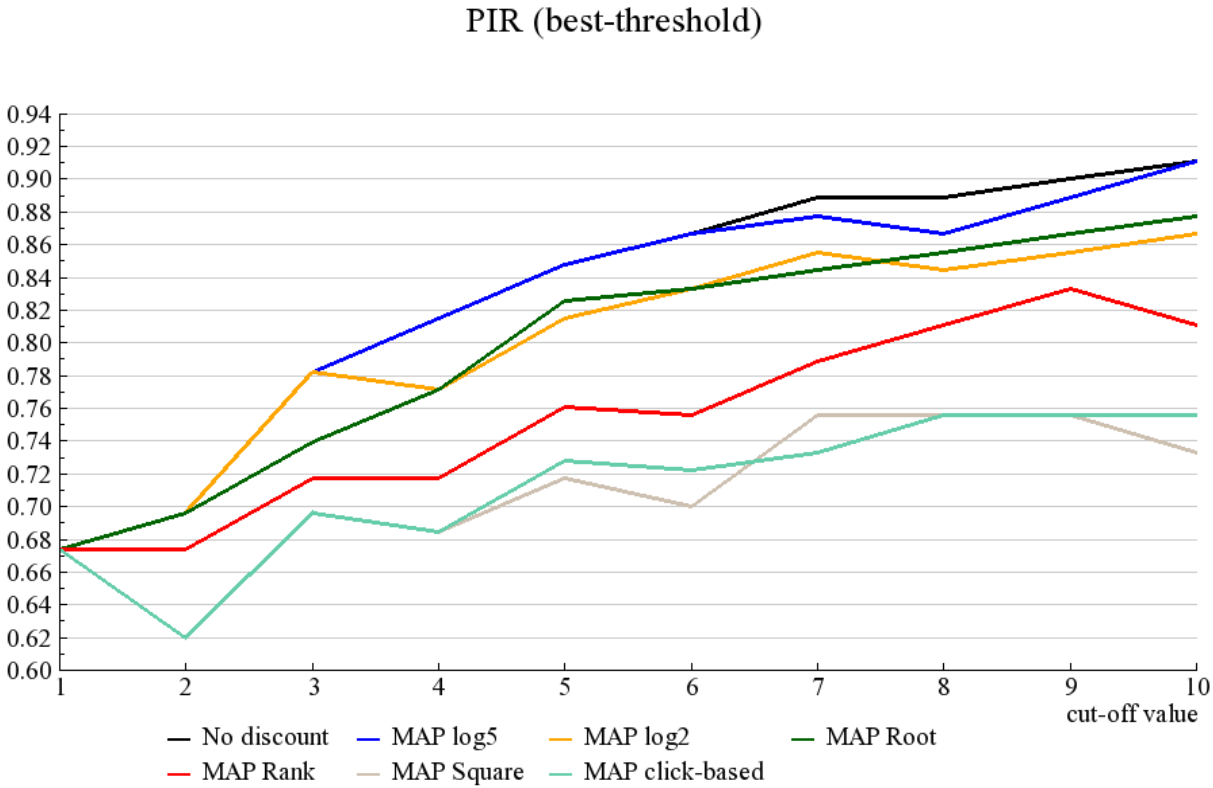

**Figure 10.57. MAP PIR scores for different discount functions, with the best-threshold approach using $R2_3$ relevance.**

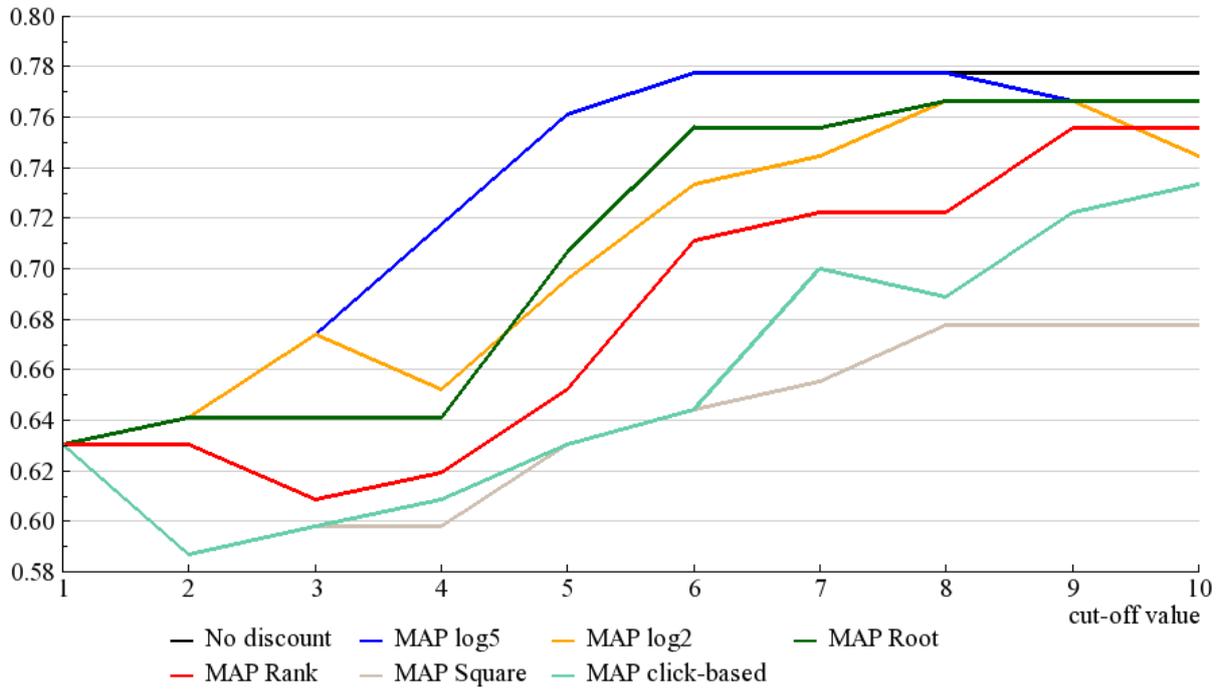

**Figure 10.58.** MAP PIR scores for different discount functions, with the best-threshold approach using $R2_5$ relevance.

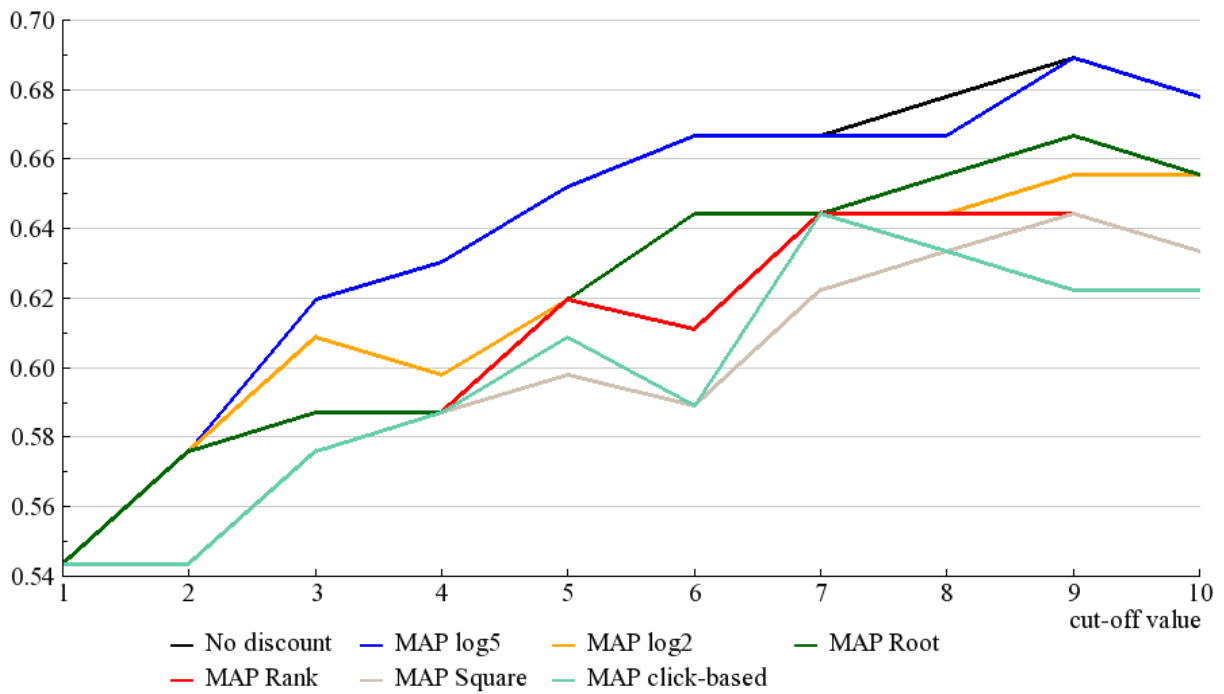

**Figure 10.59.** MAP PIR scores for different discount functions, with the best-threshold approach using $R2_1$ relevance.

The graphs for other metrics do not differ in significant or interesting ways from those for NDCG and MAP except for their absolute values; therefore, I will omit the individual metric graphs and proceed directly to the inter-metric comparison. For reference and ease of comparison, the PIR graph for six-point relevance is reproduced as Figure 10.60.

In Figure 10.61, showing PIR scores for $R2_3$ (binary relevance with the relevant results defined in the manner of "rather relevant" versus "not very relevant"), some changes are immediately apparent. While the top scores stay approximately the same, some metrics do better and some worse. MRR is the success story; rather than consistently lagging far behind the other metrics, it actually beats traditional MAP (discounted by rank) for most cut-off values. Rank-discounted MAP generally performs worse than in the six-point relevance evaluation, and so does Precision. The other metrics have results comparable to those in the six-point relevance evaluation, and no-discount MAP even improves its performance, though only by a margin. All in all, there is hardly any difference between the best-performing metrics (NDCG $\log_2$, ERR Square and ESL Rank with n=2.5), and no-discount MAP scores overtake them at the later cut-off ranks.

When we change the definition of relevance to be "not completely irrelevant" ($R2_5$), the scores sag (Figure 10.62). Instead of peaking at above 0.9, the best scores are now at just 0.8. Again, the worst-performing metrics are MRR and rank-based MAP. The best-scoring metrics are NDCG (with $\log_2$ discount) and ESL with n=2.5 and rank-based discount; however, the remaining three metrics are not far behind. It is also remarkable that, apart from Precision, no metric experiences a significant PIR decline towards the later ranks. One possible reason for that might be the generally lower PIR scores (while in the six-point relevance and $R2_3$ environments, the peak scores are high and accordingly have more ground to lose). Thus, a relatively moderate influence of latter-rank PIR may not be enough to bring the scores down.

Finally, in Figure 10.63, the peak scores fall lower still. In this condition, where "relevance" is defined as something like "a result of the most relevant kind", the PIR never reaches 0.75. MRR and rank-based MAP still constitute the rear guard, with MAP performing worst. NDCG ($\log_2$) and ERR (Square) provide the highest scores.

These evaluations seem to suggest that single result evaluations based on binary relevance perform worse than those based on six-point relevance. If the distinction between "relevant" and "non-relevant" is drawn in the middle, the decline in preference identification is moderate; if "relevant" is defined as "somehow relevant" or "highly relevant", the losses in preference prediction quality become very large indeed.

If we confine the evaluation to informational queries, the picture stays the same. Figure 10.64 through Figure 10.67 show the graphs; the lines are somewhat more jumpy (probably simply due to the reduced dataset), but all the trends remain.

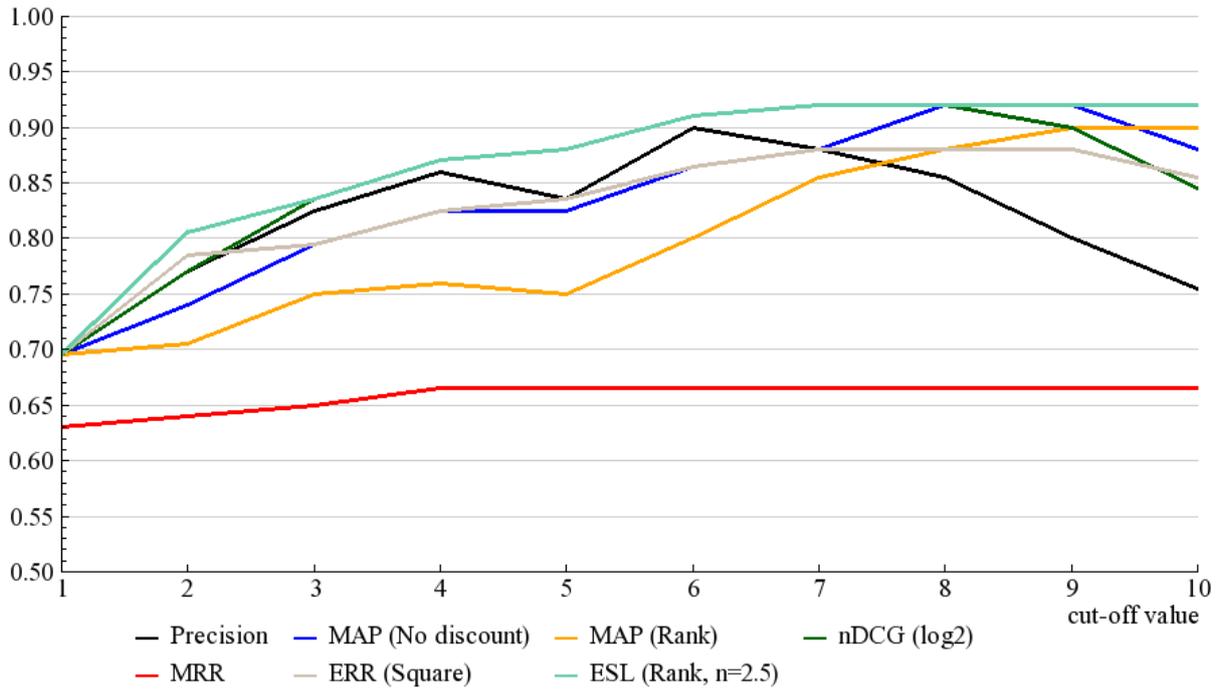

**Figure 10.60. Reproduction of Figure 10.34 with a modified Y-scale . Inter-metric PIR comparison using the best-threshold method and six-point relevance. NDCG ($log_2$) and ESL (Rank, *n*=2.5) have the same values for cut-off ranks 3 to 8.**

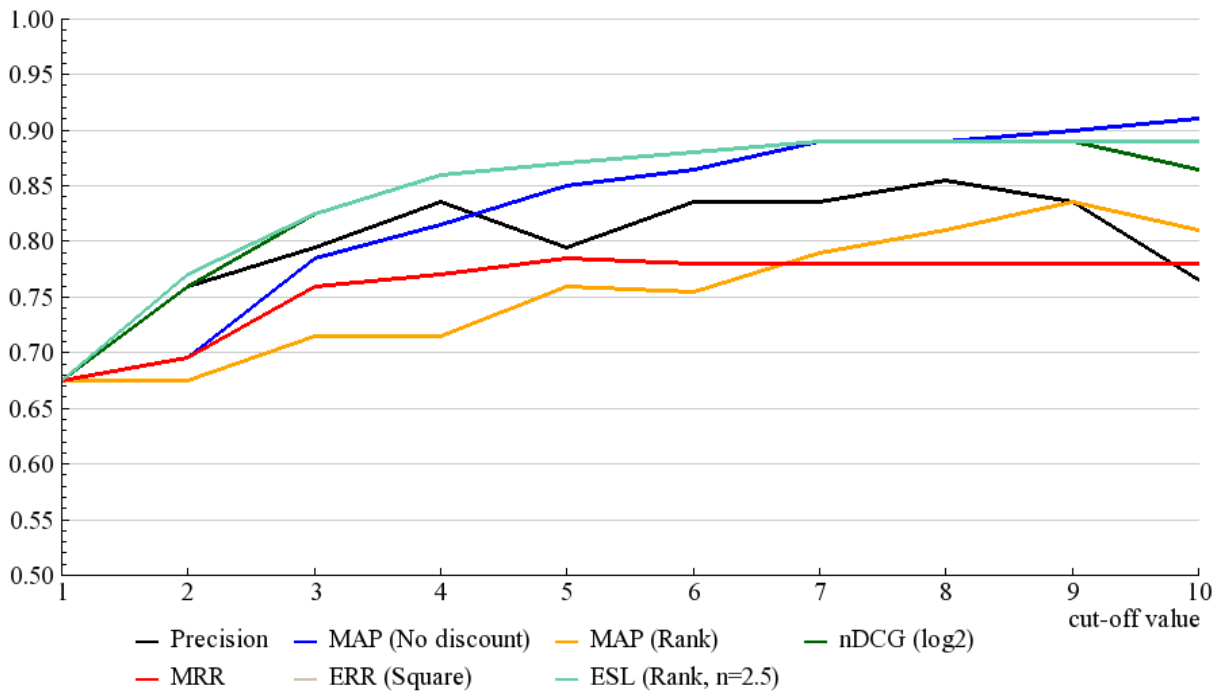

**Figure 10.61. Inter-metric PIR comparison using the best-threshold method and binary $R2_3$ relevance. NDCG ($log_2$) and ESL (Rank, *n*=2.5) have the same values for cut-off ranks 3 to 9; ERR and ESL have exactly the same scores.**

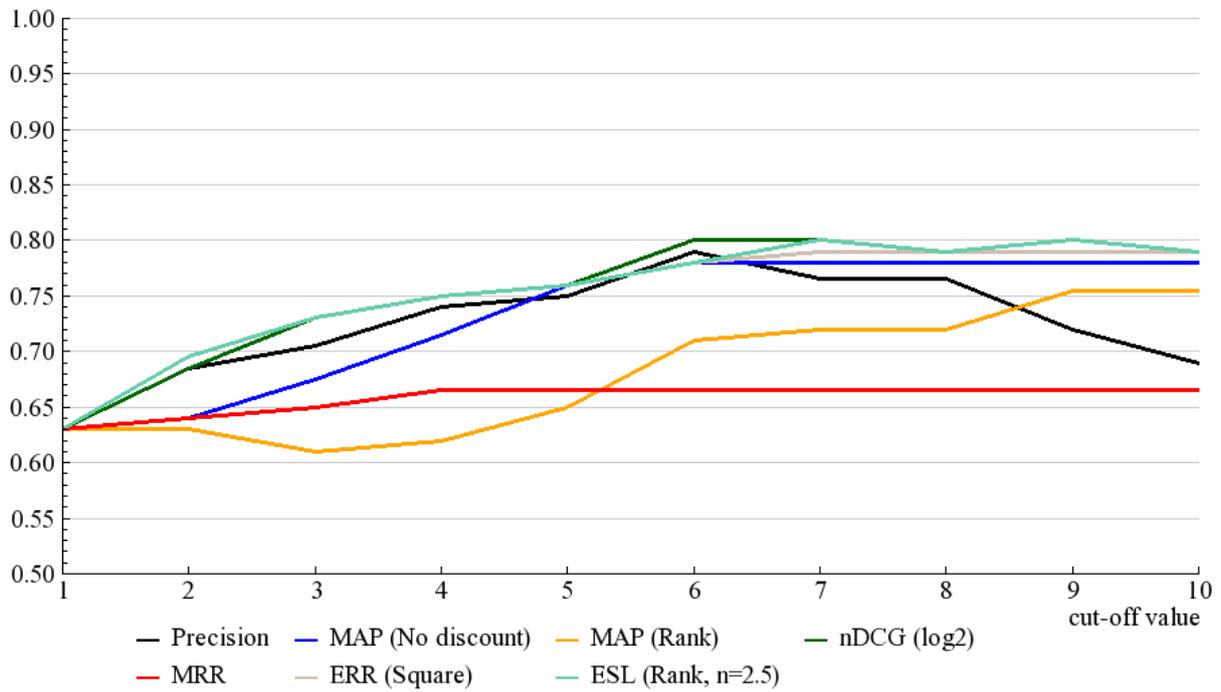

**Figure 10.62. Inter-metric PIR comparison using the best-threshold method and binary $R2_5$ relevance. NDCG ($\log_2$) and ERR (Square) have the same values for cut-off ranks 8 to 10; ERR and ESL have the same scores at ranks 1 to 6.**

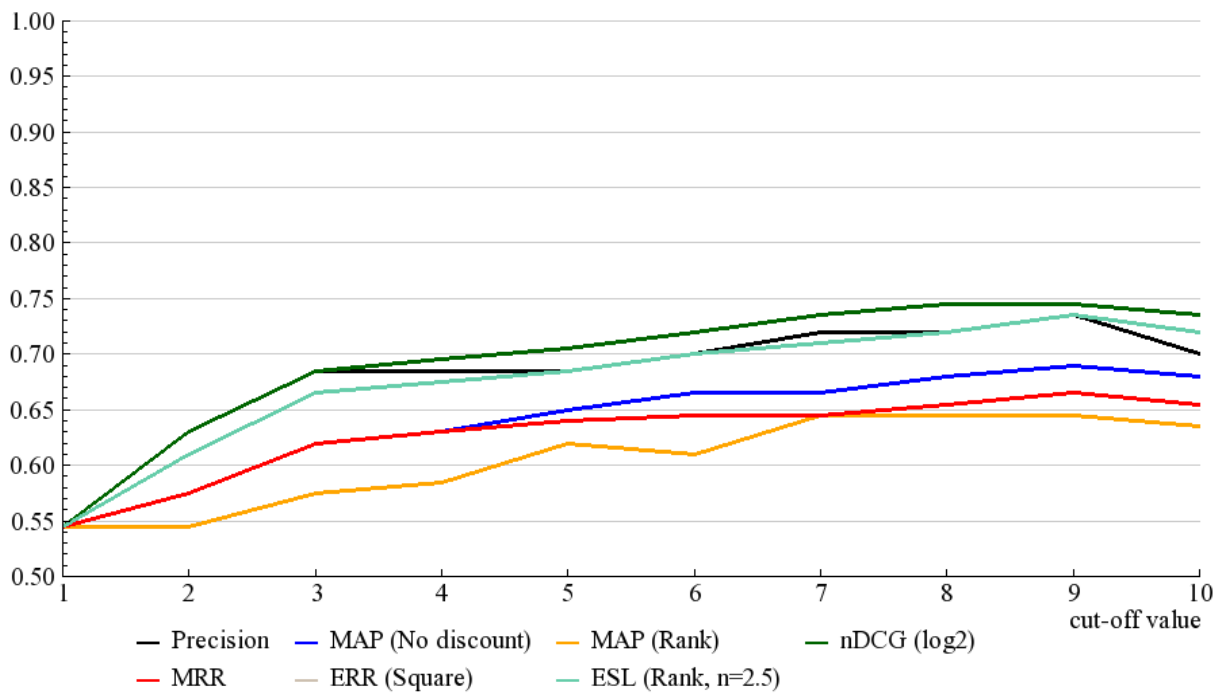

**Figure 10.63. Inter-metric PIR comparison using the best-threshold method and binary $R2_1$ relevance. ERR and ESL have exactly the same scores.**

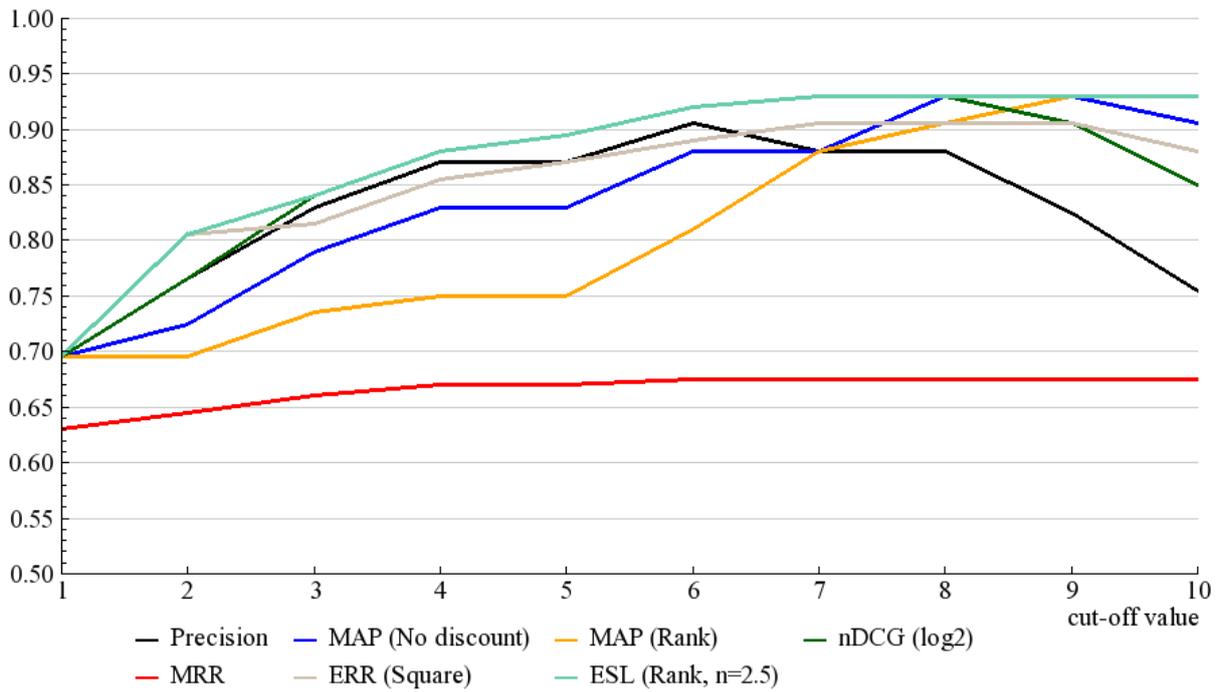

Figure 10.64. Reproduction of Figure 10.36 with a modified Y-scale . Inter-metric PIR comparison using the best-threshold method and six-point relevance, for informational queries only. NDCG (log$_2$) and ESL (Rank, *n*=2.5) have the same values for cut-off ranks 3 to 8.

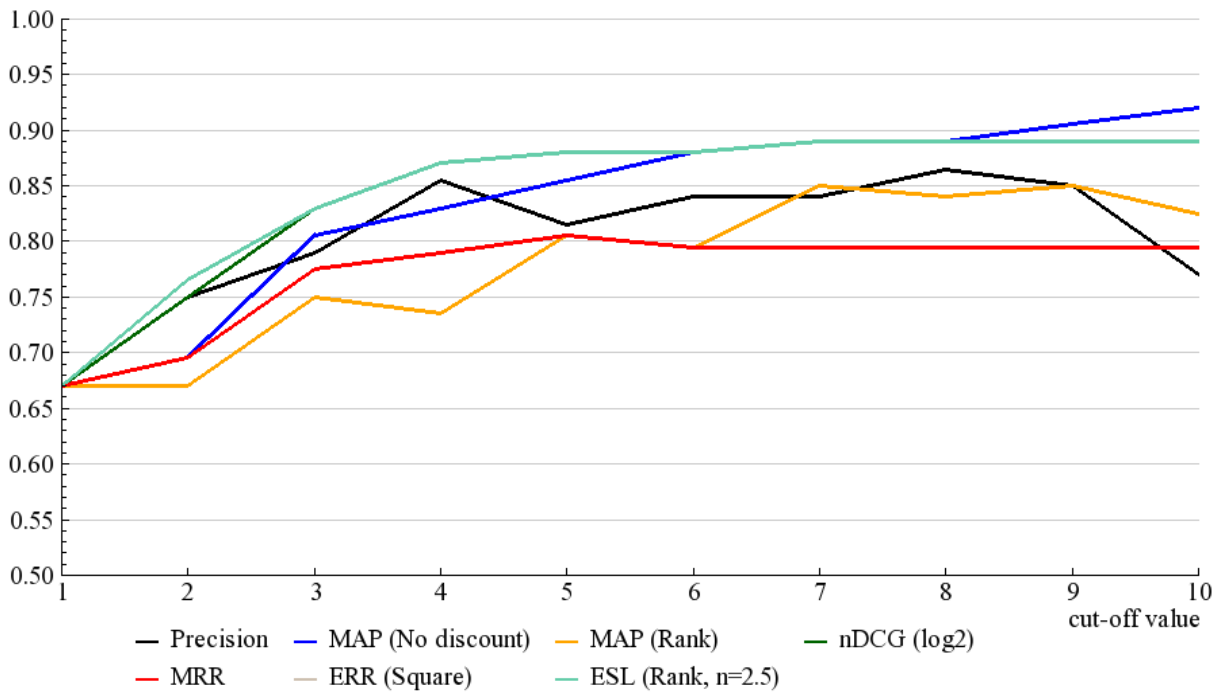

Figure 10.65. Inter-metric PIR comparison using the best-threshold method and binary R2$_3$ relevance for informational queries only.

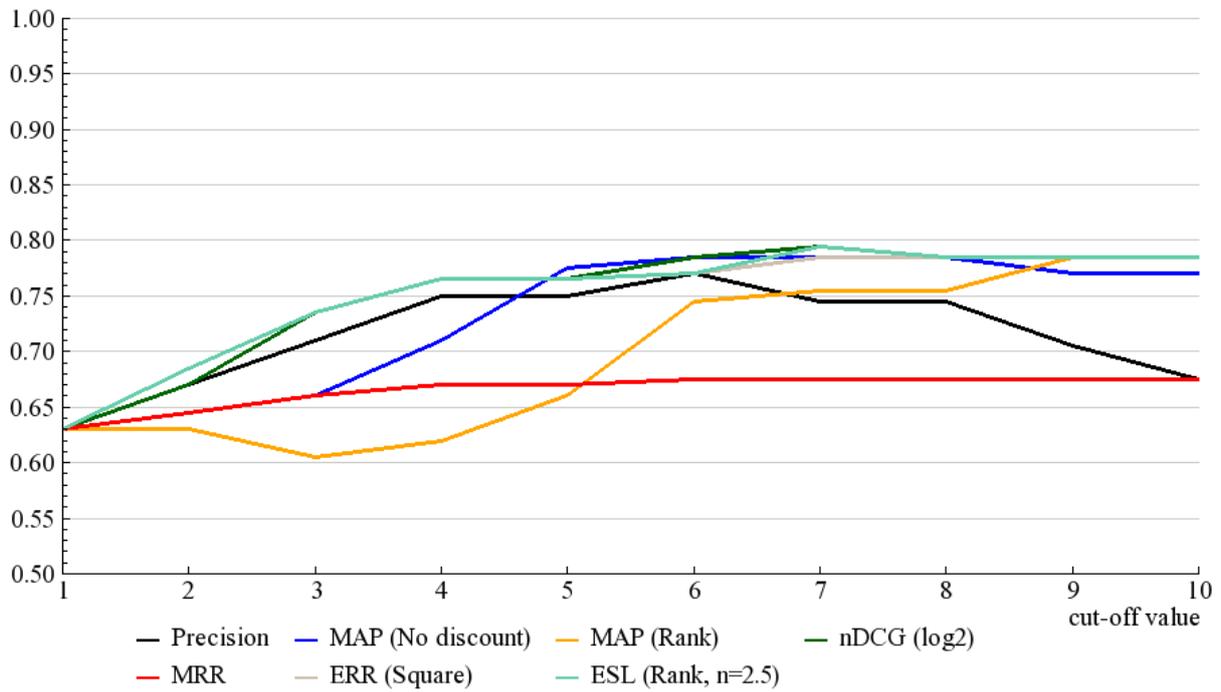

**Figure 10.66.** Inter-metric PIR comparison using the best-threshold method and binary $R2_1$ relevance for informational queries only.

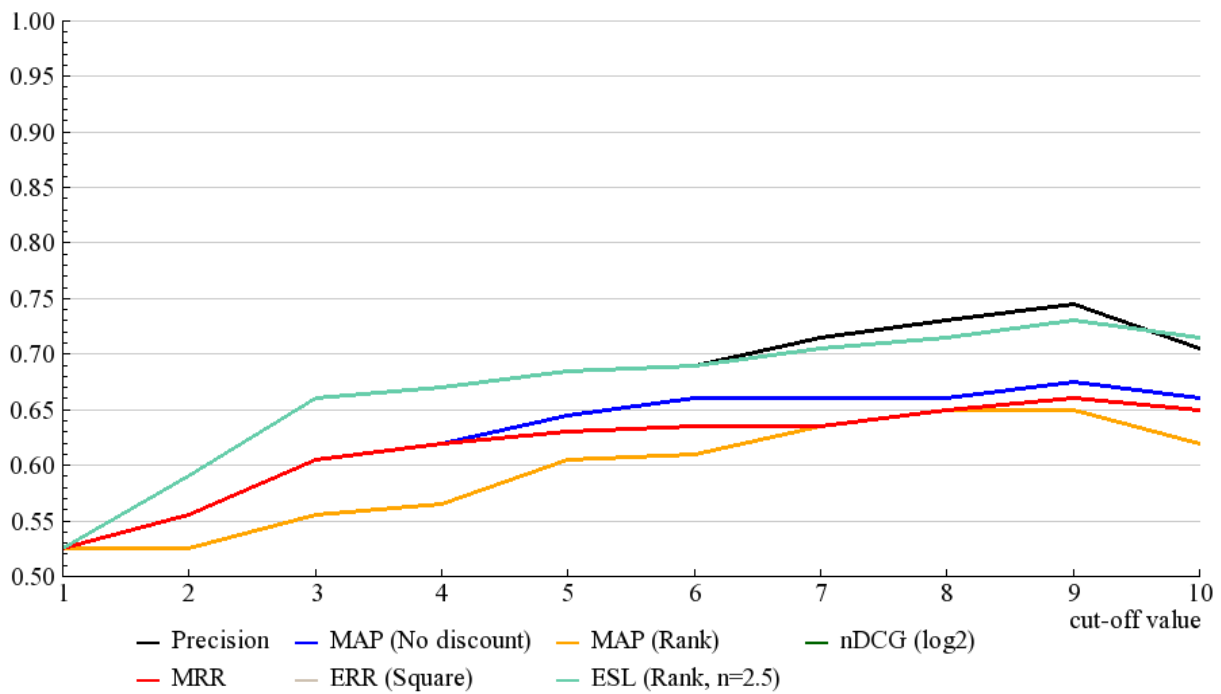

**Figure 10.67.** Inter-metric PIR comparison using the best-threshold method and binary $R2_5$ relevance for informational queries only.

The next step is to examine what happens if we use binary relevance judgments coming from users other than those providing preference ratings, thus adding the effects described in the previous section. The first graph in the next cluster (Figure 10.68) again shows the six-point relevance version, which is the reference point.[96] For the $R2_3$ binary relevance, the peak values are somewhat lower (around 0.77, down from 0.8); but not all metrics perform worse. ERR scores improve a bit, and Precision now has a top score of 0.77 as early as cut-off rank 4. The largest overall decline can be seen in both MAP metrics, which perform worst.

The situation is different for $R2_5$ relevance (Figure 10.70). At higher ranks (mostly up to rank 6), the PIR scores are even lower than for $R2_3$, apart from the MAP scores which do not perform as bad. In fact, the no-discount MAP performs best at this stage, while NDCG does worst. At about cut-off rank 7, the scores for $R2_3$ and $R2_5$ are about similar; and after that, most $R2_5$ scores improve, while those for $R2_3$ stay approximately the same or decline. The PIR scores at cut-off rank 10 are similar in both cases, except for Precision (lower for $R2_5$) and ESL. By ranks 9 and 10, ESL reaches 0.8, which is also the peak value in the six-point relevance evaluation.

How does it come about that the results at cut-off ranks 7 and 8 seem to contribute more towards the correct preference predictions than those at ranks 4 and 5? As relevance is defined as "not totally irrelevant" for $R2_5$, the only possibility to distinguish between two result lists is a situation when one has more totally irrelevant results, or (depending on the discount function) the totally irrelevant results occur earlier. One possible interpretation is that, at earlier ranks, the weak relevance criteria used carry little distinguishing power. Almost any result may be regarded as "concerning the query to some extent". After a while, either the number of totally irrelevant results changes (which might be true for the original ranking result lists – but hardly for randomized ones), or the raters' standards rise with growing acquaintance with the topic at hand and information saturation. The latter explanation cannot be true in this case since we are dealing with relevance ratings obtained from users who were not exposed to any of the result lists; they saw only single results, in random order, and mostly only two or three results from each query. However, the detailed evaluation of relevance ratings in Section 9.4 (Figure 9.19 in particular) does show some patterns similar to the issue under consideration; it is around rank 7 when the relative frequency of results judged to be absolutely irrelevant stops rising. This means that from rank 7 on, the relevance scores of the original result lists stop declining towards the levels of the randomized result lists. This is but a weak connection, and might be just coincidence – but at the present time, I do not see a more vested explanation of the results.

Evaluations using the zero-threshold approach, or informational queries only, produce very similar results. The former have slightly lower but more volatile scores (Figure 10.72), while the scores in the latter are marginally higher (Figure 10.73). Graphs are shown for $R2_3$ only since the results for $R2_5$ and $R2_1$ have the same properties.

---

[96] For better readability, the Y-scales of different-user rating evaluation graphs run not from 0.5 to 1, as in the same-user rating condition, but from 0.5 to 0.8. Since this time no metric exceeds the PIR score of 0.8, this does not constitute a problem.

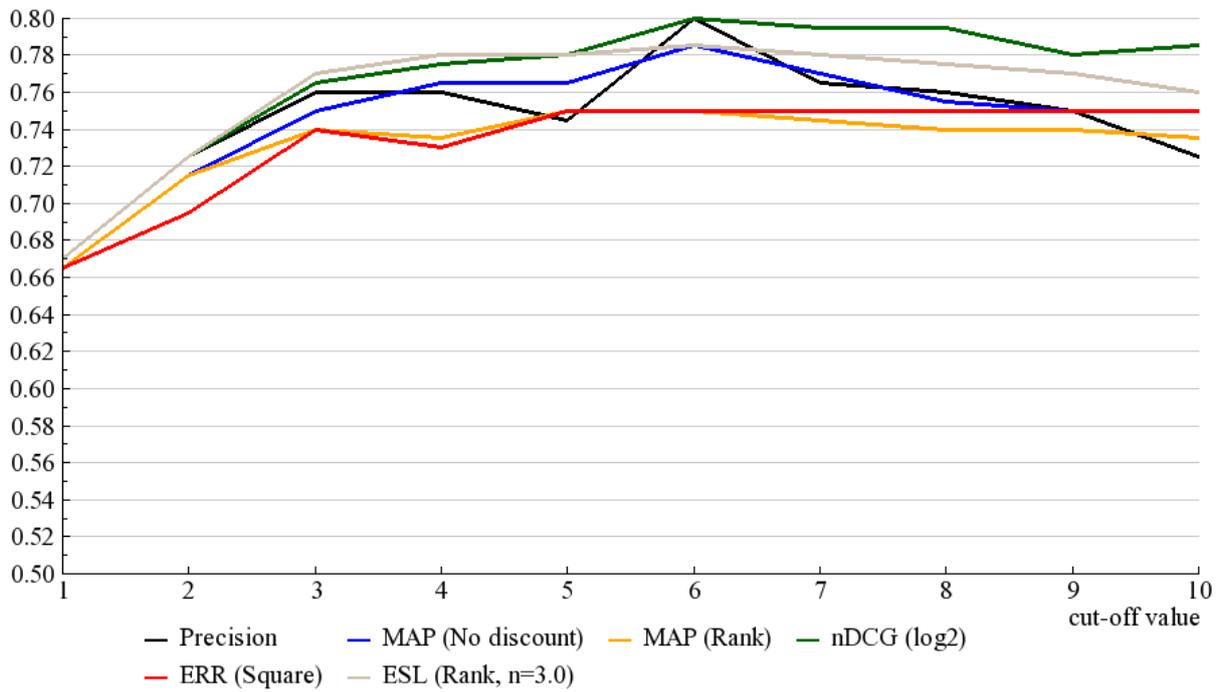

**Figure 10.68.** Reproduction of Figure 10.47 with a modified Y-scale: Inter-metric PIR comparison using the best-threshold method and six-point relevance. Result and preference ratings come from different users.

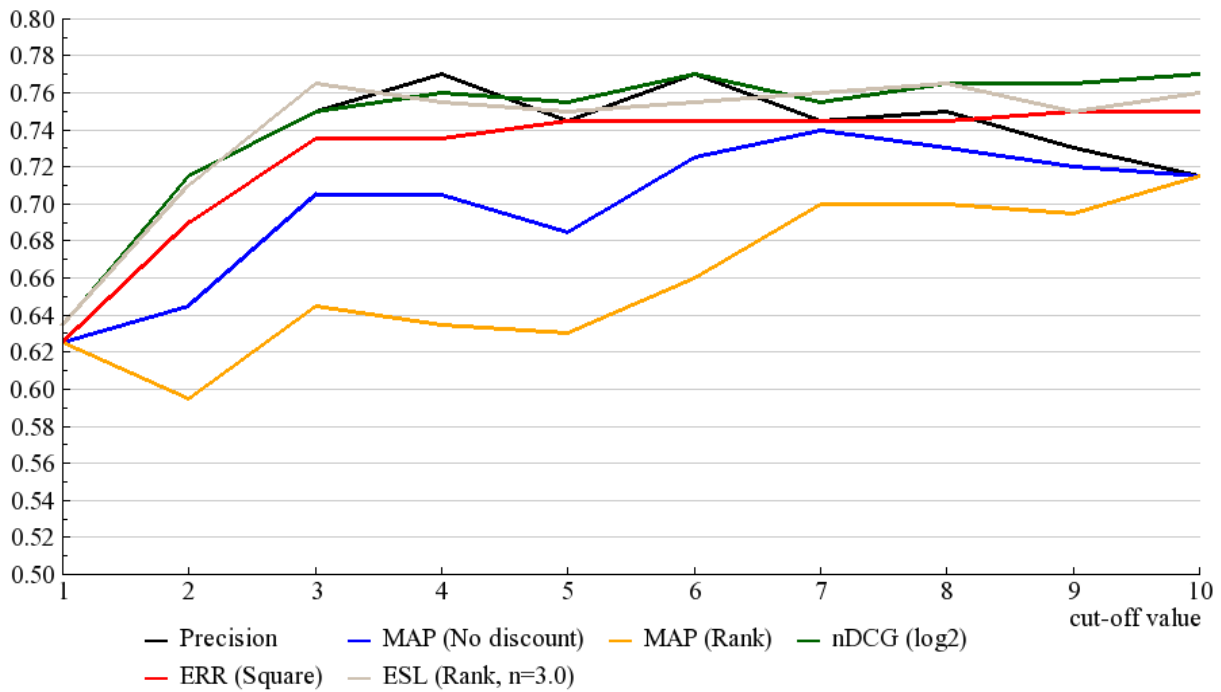

**Figure 10.69.** Inter-metric PIR comparison using the best-threshold method and binary $R2_3$ relevance. Result and preference ratings come from different users.

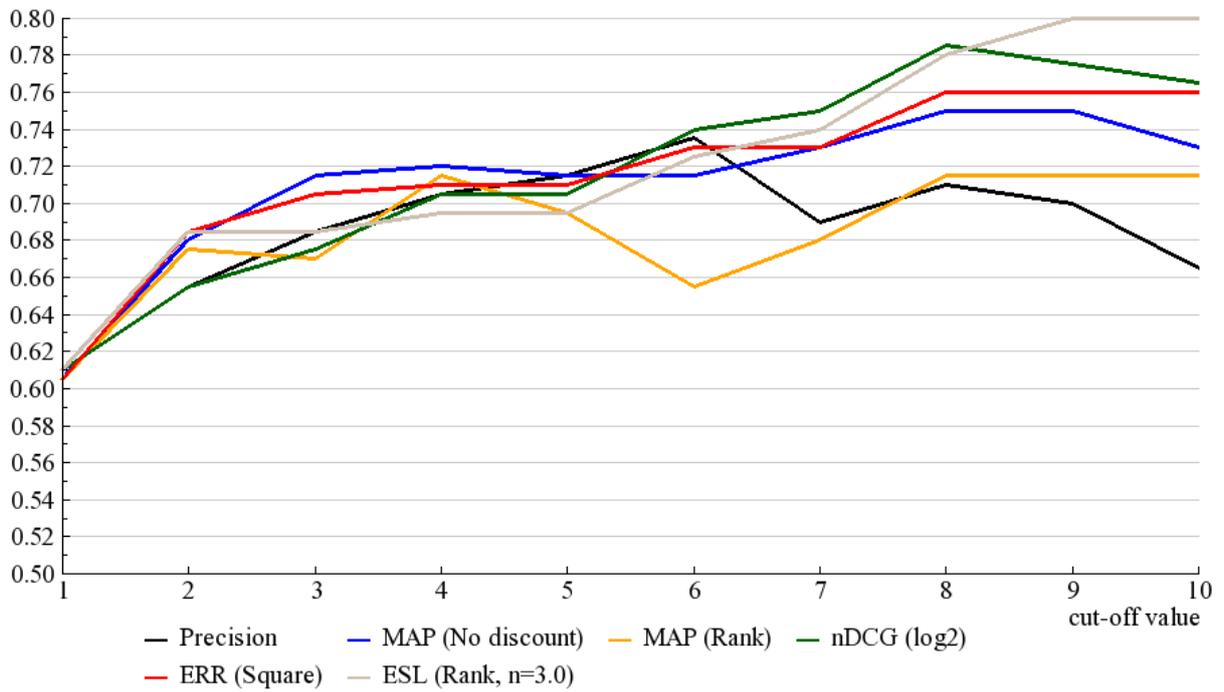

Figure 10.70. Inter-metric PIR comparison using the best-threshold method and binary $R2_5$ relevance. Result and preference ratings come from different users.

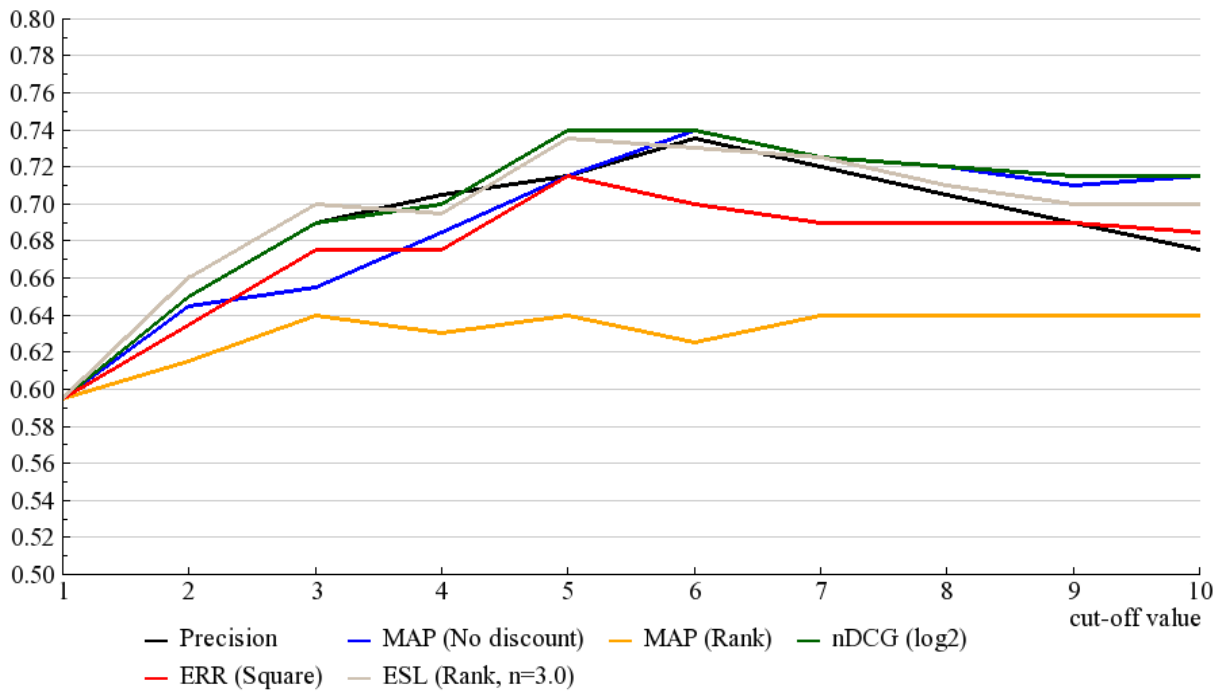

Figure 10.71. Inter-metric PIR comparison using the best-threshold method and binary $R2_1$ relevance. Result and preference ratings come from different users.

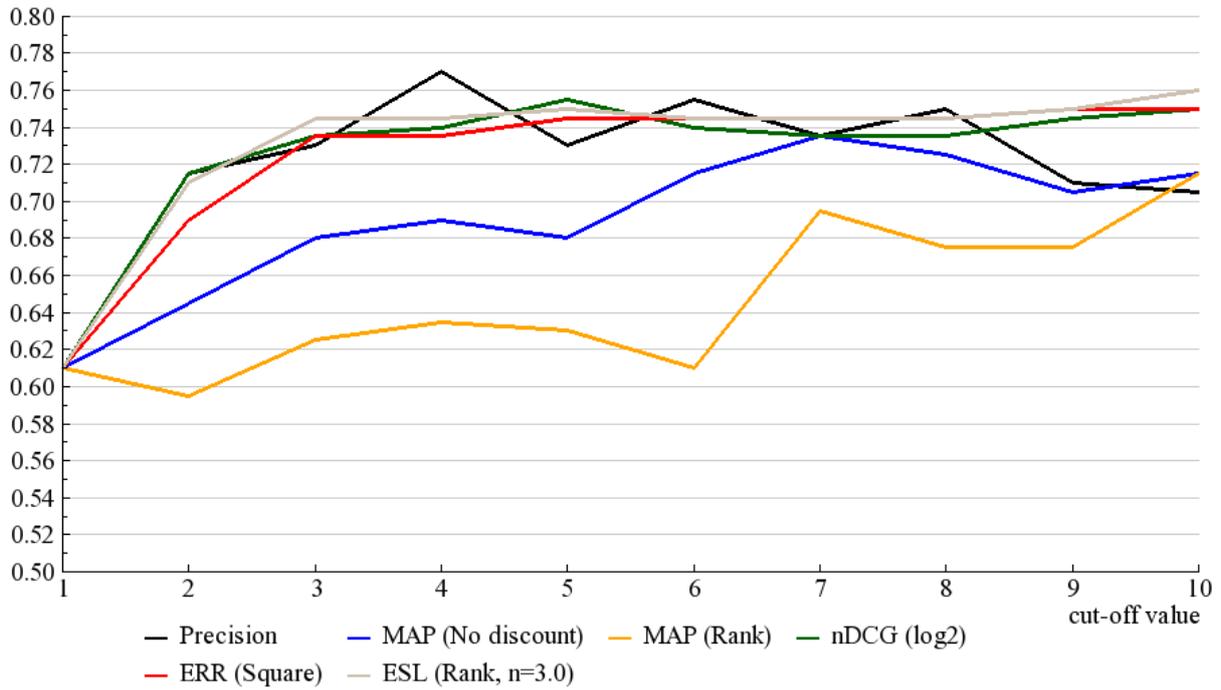

**Figure 10.72.** Inter-metric PIR comparison using the zero-threshold method and binary $R2_3$ relevance. Result and preference ratings come from different users.

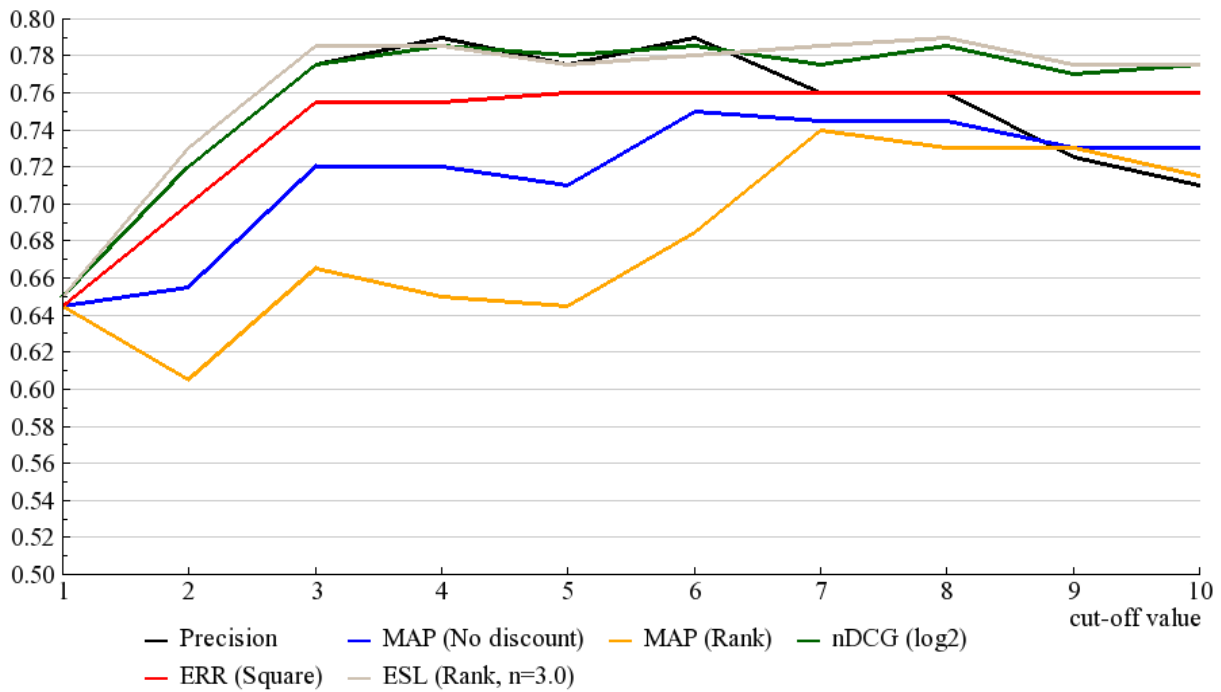

**Figure 10.73.** Inter-metric PIR comparison using the best-threshold method and binary $R2_3$ relevance for informational queries only. Result and preference ratings come from different users.

All in all, the results of this section's evaluation suggest that metrics perform better when using six-point relevance rankings than when they use binary relevance. This is less of an issue when the relevance ratings and preference judgments come from different users (which is mostly the case in real-life evaluations), when the absolute PIR values are not that high to begin with. But even in this case, there are differences. If binary relevance is employed for whatever reasons, then it is worth the time to provide the raters with guidelines as to what constitutes relevance.

### 10.7.2 Three-point Relevance

I now proceed to three-point relevance. As discussed above, the two possibilities of converting six-point relevance to three-point relevance are an equal division (1 and 2 as 1.0; 3 and 4 as 0.5; 5 and 6 as 0) or an extreme division (1 as 1.0; 2, 3, 4 and 5 as 0.5; 6 as 0). These possibilities will be examined analogous to those for binary relevance in the previous section.

The threshold-based evaluations do not show any marked differences. The rank-based NDCG scores serve as an example, but the situation is similar for other metrics and discounts. The graphs for $R3_2$ (Figure 10.74) and $R3_1$ (Figure 10.75) show some differences to each other and to the six-point relevance evaluation (Figure 10.5); but the differences are those of degree, not of kind, and consist largely of changes in absolute values, which I shall deal with immediately, in an inter-metric comparison. The same is true for discount function comparison, exemplified by NDCG in Figure 10.76 and Figure 10.77.

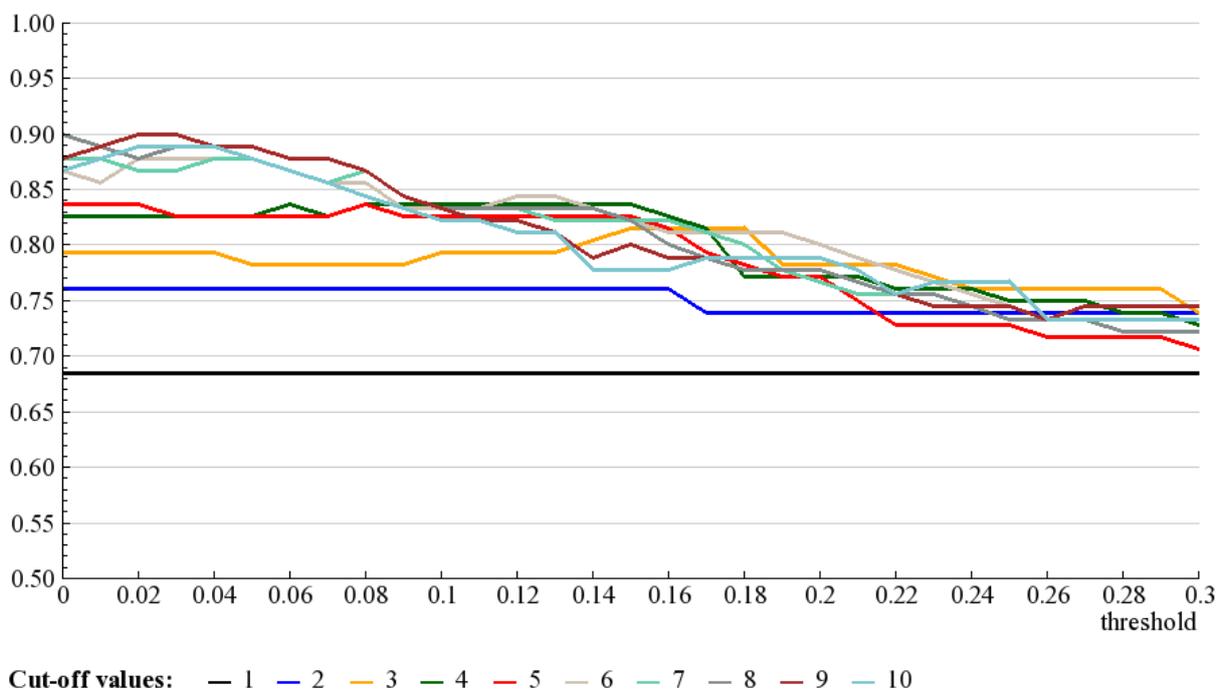

**Figure 10.74.** NDCG PIR values without any discount for results at later ranks. $R3_2$ relevance (1 and 2 as relevant, 3 and 4 as partially relevant, 5 and 6 as non-relevant) is used.

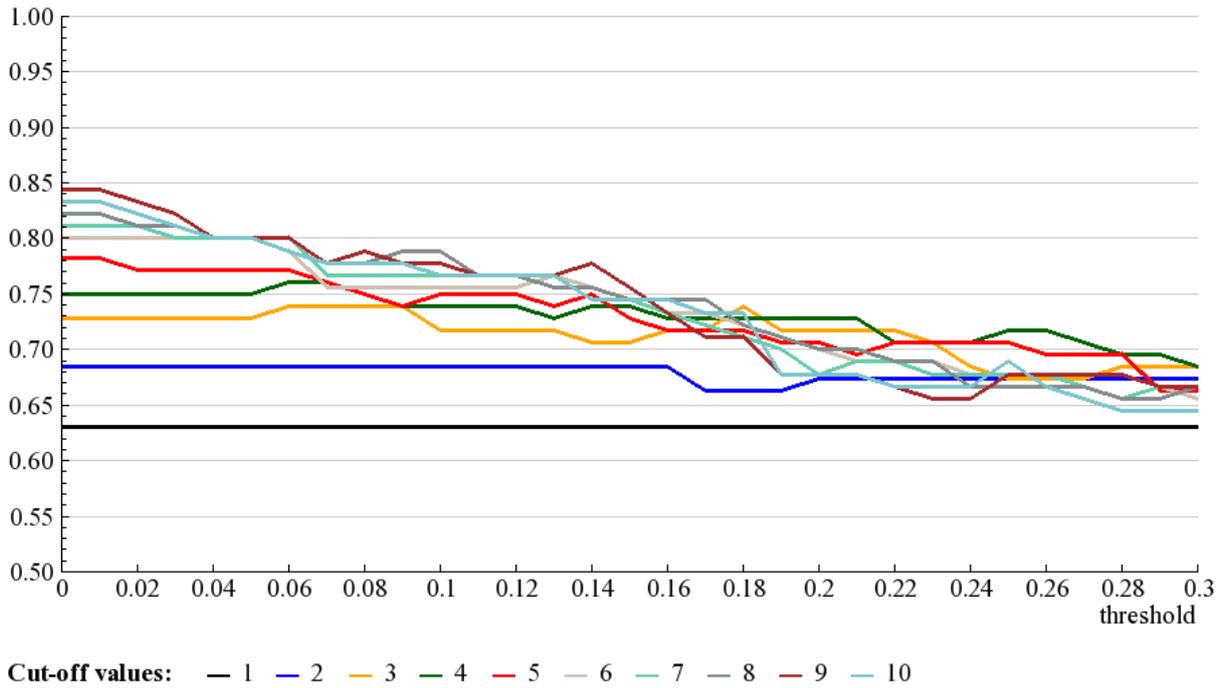

Figure 10.75. NDCG PIR values without any discount for results at later ranks. $R3_1$ relevance (1 as relevant, 2 to 5 as partially relevant, 6 as non-relevant) is used.

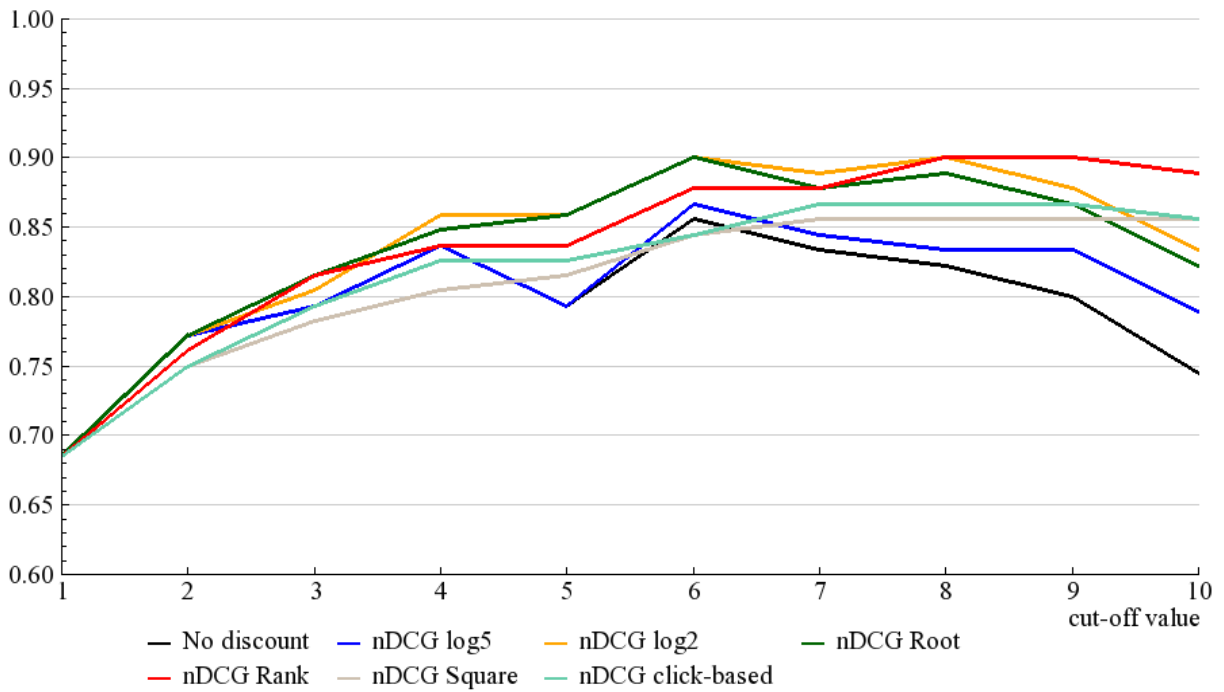

Figure 10.76. NDCG PIR scores for different discount functions, with the best-threshold approach. $R3_2$ relevance (1 and 2 as relevant, 3 and 4 as partially relevant, 5 and 6 as non-relevant) is used.

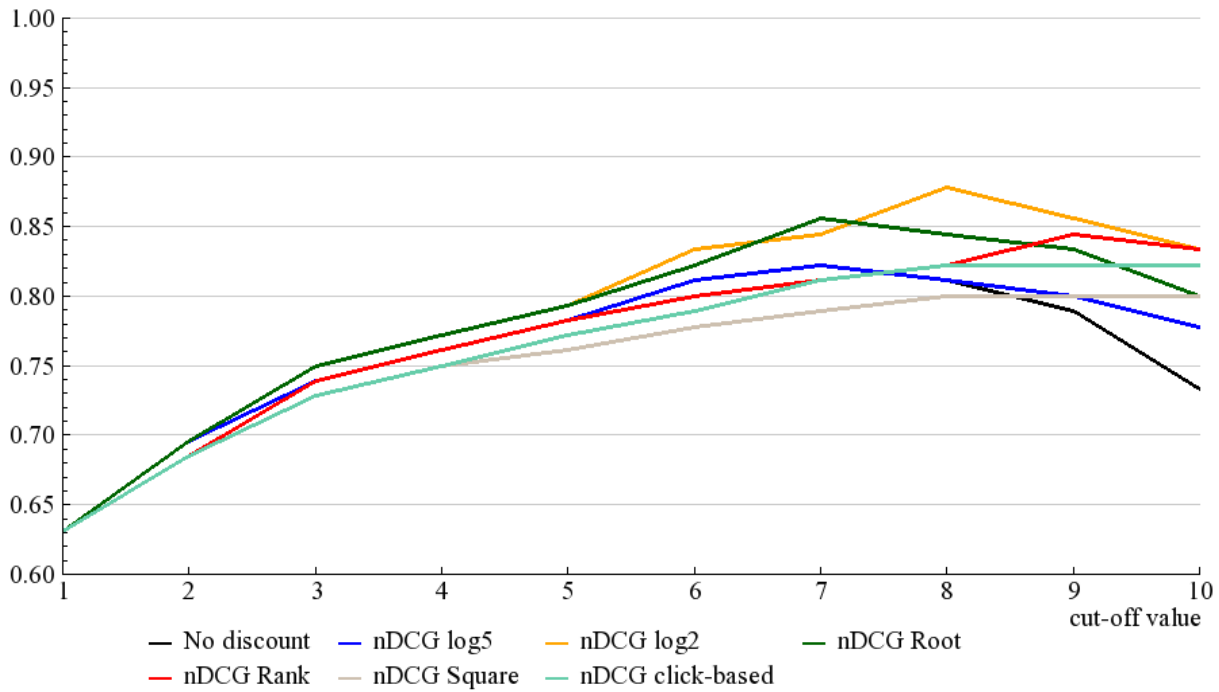

**Figure 10.77.** NDCG PIR scores for different discount functions, with the best-threshold approach. $R3_1$ relevance (1 as relevant, 2 to 5 as partially relevant, 6 as non-relevant) is used.

The small-scale issues taken care of, we can now look at the larger picture, that is, the inter-metric PIR comparisons. Once again, I start with a reminder of what the original, six-point relevance evaluation looked like (Figure 10.78). The $R3_2$ relevance evaluation shown in Figure 10.79 shows some changes reminiscent of those we saw in the binary relevance evaluations. Most metrics show a more or less pronounced decline in PIR scores, in particular those that had performed best (ESL, NDCG, no-discount MAP). MRR and (at earlier results) rank-discount MAP show slightly higher scores than in the six-point relevance evaluation. As in the case of binary relevance, the improvements in the performance of MRR can be explained easily. The MRR score is determined by the first occurrence of a result which is not irrelevant. In the six-point relevance scale, this means any result with a rating from 1 to 5; in the $R3_2$ scale this changes to the equivalent of the six-point scale's 1 to 4 ratings. Thus, the criteria are stricter, and the discriminatory power of MRR rises, albeit not to a point where it can outperform any other metric. Overall, the top PIR scores for $R3_2$ relevance are lower than those for six-point relevance by 0.03 to 0.05 points; not much, but still quite noticeable.

If we are looking for an evaluation which *does* perform much worse, we do not have to search for long; $R3_1$ relevance does the job quite nicely. As Figure 10.80 shows, almost all the scores are significantly lower than even in the $R3_2$ condition. The top PIR score (NDCG at rank 8) lies at 0.88, and the other metrics do not reach 0.85. Otherwise, the shape of the lines is quite similar to that of six-point evaluation, and (perhaps because of the broad band of "partly relevant" results) the graph is less erratic.

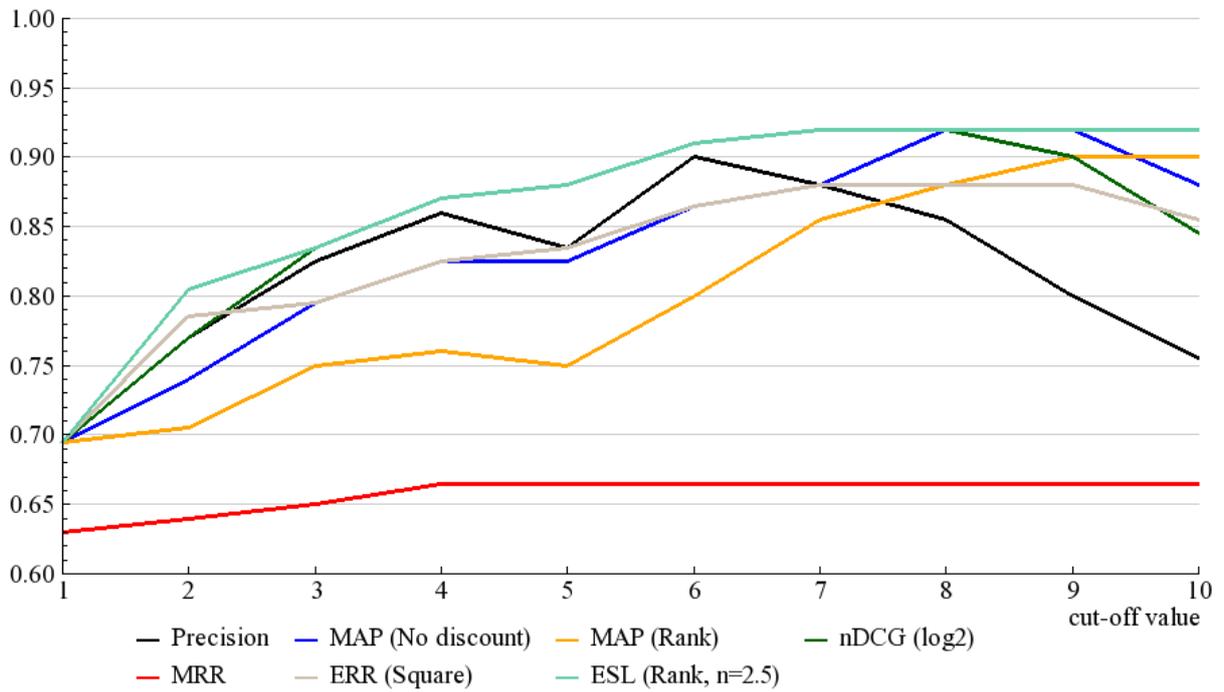

**Figure 10.78.** Reproduction of Figure 10.34 with a modified Y-scale. Inter-metric PIR comparison using the best-threshold method and six-point relevance.

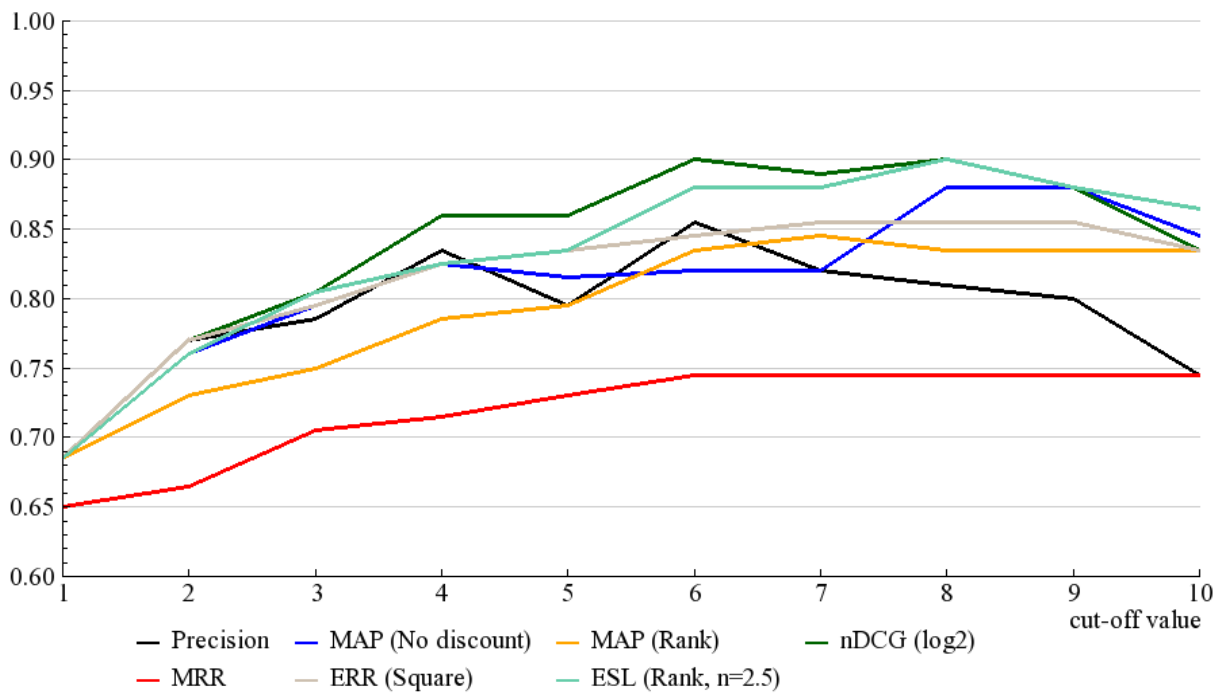

**Figure 10.79.** Inter-metric PIR comparison using the best-threshold method and three-point $R3_2$ relevance.

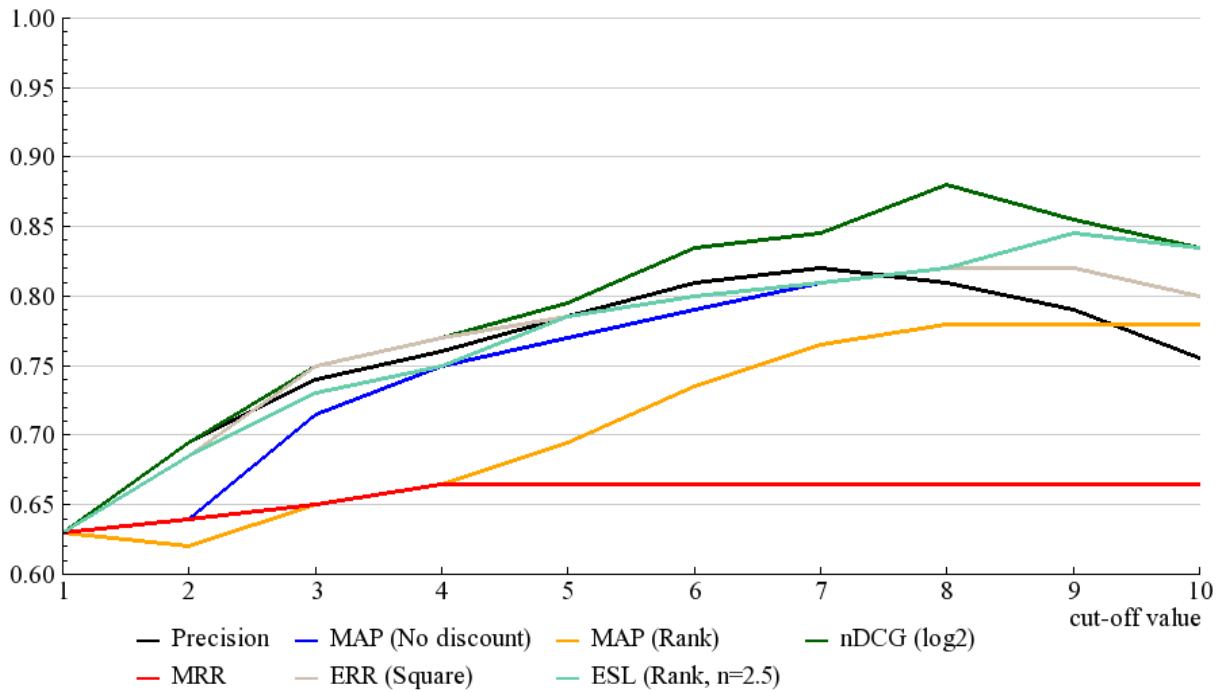

**Figure 10.80. Inter-metric PIR comparison using the best-threshold method and three-point $R3_1$ relevance.**

As the results for informational queries only are very similar to those for all queries, I proceed immediately to the next evaluation. Again, it will be the condition in which relevance ratings and preference judgments come from different users. Though the differences between metrics are lower than in the same-user condition, this time I will keep the Y-scale at PIR values 0.60 to 1, to show more clearly the differences between the two categories of evaluation.

Even the graph based on the six-point relevance evaluation (Figure 10.81) is much denser than any other previously shown in the present section. With all PIR values in the range from 0.66 to 0.80, it also gives an impression of more evenness. Again, NDCG performs best, followed by Precision and ESL, while ERR and rank-discounted MAP do worst; however, the differences between peak PIR values of the different metrics do not exceed 0.1. The discrepancies shrink further when we switch to the $R3_2$ evaluation (Figure 10.82); except for rank-discounted MAP, whose scores improve, the other metrics' PIR mostly declines.

This is, however, not the case with $R3_1$ (Figure 10.83). Some metrics' PIR scores did decline a bit, especially at earlier cut-off ranks; at cut-off rank 4, all the scores are almost equal. But starting with cut-off rank 6, all metrics' performances (with the exception of Precision) actually improve. This situation is analogous to that in the $R2_5$ evaluation with different users providing relevance ratings and preference judgments (cp. Figure 10.70). An obvious common feature of the two evaluations is the use of only the lowest possible relevance rating as "completely non-relevant". However, as has been argued in the discussion of the $R2_5$ graph, there is no clear reason why this should lead to an increase in PIR scores from this particular cut-off rank on. Thus, this issue remains a riddle for now.

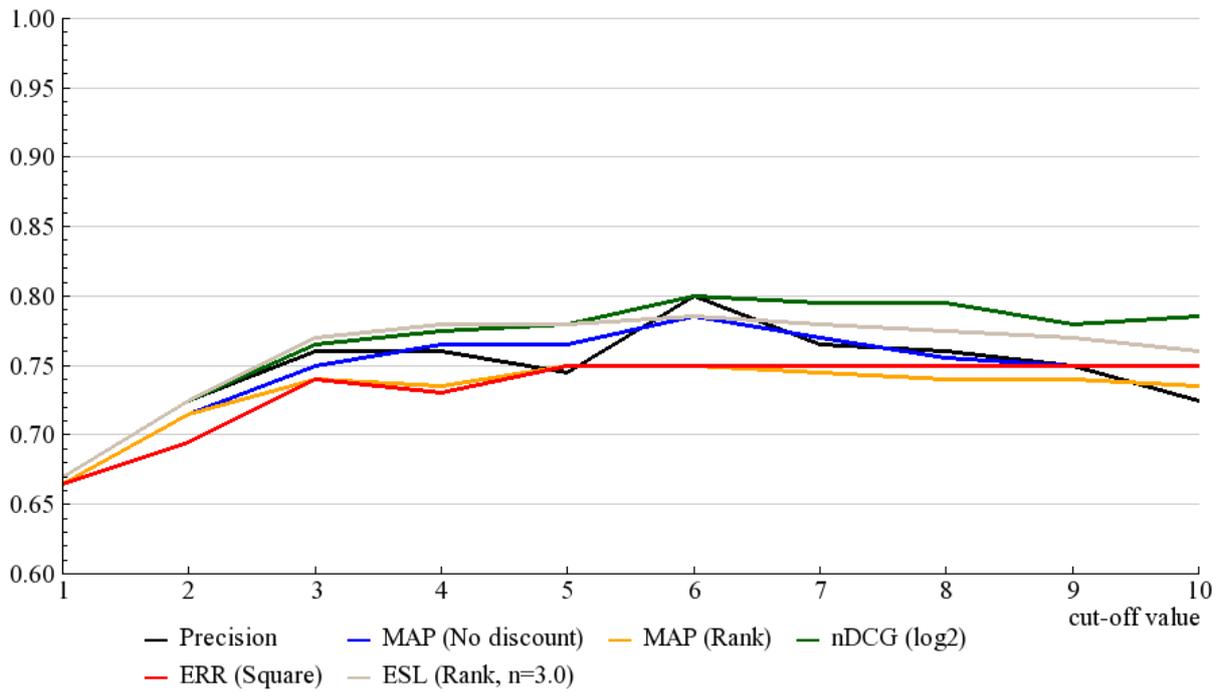

**Figure 10.81.** Reproduction of Figure 10.47 with a changed Y-scale. Inter-metric PIR comparison using the best-threshold method. Result and preference ratings come from different users.

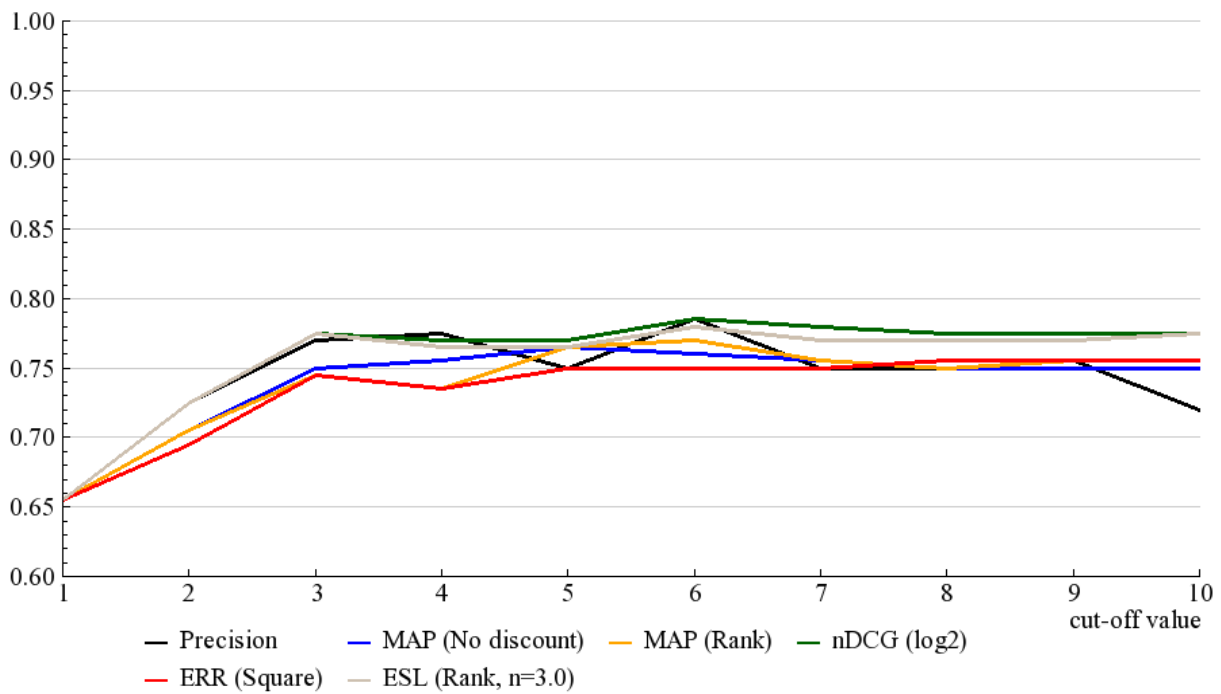

**Figure 10.82.** Inter-metric PIR comparison using the best-threshold method and three-point $R3_2$ relevance. Result and preference ratings come from different users.

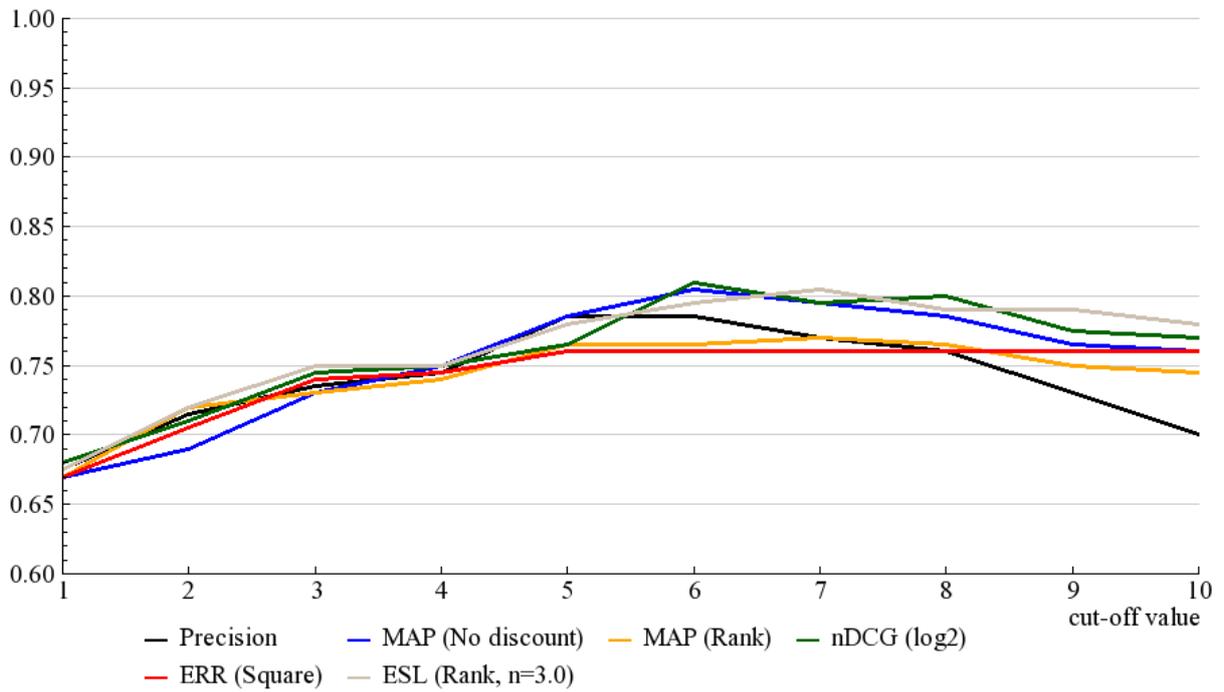

**Figure 10.83.** Inter-metric PIR comparison using the best-threshold method and three-point $R3_1$ relevance. Result and preference ratings come from different users.

# 11 Implicit Metrics

> *I have always thought the actions of men the best interpreters of their thoughts.*
>
> JOHN LOCKE,
> "ESSAY CONCERNING HUMAN UNDERSTANDING"

As I have described in Part I, explicit metrics based on users' relevance ratings of individual results are but one of the two major search engine evaluation methods. The other is based on the data provided more or less unwittingly by the users who go about their usual business of performing web searches. As mentioned in Section 9.1, some of that data has been gathered during the study. The important types of data, the ones that are going to be evaluated in this section, are session duration and clicks.

Most click metrics described in Section 5 are relative metrics; that is, they do not provide a numeric value for individual results or the result list as a whole, but rather indicate whether one result in the list is better than another. This can be a useful feature if one wants to improve a ranking algorithm, because if it works, it provides not only an indication that something is wrong, but also a concrete way of fixing the problem (for example, by swapping the two results' positions). However, relative log metrics are not easy to evaluate, at least using the methods of this study. This methods do not show which of two given result lists is better; instead, they take as their input one result list, and give as their output an improved version. Therefore, evaluating them would require three steps:

- letting users interact with a result list in the usual way;
- using the relative metric, constructing a second, presumably superior result list;
- asking (other) users which of the two result lists is preferable.

This method is not impracticable; but it requires an evaluation structure quite different from what is needed for the explicit (and log-based absolute) metrics which constitute the main body of this study. Therefore, the evaluation of relative, click-based metrics will be left to further studies to undertake.

## 11.1 Session Duration Evaluation

The amount of time a user spends on examining the results presented by the search engine has been proposed as an indicator of those results' quality. However, a problem that arises immediately is the definition and determination of "session duration". One might reasonably define it as the time between the user submitting the query[97] and his completion or abortion of the search.[98] But, while the moment of query submission is routinely recorded in search engine logs, the end point is less obvious. If we see a session as a sequence of

---

[97] Since modern web search engines return their results in a negligible time upon query submission, the time points of submitting the query and seeing the result list are the same for all practical purposes.
[98] In the present evaluation, I am concerned only with single-query sessions. Multi-query sessions constitute a significant part of all search activity, and are quite likely to be a more realistic model of user behavior than single-query sessions. There is also no obvious reason why they could not be evaluated using the same PIR

(1) submitting a query,
(2) examining the result list,
(3) selecting a result,
(4) examining the web page,

and then possibly

(5) returning to the result list and proceeding with step (2),

then the session ends after step (4), whether on the first or a subsequent iteration. But whereas the steps (1) and (3) can be (and are) recorded by search engines, there is no obvious way to know when the user has stopped examining the last result and decided he will not return to the result list. Furthermore, if the user's information need is not satisfied yet, but he looks at the result list and finds no promising results, the session might also end after step (2). Particular cases of this scenario are no-click sessions, which may have an additional twist; as described in Section 9.2, the absence of clicks can also mean a result list which provides enough information to satisfy the user's information need without delving into the results themselves, especially in the cases of factual queries or meta-queries.

In this study, two possible end points of a session can be evaluated. We happen to have definite end points for every session, as raters were required to submit a preference or satisfaction judgment when they were "done" (see Chapter 9.1 for more details). However, since this information is not usually available to search engine operators, I will also evaluate session duration with the last recordable point in the session process as described above; that is, the time of the last recorded click.

The session duration is a fairly straightforward metric; one takes the timestamp of the session end, subtracts the time of the session start, and – voila! – one has the duration. Then, it can be judged against the actual user preference, just as has been done with the explicit metrics in Chapter 10. Here also, the threshold approach can be used in a way analogous to that employed for the explicit metrics. There might be a difference in session duration for which it does not make sense to assume an underlying preference for one of the result lists (say, 65 seconds versus 63 seconds).

There is also the possibility that a longer session is not necessarily indicative of a worse result list. Longer sessions might conceivably be generally preferable. A long session can indicate that the user has found many relevant results and spent a lot of time interacting with the web pages provided. In this view, a short session might just mean there wasn't much useful information in the result list.

---

framework used here. However, multi-query sessions require a approach to study design and data gathering that different from the one employed here, and so they will be left behind, without the benefit of having been evaluated.

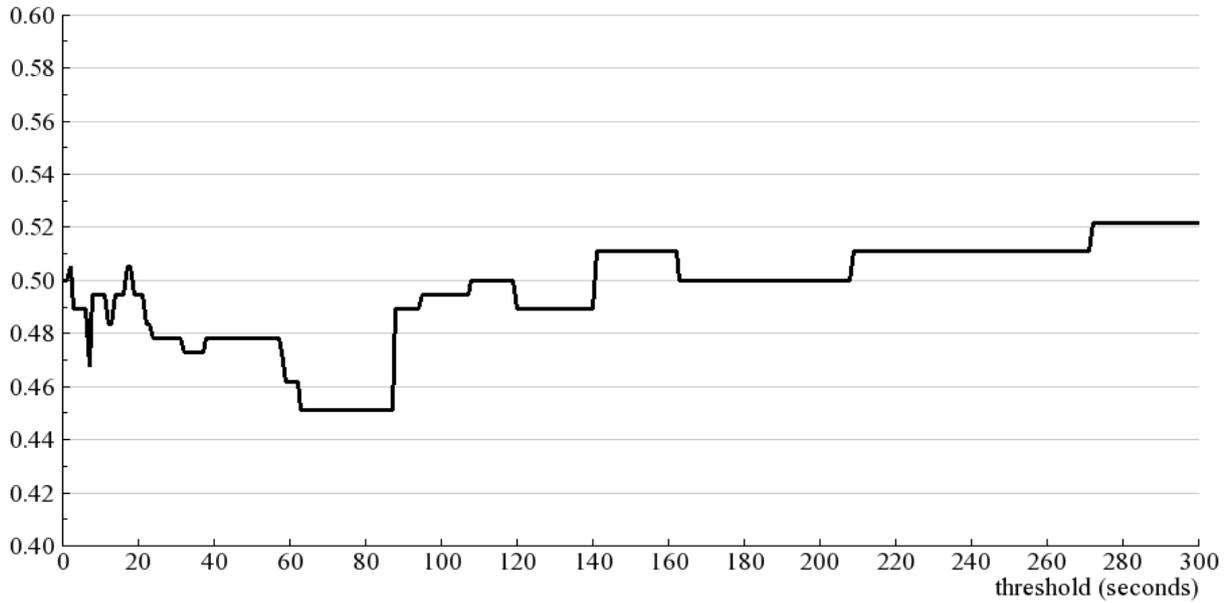

**Figure 11.1.** Preference Identification Ratio based on session duration (measured with the user-indicated session end). PIR is shown on the Y-axis and the threshold (difference in session duration, in seconds) on the X-axis. The PIR indicates the correct preference identification if shorter sessions are considered to be better. For most queries, the graph seems to indicate the opposite (better longer queries could be evaluated by mirroring the graph on 0.50 PIR), but the results are far too small to be of significance.

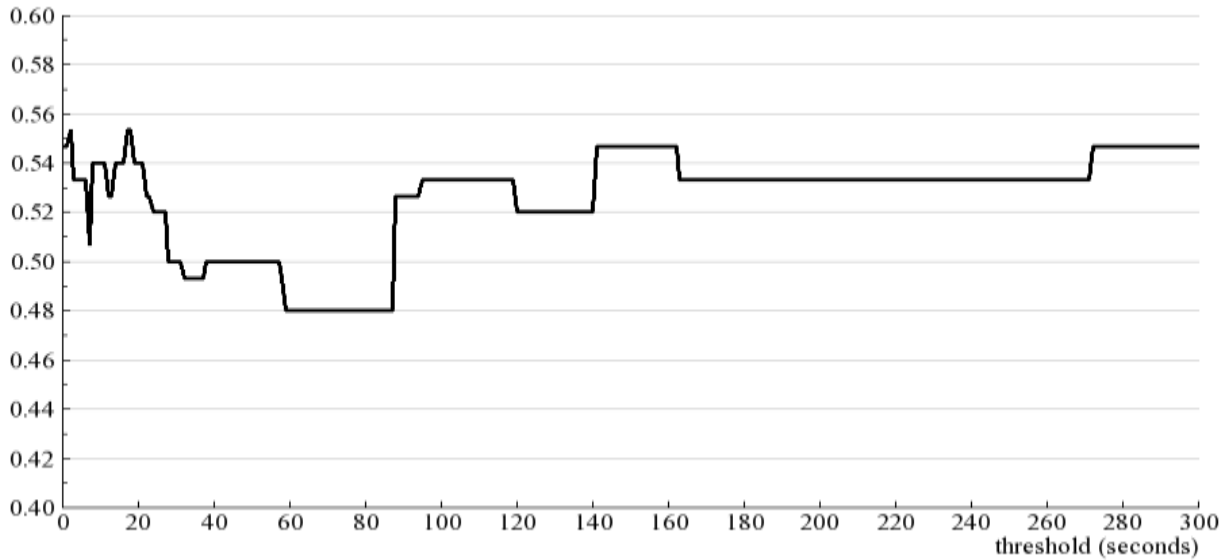

**Figure 11.2.** Preference Identification Ratio based on session duration (measured with the user-indicated session end).. PIR is shown on the Y-axis and the threshold (difference in session duration, in seconds) on the X-axis. The PIR indicates the correct preference identification if shorter sessions are considered to be better. Only informational queries have been assessed.

Figure 11.1 can put our minds to rest about the first assumption. Indeed, it shows a result that is, at least to me, quite surprising: The differences in session duration do not seem to indicate user preference in either direction. The PIR scores range approximately from 0.45 to 0.52; that is, the difference from a baseline performance is negligible. The same can be said for an evaluation of the informational queries only (Figure 11.2); while the PIR scores are slightly higher (0.48 to 0.56), the overall shape of the graph is the same as before.

Does this mean that session duration is no good at all as a predictor of user preference? Not necessarily. It might be that for certain specific durations, there are meaningful connections between the two. For example, sessions of less than five seconds may indicate a much worse result list than those from 20 seconds upwards, while for durations of more than a minute, shorter might be better.[99] Unfortunately, the more specific the situations, the narrower the duration bands become, and the less queries fall into each. In the end, the present study turns out not to be large enough to permit this kind of analysis. All that can be said is that there seems to be no direct, general and significant connection between session duration difference and user preference.

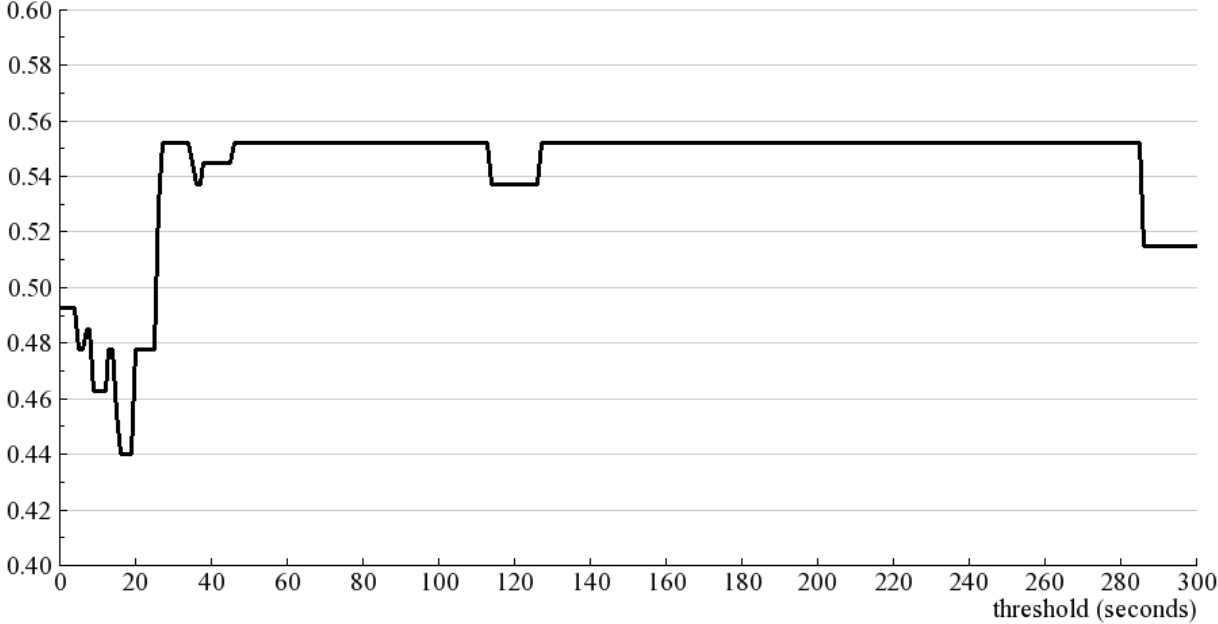

**Figure 11.3. Preference Identification Ratio based on session duration (measured with the last click as session end). PIR is shown on the Y-axis and the threshold (difference in session duration, in seconds) on the X-axis. The PIR indicates the correct preference identification if shorter sessions are considered to be better.**

---

[99] Or perhaps the opposites are true; this is meant not as a prediction of what duration means, but just of possible connections the analysis in the previous paragraphs would not have captured.

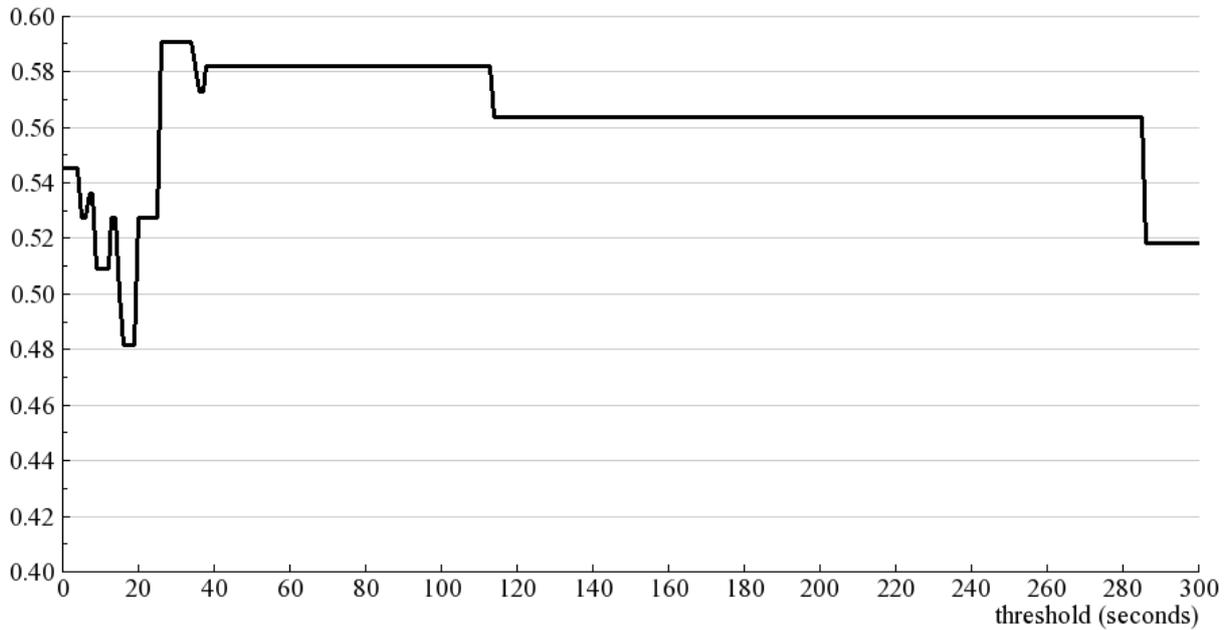

Figure 11.4. Preference Identification Ratio based on session duration (measured with the last click as session end). PIR is shown on the Y-axis and the threshold (difference in session duration, in seconds) on the X-axis. The PIR indicates the correct preference identification if shorter sessions are considered to be better. Only informational queries have been assessed.

Figure 11.3 shows the PIR graph with a different method of calculating session duration, namely, determining the end of the session by the last click made. Here, the range of PIR scores is a bit wider, from 0.44 to 0.56; however, once again this is nothing to get excited about. The scores are slightly higher than before, but still not high enough to provide meaningful clues as to user preference. A similar picture emerges in Figure 11.4, which shows the graph for informational queries only. In that case, the maximum is about 0.59 (for session duration differences of 35 seconds and above); still not enough to be useful, especially since for most session pairs, the PIR score is closer to chance.

The conclusion from this section is simple: In the present study, differences in session duration were not a good indicator for user preference; and mostly they were no indicator at all. However, this study was relatively small; in particular, it did not have enough data to evaluate sessions of particular durations only (say, sessions shorter than 10 seconds versus those taking more than half a minute). Such measures could conceivably produce better results.

## 11.2 Click-based Evaluations

Even if session duration does not cut the mustard, we can still turn to click data to provide the cutting edge. Some methods used for evaluating click data to infer the quality of a result list have been described in Chapter 5; they and a few others not widely employed in studies will be dealt with in this section The methods will be the same as those for explicit result ratings and session duration; that is, the relevant quality-indicating numbers will be calculated from clicks performed by users in the single result list condition, and then the difference in these numbers will be compared to actual user preferences using PIR. If you are still uncertain what that looks like in practice, read on; the evaluation should be a clear example of the method employed.

### 11.2.1 Click Count

The simplest way to measure something about the clicks a user performs is counting them. As is so often the case with evaluations (or at least with the approach to them taken in this study), there are at least two possible ways to interpret the number of clicks a result list gets. A low number of clicks might indicate that the user found the sought information soon; in the extreme case, no clicks can mean that the information was in the result list itself, eliminating the need to view any result.[100] Or, a low number of clicks might be a sign that most results (at least as presented in the result list) just weren't attractive, and the user was left unsatisfied. And, as usual, I will turn to evaluation to provide the conclusion as to which of the views is closer to the truth.

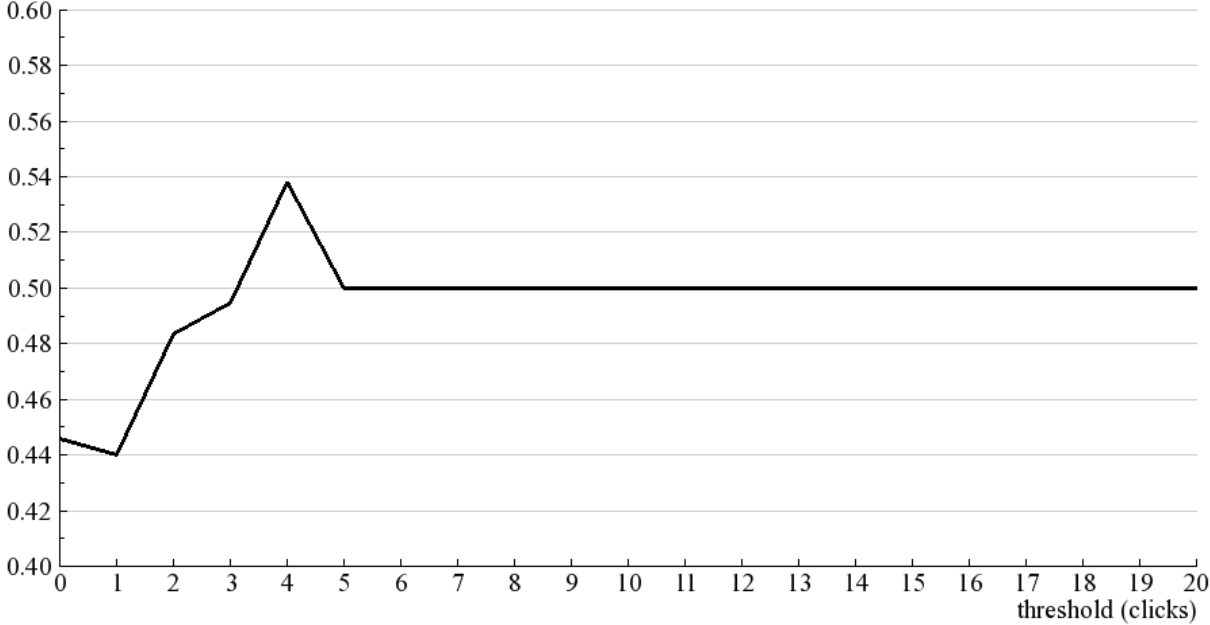

**Figure 11.5. Preference Identification Ratio based on a simple click count. PIR is shown on the Y-axis and the threshold (difference in click counts) on the X-axis. The PIR indicates the correct preference identification if less clicks are considered to be better.**

---

[100] This would necessarily be the case for meta-queries (described in Section 9.2).

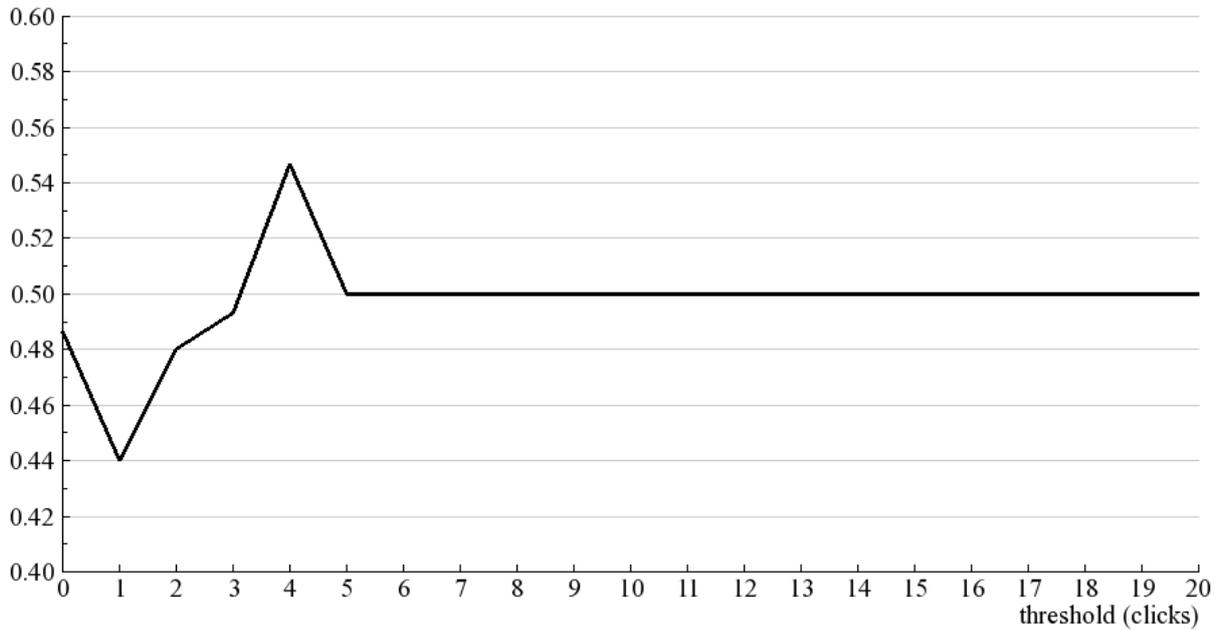

**Figure 11.6. Preference Identification Ratio based on a simple click count. PIR is shown on the Y-axis and the threshold (difference in click counts) on the X-axis. The PIR indicates the correct preference identification if less clicks are considered to be better. Only informational queries have been assessed.**

Figure 11.5 indicates that, as with the session duration, the answer is probably "none". The PIR lies between 0.44 and 0.54, never far from the baseline.[101] The picture is very similar if we restrict the evaluation to informational queries (Figure 11.6). Both show that the sheer number of clicks is not a reliable indicator of user preference; both smaller and larger click numbers can indicate superior as well as inferior result list quality.

---

[101] There were no queries where the difference in clicks for the two result lists was larger than 4; hence the stable value of 0.50 from a threshold of 5 on. Numbers for individual result lists (averaged among different sessions/users) were as high as 10 clicks, but the disparity was never too large. I have kept the larger thresholds in the graph for easier comparison to following graphs.

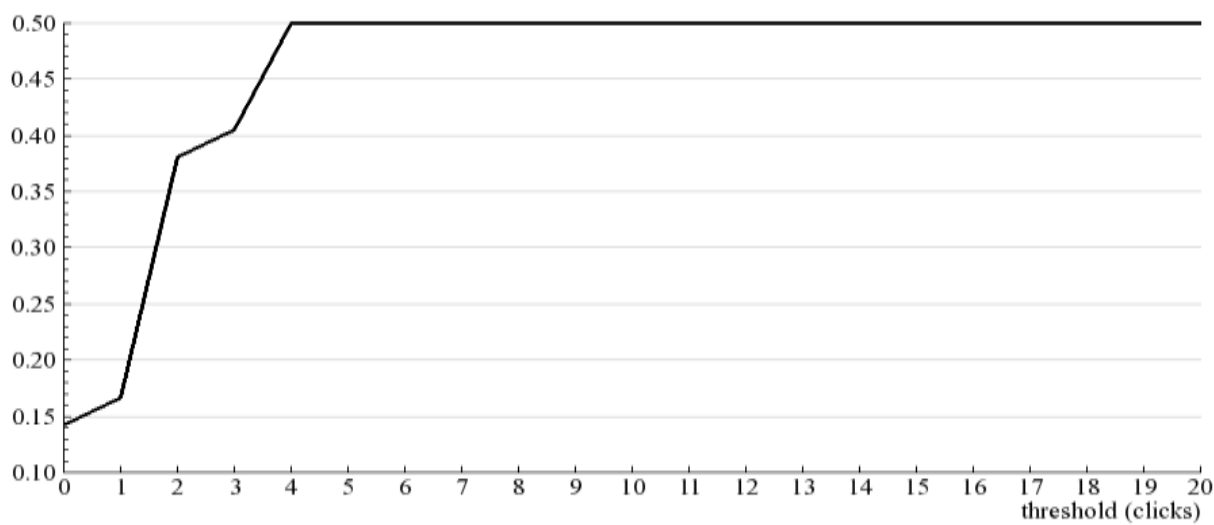

Figure 11.7. Preference Identification Ratio based on a simple click count, for queries where only one of the result lists received clicks. PIR is shown on the Y-axis and the threshold (difference in click counts) on the X-axis. The PIR indicates the correct preference identification if less clicks are considered to be better.

An additional possibility is considering that particular subset of sessions where one of the result lists does not receive any clicks. As Figure 11.7 shows, this does show a very pronounced tendency. The assumption "less clicks are better" is refuted strongly in this case; looking at the results from the other direction, we can say that the PIR for the "more clicks are better" metric is higher than 0.85. This is a high value indeed; but it has to be noted that it is not directly comparable to the other PIR scores we have seen. One reason for that is that the number of sessions fitting the criteria of one and only one of the result lists having a click count of zero is relatively small; these 35 sessions are about 1/4 of those used, for example, in the general click count evaluation. Even more importantly, the excluded cases are in large part precisely those where the metric performs poorly in terms of preference identification; for example, the sessions where neither result list receives clicks have, on their own, a PIR of 0.50.

This is not to say that the comparison of sessions with a click/no click difference is not valuable or meaningful. But previous PIR evaluations concerned the whole population of queries, which are supposed to be at least a rough approximation of real-life query variety. This result, however, stems from sessions preselected in a way that strongly favours a certain outcome. The evaluations which regard only informational queries also narrow the session population; however, they might be closer to of further from the baseline, while selecting based on the metric scores' differences necessarily nudges PIR towards a better performance.

What does this high score mean, then? Obviously, it means that if one result list has no clicks but another one does, the latter is very likely to be preferred by the user. By the way – if you wonder where the PIR evaluation for informational queries only is: all the queries that remained in this evaluation *were* informational. This also means the possibility that, especially for factual queries and meta-queries, a no-click session can be a *good* indicator, has not really been tested. But for the main body of queries, the result does not seem to be found in the result list itself. Also, it does not seem to happen with any frequency that all the clicks a user makes in a result list are for nothing; that is, here, the users seem to be quite good at scanning the result list and recognizing the absence of any useful results.

## 11.2.2 Click Rank

Another relatively straightforward metric is the average click rank. As with many of the explicit result ratings, the assumption here is that the user prefers it if good results appear at the top of the result list. If this is the case (and if the user can recognize good results from the snippets presented in the result list), the clicks in better result lists should occur at earlier ranks, while those in inferior ones would be in the lower regions.

Arguably, this approach ignores a lot of fine detail. For example, if the user clicks on all of the top five results, it could be a sign of higher result quality than if only the third (or even only the first) result is selected. However, the average click rank would be the same (or lower for the supposedly better case in the second scenario). But a metric can still be useful, irrespective of how far it goes in its attempt to capture all possible influences.

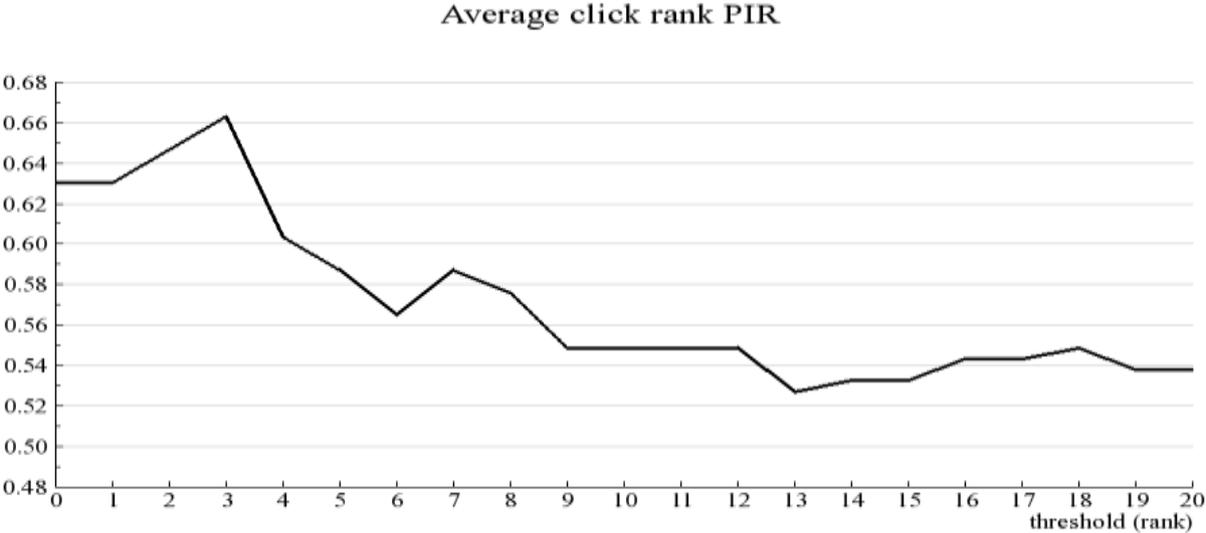

**Figure 11.8. Preference Identification Ratio based on average click rank. PIR is shown on the Y-axis and the threshold (difference in rank average) on the X-axis. The PIR indicates the correct preference identification if average click ranks which lie higher in the result list are considered to be better. No-click sessions are regarded as having an average click rank of 21, so as to have a lower score than sessions with any clicks in the first 20 results.**

And indeed, Figure 11.8 shows average click rank to perform well above chance when predicting user preferences. At low threshold values (with an average difference of 3 ranks regarded as sufficient, or even for all thresholds) the PIR scores lie at around 0.65. For thresholds of 4 ranks and more, the score declines to around 0.55 (and for thresholds starting with 10 ranks, it lies around the baseline score of 0.50).

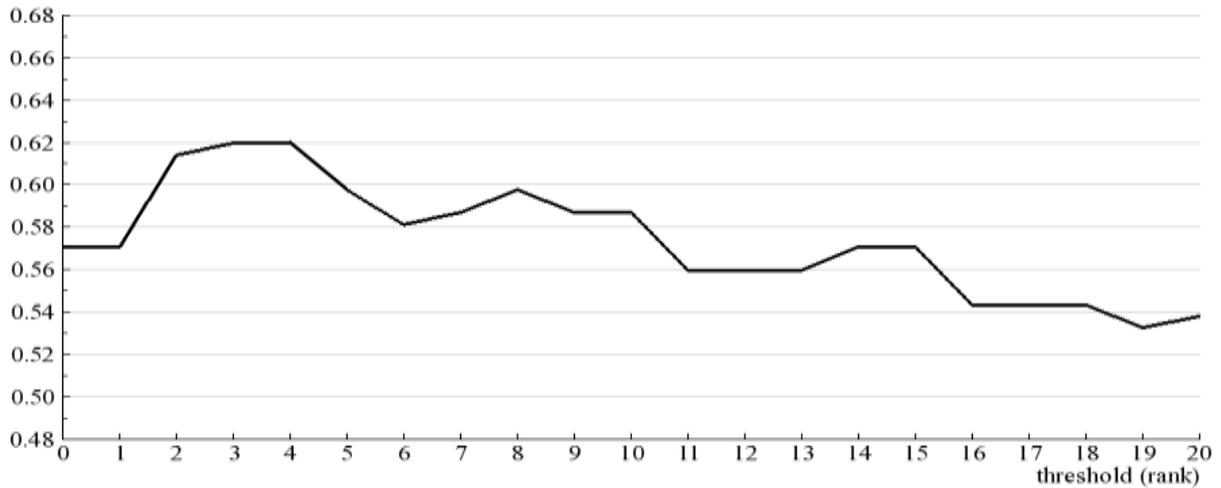

Figure 11.9. Preference Identification Ratio based on first click rank. PIR is shown on the Y-axis and the threshold (difference in first click rank) on the X-axis. The PIR indicates the correct preference identification if clicks which lie higher in the result list are considered to be better. No-click sessions are regarded as having a first click rank of 21, so as to have a lower score than sessions with any clicks in the first 20 results.

Instead of using the average click rank, we could also consider just the first click. By the same logic as before, if the first result a user clicks on is at rank 1, the result list should be preferable to one where the first click falls on rank 10. This metric is probably even more simplistic than average click rank; however, as Figure 11.9 shows, it produces at least some results.

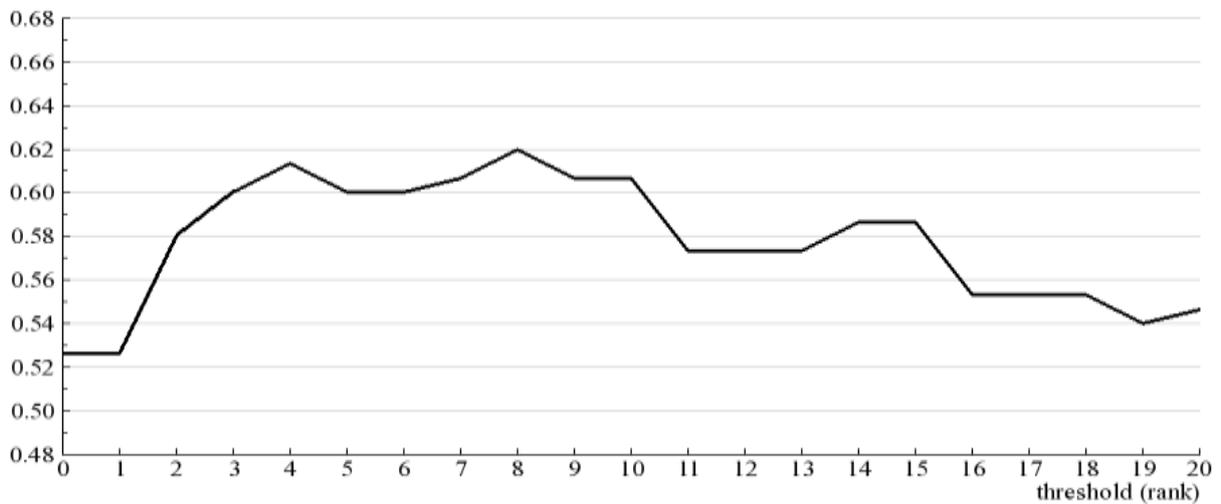

Figure 11.10. Preference Identification Ratio based on first click rank. PIR is shown on the Y-axis and the threshold (difference in first click rank) on the X-axis. The PIR indicates the correct preference identification if clicks which lie higher in the result list are considered to be better. No-click sessions are regarded as having a first click rank of 21, so as to have a lower score than sessions with any clicks in the first 20 results. Only informational queries have been evaluated.

At its best, the first-click rank method provides a PIR score of around 0.62; towards a threshold of 20, it stabilizes at 0.54 (the scores for informational results only, shown in Figure 11.10, are even more stable, with a PIR of over 0.60 up to the threshold value of 10). Thus, it is not far behind the average rank approach.

# 12 Results: A Discussion

*Results! Why, man, I have gotten a lot of results. I know several thousand things that won't work.*

**THOMAS A. EDISON**

In the previous chapters, I have presented the results of the study in some detail; perhaps sometimes in more detail than would be strictly necessary. I feel this is justified since my main goal is to show the possibilities of the approach, which is to judge metrics by comparing them to users' explicit statements regarding their preferences. I have also commented briefly on the *meaning* of most results. In this section, I want to pull together these results to provide a broader picture. I will pronounce judgment on individual metrics and types of evaluation that were studied; however, a very important caveat (which I will repeat a few times) is that this is but one study, and a not large one at that. Before anyone throws up his hands in despair and wails "Why, oh why have I ever used method X! Woe unto me!" he should attempt to reproduce the results of this study and to see if they really hold. On the other hand, most of the findings do not contradict earlier results; at most, they contradict views widely held on theoretical grounds. That is, while the results may be not as solid as I would like, they are sometimes the best empirical results on particular topics, if only because there are no others.

The discussion will span three topics. First, I will sum up the lessons gained from the general results, particularly the comparison of the two result list types. Then, I will offer an overview of and some further thoughts on the performance of individual metrics and parameters. Finally, I will turn to the methodology used and discuss its merits and shortcomings, for the evaluation presented in this study as well as for further uses.

## 12.1 Search Engines and Users

The evaluation of user judgments shows some quite important results. Those concern three areas: search engine performance, user judgment peculiarities, and the consequent appropriateness of these judgments for particular evaluation tasks.

The conclusions regarding search engine performance come from the evaluation of judgments for the original result list versus the randomized one (to recall: the latter was constructed by shuffling the top 50 results from the former). The judgments, for most evaluations, were significantly different from each other; but, also for most evaluations, much less so than could be expected for an ordered versus an unordered result list of a top search engine. In click numbers and click rank, in user satisfaction and user preference the randomized result lists performed quite acceptably.

This was most frequent for queries which have a large number of results that might be relevant. This can be the case for queries where a broad range of options is sought. Informational queries that have a wide scope ("What about climate change?") fall into that category. In this case, the user presumably wants to look at a variety of documents, and there are also likely to be more than enough pages in the search engine index to fill the top 50 ranks with more or less relevant results. Another possibility is that the user looks for something concrete which can be found at any number of pages. This can be a factual query ("When was Lenin born?"), or a large subset of transactional queries (there are hundreds of online shops selling similar products at similar prices, or offering the same downloads). All in all, those options cover a significant part of all queries. The queries for which there is likely to be a large difference between the performance of a normally ranked and a randomized list are those which have a small number of relevant results; these can be narrow informational or transactional queries ("Teaching music to the hard of hearing"), or, most obviously, navigational queries, which tend to have only one relevant result. In the present study, only 2 of the 42 queries were navigational, compared to the 10% to 25% given in the literature (see Section 2.2); a higher proportion may well have resulted in more differences between result list performance.

One evaluation where the two result list types performed very alike was the number of result lists that failed to satisfy any rater (17% versus 18%). This and the relatively high percentage of queries rated as very satisfactory in the randomized result list (57%) indicate that user satisfaction might not be a very discriminating measure. Of course, the satisfaction was binary in this case, and a broader rating scale might produce more diverse results.

For clicks, the results are twofold. On the one hand, the original result lists had many more clicks at earlier ranks than the randomized one. But after rank five, the numbers are virtually indistinguishable. They generally decline at the same pace, apparently due to the users' growing satisfaction with the information they already got, or to their growing tiredness, or just to good old position bias; but result quality seems to plays no role. Almost the same picture arises when we consider relevance, whether macro or micro; at first, the original result list has more relevant results, but, halfway through the first result page, the quality difference between it and the randomized result list all but disappears.

An important lesson from these findings is that authors should be very careful when, in their studies, they use result lists of different quality to assess a metric. The a priori difference in quality can turn out to be much smaller than expected, and any correlation it shows with other values might turn out to be, in the worst case, worthless. The solution is to define the quality criteria in an unambiguous form; in most cases, a definition in terms of explicit user ratings (or perhaps user behavior) will be appropriate.

## 12.2 Parameters and Metrics

Which metric performs best? That is, which metric is best at capturing the real user preferences between two result lists? This is a relatively clear-cut question; unfortunately, there is no equally unequivocal answer. However, if we look closer and move away from unrealistically perfect conditions, differences between the metrics' performances become visible. Then, towards the end of this section, we will return to the question of general metric quality.

### 12.2.1 Discount Functions

One important point regards the discount functions. This is especially important in the case of MAP, where classical, widely-used MAP (with result scores discounted by rank) is consistently outperformed by a no-discount version.[102] However, this by no means means that no discount is the best discount. In NDCG and ESL, for example, the no-discount versions tend to perform worst. All in all, the hardly world-changing result is that different discount functions go down well with different metrics; though once again a larger study may find more regularities.

However, this lack of clarity does not extend to different usage situations of most metrics. That is, no-discount MAP outperforms rank-discounted MAP for all queries lumped together and for informational queries on their own, for result ratings given by the same user who made the preference judgment and for those made by another one, for ratings made on a six-point scale and for binary ratings. If this finding is confirmed, it would mean that you only need to determine the best discount function for a given metric in one situation, and then would be able to use it in any.

Another point to be noted is that there is rarely a single discount function for any metric and situation performing significantly better than all others. Rather, there tend to be some well-performing and some not-so-well-performing discounts, with perhaps one or two not-so-well-but-also-not-so-badly-performing in between. For NDCG, for example, the shallowest discounts (no discount and $\log_5$) perform worst, the steepest (square and click-based) do better, with the moderately discounting functions providing the highest results.

All of these results are hard to explain. One thing that seems to be clear is that the properties of a metric make it work better with a certain kind of discount; however, this is a reformulated description rather than an explanation. If pressed, I think I could come up with a non-contradictory explanation of why MAP works best without discount, NDCG with a slight one, and ERR likes it steep. For example: MAP already regards earlier results to be more important, even without an explicit discount function. As it averages over precision at ranks 1 to 3, say, the relevance of the first result influences all three precision values, while that of the third result has an effect on the third precision value only. Thus, using an additional, explicit discount by rank turns the overall discount into something more like a squared-rank function. However, the other cases are less clear-cut; and in all of them, the *a posteriori* nature of possible explanations should make us cautious not to put too much trust in them until they are corroborated by further results.

### 12.2.2 Thresholds

The thresholds, as employed in this study, have the function of providing some extra sensitivity to preference predictions. These predictions depend on differences between two values; and it should make immediate sense that a difference of 300% should be a pretty strong predictor, while a difference of 1% might or might not mean anything. Indeed, the smaller the difference, the larger the chance that the user preference actually goes in the opposite direction.[103]

---

[102] At later ranks, no-discount MAP sometimes has lower scores than rank-discounted MAP. However, this is not really a case of the latter improving with more data, but rather the former falling. This is an important difference, since it means that no-discount MAP generally has a higher peak score (that is, its highest score is almost always higher than the highest score of rank-discounted MAP).

The good news is that in many cases, the best predictions can be made with the threshold set to zero. For example, the inter-metric evaluation results given in Section 10.5 look very similar, whether we use the zero-threshold approach, or take the best-performing thresholds for every individual metric, discount and cut-off value. If there are changes, they generally lie in the margin of 0.01-0.02 PIR points.

The bad news is that this is not the case when we use single-result and preference judgments by different users – the most likely situation for a real-life evaluation. For the six-point relevance scale as well as for binary and three-point versions, the zero-threshold scores are lower and more volatile. While the peak values are relatively stable, it becomes more important to either use precisely the right cut-off value, or go at least some way towards determining good threshold values; else, one can land at a point where PIR scores take a dip, and get significantly worse predictions than those using a slightly different parameter would yield.

### 12.2.2.1 Detailed Preference Identification

There is an additional twist to threshold values. Throughout this study, I have assumed that the goal, the performance we were trying to capture, was to provide a maximal rate of preference recognition; this is what PIR scores are about. Sensible as this assumption is, there can be others. One situation where PIR scores are not everything occurs when we are trying to improve our algorithm without worsening anyone's search experience.

It is no revelation that most changes, in search as elsewhere, make some things better and some worse. Unless we reach a PIR of 1 (and we won't), some user's preferences are not going to be recognized, or will even be inverted. If we have a PIR of 0.7, than 70% of user preferences are recognized correctly, while 30% are reversed; or 40% are recognized correctly, and 60% are treated as no preference at all; or anything in between. For a search engine operator, it is a big difference whether you improve results for two fifth of your users, or improve it for about two thirds while antagonizing the remaining third. Users may be content with stagnating quality, but react badly if it actually deteriorates.

Consider the example of one of the best-performing metrics, NDCG with a default $\log_2$ discount[104] (the graph is reproduced for your convenience as Figure 12.1). Let us assume that you've gone for the cut-off value of 10 to use as much information as possible. You can see in the graph that the peak PIR value is reached at thresholds 0.01-0.03 and again at 0.08. Which of those should you choose if you want to maximize benefits, but only provided you cause a minimum of harm? Or is there another threshold which is best suited for that?

---

[103] This example suggests that another method would be to determine thresholds by relative, not absolute values. However, since the PIR scores are unlikely to differ by orders of magnitude, large differences in the results seem improbable.
[104] This example has been picked at random; obviously, the same methods can be used to evaluate in detail any other metric and discount.

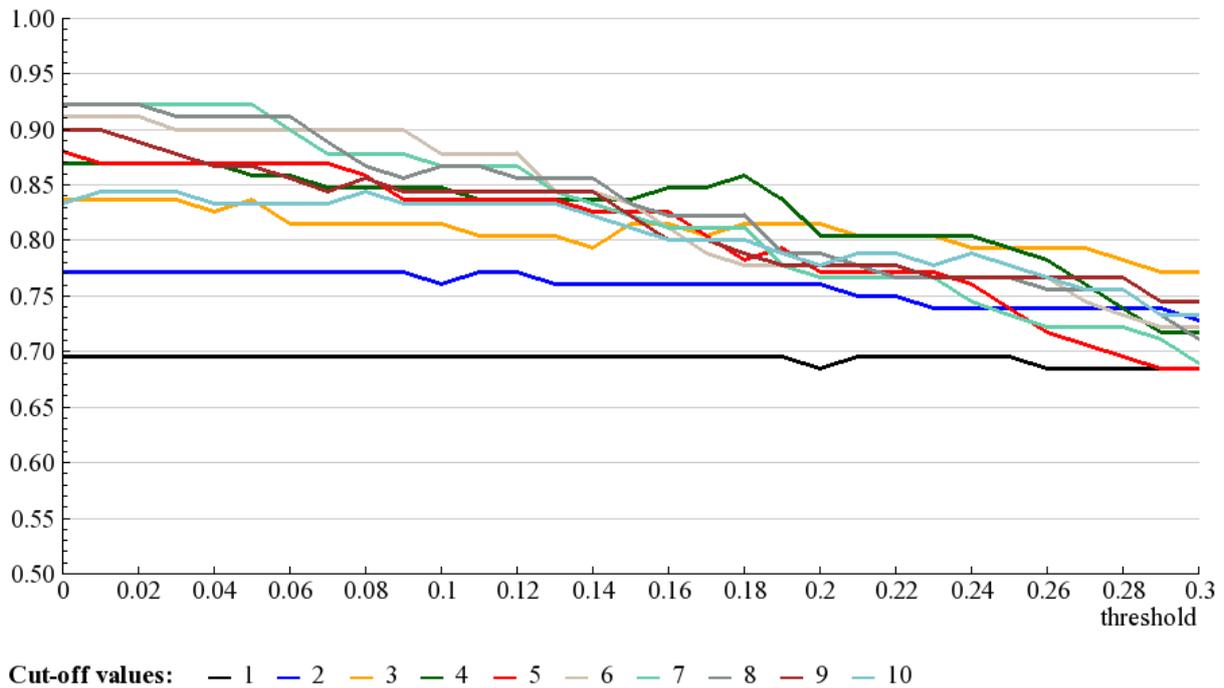

Figure 12.1. NDCG PIR values discounted by $\log_2$ of result rank (reproduction of Figure 10.3).

In this case, it might make sense to go beyond PIR. Figure 12.2 shows the detailed distribution of judgments at the first peak point (threshold 0.01). I will take a moment to explain what the colours mean.

- Dark green shows the proportion of queries where result list preference is identified correctly by the difference in NDCG scores.
- Light green means that the user has no preference, and the NDCG scores' difference is below the threshold, so that the equal quality of the result lists is correctly inferred. These queries are not used in PIR calculation for reasons stated below.
- The yellow sector is for the queries where the user considers the result lists to be equally good, but the difference in NDCG scores is larger than the threshold, falsely showing a preference. This is the second kind of query not used for PIR calculation. As explained in Section 8.2.1, if a user considers two result lists to be of equal use, than recognizing this (non-)preference or failing to do that has no direct influence on the user's experience since he will be neither better no worse off, whichever result list he gets.
- Orange means that the NDCG difference is once again below the threshold, but this time, there *is* a user preference, which is thus not captured.
- Red means that there is a user preference, but the difference in NDCG scores actually predicts the opposite preference.

As an easy guide, the colours go from dark green (good) through yellow (relatively harmless) towards red (really bad).

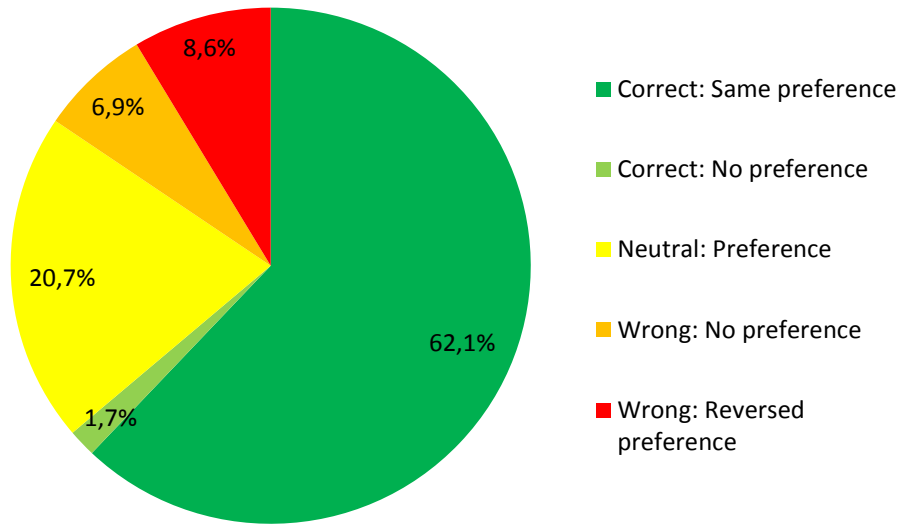

**Figure 12.2. Preference identifications for NDCG discounted by log$_2$ of result rank, at a cut-off value of ten and a threshold of 0.01.**

The PIR value for the point shown in Figure 12.2 is 0.84, as can be read from Figure 12.1. However, the pie graph provides more information. We see that for over 60% of all queries, user have preferences which are correctly identified by NDCG (dark green); that of the 23% of queries where the user does not have a preference, NDCG scores reflect this in just 2% (light green), while for 21%, they suggest a non-existing preference (yellow); that in 7% of cases, NDCG scores miss out on existing preferences (orange); and that for 8% of all queries, NDCG actually picks the result list the user likes less. It is this last number that you are interested in minimizing in this scenario, while keeping the first one as high as possible.[105] But 8% seems quite large; one in thirteen users will actually experience a decline in result list quality. Therefore, we need a method to look at the number of reversed preference judgments depending on the threshold used.

---

[105] Of course, the erroneous "No preference" statement should also be kept as low as possible; but while this would mean missed chances to improve the user experience, reversed preference actually impairs it.

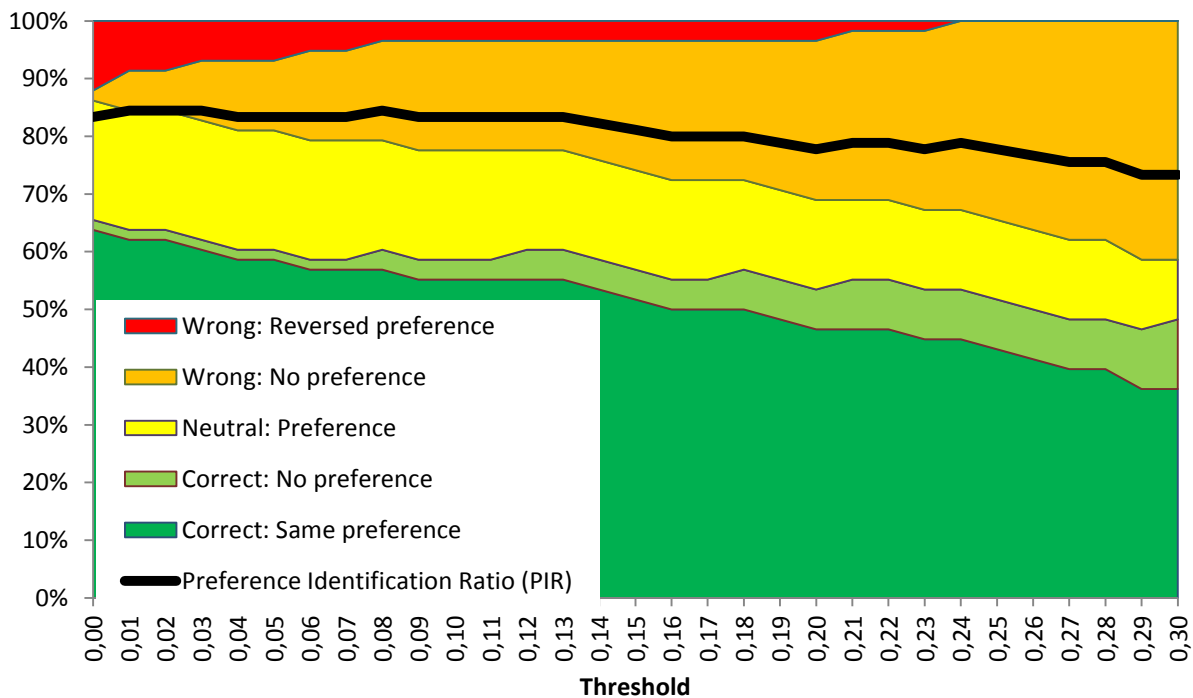

**Figure 12.3.** Preference identifications for NDCG discounted by $\log_2$ of result rank, at a cut-off value of 10.

This is where Figure 12.3 comes into play, and I suspect that if you regarded PIR graphs as too immersed in specifics, you are going to hate this one. It shows the relative development of the five categories introduced above as a function of threshold values. As above, we would like the green areas to be as large as possible, while keeping orange and red small. Additionally, the black line shows the PIR at each threshold.

If we look at the PIR line, we will see that the peak scores described above are indeed quite different with regard to the number of queries which will have their user preference reversed. With a threshold of 0.01, more than 8% of all users will experience a decline in result list quality; at 0.03, the number falls to about 6%, and at 0.08 to 4%. If our aim is to minimize wrong predictions, we should go with the latter threshold; or even raise it to as much as 0.24, when PIR is 0.05 points below its maximum, but no user will be worse off than before.

Figure 12.3 also illustrates an inherent property of PIR development. With a threshold of 0, all queries (except for those where metric scores are absolutely identical) are regarded as having a preferred result list. If we increase the threshold, the only possible changes are that results that some of these queries move into one of the "no preference" categories. Thus, if PIR falls, it means that some queries moved from correct preference identification (dark green) to mistaken equality (orange); and if PIR increases, some queries must have gone from wrong preference (red).[106] This gives us an easy possibility to discover purged mistakes: whenever PIR goes up when thresholds increase, the red faction has shrunk. We cannot assume that PIR always increases when errors are corrected, however, since correct judgments may be dropped at the same point.

---

[106] To be more precise, more queries must have moved in the mentioned direction than in the opposite.

### 12.2.3 Rating Sources

A large difference exists between results depending on whether the explicit result ratings come from the same raters as the preference judgments. Not surprisingly, if a user's ratings are used to predict his own preference, the PIR scores are significantly higher. While the highest attained PIR score in this condition is about 0.92, it drops to just 0.80 if the preference is inferred from other users' judgements. This means that with all other parameters being optimal, one in five users would actually experience worse results than in the absence of any judgments.

This result holds broadly across most parameters. It can be seen at any cut-off rank, with any metric, threshold and scale (with an exception mentioned in the next section). Moreover, the results are not only lower in absolute PIR scores; they also get much less stable. The increased fluctuations can be seen particularly well in threshold evaluations, though most parameters are affected.

What are we to make of these results? To me, they mainly suggest the importance of search personalization. I stated that explicit rating results from the real-life user issuing a query are a luxury hardly imaginable for a search engine. But neither is it necessary to base preference assessments on all (or a random subset) of other users' judgements. The personalization of web search is a well-established topic which has grown in importance in the last decade (see, for example, Speretta and Gauch 2005; Teevan, Dumais and Horvitz 2005; Sieg, Mobasher and Burke 2007). Usually, the research focuses on generating useful changes to a result list drawing upon the user's previous actions. For our purposes, some kind of user profile could be generated, and a user's preference could be induced from the preferences of raters or real users with similar profiles. The resulting evaluation scores would then fall in between the conditions of same-user and different-user ratings.

### 12.2.4 Relevance Scales

The evaluation of relevance scales is somewhat less satisfactory than that of other parameters in that it is not direct. The better method would certainly have been to provide different rater groups with different scales, and then give different sub-groups different instructions as to what the individual relevance ratings are supposed to signify. However, for the relevance scales tested here this would mean a total of six subgroups, in which case each subgroup would be too small to provide reliable results. Therefore, I had to make do with conflating a six-point scale; and a more direct evaluation would be appropriate before taking any drastic measures.[107]

Generally, the six-point scale performs better than any binary or three-point scale. However, there are large differences among the subgroups as well as for different situations. For binary scales, the subgroup where the division between relevant and non-relevant was drawn in the middle (between the six-point scale's ratings 3 and 4) performed only slightly worse than the six-point scale itself, while the subgroups where the relevance criteria were either very strict or very lax did far worse. This is important since users tend to be generous with their ratings in binary evaluations. This is also in agreement with findings that users are more confident in more detailed ratings, but the effect is negligible for scales with more than 6 points (Tang, Shaw and Vevea 1999).

---

[107] Although I do not think any of the results are very revolutionary in the first place.

For three-point relevance scales, an equal distribution of relevance ratings also provided results not far below those of the six-point scale. If "relevant" and "non-relevant" categories were narrower, with more ratings falling into the "partially relevant" category, the PIR scores dropped. However, there was a noticeable exception: in one condition (with result ratings and preference judgements coming from different users), the latter three-point scale was actually more successful than the six-point scale. The result is not easily explained, and awaits further examination.

**12.2.5 Cut-off Ranks**
The evaluation of cut-off ranks and their influence on the metrics' performances was a surprise for me and could lead to a change in the methodology of search engine evaluation if it is confirmed by further studies. Usually, the necessity of introducing a cut-off rank is regarded as just that – a necessary compromise between optimal quality (evaluating all results) and reality (having a limited amount of resources). Accordingly, my initial aim regarding cut-offs was to see if there are certain points at which the marginal gain of evaluating one more rank would fall significantly, therefore making them good candidates for cut-off-rankdom.

I did not expect to find that for most metrics and discount functions, as well as other parameters, the metrics' performance does not rise steadily from rank one to ten. Instead, it rises initially, levels out, and then declines. The peak cut-off rank differs, and I will come to the particulars shortly. But before that, I would like to illustrate how more information can actually decrease the correspondence between an explicit metric and user preference.

As an extreme example, consider a navigational query. In the first result list, the desired result is at rank one; in the second, it is at rank 10. It is reasonable to expect that users will prefer the first result list. However, with precision and a cut-off rank of 10, the metrics will have identical scores. As soon as we lower the cut-off rank, the "right" result list takes the lead.

Of course, it is also possible that the first result list has another really good result at rank 10, while the second result has none at all. In this case, going all the way to cut-off rank 10 would increase the chance of getting the right prediction. But if we assume that both scenarios are equally likely,[108] than we have a fifty-fifty chance of improving or decreasing PIR, all other things being equal. But usually, all other things are not equal. PIR, in all cases we have seen, lies above the baseline of 0.5, and mostly well above it, sometimes in regions around 0.9. This means that it has much more room for decline than for gain; and, in our example, if the first result list had the good result at rank ten, the PIR would not have improved.

There is also another tendency working in the same direction. As shown in Section 9.4, the average relevance of the results up to rank 5 is higher for the original result lists than for the randomized ones. After that, the ratings become very similar. Given this distribution, the evaluation of later results would indeed make them more similar, and provide less discriminative power.

---

[108] That is not necessarily true; but the general logic holds in any case.

If we accept this "regression towards the mean" explanation, what parameters would determine a metric's proneness to this cut-off decay? It would be the discount function and the PIR score itself.

The discount function's influence is easy to see: the steeper the discount, the less heed is taken of later results. With square rank discount, for example, the tenth result has hundred times less influence then the first; for most purposes, it might as well not have been included at all. This effect can be seen clearly in Figure 12.4, which shows the performance of NDCG with different discount functions. The legend orders them in order of increasing steepness, and it is easy to see that no-discount DCG is the first to go into decline, and has the largest fall. $\log_5$ starts to decline at the same point (cut-off rank 5), but does so less dramatically. $\log_2$ and root are next to go, at cut-off rank 8. The steeper discounts (rank and squared rank) do not decline at all, and the click-based discount actually manages a marginal rise as late as cut-off rank 10.

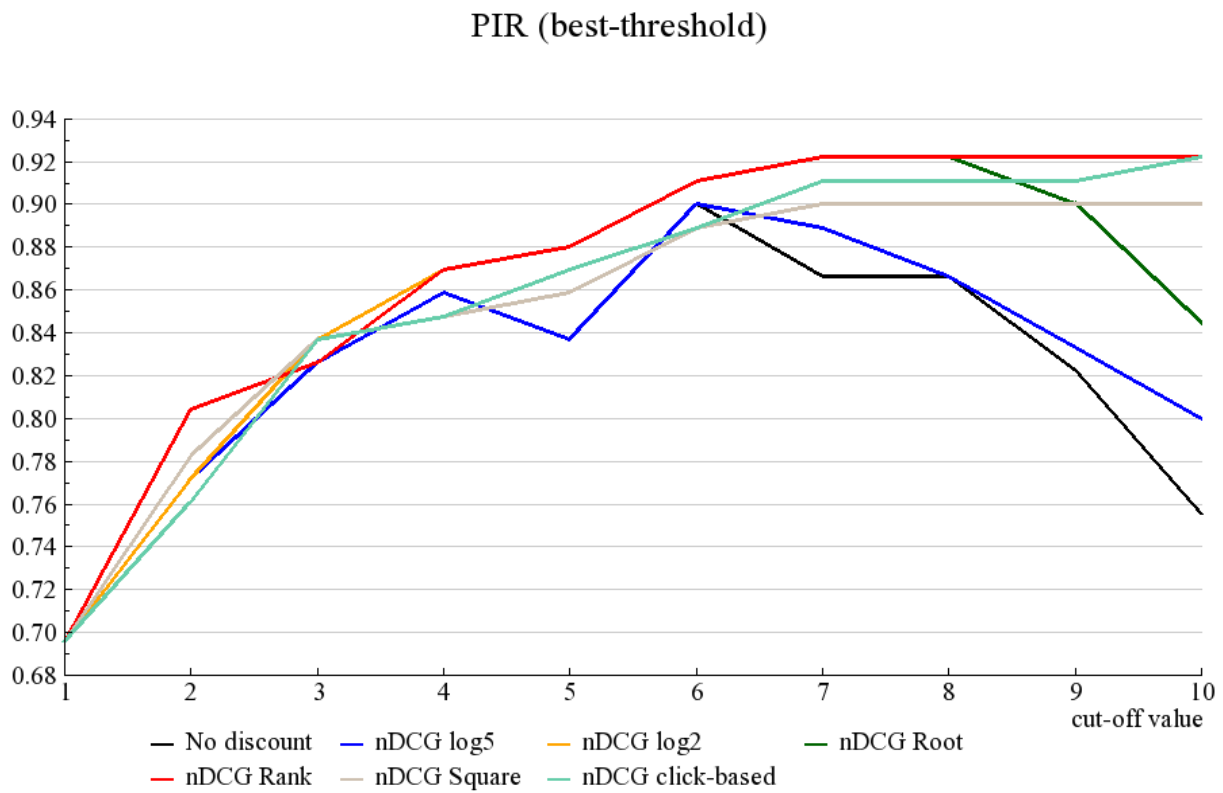

Figure 12.4. NDCG PIR scores for different discount functions, with the best-threshold approach. As many of the discount functions often produce identical PIR scores, a description is needed. Due to the discount function definitions, the "No discount", "$\log_5$" and "$\log_2$" conditions produce the same results for cut-off values 1 and 2, and "No discount" and "$\log_5$" have the same values for cut-off values 1 to 5. Furthermore, the "Root" and "Rank" conditions have the same scores for cut-off ranks 1 to 8, and "Square" and "Click-based" coincide for cut-off values 3 to 4. "$\log_2$" has the same results as "Root" for cut-off values 4 to 10 and the same results as "Rank" for cut-off values 4 to 8. Reproduction of Figure 10.8.

Of course, the picture is not always as clear as that. When the curves become more volatile (for example, when result ratings and preferences come from different users), it is harder to discern patterns, but mostly, they are still visible. But the second parameter which can influence the decline in PIR scores is harder to observe in an actual evaluation.

It seems logical to assume that if the explanation for the tendency of PIR scores to decline after some rank given above is true, then the peak PIR score should play a role in the extent of this decline. After all, if the PIR is at 0.5 to begin with, than it has nothing to fear from results that randomize its predictive power. Conversely, the higher the peak score, the higher the fall possibilities (and the fall probabilities) could be expected to be. However, this tendency is hard to see in the data. A possible explanation is that a metric that has a high PIR peak is, on some level, a "good" metric. And a "good" metric would also be more likely than a "bad" metric to provide a better preference recognition at later ranks, which counterbalances any possible "the higher the fall" effect.[109] This argument is not very satisfactory since it would imply that testing the effect is generally not possible. However, a more confident explanation will require more data.

### 12.2.6 Metric Performance

We are now approaching the results which are probably of most immediate importance for search engine evaluation. Which metric should be employed? I hope that it has by now become abundantly clear that there is no single answer to this question (or perhaps even not much sense in the question itself, at least in such an oversimplified form). However, there are doubtless some metrics more suitable than others for certain evaluations, and even cases of metrics that seem to perform significantly better (or worse) than others in a wide range of situations.

Two of the log-based metrics that have been examined, Session Duration and Click Count, did not perform much better than chance. They might be more useful in specific situations (e.g. when considering result lists which received clicks versus no-click result lists). The third, Click Rank, had PIR scores significantly above the baseline, but its peak PIR values of 0.66 were only as good as those of the worst-performing explicit metrics.

Of the explicit metrics evaluated in this study, the one which performed by far the worst was Mean Reciprocal Rank. This is not too surprising; it is a very simple metric, considering only the rank of the first relevant result. But for other metrics, the picture is less clear.

ERR was a metric I expected to do well, as it tries to incorporate more facets of user behaviour than others. Perhaps it makes too many assumptions; but it is never best of the class. In some cases it can keep up with the best-performing metrics, and sometimes, as in the general same-user evaluation (Figure 10.34), its peak value lies significantly below all others but MRR.

---

[109] "Good" and "bad", in this case, can be taken to mean "tend to produce high PIR scores" and "tend to produce low PIR scores". The argument says only that if a metric produces a higher PIR score at its peak value than another metric, it can also be expected to produce a higher PIR score at a later rank.

Precision did not produce any miracles; but nevertheless, its performance was better than I expected considering its relatively simple nature.[110] Its peak values have been not much lower than those of most other metrics in same-user evaluations, and actually lead the field when the explicit result ratings and preference judgments came from different users. Precision's weak point is its bad performance at later cut-off ranks; as it lacks a discounting function, the results at rank 10 are almost always significantly worse than those at an (earlier) peak rank.

Another positive surprise was the performance of ESL. The metric itself is almost as old as precision; and it performed significantly better. It is not quite the traditional version of ESL that has been evaluated; in Section 10.4, I described some changes I have made to allow for non-binary relevance and a discount factor, as well as for normalization. However, I think that its essence is still the same. Be that as it may, ESL's PIR scores are impressive. In many situations it is one of the best-performing metrics. However, its scores are somewhat lower when explicit result ratings and preference judgments come from different users. Also, it is the metric most susceptible to threshold changes; that is, taking a simple zero-threshold approach as opposed to a best-threshold one has slightly more of a detrimental effect on ESL than on other metrics.

MAP, on the other hand, is performing surprisingly poorly. As explained in Section 10.3, two versions of MAP have been evaluated, one with the traditional rank-based discount, and another without a discount. The no-discount MAP performed unspectacularly, having mostly average PIR scores, and reaching other metrics' peak scores in some circumstances (although at later cut-off ranks). The PIR scores of traditional MAP, discounted by rank, were outright poor. It lagged behind the respective top metrics in every evaluation. It came as a bit of a shock to me that one of the most widely used evaluation metrics performed so poorly, even though there have been other studies pointing in the same direction (see Section 4.1 for an overview). However, taken together with the current study, those results seem convincing. While in most cases, I would advocate attempting to replicate or substantiate the findings of the current study before taking any decisions, for MAP (at least, for the usual version discounted by rank), I recommend further research before continuing to use the metric.

Finally, there is (N)DCG. In earlier sections I pointed out some theoretical reservations about the metric itself and the lack of reflection when selecting its parameter (discount function). However, these potential shortcomings did not have a visible effect on the metric's performance. If there is a metric which can be recommended for default use as a result of this study, it is NDCG, and in particular NDCG with its customary $\log_2$ discount. It has been in the best-performing group in virtually every single evaluation; not always providing the very best scores, but being close behind in the worst of cases.

If saying which metrics perform better and which do worse can only be done with significant reservations, explaining the *why* is much harder still. Why does NDCG generally perform well, why does ERR perform worse? To find general tendencies, we can try to employ the Web Evaluation Relevance model described in Chapter 7.

---

[110] I do not think I was alone in my low expectations, either, as the use of precision in evaluations has steadily declined over time.

The WER model had three dimensions. The first of those was representation. While different evaluation situations had different representation styles (information need or request), they are identical for all metrics. The same can be said about the information resources; all explicit metrics are document-based. The only differences, then, lie in the third category: context. Context is somewhat of a catch-all concept, encompassing many aspects of the evaluation environment. On an intuitive level, it reflects the complexity of the user model used by a metric.

One component of context that is present in most of the evaluated metrics is a discount function. However, as described above, the same discount functions have been tested for all metrics. A second component is the taking into the consideration the quality of earlier results. This happens, in different ways, with ERR (where a relevant result contributes less if previous results were also relevant), and ESL (where the number of relevant results up to a rank is the defining feature).[111] But this does not shed too much light on the reasons for the metrics' performances, either. The two metrics handle the additional input in quite different ways that to not constitute parts of a continuum to be rated for performance improvement. And the use of additional input itself does not provide any clues, either; ESL generally performs well, while ERR's PIR scores are average. This might be explained by the fact that ESL has an extra parameter, the number of required relevant results,[112] which can be manipulated to select the best-performing value for any evaluation environment. All in all, it can be said that there is not enough evidence for metrics with a more complex user model (as exemplified by ERR and ESL) to provide better results.

## 12.3 The Methodology and Its Potential

The results presented in the two previous sections, if confirmed, can provide valuable information for search engine operators and evaluators. The different impacts of user satisfaction and user preference or the performance differences between MAP and NDCG are important results; however, I regard the methodology developed and used in this study as its most important contribution.

In my eyes, one of the most significant drawbacks of many previous studies was a lack of clear definitions of what it was the study was actually going to measure in the real world. Take as example the many evaluations where the standard against which a metric is evaluated is another metric, or result lists postulated to be of significantly different quality. The question which these metrics attempt to answer is "How similar are this metric's results to those provided by another metric", which is of limited use if the reference metric does not have a well-defined meaning itself; or "If we are right that the result lists are of different quality, how close does this metric reflect the difference", which depends on the satisfaction of the condition. Moreover, even if the reference values are meaningful, such evaluations provide results which are at best twice removed from actual, real-life users and use cases.

---

[111] MAP could be argued to actually rate results higher if previous results have been more relevant (since the addend for each rank partly depends on the sum of previous relevancies). However, as the addends are then averaged, the higher scores at earlier ranks means that the MAP itself declines. For an illustration, you might return to the example in Table 4.2 and Figure 4.2, and consider the opposed impacts of the results at rank 4 in both result lists.

[112] Or rather, in the form used here, the total amount of accumulated relevance (see Formula 10.5).

This is the reason for the Preference Identification Ratio and the methodology that goes with it. The idea behind PIR is defining a real-life, user-based goal (in this case, providing users with better result lists by predicting user preferences), and measuring directly how well a metric performs. I would like to make it clear that PIR is by no means the only or the best measure imaginable. One could concentrate on user satisfaction instead of preference, for example; or on reducing (or increasing) the number of clicks, or on any number of other parameters important for a search engine operator. However, the choice of measure does not depend on theoretical considerations, but only on the goal of the evaluation. The question of how well PIR reflects a metric's ability to predict user preference is answered by the very definition of PIR.

Another important method employed in this study is parameter variation. Metric parameters are routinely used in studies without much reflection. I have mostly used the $\log_2$ discount of NDCG as an example, but it is only one instance, and probably not the most significant one. I am not aware of any research as to the question of whether the discount by rank used in MAP for decades is really the best one; and the results of this evaluation suggest it emphatically is not. I am also not aware of a systematic evaluation of different cut-off ranks to determine what cut-off ranks produce the best effort/benefit ratio, or just the best results.

I hope I have shown a few important facts about parameters. They should not be taken for granted and used without reflection; you may be missing an opportunity to reduce effort or improve performance, or just providing bad results. But also important is the realization that it is not necessarily outlandishly hard to find the best values. In this study, there were two main sets of data; user preferences and result judgments. But with just these two input sets, it was possible to evaluate a large range of parameters: half a dozen discount functions, ten cut-off ranks, thirty threshold values, two types of rating input, three rating scales (with multiple subscales), and five main metrics, which also can be viewed as just another parameter.

Granted, the evaluation's results are not necessarily as robust as I would have liked. However, that is not a result of the large number of evaluated parameters, but of the small scale of the study itself. I expect that a study with, say, fifty participants and a hundred queries would have stronger evidence for its findings; and, should its findings corroborate those of the present evaluation, be conclusive enough to act upon the results.

## 12.4 Further Research Possibilities

What could be the next steps in the research framework described here? There are quite a lot; while some could revisit or deepen research presented in this study, other might consider new questions or parameters. Below, I will suggest a few possibilities going further than the obvious (and needed) one of attempting to duplicate this evaluation's results.

One avenue of research discussed in Section 12.2.2.1 is detailed preference identification. It goes beyond a simple, single score to provide a faceted view of a metric's performance. By including all sessions, not just the ones where a user preference exists, and providing more details, it allows for a finer evaluation, and consideration of more precise goals. In Section 12.2.3, I mentioned personalization efforts; if such mechanisms could be implemented and tested, they might provide an important bridge between the relatively low results of different-user evaluation and same-user evaluation. In Section 12.2.4, a useful possible study was suggested, which would provide user groups with different relevance scales and different instructions, and could show if an intuitive six-point relevance scale is really better than a binary or three-point one; and which subcategory of the latter scales produces the most accurate preference predictions.

Another interesting topic is constituted by log-based, and in particular click-based, metrics. In Chapter 11, I have explained that the study layout was not suited for most click metrics proposed and used in the last years. They do not provide absolute scores for a result or result list, but rather, given a result list and click data, construct a result list that is postulated to be of better quality. Thus, we cannot take two result lists and compare them using, say, the "click > skip above" model (see Chapter 5); instead, we have to take one result list, collect the log data, then construct a second result list, and obtain a user preference between the two lists.

Obviously, there also possible research topics apart from those already mentioned in this study. Perhaps the most intriguing of them concerns result snippets. If you recall, one of the ratings required from the evaluators in the present study was whether the result descriptions they encountered in the result lists were "good", that is, whether the user would click on this result for this query. I have not performed the evaluation (yet), as this is a large, separate task with many possibilities of its own.

Snippets have been the object of some research lately; however, it mostly focused on snippet creation (e.g. Turpin et al. 2007; Teevan et al. 2009). There has also been research on snippet-based evaluation (Lewandowski 2008; Höchstötter and Lewandowski 2009), which mostly focused on the evaluation of snippets on their own, as well as comparing ratings for snippets and documents. But there is more to be done. One possibility for which the PIR framework seems well suited is combining snippet and document judgment.

It is well established that the user only examines a small subset of available results, even within a cut-off rank of, say, ten. One method for determining which results will be clicked on is evaluating the snippets. It stands to reason that if a user does not regard a snippet as relevant, he will not click on it; he will not see the document itself; and he will not gain anything from the result.[113] Thus, the results with unattractive snippets will be assigned a relevance score of zero. This is an approach that has received attention (Turpin et al. 2009), although not as much as its possible impact would warrant.

In the next step, there are at least two variants of dealing with unattractive snippets. We can consider them as having no influence on the user at all, and discard all documents which will not be seen; in this case, the rank of the following documents will move up. Or we can assume that the results do not provide any benefits to the user, and furthermore distract him from the possibly more useful results; then, we would just set the relevance scores of the documents to zero. PIR should be able to provide an answer which of these models (or perhaps some other) can better predict user preferences, and whether any of them performs better than the current model which does not consider snippets at all.

---

[113] Except for the case where the snippet actually contains the sought-after information, as may be the case with factual queries. This case can be accounted for by using three snippet relevance categories: "unpromising", "promising" and "useful on its own".

There are, of course, many more open questions out there, such as different result types (e.g. image or news search). I think that the framework introduced in this thesis, combining direct, session-based user opinion as a reference point and variation of as much data as possible to assess not only one metric, but as many parameters as needed, can help answer some of them.

# Executive Summary

> *Let thy speech be short, comprehending much in a few words.*
> **APOCRYPHA**

*In this thesis, I describe...*

- an overview of the metrics used in search engine evaluation;
- the theoretical issues which can be raised about them;
- previous studies which evaluate evaluation metrics.

*I introduce...*

- a meta-evaluation measure, the Preference Identification Ratio (PIR), which captures a metric's ability to correctly recognize explicitly stated user preferences;
- an evaluation method which varies metrics and parameters to allow to use one set of data to run dozens or hundreds of different evaluations;
- a new category of query, called "meta-query", which requires information on the search engine itself and cannot produce a "good" or "bad" result list.

*I find that...*

- randomizing the top 50 results of a leading search engine (Yahoo) leads to result lists that are regarded to be equal or superior to the original ones for over 40% of queries;
- after the first five ranks, further results of a leading search engine (Yahoo) are on average no better than the average of the first 50 results;
- a cut-off rank slightly smaller than 10 does not only reduce the data-gathering effort but in most cases actually improves the evaluation accuracy;
- the widely-used Mean Average Precision metric is in most cases a poor predictor of user preference, worse than Discounted Cumulated Gain and not better than Precision;
- the kind of discount function employed in a metric can be crucial for its ability to predict user preference;
- a six-point relevance scale intuitively understandable to raters produces better results than a binary or three-point scale;
- session duration and click count on their own are not useful in predicting user preference

*I find preliminary evidence that...*

- (normalized) Discounted Cumulative Gain and a variety of Estimated Search Length are, with appropriate discount functions, the best predictors of user preference;
- if a binary or three-point relevance scale is used, certain rater instructions as to what the single ratings mean can significantly influence user preference prediction quality;
- depending on the metric and other evaluation parameters, cut-off ranks as low as 4 may provide the best effort-quality ratio.

# Bibliography

*The surest way to make a monkey of a man is to quote him. That remark in itself wouldn't make any sense if quoted as it stands.*
ROBERT BENCHLEY,
"MY TEN YEARS IN A QUANDARY "

# Appendix: Metrics Evaluated in Part II

*ERR – Expected Reciprocal Rank*

$$ERR = \sum_{r=1}^{n} \frac{1}{disc(r)} \prod_{i=1}^{r-1} (1 - \frac{2^{rel(d_i)} - 1}{2^{rel_{max}}}) \frac{2^{rel(d_r)} - 1}{2^{rel_{max}}}$$

**ERR formula. For each rank *r*, the probabilities of a relevant result at each earlier rank *i* are multiplied; the inverse probability is used as a damping factor for the gain at the current rank.**

ERR is a relatively new and user-centric metric. The importance of each later result is assumed to be negatively correlated with the usefulness of previous results; that is, the better earlier results, the less important the later ones.

Theory: Section 4.3, page 33.

Study: Section 10.4, page 104.

In the evaluation, ERR showed a relatively stable discriminative performance, albeit on a below-average level. It performed best with a discount by square rank, though the differences between different discount functions were small.

*ESL – Expected Search Length*

$$ESL = 1 - \frac{r_n - \sum_{i=1}^{r_n} \frac{rel(i)}{disc(i)}}{c}$$

**Expected Search Length adapted for graded relevance and discount. *r* is the rank at which the sum of single result relevance scores reaches a threshold *n*. The relevance scores are obtained by dividing the user relevance score for rank *i* by a discount *disc(i)* dependant on the same rank. *c* is the cut-off value used.**

ESL scores depend on the number of results a user has to examine before satisfying his information need; this is to be modelled as seeing a set minimal number of relevant results. Because of non-binary relevance used in much of the present study, the satisfaction of the information need has been instead regarded to occur by having seen results with a set cumulative relevance.

Theory: Section 4.3, page 32.

Study: Section 10.4, page 108.

In the evaluation, ESL performed quite well. The best parameters for a six-point relevance evaluation were a cumulative relevance threshold *n* of around 2.5 with a high-discount function (e.g. a rank-based discount).

## MAP – Mean Average Precision

$$\text{MAP} = \frac{1}{|Q|} \left( \sum_{Q_i} \frac{1}{|D|} \left( \sum_{r=1}^{|D|} rel(d_r) \frac{\sum_{k=1}^{r} rel(d_k)}{disc(r)} \right) \right)$$

**Modified MAP formula with queries *Q*, relevant documents *R* and documents *D* (at rank *r*). *rel* is a relevance function assigning *1* to relevant and *0* to non-relevant results, or, in this study, one of six values in the range from 1 to 0. *disc(r)* is a discount function depending on the rank *r*.**

As its name suggests, MAP calculates average precision scores (at each relevant results), and then calculates the means of those over multiple queries. Usually, later addends are discounted by rank (*disc(r)=r)*.

Theory: Section 4.1, page 25.

Study: Section 10.3.

An important (and relatively stable) result is that the routinely used discount by rank produces results far below other metrics like (N)DCG, or even MAP without any discount at all, which performs quite well.

## MRR – Mean Reciprocal Rank

$$RR = \begin{cases} 0 & \text{if no relevant results} \\ \frac{1}{disc(r)} & else \end{cases}$$

**Reciprocal Rank (RR), with *r* the first rank where a relevant result is found. *disc(r)* is a discount function depending on the rank *r*.**

$$MRR = \frac{1}{|Q|} \sum_{q \in Q} RR_q$$

**Mean Reciprocal Rank (MRR), *Q* being the set of queries and *$RR_q$* the Reciprocal Rank measured for query *q*.**

MRR is a straightforward metric which depends solely on the rank of the first relevant result.

Theory: Section 4.2, page 27.

Study: Section 10.4, page 107.

In the evaluation, MRR performed significantly worse than most other metrics.

### (N)DCG – (Normalized) Discounted Cumulative Gain

$$DCG_r = \begin{cases} CG_r & if\ r < b \\ DCG_{r-1} + \dfrac{rel(r)}{disc(r)} & if\ r \geq b \end{cases}$$

**DCG with logarithm base *b*. $CG_r$ is the Cumulated Gain at rank *r*, and rel(r) a a relevance function assigning *1* to relevant and *0* to non-relevant results. *disc(r)* is a discount function depending on the rank *r*.**

$$NDCG_r = \frac{DCG_r}{iDCG_r}$$

**Normalized DCG is calculated by dividing DCG at rank *r* by the DCG of an ideal result ranking iDCG at the same rank.**

A metric derived from *Discounted Cumulative Gain (DCG)*. It adds up the relevance scores of all documents up to a cut-off value, with each later result discounted by $\log_2$ of its rank. Then, it is normalized by dividing it by the DCG of an idealized result list (*iDCG*) constructed from the pool of all available results.

Theory: Section 4.2, page 29.

Study: Section 10.1.

Results: NDCG is found to perform well in most circumstances, with the usually employed $\log_2$ discount.

### Precision

$$Precision = P = \frac{relevant\ retrieved\ results}{overall\ retrieved\ results}$$

One of the oldest and simplest metrics, Precision is just the share of relevant results in the retrieved list. For reasons described in the study, it is very similar to (N)DCG without a discount for later ranks.

Theory: Section 4.1, page 24.

Study: Section 10.2.

Mostly, Precision performed acceptably; the most startling feature was a strong decline in discriminative power at later cut-off ranks.